\documentclass[3p, number,sort&compress,times,fleqn]{elsarticle}      

\usepackage[colorlinks = true,
linkcolor = blue,
urlcolor  = blue,
citecolor = blue,
anchorcolor = blue]{hyperref}

\graphicspath{{./figs/}}
\usepackage{url}
\usepackage{booktabs}
\usepackage[table]{xcolor}

\usepackage{tikz}
\usetikzlibrary{arrows,calc,positioning,shapes}

\usepackage{csml}

\begin{document}
	
\begin{frontmatter}
	
\title{The statistical finite element method (statFEM) for coherent synthesis of observation data and model predictions}

\author[1,2]{Mark Girolami}
\author[1,2]{Eky Febrianto}
\author[1,2]{Ge Yin}
\author[1,2]{Fehmi Cirak\corref{cor1}}
\ead{f.cirak@eng.cam.ac.uk}

\cortext[cor1]{Corresponding author}

\address[1]{Department of Engineering, University of Cambridge, Cambridge, CB2 1PZ, UK }
\address[2]{The Alan Turing Institute, London, NW1 2DB, UK }

\begin{abstract}
The increased availability of observation data from engineering systems in operation poses the question of how to incorporate this data into finite element models. To this end, we propose a novel statistical construction of the finite element method that provides the means of synthesising measurement data and finite element models. The Bayesian statistical framework is adopted to treat all the uncertainties present in the data, the mathematical model and its finite element discretisation. From the outset, we postulate a statistical generating model which additively decomposes data into a finite element, a model misspecification and a noise component. Each of the components may be uncertain and is considered as a random variable with a respective prior probability density. The prior of the finite element component is given by a conventional stochastic forward problem. The prior probabilities of the model misspecification and measurement noise, without loss of generality, are assumed to have a zero-mean and a known covariance structure. Our proposed statistical model is hierarchical in the sense that each of the three random components may depend on one or more non-observable random hyperparameters with their own corresponding probability densities. We use Bayes rule to infer the posterior densities of the three random components and the hyperparameters from their known prior densities and a data dependent likelihood function. Because of the hierarchical structure of our statistical model, Bayes rule is applied on three different levels in turn. On level one, we determine the posterior densities of the finite element component and the true system response using the prior finite element density given by the forward problem and the data likelihood. In this step, approximating the prior finite element density with a multivariate Gaussian distribution allows us to obtain a closed-form expression for the posterior. On the next level, we infer the hyperparameter posterior densities from their respective priors and the marginal likelihood of the first inference problem. These posteriors are sampled numerically using the Markov chain Monte Carlo (MCMC) method. Finally, on level three we use Bayes rule to choose the most suitable finite element model in light of the observed data by computing the respective model posteriors. We demonstrate the application and versatility of statFEM with one and two-dimensional examples. \end{abstract}
	

\end{frontmatter}

\newpage

%
\section{Introduction \label{sec:introduction}}
%
Most engineering systems, such as structures and machines, are designed using deterministic mathematical models, indeed their finite element discretisations, which depend on material, geometry, loading and other parameters with significant uncertainties. Traditionally, these uncertainties have been taken into account through codified safety factors. Although this approach has been perfected over the centuries, it is known, for instance in structural engineering, that the response of the actual system and the model often bear no resemblance to each other \cite[Ch. 9]{heyman1998structural}. More accurate and reliable predictions are essential towards the design of more efficient systems and for making rational decisions about their operation and maintenance. To achieve this, the uncertainties in the model parameters must be taken into account \cite{oden2010computerI,roy2011comprehensive}; even so, the predicted uncertainties in the model response can be significant rendering them practically useless.  Fortunately, modern engineering systems are more and more equipped with sensor networks that continuously collect data for their in-situ monitoring, see e.g.~\cite{lin2018performance,everton2016review}. The available data includes, for instance, strains from fibre-optic Bragg sensor networks or digital image correlation, temperatures from infrared thermography and accelerations. Incorporating this readily available measurement data into finite element models provides a means to infer the uncertain true system behaviour. 

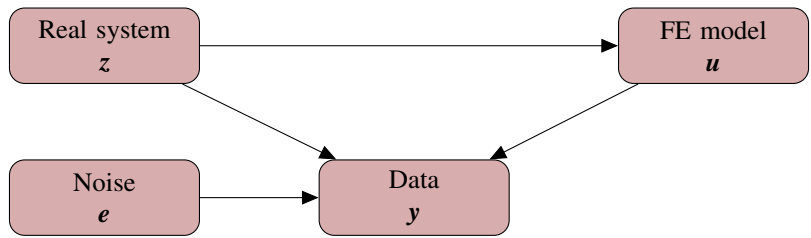
\begin{figure}[b]
\centering
\begin{tikzpicture}[
 ->, >=triangle 45, shorten >=0pt, auto,node distance=55mm, 
 main node/.style={rectangle, rounded corners =2mm, fill=csmlPink, draw=black, minimum width = 25mm, minimum height = 10.mm, align=center, text width=20mm}, 
 scale=.5] 
\node[main node]    (1)             {Real system \\ $\vec z$}; 
\node[main node]    (2)            [right= of 1]     {FE model \\ $\vec u$}; 
\node[main node]    (3)            [below left= 1 cm  and 14.25mm of 2]     {Data \\ \vec y}; 
\node[main node]    (4)            [below= 1cm of 1]     {Noise \\ \vec e}; 

\path[every node/.style={font=\sffamily\small}] 
(1) edge node [] {} (3)
(2) edge node [] {} (3)
(1) edge node [] {} (2) 
(4) edge node [] {} (3) ; 
\end{tikzpicture} 
\caption{Statistical model underlying statFEM. The observation data~$\vec y$  is decomposed as $\vec y= \vec z + \vec e = \rho \vec P \vec u + \vec d + \vec e $, where~$\rho$ is a random hyperparameter, $\vec P$ is a suitable projection operator and  other variables are all random vectors. Each of the random vectors depend on additional random hyperparameters which have been omitted in this diagram.  \label{fig:statFEMintro}}
\end{figure}

To this end, we propose a statistical construction of the finite element method, dubbed as statFEM, which allows one to make predictions about the true system behaviour in light of measurement data. Adopting a Bayesian viewpoint, all uncertainties in the data and model parameters are treated as random variables with suitably chosen prior probability densities, which consolidate any knowledge at hand. Starting from the prior probability densities, Bayes rule provides a coherent formalism to determine their respective posterior densities while making use of the likelihood of the observations. The selected model determines the probability, or the likelihood, that the observed data was produced by the model. See, e.g., the books~\cite{mackay2003information,sivia2006data, murphy2012machine,gelman2013bayesian,rogers2016first} for an introduction to Bayesian statistics and data analysis. Following Kennedy and O'Hagan's seminal work on calibration of computer models \cite{kennedy2001bayesian}, we decompose the observed data~$\vec y$ into three random components, namely a finite element component~$\vec u$, a model misspecification component~$\vec d$ and a measurement noise component~$\vec e$, see Figure~\ref{fig:statFEMintro}. We refer to this decomposition as the {\em statistical generating model}, or in short the {\em statistical model}, and have additional models corresponding to each of the random variables, i.e.~$\vec u$, $\vec d$ and $\vec e$. The three random variables depend in turn on random parameters with corresponding probability densities. Following standard statistics terminology, we refer to the  unknown random parameters as {\em hyperparameters}. Evidently, the proposed statistical construction has an inherent hierarchical structure, see e.g. \cite[Ch. 5]{gelman2013bayesian}. That is, each of the random variables~$\vec u$, $\vec d$ and~$\vec e$ depend in turn on a set of random hyperparameters. 

We obtain the prior probability density for the finite element component~$\vec u$ by solving a traditional probabilistic forward problem in the form of a stochastic partial differential equation.  The source, or forcing, term and the coefficients of the differential operator can all be random. Any unknown variables used for parameterising the respective random fields, e.g., for describing their covariance structure, are treated as hyperparameters. We solve the finite element discretised forward problem with a classical first-order perturbation method~\cite{liu1986random}. For the misspecification component~$\vec d$, we assume a Gaussian process prior, which for the purposes of illustration is assigned a square exponential kernel, and treat the respective covariance parameters as hyperparameters. The measurement noise~$\vec e$ is, as usual, assumed to be independent and identically distributed. It is possible to determine with Bayes rule the joint posterior density of all random variables, i.e.~$\vec u$, $\vec d$ and $\vec e$, and their hyperparameters, and to obtain subsequently the posterior densities of the individual random variables by marginalisation. This leads, however, to a costly high-dimensional inference problem for which advanced sampling schemes are being developed, see e.g. \cite{marzouk2009dimensionality,robert2013monte,petra2014computational,beskosetal2017}. We circumvent the need for costly sampling by using approximate inference, that is, by exploiting the hierarchical structure of the proposed construction and applying Bayes rule on three different levels in turn. The overall approach is akin to the empirical Bayes or evidence approximation techniques prevalent in machine learning, see~\cite{mackay1992bayesian,mackay1999comparison}, \cite[Ch. 3]{bishop2006pattern} and \cite[Ch. 5]{murphy2012machine}. Ultimately, we infer, or in other words learn, from the data~$\vec y$ the posterior densities of~$\vec u$,~$\vec d$ and~$\vec e$ and their respective hyperparameters. Moreover, we assess the suitability of different finite element models by computing their Bayes factors~\cite{kass1995bayes,mackay2003information}. The posteriors are, depending on the level, either analytically approximated or numerically sampled with MCMC. We refer to~\cite{stuart2010inverse} on mathematical foundations of Bayesian inverse problems and for the necessary theory in defining the probability measures required in statFEM. 

The seemingly innocuous decomposition of the data~$\vec y$ according to Figure~\ref{fig:statFEMintro}, as proposed in~\cite{ kennedy2001bayesian}, provides a versatile framework for statistical inference and has been extensively used in the past. The choice of the models for each of the components~$\vec u$, $\vec d$, and~$\vec e$ and the numerical techniques for the treatment of the resulting inference problem leads to a rich set of approaches. In~\cite{kennedy2001bayesian} and most subsequent papers, including~\cite{ bayarri2007framework, higdon2004combining}, the component~$\vec u$ representing the simulation model is obtained from a given black-box computer code (simulator). The deterministic model response~$\vec u$ is approximated with a standard Gaussian process emulator~\cite{ williams2006gaussian}, or surrogate model, using multiple evaluations of the simulator.  Obviously, instead of Gaussian process regression any other technique for creating surrogates can be used~\cite{peherstorfer2018survey, forrester2008engineering}. The essential advantage of using a surrogate model as a forward model for~$\vec u$ is that the inferential framework becomes independent from the complexities of the specific simulator used. Calibration aims to determine the parameters of the deterministic forward model, including its constitutive parameters and forcing and their spatial distributions. However, in many engineering systems the forward model is not deterministic. For instance, the loading of a bridge under operation is inherently random. Similarly, the constitutive parameters of a mechanical part, say a connecting rod in an engine, will indeed be random over its different realisations. The aim of calibration in such cases is, as proposed in this paper, to determine the hyperparameters characterising the random loading or the constitutive parameters. The importance of a random misspecification component~$\vec d$ in calibration is meanwhile well-established ~\cite{brynjarsdottir2014learning, ling2014selection,jiang2020sequential}. 

StatFEM is complementary to the conventional probabilistic finite element method and the Bayesian treatment of inverse problems. As mentioned, we solve the forward problem via a probabilistic finite element method. Over the years, a wide range of approaches have been proposed to solve stochastic partial differential equations using finite elements. They differ in terms of discretisation of the prescribed random inputs, like the forcing or the constitutive parameters, and the approximation of the solution in the probability domain. For insightful reviews  see~\cite{matthies1997uncertainties,sudret2000stochastic,stefanou2009stochastic,xiu2010numerical,lord2014introduction,aldosary2018structural}. In this paper, we assume that the random forcing term and the coefficients  of the differential operator are both Gaussian processes. When only the forcing is random, the resulting finite element solution is also Gaussian and the solution probability measure can be readily obtained.  However, when the coefficients of the differential operator are random, the solution is usually not a Gaussian and it becomes more taxing to solve the forward problem. In addition to the spatial domain the probability domain has to be also discretised or in some way approximated. We choose a perturbation method to approximate the solution in the probability domain~\cite{liu1986random}, but could use any one of the other well-known techniques, like Monte-Carlo~\cite{robert2013monte}, Neumann-series~\cite{yamazaki1988neumann} or polynomial-chaos expansions with their Galerkin and collocation implementations~\cite{ghanem1991stochastic, xiu2002modeling,babuska2004galerkin,xiu2005high}. In contrast to forward problems, inverse problems are considerably more challenging to formulate and to solve, because the same observation can usually be generated by different sets of model parameters. Adopting a Bayesian viewpoint and treating the model parameters as random can resolve the ill-posedness of the inverse problem. See~\cite{oden2010computerII,tarantola2005inverse,kaipio2006statistical} for an introduction to Bayesian inversion and~\cite{ stuart2010inverse} for a detailed mathematical analysis. Bayesian inversion is closely related to the calibration framework proposed by Kennedy and O'Hagan~\cite{kennedy2001bayesian}. However, in Bayesian inversion literature the model misspecification errors are usually not considered. Different from statFEM, in both Bayesian inversion and~\cite{kennedy2001bayesian} the forward model mapping the random parameters to the observations is usually deterministic. Despite the principal differences between statFEM and Bayesian inversion they share a number of algorithmic similarities in terms of implementation. Therefore, many of the numerical techniques developed for efficiently solving large-scale inverse problems, like the treatment of non-local covariance operators~\cite{bui2013computational,lindgren2011explicit}, representation of random fields~\cite{marzouk2009dimensionality} or the sampling of posteriors~\cite{vigliotti2018bayesian}, can all be adapted to statFEM.

The outline of this paper is as follows. In Section~\ref{sec:forward} we review the solution of stochastic partial differential equations with random forcing and coefficients with the finite element method. We then introduce in Section~\ref{sec:inference} the proposed statistical construction of the finite element method. After introducing the underlying statistical generating model we detail the computation of the posterior densities of the finite element solution, the true system response, the hyperparameters and the finite element model itself. This is followed in Section~\ref{sec:examples} by the study of one and two-dimensional Poisson problems with the proposed approach.  Amongst others, we study the convergence of the computed posterior densities to the true densities used for generating the synthetic observation data with an increasing number of observations.  We also illustrate how statFEM reduces to a conventional Bayesian inverse problem when the assumed statistical model is simplified.

%
\section{Probabilistic forward model \label{sec:forward}}
%
The forward problem consists of a stochastic partial differential equation with both a random coefficient and forcing. The coefficient and forcing fields are both assumed to be Gaussian processes. We solve the finite element discretised forward problem with a classical  first-order perturbation method so that its solution is also a Gaussian~\cite{liu1986random}. However, we can use in statFEM any of the established discretisation techniques which can yield the second-order statistics, i.e. mean and covariance, of the solution field; see e.g. the reviews~~\cite{matthies1997uncertainties,sudret2000stochastic,stefanou2009stochastic,xiu2010numerical, aldosary2018structural}.

\subsection{Governing equations \label{sec:governingEqs}}
%
As a representative stochastic partial differential equation we consider on a domain~\mbox{$\Omega \subset \mathbb R^d$} with~\mbox{$d \in \{ 1, \, 2, \, 3\}$}  and the boundary~$ \partial \Omega$ the Poisson equation 
\begin{subequations}	 \label{eq:poisson}
\begin{alignat}{2}
	- \vec \nabla \cdot  \left ( \mu (\vec x)   \vec \nabla u(\vec x) \right ) &=  f(\vec x)   \qquad &&\text{in }  \Omega    \\ 
	u (\vec x ) &= 0 &&\text{on } \partial \Omega  \, ,
\end{alignat}
\end{subequations}
where~$u(\vec x) \in \mathbb R$ is the unknown,~$\mu (\vec x) \in \mathbb R^+ $ is a random diffusion coefficient and~$ f(\vec x) \in \mathbb R$ is a random 
source term. 

To ensure that the diffusion coefficient~$ \mu (\vec x)$ is positive we introduce~$\kappa(\vec x)$ such that~\mbox{$\mu (\vec x) = \exp(\kappa(\vec x))$.} In the following, this new function~$\kappa(\vec x)$  is referred to as the diffusion coefficient with a slight abuse of terminology. The diffusion coefficient~$\kappa(\vec x)$ is modelled as a Gaussian process 
\begin{equation} \label{eq:gprKappa}
	\kappa (\vec x ) \sim \set {GP} \left ( \overline {\kappa}(\vec x), \, c_\kappa(\vec x, \, \vec x' ) \right ) 
\end{equation}
with the mean
\begin{equation} \label{eq:meanKappa}
	\expect [\kappa (\vec x)]= \overline \kappa(\vec x) 
\end{equation}
and the covariance 
\begin{equation} \label{eq:covarKappa}
	 \cov \left (  \kappa(\vec x), \, \kappa(\vec x') \right )  \coloneqq  \expect \left [ \left (\kappa(\vec x) - \overline \kappa(\vec x) \right )  \left (\kappa(\vec x') - \overline \kappa(\vec x') \right ) \right  ]  =  c_\kappa(\vec x, \, \vec x' )  \, .
\end{equation}
Although the specific form of the kernel~$c_\kappa(\vec x, \, \vec x' )$ is inconsequential for the presented approach, we assume for the sake of concreteness a squared exponential kernel of the form
\begin{equation}
	c _\kappa (\vec x, \, \vec x') = \sigma^2_\kappa \exp \left ( - \frac{\| \vec x - \vec x'  \|^2}{2 \ell^2_\kappa }  \right )  
\end{equation}
with the scaling parameter~$\sigma_\kappa \in \mathbb R^+$  and lengthscale parameter~$\ell_\kappa \in \mathbb R^+$. However, we stress that this will be a modelling choice in an actual application based on what is understood about the structure and form of the diffusion coefficient.

Similarly, the random source term~$f(\vec x)$ is modelled as a Gaussian process
\begin{equation} \label{eq:gprR}
	f (\vec x ) \sim \set {GP} \left ( \, \overline{f} (\vec x), \, c_f(\vec x, \, \vec x' ) \right )  
\end{equation}
with the mean
\begin{equation} \label{eq:meanR}
	\expect [f (\vec x)]= \overline{f} (\vec x)
\end{equation}
and the covariance 
\begin{equation}\label{eq:covarR}
	c_f (\vec x, \, \vec x' )    =  \cov \left (  f (\vec x), \, f (\vec x') \right )  = \expect \left [ \left ( f(\vec x) - \overline {f}(\vec x) \right ) \left (  f(\vec x') - \overline {f}(\vec x')  \right ) \right ] =  \sigma^2_f \exp \left ( - \frac{\| \vec x - \vec x'  \|^2}{2 \ell^2_f}  \right )  
\end{equation}
with the respective scaling and lengthscale parameters~$\sigma_f \in \mathbb R^+$  and~$\ell_f \in \mathbb R^+$. 

\begin{figure}
	\centering
	\subfloat[][Source $f(x)$] {
		\includegraphics[width=0.425\textwidth]{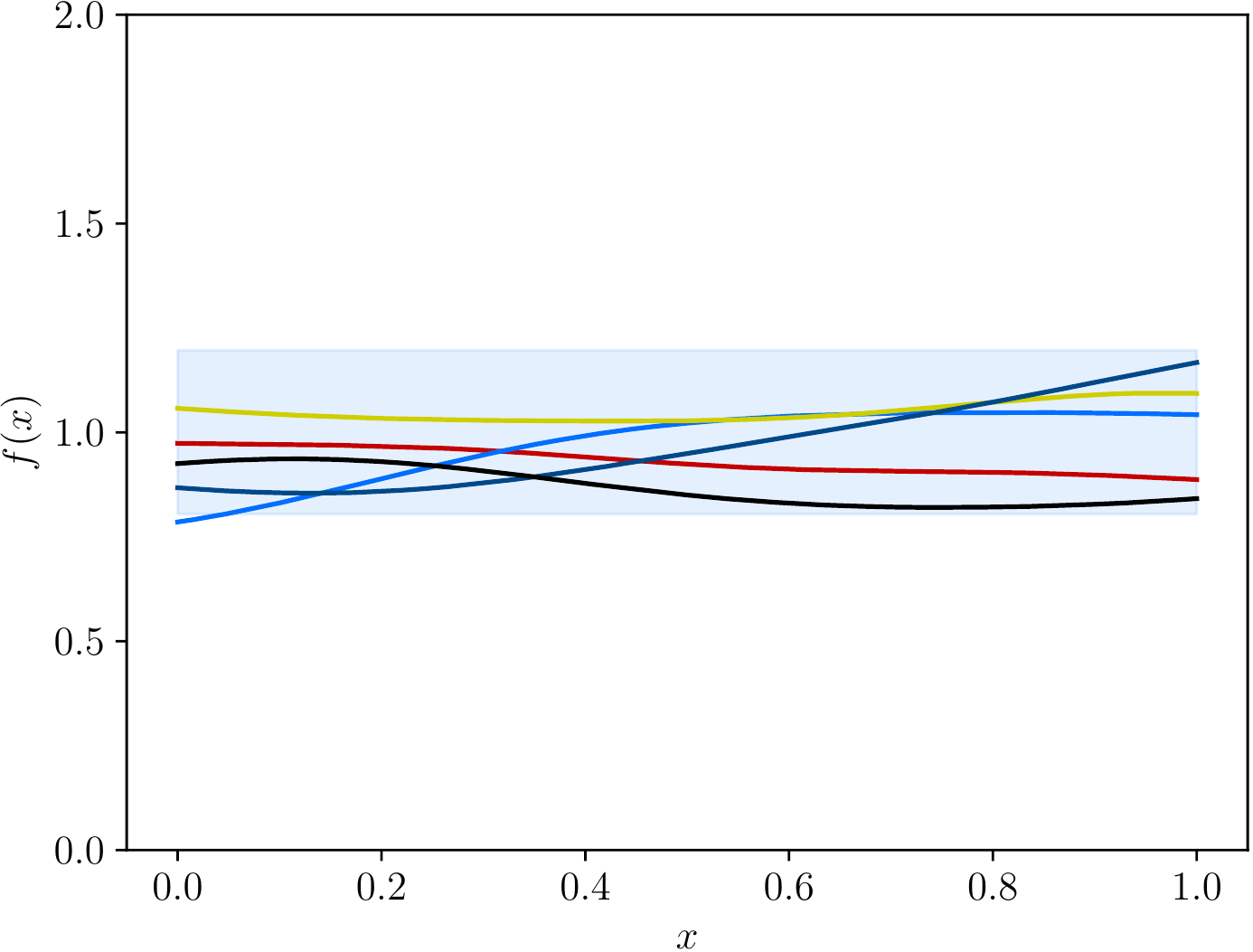}
	}
	\hspace{0.05\textwidth}
	\subfloat[][Solution $u(x)$] {
		\includegraphics[width=0.425\textwidth]{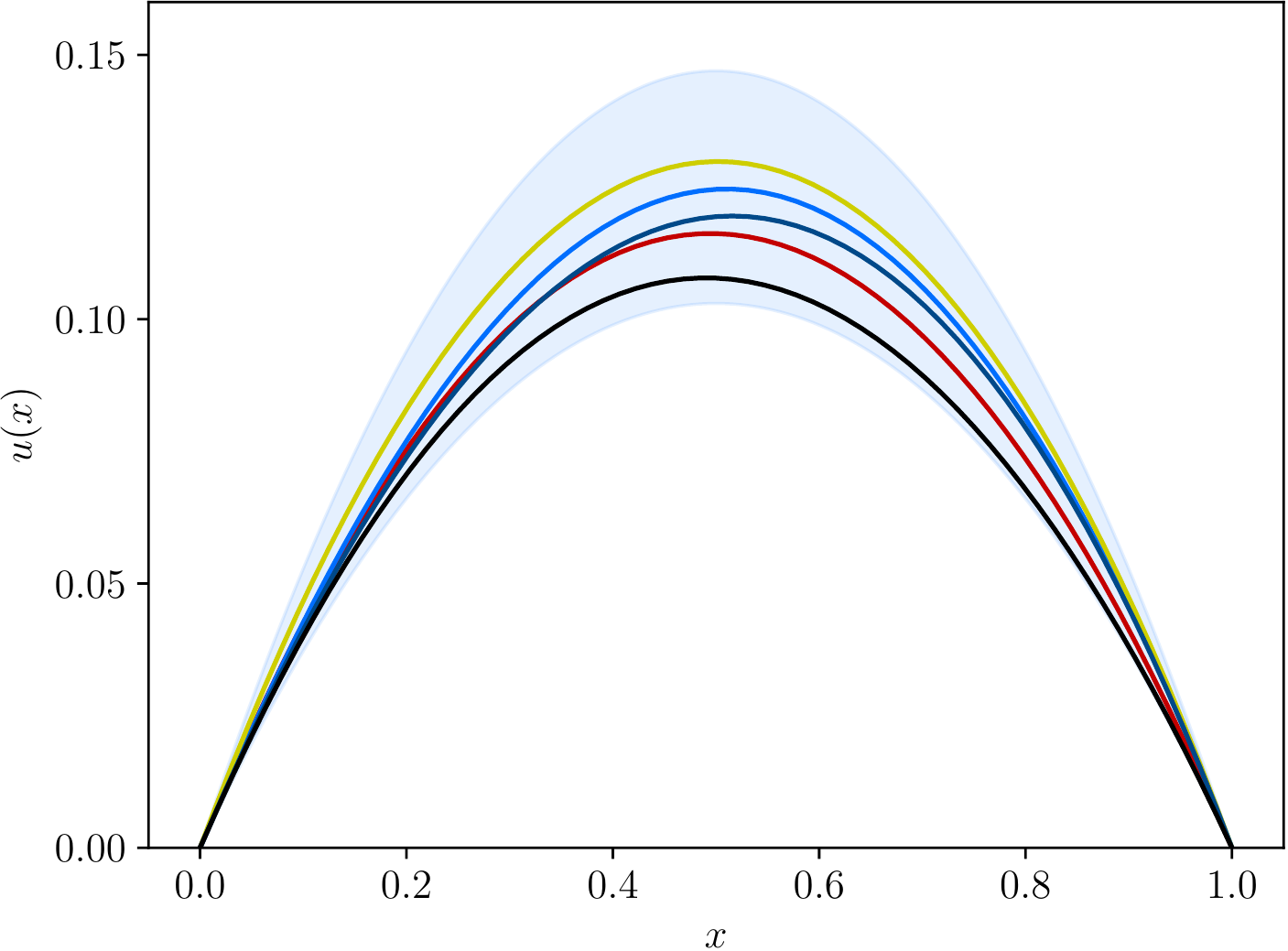}
	}
	\caption{One-dimensional Poisson problem $\D^2 u(x) / \D x^2 = f(x)$ with a random source with a mean~\mbox{$\overline f(x)=1$} and covariance kernel parameters~\mbox{$\sigma_f = 0.1$} and \mbox{$\ell_f=0.4$.}  The five lines in (a) represent samples drawn from~\eqref{eq:gprR} with the corresponding solutions shown in (b). The shaded areas are the $95\%$ confidence regions obtained from~\eqref{eq:gprR} and~\eqref{eq:uExact}, respectively. \label{fig:poisson1D-RHS}} 
\end{figure}

In Figure~\ref{fig:poisson1D-RHS} an illustrative one-dimensional Poisson example,~$\D^2 u(x) / \D x^2 = f(x)$, with a random source and corresponding solution are shown. The diffusion coefficient~$\kappa(x)=1$ is chosen as non-random. The source has the mean~$\overline f(x)=1$ and the parameters of the exponential covariance kernel are chosen with $\sigma_f = 0.1$ and $\ell_f=0.4$. The solution~$u(\vec x)$ is given by the push forward measure of the Gaussian process on the source term, which due to the linearity of the differential operator is also a Gaussian process,  
\begin{equation} \label{eq:uExact}
	u(\vec x) \sim   \set{GP} \left ( \overline{ u}(x), \,    c_u(\vec x, \, \vec x') \right ) =  \set{GP} \left ( g(\vec x, \, \vec x')*\overline{f}(\vec x'), \,   g(\vec x, \, \vec x'') *  c_f(\vec x'', \, \vec x''' ) *  g(\vec x''', \, \vec x') \right ) \, ,
\end{equation}
where~$g(\vec x, \, \vec x')$ is the Greens function of the Poisson problem and~$*$ denotes convolution,  see e.g.~\cite{stuart2010inverse,owhadi2015}. Due to the  smoothing property of the convolution operation the lengthscale of the kernel~$ c_u(\vec x, \, \vec x')$ is larger than~$\ell_f$. Furthermore, the source and the solution are both~$C^\infty$ smooth owing to the squared exponential kernel used for the source~$f(\vec x)$ with the constant mean~$\overline f(x) =1$. 

%
\subsection{Finite element discretisation \label{sec:forwardFE}}
%
We discretise the weak form of the Poisson equation~\eqref{eq:poisson} with a standard finite element approach. Specifically, the domain~$\Omega$ is subdivided into a set~$\{ \omega_e \}$ of non-overlapping elements
\begin{equation}  
	\Omega = \bigcup_{e=1}^{n_e} \omega_e  
\end{equation} 
of maximum size
\begin{equation}
	h = \max_e \diam (\omega_e) \, .
\end{equation}
The unknown field~$ u(\vec x)$  is approximated with Lagrange basis functions~$\phi_i (\vec x)$ and the respective nodal coefficients~$u_i$ of the~$n_u$ non-Dirichlet boundary mesh nodes  by
\begin{equation} \label{eq:LagIntp}
	u_h (\vec x) = \sum_{i=1}^{n_{u}} \phi_i  (\vec x) u_i \, .
\end{equation}
The discretisation of the weak form of the Poisson equation yields the discrete system of equations
\begin{equation} \label{eq:sysEqs}
	\vec A (\vec  \kappa )  \vec u = \vec f    \, ,
\end{equation}
where~\mbox{$\vec A  ( \vec \kappa) \in \mathbb R^{{n_u} \times {n_u}} $}  is the system matrix,~\mbox{$\vec \kappa \in \mathbb R^{n_e}$} is the vector of diffusion coefficients,~\mbox{$\vec u \in \mathbb R^{n_u}$} is the vector of nodal coefficients and~\mbox{$\vec f \in \mathbb  R^{n_u}$} is the nodal source vector. 

The diffusion coefficient vector~$\vec \kappa$ is given by the Gaussian process~\eqref{eq:gprKappa}. We assume that the diffusion coefficient is constant within each element and collect all element barycentre coordinates in the matrix~$\vec X^{(c)} = \left \{ \vec x^{(c)}_e \right \}_{e=1}^{n_e}$. Thus, the diffusion coefficient vector is given by the multivariate Gaussian density 
\begin{equation} \label{eq:kappaMvG}
	\vec \kappa \sim p( \vec \kappa) = \set N \left (\overline{\vec \kappa} \left(  \vec X^{(c)} \right ), \, \vec C_{\kappa} \left ( \vec X^{(c)} , \, \vec X^{(c)} \right ) \right ) \, , 
\end{equation}
where the mean vector~$\overline{\vec \kappa}$ and the covariance matrix~\mbox{$\vec C_\kappa$} are obtained by evaluating the mean~\eqref{eq:meanKappa} and covariance kernel~\eqref{eq:covarKappa} at the respective element barycentres.
 The system matrix~$\vec A$ is assembled from the element system matrices given by 
\begin{equation}
	A_{ij}^e \left ( \kappa_e \right ) = \int_{\omega_e}  \exp (\kappa_e)  \frac{\partial \phi_i (\vec x)}{\partial \vec x} \cdot \frac{\partial \phi_ j(\vec x) }{\partial \vec x} \D \omega_e \, ,
\end{equation}
where~$\kappa_e$ is the diffusion coefficient of the element with the index~$e$. And, the components of the source vector are given by
\begin{equation} \label{eq:sourceF}
	f_{i} = \int_{\Omega}   f(\vec x)   \phi_i(\vec x)   \D \Omega \, .
\end{equation}
As introduced in~\eqref{eq:gprR}, the source is a Gaussian process with the expectation~\eqref{eq:meanR} and covariance~\eqref{eq:covarR}. This implies for the components of the source vector, owing to the linearity of expectation (see e.g.~\cite{murphy2012machine}) and Fubini's theorem,  
\begin{subequations} \label{eq:gprComps}
\begin{align}
	\expect [ f_i ] &=  \expect \left [ \int_\Omega f(\vec x) \phi_i (\vec x)  \D \Omega  \right]   =   \int_\Omega \expect \left [ f( \vec x) \right ]  \phi_i (\vec x)  \D \Omega   =   \int_\Omega   \overline{f}( \vec x)  \phi_i (\vec x)  \D \Omega = \overline{f}_i \label{eq:gprCompsA} \\ 
	 \cov ( f_i, \, f_j) &=  \expect \left [  \int_\Omega \left (  f(\vec x)  - \overline f(\vec x) \right )\phi_i (\vec x)  \D \Omega  \int_\Omega \left ( f(\vec x')  - \overline f (\vec x') \right )\phi_j (\vec x')  \D \Omega (\vec x')   \right]  \nonumber \\
	 & =  \int_\Omega \int_\Omega  \phi_i(\vec x) \expect \left [ \left ( f(\vec x ) -  \overline{f}(\vec x) \right ) \left ( f(\vec x') - \overline{f} (\vec x')\right ) \right ] \phi_j(\vec x' ) \D \Omega (\vec x) \D \Omega (\vec x') \nonumber \\
	&  =  \int_\Omega \int_\Omega  \phi_i(\vec x) c_f (\vec x , \, \vec x') \phi_j(\vec x' ) \D \Omega (\vec x) \D \Omega (\vec x')  \label{eq:gprCompsB} \, .
\end{align}
\end{subequations}
The components of the covariance matrix~$~\mbox{$( \vec C_f) {_{ij}}  = \cov ( f_i, \, f_j)$}$ are obtained by interpolating the covariance kernel with finite element basis functions, i.e.,  
\begin{equation} \label{eq:gprCompsM}
	 ( \vec C_f)_{ij} = \sum_k \sum_l    \int_\Omega \int_\Omega  \phi_i(\vec x)  \phi_k(\vec x) c_f(\vec x_k , \, \vec x'_l)  \phi_l(\vec x' )  \phi_j(\vec x' ) \D \Omega (\vec x) \D \Omega (\vec x')  \, .
\end{equation}
Hence, the source vector is given by the multivariate Gaussian density
\begin{equation} \label{eq:randomVec}
	\vec f \sim p(\vec f )  = \set N \left (\overline{\vec f}, \, \vec C_f  \right )  \, .
\end{equation}

For a globally supported covariance kernel~$c_f(\vec x, \, \vec x')$, such as the used squared exponential kernel, the covariance matrix~$\vec C_f$ is dense. It is non-trivial to efficiently compute and assemble its components.  To obtain a more easily computable covariance matrix, notice that in~\eqref{eq:gprCompsM} the two integrals over the products of basis functions denote indeed two mass matrices. Replacing the two mass matrices with their lumped versions we obtain the approximation 
\begin{align} \label{eq:CRapprox}
	 ( \vec C_f)_{ij}  \approx     \left (  \int_\Omega   \phi_i(\vec x) \D \Omega \right ) c_f(\vec x_i , \, \vec x_j)  \left ( \int_\Omega \phi_j(\vec x )  \D \Omega  \right )  \, .		
\end{align}
Each of the brackets here corresponds to a source vector with a uniform prescribed source~$f(\vec x) =1$, c.f.~\eqref{eq:sourceF}. The covariance kernel~$c_f(\vec x_i , \, \vec x_j) $  is evaluated at the nodes corresponding to basis functions~$\phi_i(\vec x)$ and~$\phi_j(\vec x)$. The introduced approximation makes it possible to assemble and compute the covariance matrix~$\vec C_f$ with standard finite element data structures. For a review on similar and other approximation techniques for computing finite element covariance matrices see~\cite{matthies1997uncertainties, sudret2000stochastic}.

Finally, we can write the probability density for the finite element solution vector~$\vec u$ for a given diffusion coefficient vector~$\vec \kappa$. Solving the discrete system of equations~\eqref{eq:sysEqs} gives
\begin{equation}
		  \vec u =  \vec A ( \vec \kappa )^{-1} \vec f   \, .
\end{equation}  
The right-hand side  represents an affine transformation of the source vector~$\vec f$ with the  multivariate Gaussian density~\eqref{eq:randomVec} such that
\begin{equation} \label{eq:feDensity2}
	\vec u \sim p( \vec u  | \vec \kappa) = \set N \left (\vec  A(\vec \kappa)^{-1} \overline{\vec f} ,\,  \vec A(\vec \kappa)^{-1} \vec C_f \vec A(\vec \kappa)^{-\trans} \right )  \, .
\end{equation}
The corresponding unconditional density~$p (\vec u)$ is obtained by marginalising the joint density~\mbox{$p(\vec u, \, \vec \kappa) = p(\vec u |  \vec \kappa) p(\vec \kappa)$}, which yields
\begin{equation} \label{eq:uDensity}
	p(\vec u) = \int p(\vec u | \vec \kappa ) p(\vec \kappa ) \D \vec \kappa \, .
\end{equation}
It is possible to evaluate this integral numerically using, e.g., MC, MCMC or (sparse) quadrature, however it is impractical for large scale problems given that~$\vec \kappa \in \mathbb R^{n_e}$ is usually a high-dimensional vector. Instead, we use a perturbation method to compute a first order approximation to the density~$p(\vec u)$ \cite{liu1986random}. By explicitly denoting the dependence of the solution~$\vec u$ on the random diffusion coefficient~$\vec \kappa$ and source~$\vec f$,  we can write the series expansion
\begin{align}
\begin{split}
	\vec u (\vec \kappa, \, \vec f) &= \vec u ( \overline{\vec \kappa}, \, \vec f ) + \sum_{e=1}^{n_e} \frac{\partial  \vec u ( \overline{\vec \kappa}, \, \vec f ) }{\partial \kappa_e} ( \kappa_e  - \overline{\kappa}_e )  + \dotsc = \vec u^{(0)} + \sum_{e=1}^{n_e}  \vec u_e^{(1)} \lambda_e  + \dotsc \, , 
\end{split}
\end{align}
where the coefficients~$\vec u^{(0)}$,~$\vec u_e^{(1)}$, $\dotsc$ are obtained by successively differentiating  the system equation~\eqref{eq:sysEqs}  with respect to the element diffusion coefficients. The two coefficients relevant for a first order approximation are given by
\begin{subequations}
\begin{align}
	\vec u^{(0)} &= \vec A( \overline{\vec \kappa})^{-1} \vec f \\
	\vec u_e^{(1)} &= -  \vec A( \overline{\vec \kappa})^{-1}  \frac{\partial \vec A ( \overline{\vec \kappa})}{\partial \kappa_e} \vec u^{(0)} \, .
\end{align}
\end{subequations}
Hence, we obtain for the approximate mean and covariance 
\begin{subequations} \label{eq:approxUcU}
\begin{align}
	\overline {\vec u} &= \expect  \left [   \vec u^{(0)} + \sum_e \vec u^{(1)}_e \lambda_e   \right ]  =   \vec A(\overline {\vec \kappa})^{-1}  \overline{\vec f}  \\
	\vec C_u &= \expect \left [  \left (  \vec u^{(0)} +  \sum_e \vec u_e^{(1)} \lambda_e  \right )   \otimes  \left ( \vec u^{(0)} +  \sum_e \vec u^{(1)}_e \lambda_e  \right ) \right ] - \overline {\vec u} \otimes \overline {\vec u}   \, .
\end{align}
\end{subequations}
Note that the expectation is over both~$\vec \kappa$ and~$\vec f$. After lengthy but straightforward algebraic manipulations we obtain  
\begin{equation} \label{eq:cUperturb}
	\vec C_u  = \vec A(\overline {\vec \kappa})^{-1} {\vec C}_f  \vec A(\overline {\vec \kappa})^{-\trans}  + \sum_e \sum_d { ( \vec C_\kappa})_{ed } \vec A(\overline {\vec \kappa})^{-1} \frac{\partial \vec A(\overline {\vec \kappa})}{\partial \kappa_e} \vec A(\overline {\vec \kappa})^{-1} \left (  {\vec C}_f + \overline{\vec f} \otimes  \overline{\vec f}  \right )   \vec A(\overline {\vec \kappa})^{-\trans} \frac{\partial \vec A(\overline {\vec \kappa})^{\trans}}{\partial \kappa_d} \vec A(\overline {\vec \kappa})^{-\trans} \, .
\end{equation}
Finally, we can approximate the density of the finite element solution~\eqref{eq:uDensity} with the multivariate Gaussian density
\begin{equation} \label{eq:forwardDensityFinal}
  	p(\vec u) = \set N \left (\overline{\vec u} , \, \vec C_u \right ) \, .
 \end{equation} 

It is clear that the true density~$p(\vec u)$ according to~\eqref{eq:uDensity} is usually not a Gaussian. The approximation~\eqref{eq:forwardDensityFinal} is only valid when the scaling parameter~$\sigma_\kappa$ of the variance is relatively small.  However, when the diffusion coefficient~$\vec \kappa$ is deterministic  the density~$p(\vec u)$  is a Gaussian as can be seen in~\eqref{eq:feDensity2}. Furthermore, we can deduce from~\eqref{eq:cUperturb} that there is a fundamental difference in how the source and diffusivity covariance matrices~$\vec C_f$ and~$\vec C_\kappa$ contribute to~$\vec C_u$. The source covariance~$\vec C_f$ is always multiplied twice with the inverse of the system matrix, which increases the smoothness of the covariance operator.  In contrast, the diffusivity covariance~$\vec C_\kappa$ contributes directly with no such smoothing. 

\begin{figure}[]
	\centering
	\subfloat[][Diffusion coefficient $\mu(x)$] {
		\includegraphics[width=0.425\textwidth]{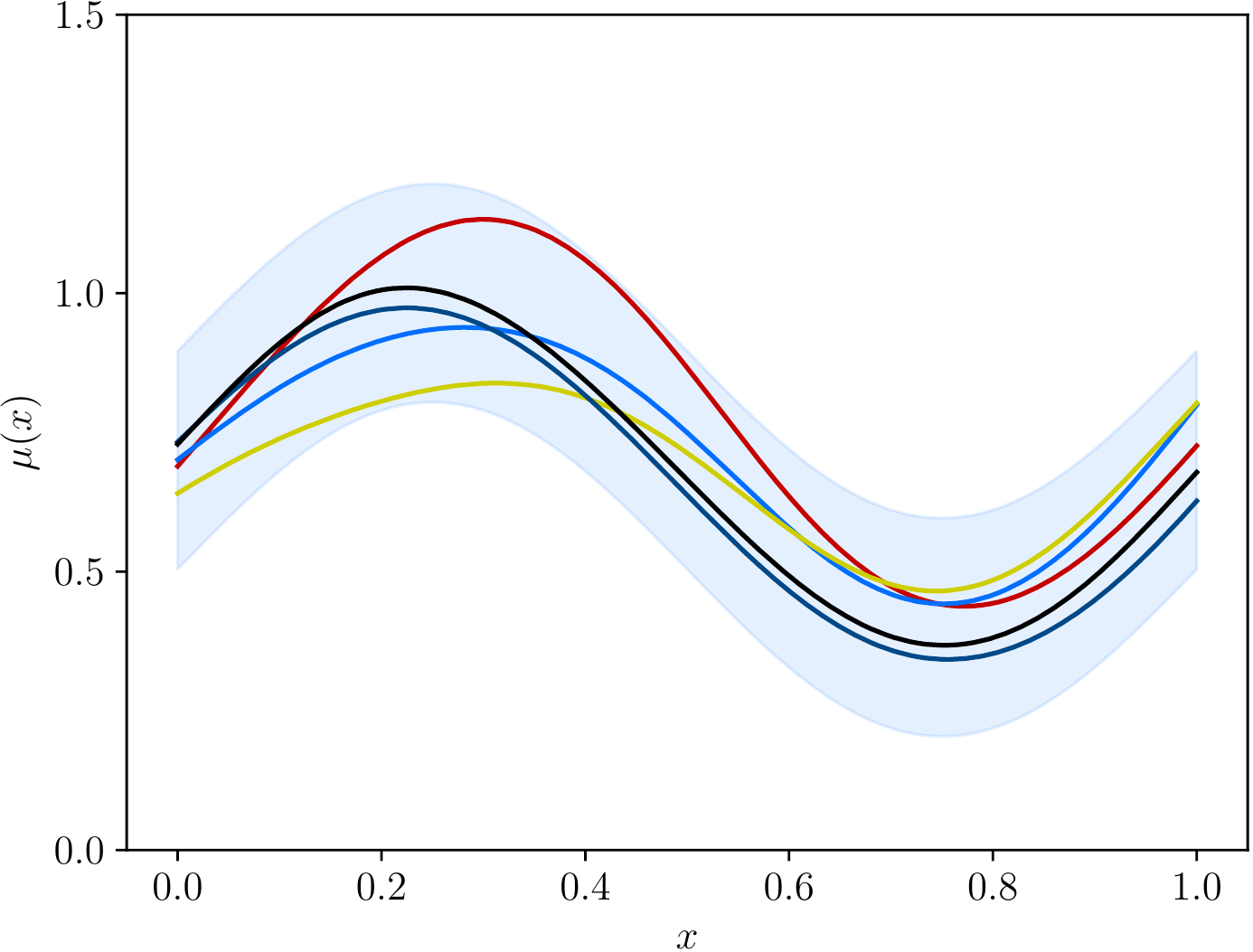}
	}
	\hspace{0.05\textwidth}
	\subfloat[][Solution $u(x)$] {
		\includegraphics[width=0.425\textwidth]{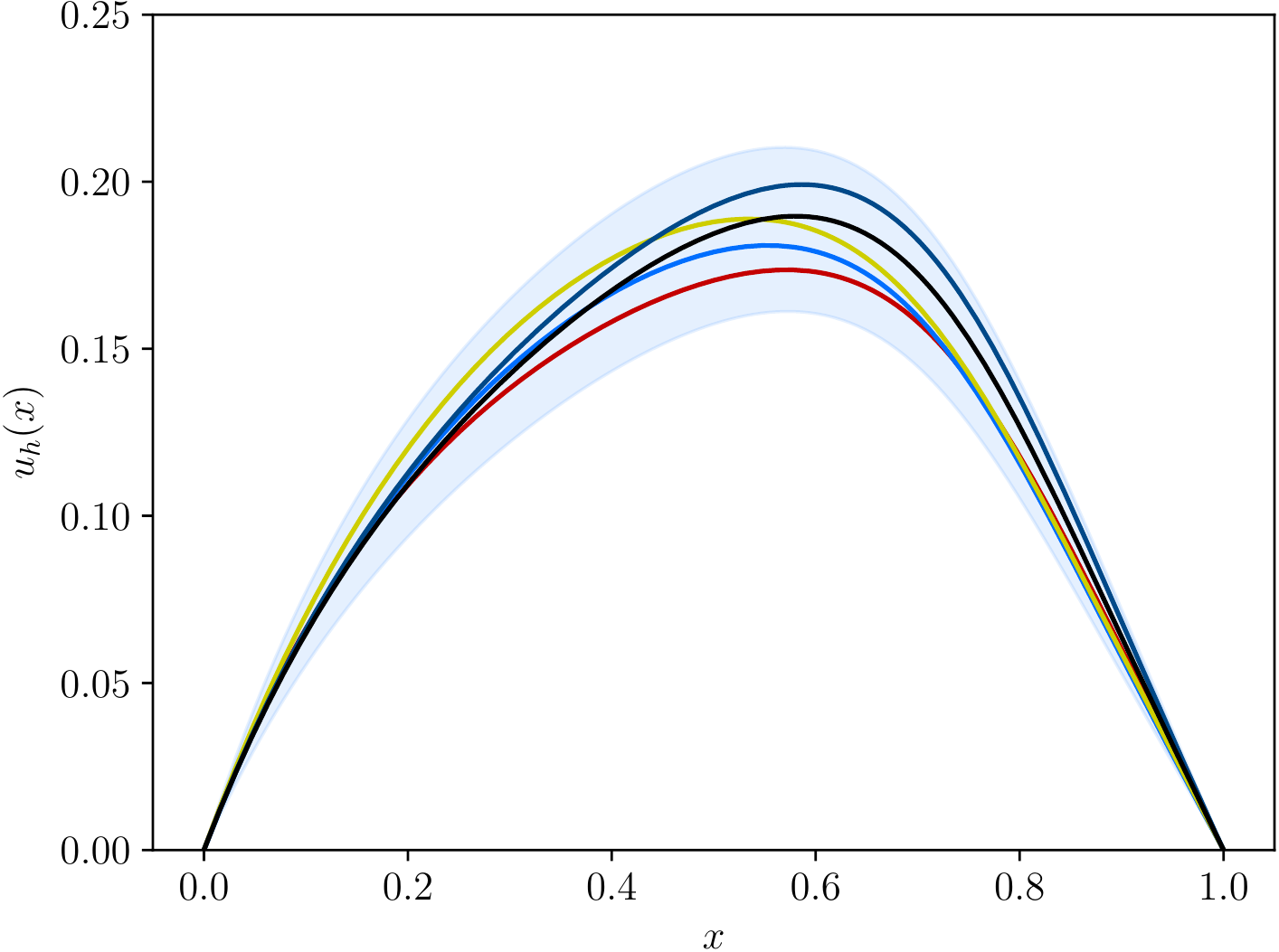}
	}
	\caption{One-dimensional Poisson problem \mbox{$-\D \left ( \mu(x)  \D u(x) / \D x \right )/\D x  =  1$} with a random diffusion coefficient~\mbox{$\mu(x) = \exp (\kappa(x))$} with a mean~\mbox{$\overline \kappa (x) = \ln \left( 0.7 + 0.3 \sin ( 2 \pi x ) \right)$} and the covariance kernel parameters~$\sigma_\kappa = 0.1 $ and $\ell_\kappa= 0.25$.  The five lines in (a) represent samples drawn from~\eqref{eq:kappaMvG} with the corresponding solutions shown in (b). The shaded areas are the $95\%$ confidence regions obtained from~\eqref{eq:kappaMvG} and Monte Carlo sampling. \label{fig:poisson1D-LHS}} 
\end{figure}

\begin{figure}[h!] 
	\centering
		\includegraphics[width=0.485\textwidth]{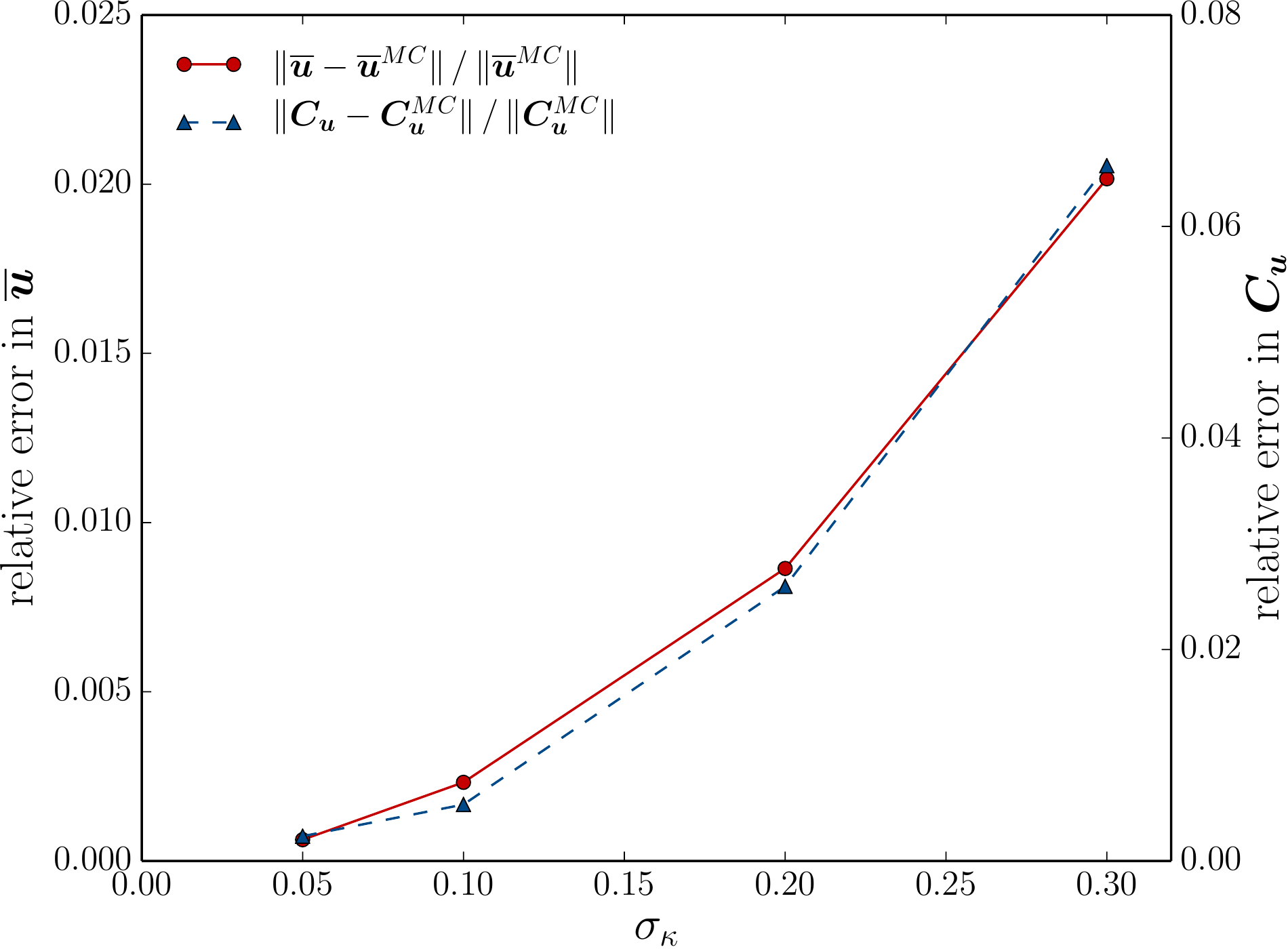}
	\caption{Relative perturbation error of the approximate mean~$\overline{\vec u}$ and covariance~$\vec C_u$ obtained from~\eqref{eq:approxUcU} for the one-dimensional Poisson problem with a random diffusion coefficient shown in Figure~\ref{fig:poisson1D-LHS}. Only the covariance scaling parameter~$\sigma_k$ is varied. The empirical mean~$\overline{\vec u}^{MC}$ and covariance~$\vec C_u^{MC}$ are obtained by Monte Carlo sampling~\eqref{eq:uDensity} and can be considered as exact. The errors are measured in the Frobenius norm. \label{fig:convCuLHS}} 
\end{figure}

In Figure~\ref{fig:poisson1D-LHS} an illustrative one-dimensional Poisson problem,~ $- \D \left ( \mu(x)  \D u(x) / \D x \right ) / \D x  =  f(x)$, with a random diffusion coefficient and corresponding finite element solution are shown. The source~$f(x)=1$ is chosen as non-random. The diffusion coefficient~$\mu(x) = \exp (\kappa(x))$ has the mean~$\overline \kappa (x) = \ln \left( 0.7 + 0.3 \sin ( 2 \pi x ) \right)$ and the parameters of the exponential covariance kernel are chosen with~$\sigma_\kappa = 0.1 $ and $\ell_\kappa= 0.25$. The one-dimensional problem is discretised with~$128$ linear finite elements. To assess the accuracy of the approximate mean~$\overline {\vec u}$  and covariance~$\vec C_u$ according to~\eqref{eq:approxUcU}, we compare both with the empirical mean~$\overline{\vec u}^{MC}$ and covariance~$\vec C_u^{MC}$ obtained by Monte Carlo sampling~\eqref{eq:uDensity}. As depicted in Figure~\ref{fig:convCuLHS} the first-order perturbation and the Monte Carlo results are in good agreement for relatively large~$\sigma_\kappa$.

%
\section{Bayesian inference \label{sec:inference}}
%
In this section, we first introduce following~\cite{kennedy2001bayesian,higdon2004combining,bayarri2007framework} the statistical generating model for the true process underlying the observed data.  Each of the random variables in this model depends on parameters, like the covariance lengthscale and scaling parameters in the forward model, which may be chosen to be known or unknown with only their prior densities given. The unknown random parameters are referred to as the hyperparameters. We sequentially apply Bayes rule on three different levels to infer, or learn, from the observed data all the random variables and hyperparameters. On level one, in Section~\ref{sec:posteriorFE}, we derive the posterior finite element and true system response densities with the finite element density derived in Section~\ref{sec:forwardFE} serving as a prior. On level two, in Section~\ref{sec:hyperPlearning}, the posterior densities and point estimates of the hyperparameters are obtained. Finally, on level three,  in Section~\ref{sec:modCompare}, we rank different finite element models, e.g., with different mesh sizes or modelling assumptions, based on their ability to explain the observed data.  

%
\subsection{Statistical generating model for the observations  \label{sec:generatingModel}}
\begin{figure}[htb] 
\centering
\newlength{\myTikzDist}
\setlength{\myTikzDist}{1.1cm}
\begin{tikzpicture}[
 ->,>=latex,shorten >=0pt,auto,node distance=\myTikzDist,  thin, align=center,
 main node/.style={circle, draw=black, minimum size = 12mm}, 
 scale=.5] 
\node[main node, fill=csmlLightBlue]    (2)                                   {$\{ \omega_e \}$}; 
\node[main node]                                  (5)            [right= of 2]     {$\vec {A}$}; 
\node[main node]                                  (1)            [right= of 5]     {$\vec \kappa$}; 
\node[main node, fill=csmlLightBlue]  (15)            [right= of 1, yshift=0.7\myTikzDist]    {$\overline {\vec \kappa}$}; 
\node[main node, fill=csmlLightBlue]  (16)            [right= of 1, yshift=-0.7\myTikzDist]   {$\sigma_\kappa$, $\ell_\kappa$}; 
\node[main node]                                  (6)            [below= of 5] {$\vec{u}$ }; 
\node[main node]                                  (4)            [left= of 6]     {$\vec f$}; 
\node[main node, fill=csmlLightBlue]    (3)            [left= of 4, yshift=0.7\myTikzDist]    {$\overline {\vec f}$}; 
\node[main node, fill=csmlLightBlue]   (17)            [left= of 4,  yshift=-0.7\myTikzDist] {$\sigma_f$, $\ell_f$}; 
\node[main node]                                  (7)            [below= of 6]   {$\vec{y}$}; 
\node[main node]                                 (9)             [right= of 7]      {${\vec d}$}; 
\node[main node, fill=csmlPink]          (10)            [below= of 7]    {$\rho$}; 
\node[main node]                                (11)            [left= of 7]         {$\vec e$}; 
\node[main node, fill=csmlPink]          (12)            [left= of 11]         {$\sigma_e$}; 
\node[main node, fill=csmlPink]          (14)            [right= of 9]      {$\sigma_d, \, \ell_d$}; 

\path[every node/.style={font=\sffamily\small}] 
(15) edge node [] {} (1) 
(16) edge node [] {} (1) 
(1) edge node [] {} (5) 
(2) edge node [] {} (5) 
(3) edge node [] {} (4) 
(4) edge node [] {} (6) 
(5) edge node [] {} (6) 
(6) edge node [] {} (7) 
(17) edge node [] {} (4) 
(9) edge node [] {} (7) 
(10) edge node [] {} (7) 
(11) edge node [] {} (7) 
(12) edge node [] {} (11) 
(14) edge node [] {} (9); 

\draw[csmlRed,thick,dashed] ($(6.north west)+(-0.8,0.8)$) rectangle ($(7.south east)+(0.8,-0.8)$); 

\end{tikzpicture} 
\caption{Graphical model of statFEM. The coloured circles represent the (possibly unknown) parameters and the empty circles denote the random variables that are either observed or derived. The parameters of the forward problem introduced in Section~\ref{sec:forward} are shown in blue and the parameters of the statistical generating model~\eqref{eq:decomposition} are shown in red. Usually, some or all of the parameters in the shaded circles are known and need not be inferred. The remaining unknown random hyperparameters are inferred from the observations~$\vec y$ and the finite element solution~$\vec u$ using the statistical generating model.  \label{fig:graphStatFEM}}
\end{figure}
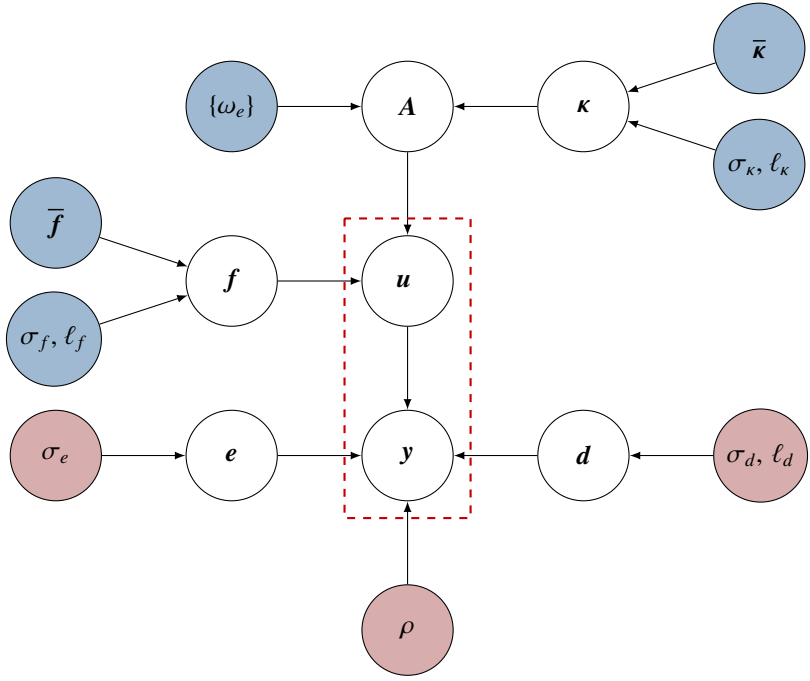

%
In statFEM the observed data vector~$\vec y \in \mathbb R^{n_y}$  is, as graphically illustrated in Figure~\ref{fig:graphStatFEM}, additively composed in three components 
\begin{equation} \label{eq:decomposition} 
	\vec y = \vec z + \vec e = \rho \vec P \vec u + \vec d  + \vec e \, .
 \end{equation}
That is, the observed data vector is equal to the unknown true system response~\mbox{$\vec z  \in \mathbb R^{n_y}$} and the random observation error, i.e. noise,~\mbox{$\vec e  \in \mathbb R^{n_y}$.} In turn, the  true system response is characterised with the parameter~$\rho \in \mathbb R^+$ scaled projected finite element solution~$\vec P \vec u$ and the  mismatch error, or model inadequacy,~\mbox{$\vec d \in \mathbb R^{n_y}$}. The matrix~\mbox{$\vec P \in \mathbb R^{n_y \times n_u}$} projects the finite element solution to the observed data space and consists of the finite element basis functions~$\phi_i (\vec x)$ evaluated at the~$n_y$ observation points. Of course, the observations in~$\vec y$ can correspond to almost any physical quantity of interest, like the flux~$\nabla u_h(\vec x)$, which can be obtained from the solution~$u_h(\vec x)$ by applying a linear operator. In such cases the projection matrix~$\vec P$  is the discretisation of the linear operator in question. 

Although the mismatch error~$\vec d$ is not known, we approximate its distribution using a Gaussian Process
\begin{equation} \label{eq:densityD}
	\vec d \sim p (\vec d | \sigma_d \,  , \ell_d) = \set N(\vec 0, \, \vec C_d) \, ,
\end{equation}
and choose as a kernel the squared exponential kernel
\begin{equation} \label{eq:mismatchCovKernel}
	c _d (\vec x, \, \vec x') = \sigma^2_d \exp \left ( - \frac{\| \vec x - \vec x'  \|^2}{2 \ell^2_d }  \right )  
\end{equation}
with the parameters~$\sigma_d \in \mathbb R^+$ and~$\ell_d \in \mathbb R^+$. The covariance matrix~$\vec C_d \in \mathbb R^{n_y}  \times  \mathbb R^{n_y}$ is obtained by evaluating the kernel at the~$n_y$ observation locations. It is straightforward to consider other covariance kernels or a linear combination of covariance kernels, see e.g.~\cite[Ch. 4]{williams2006gaussian}.
 
Furthermore, as usual, we assume that the observation error~$\vec e$  has the multivariate Gaussian density
\begin{equation} \label{eq:densityE}
	\vec e \sim p (\vec e  ) = \set N(\vec 0, \, \vec  C_e ) 
\end{equation}
with the diagonal covariance matrix $ \vec C_e=  \sigma^2_e \vec I $.

All the variables in the decomposition~\eqref{eq:decomposition} are Gaussians so that the observed data vector~$\vec y$ has the conditional density 
\begin{equation}\label{eq:likelihood}
	p(\vec y |   \vec u) = \set N( \rho \vec P \vec u, \, \vec C_d + \vec C_e )  \, .
\end{equation}
This density is the likelihood of observing the data~$\vec y$ for a given finite element solution~$\vec u$. The likelihood depends in addition~ to the scaling parameter~$\rho$ on a number of parameters, including the introduced kernel scaling and lengthscale parameters, not all of which are known from the outset, see Figure~\ref{fig:graphStatFEM}. The linear summation~$\vec{C}_d + \vec{C}_e$ in the likelihood  indicates an identifiability issue, as discussed in detail in~\cite{bayarri2007framework,kennedy2001bayesian}. We enforce weak identifiability by employing prior distributions on the hyperparameters defining each~$\vec C_d$ and~$\vec C_e$, see also Section~\ref{sec:hyperPlearning}.

In passing, we note that in a non-Bayesian context the unknown hyperparameters are determined by maximising the likelihood~\eqref{eq:likelihood} and the obtained values are referred to as MLE estimates. Clearly, the likelihood 
\begin{equation}
	p(\vec y |   \vec u )  \propto \exp \left ( -\frac{1}{2} \left (\rho \vec P \vec u  - \vec y \right )^{\trans} (\vec C_d + \vec C_e)^{-1}  \left (\rho \vec P \vec u  - \vec y \right )    \right ) 
\end{equation}
 has its maximum at~$\vec y = \rho \vec P \vec u$. Hence, MLE  prefers models which match the observation vector~$\vec y$ as closely as possible  irrespective of the true system response and measurement errors. As widely discussed in the literature, this gives rise to overly complex models prone to overfitting, see~\cite{mackay2003information,bishop2006pattern,kaipio2006statistical,murphy2012machine,vigliotti2018bayesian}.  

\subsection{Posterior finite element and true system response densities~\label{sec:posteriorFE}}
%
We use Bayes rule to update the finite element density of the forward problem~\eqref{eq:forwardDensityFinal} with the available observed data in line with the postulated statistical model~\eqref{eq:decomposition}. All the hyperparameters are assumed to be known or, in other words, all the mentioned densities are conditioned on the hyperparameters.

\subsubsection{Single observation vector \label{sec:posteriorFEsingle}}
%
To begin with, we consider only one single observation vector~$\vec y$. The  posterior finite element density~$p(\vec u | \vec y)$ conditioned on observed data~$\vec y$ is given by 
\begin{equation} \label{eq:bayesU}
	p(  \vec u | \vec y ) = \frac{ p(\vec y | \vec u )  p  (\vec u )}{ p (\vec y)}
\end{equation}
with the likelihood~\mbox{$p(\vec y | \vec u )$} in~\eqref{eq:likelihood}, the prior~$p(\vec u )$ in~\eqref{eq:forwardDensityFinal} and the marginal likelihood, or the evidence,  
\begin{equation} \label{eq:magrLikely}
	p(\vec y ) = \int p(\vec y | \vec u)  p  ( \vec u) \D  \vec u   \, , 
\end{equation}
which ensures that the posterior~$p(\vec u  | \vec y)$ is a probability distribution integrating to one.  The likelihood~\mbox{$p(\vec y | \vec u )$} is a function of~$\vec u $ and measures the fit of the statistical generating model~\eqref{eq:decomposition} to the given observation~$\vec y$.  On the other hand, the prior~$p  ( \vec u )$ reflects our knowledge of the system before any observations are made. The marginal likelihood~$p(\vec y )$ is the probability of observing the known (fixed) observation~$\vec y$ averaged over all possible finite element solutions~$\vec u$.  Marginal likelihood plays a key role in Bayesian statistics as will be detailed in the following sections. As shown in the~\ref{sec:appPuCy}, the posterior density is a multivariate Gaussian and is given by 
\begin{subequations}  \label{eq:postUcY}
\begin{align} 
	& p(  \vec u | \vec y )  = \set N ( \overline {\vec u}_{|y} , \, \vec C_{u|y}) \, , 
	\intertext{where}    
	& \overline {\vec u}_{|y} = \vec C_{u|y} \left (  \rho  \vec P^\trans  \left ( \vec C_d + \vec C_e \right )^{-1} \vec y   +  \vec C_u^{-1} \overline{\vec u}  \right )  \quad \text {and} \quad 
	\vec C_{u|y}  =  \left  ( \rho^2 \vec P^\trans (\vec C_d + \vec C_e )^{-1} \vec P + \vec C_u^{-1}  \right )^{-1} \, .
\end{align}
\end{subequations}
Likewise, the marginal likelihood is a Gaussian and can be obtained by analytically evaluating~\eqref{eq:magrLikely}, or more easily by revisiting the decomposition~\eqref{eq:decomposition}. All the variables in~\eqref{eq:decomposition} are multivariate Gaussians so that the marginal likelihood simply reads
\begin{equation} \label{eq:marginalLevald}
	p(\vec y ) = \set N \left  ( \rho \vec P \overline{\vec u} , \, \vec C_d + \vec C_e + \rho^2 \vec P \vec C_u \vec P^\trans \right )  \, .
\end{equation}

In~\eqref{eq:postUcY}, we can see when  $\vec C_d + \vec C_e$ is small in comparison to $\vec C_u$ (in some norm) the  mean~$\overline {\vec u}_{|y}$  tends to~$\vec y / \rho$  and the covariance~$\vec C_{u|y}$  to~$(\vec C_d + \vec C_e) / \rho^2$. However, when  $\vec C_d + \vec C_e$ is relatively large  the  mean~$\overline {\vec u}_{|y}$  tends to~$\overline{\vec u}$  and the covariance~$\vec C_{u|y}$  to~$\vec C_u$. These bounds are reasonable reminding ourselves that the density of the unobserved true system response~\mbox{$\vec z = \rho \vec P \vec u + \vec d$} is given by
\begin{equation} \label{eq:trueSystemPosterior}
	p(  \vec z | \vec y ) =  \set N \left ( \rho \vec P \overline {\vec u}_{|y} , \, \rho^2 \vec P \vec C_{u|y}  \vec P^\trans + \vec C_d \right ) \, .
\end{equation}
With the mentioned bounds, for relatively small $\vec C_d+ \vec C_e$ the mean of~$\vec z$ tends to~$\vec y$ and its covariance to~$2 \vec C_d + \vec C_e$. In contrast, for large $\vec C_d + \vec C_e$ the mean tends to~$\rho \overline{\vec u}$ and the covariance to~$\vec C_u + \vec C_d$.

In terms of implementation, usually the number of observations points is significantly smaller than the unknowns in the finite element method, i.e.~\mbox{$n_y \ll n_u$}. Hence, the required inversion of large dense matrices of size~$n_u \times n_u$ in~\eqref{eq:postUcY} can be avoided by using the Sherman-Morrison-Woodbury matrix identity, see~\ref{sec:appPuCy}. 

%
\subsubsection{Multiple observation vectors~\label{sec:multilePY}}
%
In engineering applications usually the same sensors are used to repeatedly sample a set of observation vectors~$\{ \vec y_i\}_{i=1}^{n_{o}}$. For notational convenience we collect the set of observation vectors in a matrix~$\vec Y \in \mathbb R^{n_y \times n_o}$. The posterior finite element density~$p(\vec u | \vec Y)$ conditioned on all the observations~$\vec Y$ is once again given by 
\begin{equation} \label{eq:bayesUMult}
	p(  \vec u | \vec Y ) = \frac{ p(\vec Y | \vec u)  p  (\vec u)}{ p (\vec Y)}   \, .
\end{equation}
The physically sensible assumption of statistical independence between the~$n_o$ observations yields the likelihood
\begin{equation}
	p(\vec Y | \vec u)  =  p(\vec y_1 | \vec u) p(\vec y_2 | \vec u) \cdots  p(\vec y_{n_o} | \vec u)    =  \prod_{i=1}^{n_o} p(\vec y_i | \vec u)
\end{equation}
and the marginal likelihood, or the evidence, 
\begin{equation} \label{eq:magrLikelyMult1}
	p(\vec Y ) = \int  \prod_{i=1}^{n_o} p(\vec y_i | \vec u)    p  ( \vec u) \D \vec  u   \, .
\end{equation}
Following the same steps as in the preceding Section~\ref{sec:posteriorFEsingle} we obtain the posterior density  
\begin{subequations}  \label{eq:postUcYMult}
\begin{align}
	& p(  \vec u | \vec Y )  = \set N ( \overline {\vec u}_{|Y} , \, \vec C_{u|Y}) \, , 
	\intertext{where}    
	& \overline {\vec u}_{|Y} = \vec C_{u|Y} \left (  \rho \vec P^\trans  \left ( \vec C_d + \vec C_e \right )^{-1} \sum_{i=1}^{n_o} \vec y_i   +  \vec C_u^{-1} \overline{\vec u}  \right )  \quad \text{and} \quad
	\vec C_{u|Y} =  \left  ( \rho^2 n_o \vec P^\trans (\vec C_d + \vec C_e )^{-1} \vec P + \vec C_u^{-1}  \right )^{-1} \, .
\end{align}
\end{subequations}
Ostensibly, with an increase in the number of observation vectors~$n_o$ the covariance~$\vec C_{u|Y}$ tends to zero and, in turn, the mean~$\overline {\vec u}_{|Y}$ tends to the empirical mean of the observations~$\sum \vec y_i / n_o$. For later reference, we note that the marginal likelihood is because of the statistical independence assumption between the observations given by 
\begin{equation} \label{eq:magrLikelyMult}
	p(\vec Y) = \prod_{i=1}^{n_o} p(\vec y_i ) \, ,
\end{equation}
where~$p(\vec y_i)$ is the marginal likelihood~\eqref{eq:magrLikely} of each of the readings.

%
\subsection{Hyperparameter learning \label{sec:hyperPlearning}}
%
The marginal likelihood~$p (\vec Y) $ given in~\eqref{eq:magrLikelyMult}, or~$p(\vec y)$ in~\eqref{eq:magrLikely}, is critical for determining the hyperparameters of the statistical model~\eqref{eq:decomposition}.  To begin with, we collect the parameters introduced so far in the vector
\begin{equation} \label{eq:wVecLong}
	\setcounter{MaxMatrixCols}{20}    
	\vec w \coloneqq \begin{pmatrix}
	\rho & \sigma_\kappa & \ell_\kappa & \sigma_f & \ell_f & \kappa_1 & \kappa_2 & \dotsc & \kappa_{n_e} &  \sigma_d  &  \ell_d &  \sigma_e 
	\end{pmatrix}^\trans   
	\,  \in \mathbb R^{n_w} \, .
\end{equation}
Some of these parameters may be known or unknown with only their prior densities given. To sidestep the issue of non-identifiability of the parameters it is important that the priors are informative. In the following,~$\vec w$ includes only the {\em unknown} hyperparameters so that its dimension varies depending on the considered problem. The hyperparameters are estimated from the observed data by applying the Bayes formula one more time. To this end, note that the marginal likelihood~$p(\vec Y)$ in~\eqref{eq:magrLikelyMult} is indeed conditioned on the hyperparameter vector~$\vec w$  so that we write more succinctly~$p(\vec Y | \vec w)$.  Consequently, the Bayes formula for  obtaining the posterior density of the hyperparameter vector reads 
\begin{equation}  \label{eq:bayesHypers}
	p( \vec w  |  \vec Y )  = \frac{p ( \vec Y | \vec w ) p( \vec w )}{\int p ( \vec Y | \vec w ) p( \vec w ) \D \vec w}  \, .
\end{equation}
As usual, the prior~$p(\vec w)$ encodes any information that we might have prior to making the observation~$\vec Y$. In choosing~$p(\vec w)$ it is justified to assume that all the hyperparameters are statistically independent  such that 
\begin{equation}
	p(\vec w) = \prod_{i=1}^{n_w} p(w_i)    \, .
\end{equation}
Moreover, the normalisation constant in the denominator of~\eqref{eq:bayesHypers} can be omitted when only a point-estimate is needed or when the posterior is sampled with MCMC. In that case it is sufficient to consider just
\begin{equation} \label{eq:bayesParams}
	p( \vec w  |  \vec Y )  \propto p( \vec Y | \vec w ) p( \vec w ) \, .
\end{equation}
It bears emphasis that this posterior is in contrast to the posterior~\eqref{eq:bayesUMult} analytically intractable. The optimal hyperparameter vector~$\vec w^*$ referred to as the  maximum  posteriori (MAP) estimate is given by 
\begin{equation} \label{eq:hypMAP}
	\vec w^*  = \argmax_{\vec w }  p( \vec Y |  \vec w  ) p( \vec w ) \, .
\end{equation}
This, often non-convex, optimisation problem can be numerically solved with conventional algorithms, see e.g.~\cite{zhu1997algorithm}. Alternatively, as in this paper, we can use the expectation of the parameter vector 
\begin{equation}
	\overline{\vec w} = \mathbb E [\vec w] = \int \vec w p (\vec w | \vec Y ) \D \vec w \, ,
\end{equation}
as a point estimate. The expectation is obtained by  sampling~$p (\vec w | \vec Y )$ using MCMC, see~\ref{sec:appMCMC}, and then calculating the empirical mean 
\begin{equation}  \label{eq:hypMCMC}
	\overline{\vec w}   \approx \frac{1}{N} \sum_i \vec w^{(i)} \quad \text{with }   \vec w^{(i)} \sim p( \vec w |  \vec Y  ) \, .
\end{equation}
The empirical variance of the samples~$\vec w^{(i)}$ represents the uncertainty in the obtained point estimate. In contrast to sampling, the optimisation problem~\eqref{eq:hypMAP} does not yield such a variance estimate. Evidently, the two estimates~\eqref{eq:hypMAP} and~\eqref{eq:hypMCMC} will have different values and either one or both of them might be sufficient to characterise~$p(\vec w| \vec Y)$. When~$p(\vec w | \vec Y)$ is multimodal, usually, both are inadequate. In such cases it is better to take the entire distribution~$p(\vec w | \vec Y)$ and to marginalise out~$\vec w$ whenever a distribution depends, i.e., is conditioned, on~$\vec w$. 
  
%
\subsection{Model comparison and hypothesis testing \label{sec:modCompare}}
%
The marginal likelihood~$p (\vec Y) $ in~\eqref{eq:magrLikelyMult}, or~$p(\vec y)$  in~\eqref{eq:magrLikely}, plays also a key role in Bayesian model comparison, see e.g.~\cite{kass1995bayes,mackay2003information}. Without loss of generality, we consider in the following finite element models which differ only in terms of mesh resolution, specifically, maximum element size~$h$. However, the same approach can be applied to finite element models that have, for instance, differing domain geometries or boundary conditions or are based on fundamentally different mathematical models.   

We aim to compare the fidelity of two finite element models, i.e. the two meshes~\mbox{$\set M_1$} and~\mbox{$\set M_2$}, in explaining the observed data~$\vec Y$. As previously defined, the marginal likelihood~$p(\vec Y)$ is obtained by averaging the likelihood~$p(\vec Y| \vec u)$ over all possible finite element solutions~$\vec u$. The marginal likelihood is conditioned on the specific mesh used for computing~$\vec u$ so that we denote it either with~\mbox{$p(\vec Y | \set M_1)$} or~\mbox{$p(\vec Y | \set M_2)$.} Because the true system response~$\vec z$ is unknown it is from the outset unclear which of the two models is sufficient to explain the observed data~$\vec Y$. Obviously, the model with the finer mesh yields the more accurate finite element solution. However, in light of inherent observation errors and model inadequacy the coarser mesh may indeed be sufficient to explain the observed data. To quantify the fidelity of the two models, we compute their posterior densities using Bayes formula, that is, 
\begin{subequations}
\begin{align}
	p(\set M_1 | \vec Y) &= \frac{p(\vec Y | \set M_1 ) p(\set M_1 )}{p(\vec y | \set M_1 ) p(\set M_1 ) + p(\vec Y | \set M_2 ) p(\set M_2 )} \,  , \\
	p(\set M_2 | \vec Y) &= \frac{p(\vec Y | \set M_2 ) p(\set M_2 )}{p(\vec y | \set M_1 ) p(\set M_1 ) + p(\vec Y | \set M_2 ) p(\set M_2 )} \, , 
\end{align}
\end{subequations}
where the two priors~$ p(\set M_1 )$  and~$ p(\set M_2 )$ encode our subjective preference for either one of the models or any prior information available. If we do not have any prior information, we can choose them with~\mbox{$ p(\set M_1 ) \propto 1$} and \mbox{$ p(\set M_2 ) \propto 1$.} Furthermore, for model comparison rather than the absolute values of the posteriors their ratio, referred to as the Bayes factor,  
\begin{equation}
	\frac{p(\set M_1 | \vec Y)}{p(\set M_2 | \vec Y) } =  \frac{p(\vec Y | \set M_1 ) p(\set M_1 )}{ p(\vec Y | \set M_2 ) p(\set M_2 )}
\end{equation}	
is more meaningful. A ratio larger than one indicates a preference for~\mbox{$\set M_1$} and a value smaller a preference for~\mbox{$\set M_2$}~\cite{kass1995bayes}.   

\subsection{Predictive observation density~\label{sec:predictY}}
%
We can use the posterior finite element density derived in Section~\ref{sec:posteriorFE}  and the likelihood according to our statistical model to compute the predictive observation density at locations where there are no observations. The unknown predictive distribution at the~$n_{\tilde y}$ non-observed locations of interest are collected in a vector~$\tilde{\vec y} \in \mathbb R^{n_{\tilde y}}$.  First, we consider the conditional joint distribution
\begin{equation}
	p ( \tilde{ \vec y},  \, \vec u | \vec Y ) = p ( \tilde{\vec y}  | \vec u , \,  \vec Y ) p (  \, \vec u | \vec Y ) =  p ( \tilde {\vec y}  | \vec u  ) p (  \, \vec u | \vec Y ) \, ,
\end{equation}
where we used  the statistical independence of the random vector~$\tilde {\vec y}$ and the observation matrix~$\vec Y$. Then, marginalising out the finite element solution we obtain for the predictive observation density
\begin{equation} \label{eq:predDensity}
	p ( \tilde{\vec y} | \vec Y )  = \int  p ( \tilde{\vec y}  | \vec u  ) p (  \, \vec u | \vec Y ) \D \vec u \, .
\end{equation}
The likelihood is given in~\eqref{eq:likelihood} and the posterior finite element density in~\eqref{eq:postUcYMult}. Denoting the matrices corresponding to the~$n_{\tilde y}$ non-observed locations with~$\tilde {\vec P} \in \mathbb R^{n_{\tilde y} \times n_u } $, $\tilde {\vec C}_d  \in \mathbb R^{n_{\tilde y} \times n_{\tilde y} }$ and $\tilde {\vec C}_e  \in \mathbb R^{n_{\tilde y} \times n_{\tilde y} }$,  the likelihood reads
\begin{equation}\label{eq:likelihoodYtilde}
	p(\tilde{\vec y} |   \vec u) = \set N \left ( \rho \tilde{\vec P}  \vec u , \, \tilde{\vec C}_d + \tilde {\vec C}_e \right )  \, .
\end{equation}
In~$\eqref{eq:predDensity}$, both terms in the integrand are Gaussians so that the integral can be analytically evaluated, c.f.~\ref{sec:appPuCy} and the references therein, yielding
\begin{equation}
	p(\tilde{\vec y} |   \vec Y) = \set N \left (   \rho \tilde{\vec P} \overline {\vec u}_{|Y}  , \, \tilde{\vec C}_d + \tilde {\vec C}_e  + \rho^2  \tilde{\vec P} \vec C_{u|Y}  \tilde{\vec P}^\trans \right )  \, .
\end{equation}
The mean is the with~$\rho$ scaled mean of the posterior finite element density and the covariance is the sum of all contributions to the overall uncertainty in prediction.

%
\section{Examples \label{sec:examples}}
%
In this section, we apply statFEM to one- and two-dimensional Poisson problems. All examples are discretised with a standard finite element approach using linear Lagrange basis functions. After establishing the convergence of the probabilistic forward problem we study the convergence of the posterior densities with an increasing number of observation points~$n_y$  and readings~$n_o$. In addition, we illustrate how statFEM reduces to conventional Bayesian inversion when the statistical model~\eqref{eq:decomposition} is simplified to~$\vec y = \vec u+ \vec e$  and the mapping of the model parameters to the finite element solution becomes deterministic. 
Usually, the finite element covariance matrix~$\vec C_u$ given by~\eqref{eq:approxUcU} becomes ill-conditioned when the covariance lengthscale for the source~$\ell_f$ or for the diffusion coefficient~$\ell_\kappa$ is larger than the characteristic element size~$h$. Therefore, in all the examples, we use instead of~$\vec C_u$ the stabilised covariance matrix \mbox{$\vec C_u + 0.001 (\sigma_f^2 + \sigma_\kappa^2) \vec I$.}

%
\subsection{One-dimensional problem \label{sec:oneD}}
%
We seek the solution of the one-dimensional Poisson-Dirichlet problem  
\begin{subequations} \label{eq:oneDex}
\begin{alignat}{2}
	- \frac{\D }{\D x}  \left ( \mu (x)    \frac{\D u }{\D x}  \right ) &=  f(x)  \qquad &&\text{in }  \Omega = (0, \,1)    \\ 
	u ( x ) &= 0 &&\text{on } x=0 \text{ and }  x=1\,  \, , 
\end{alignat}
\end{subequations}
where either the diffusion coefficient~$\mu(x)$ or the source~$f(x)$ is random. As mentioned, to ensure that the diffusion coefficient~$\mu(x)$ is positive we consider the auxiliary variable~$\kappa(x) = \ln \left (  \mu(x) \right )$. That is, in MCMC sampling~\mbox{$\kappa(x) \in \mathbb R$} is the unknown variable and the positive diffusion coefficient is~$\mu(x) = \exp(\kappa(x))$, see \ref{sec:appMCMC}. 

%
\subsubsection{Convergence of the forward finite element density for random source}
%
To establish the convergence of the discretised probabilistic forward problem, we  consider the  Poisson-Dirichlet problem~\eqref{eq:oneDex} with a deterministic diffusion coefficient~\mbox{$\mu(x) = 1$} and a random source with a mean~\mbox{$\overline f(x) =1$,} covariance scaling parameter~$\sigma_f=0.2$ and a lengthscale parameter~$\ell_f \in \{ 0.25, \, 0.5, \, 1.0\}$. The exact solution~$u(x)$ is a Gaussian process~\eqref{eq:uExact} with a mean~$\overline u (x)$ and covariance~$c_u(x, \, x')$. The required Greens function~$g(x, \, x')$ can be easily analytically obtained. 

\begin{figure}[] 
\centering
	\subfloat[][Relative $L_2$ norm error\label{fig:1dconv}] {
		\includegraphics[width=0.435\textwidth]{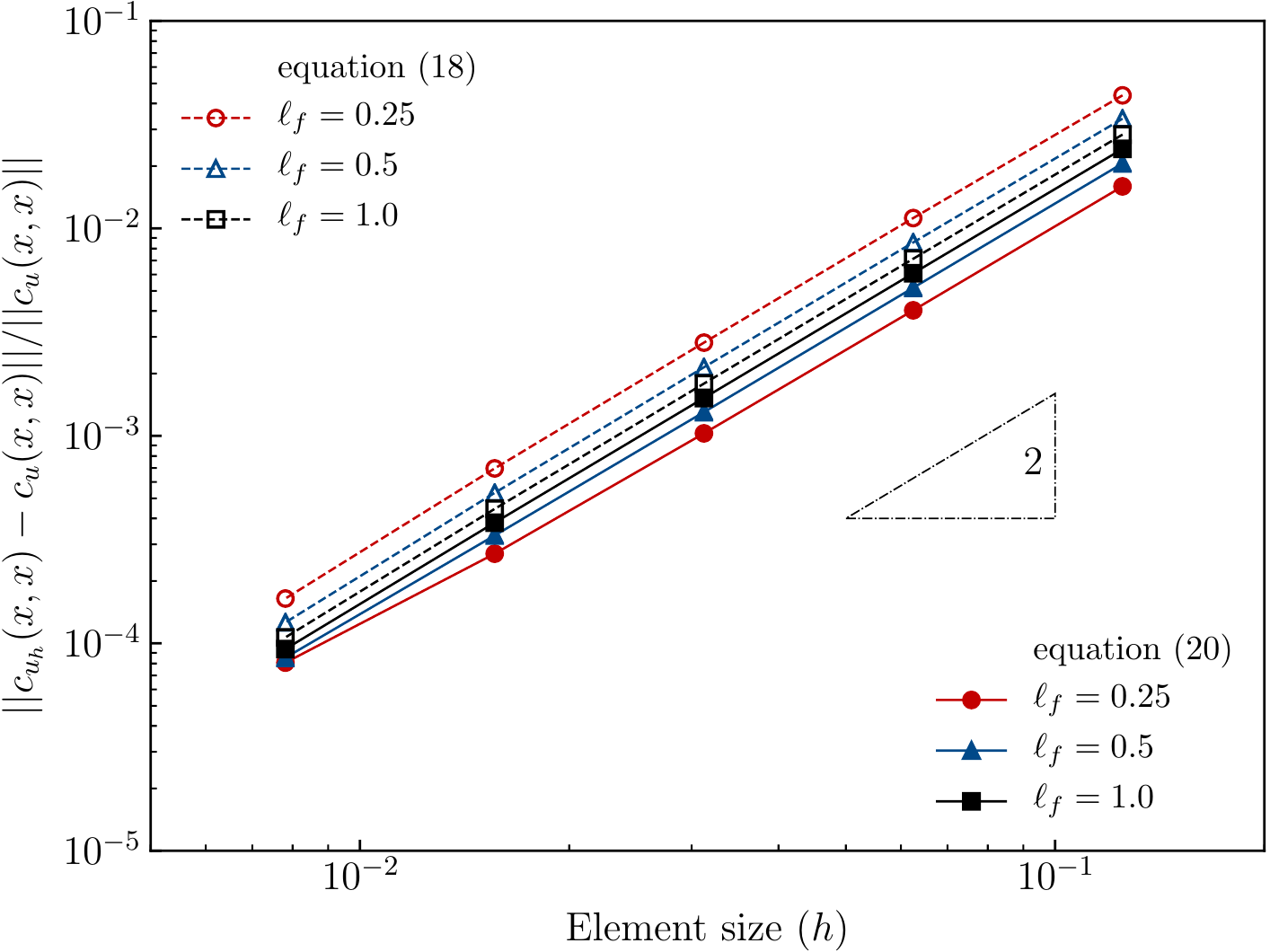}
	}
	\hspace{0.05\textwidth}
	\subfloat[][$c_{u} (x, \, x) $  and $c_{u_h} (x, \, x)$ \label{fig:cuH}] {
		\includegraphics[width=0.425\textwidth]{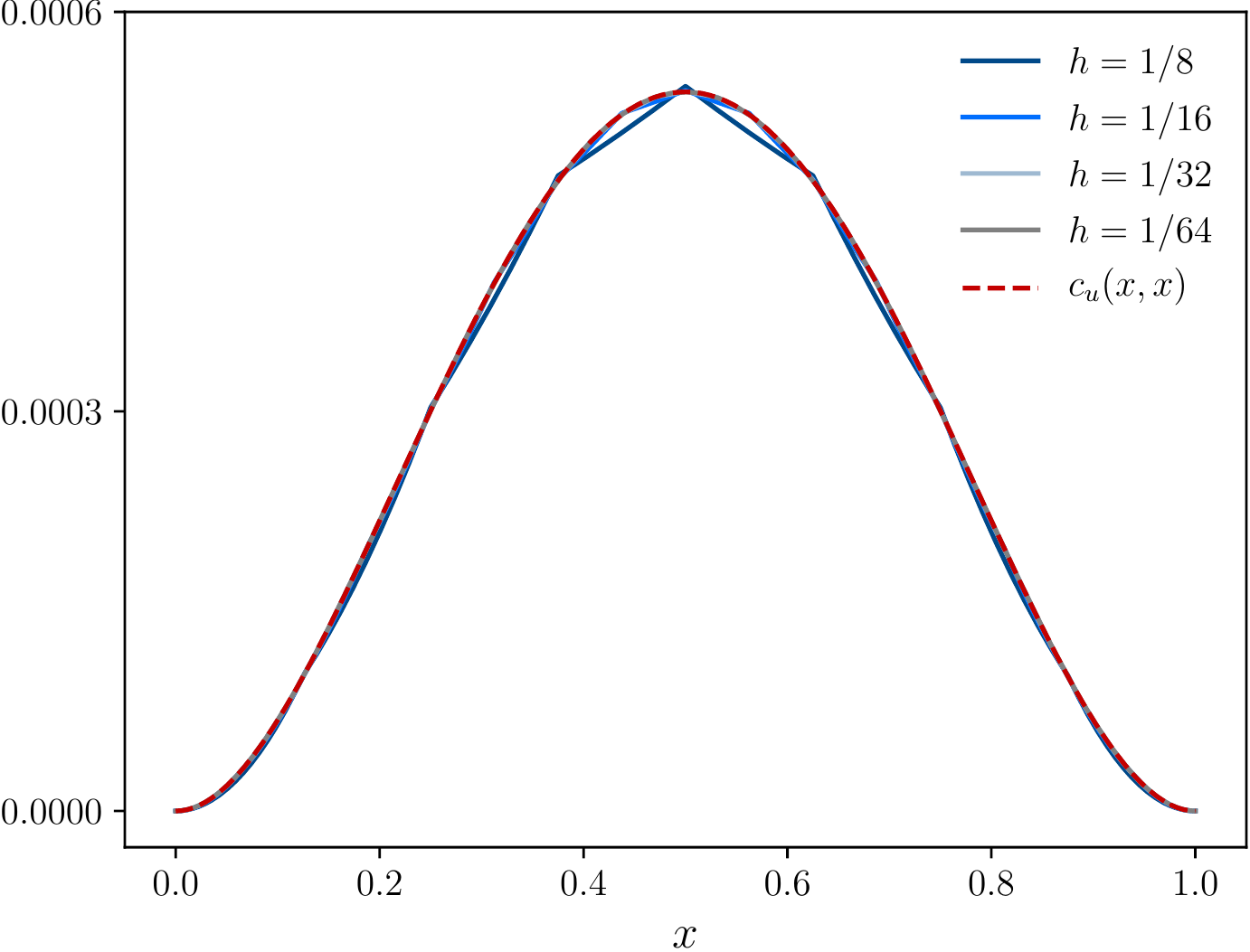}
	} 
\caption{One-dimensional problem. Convergence of the finite element variance~$c_{u_h} (x, \, x)$ of the forward problem. Source covariance matrix~$\vec C_f$ is either according to~\eqref{eq:gprCompsM} or~its approximation~\eqref{eq:CRapprox}.}
\end{figure}

According to~\eqref{eq:feDensity2} the density of the finite element solution  is a multivariate Gaussian~\mbox{$ p(\vec u) = \set N (\overline{\vec u}, \, \vec C_u )$}  with a mean~\mbox{$\overline{\vec u} = \vec A^{-1} \overline{\vec f}$} and covariance \mbox{$\vec C_u = \vec A^{-1} \vec C_f \vec A^{-\trans}$.} The mean is identical to the solution of a Poisson-Dirichlet problem with a deterministic source and as such its convergence characteristics are well studied, see e.g.~\cite{Ern:2003aa}. Therefore, we focus here on the convergence of the covariance. The  covariance of the finite element approximation~\mbox{$u_h( x) = \sum_i \phi_i(x) u_i = \vec \phi (x)^\trans \vec u$} is given by
\begin{equation}
\begin{split}
	c_{u_h} (x, \, x') =  \cov (u_h (x) , \, u_h(x') ) & =  \mathbb E \left  [ \left ( \sum_i \phi_i(x) \left ( u_i - \overline{u}_i \right )  \right ) \left ( \sum_j \phi_j(x')  (  u_j - \overline{u}_j  ) \right ) \right ]   \\
	& =   \sum_i \sum_j  \phi_i (x) \mathbb E \left  [  ( u_i - \overline u_i   )( \,   u_j - \overline{u}_j )  \right ] \phi_j(x')  = \vec \phi(x)^\trans  \vec C_u \vec \phi(x')  \, .
\end{split} 
\end{equation}
In Figure~\ref{fig:1dconv} the $L_2$ norm  of the approximation error of the variance~\mbox{$ \| c_{u_h} (x, \, x)  - c_{u} (x, \, x)  \|$} as a function of the element size~$h$ is shown. The covariance~$\vec C_f$ of the source is computed either according to~\eqref{eq:gprCompsM} or its approximation~\eqref{eq:CRapprox} and using~$10$ quadrature points per element. In both cases, the error converges with a rate of around two irrespective of the lengthscale parameter~$\ell_f$. However, the approximation~\eqref{eq:CRapprox} leads to a smaller error than the exact expression~\eqref{eq:gprCompsM}, especially for smaller~$\ell_f$. Figure~\ref{fig:cuH} depicts the variance~$c_{u} (x, \, x) $ and its finite element approximation~$c_{u_h} (x, \, x) $ for~\mbox{$\ell_f = 0.5$} on successively finer meshes. As expected, the finite element approximation converges to the exact solution when the mesh is refined. 
%

%
\subsubsection{Posterior finite element density and system response for random source \label{sec:1dUcondY}}
\begin{figure}[]
	\centering
	\includegraphics[width=0.425\textwidth]{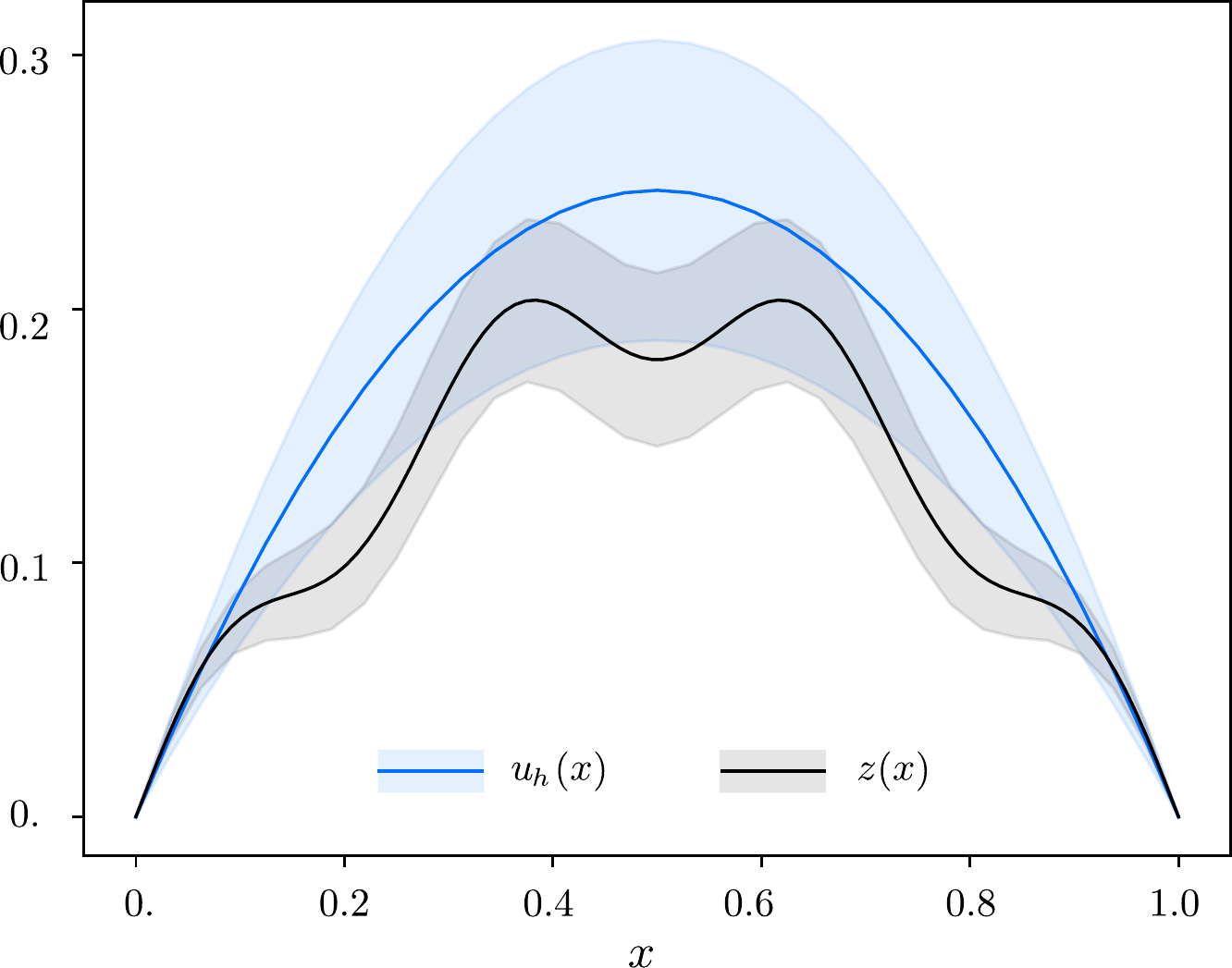}
	\caption{One-dimensional problem. Finite element solution $u_h(x)$ and the true system response $z(x)$.  The solid lines represent the respective means and the shaded areas the~$95\%$ confidence regions.  \label{fig:trueVSsoln}}
\end{figure}

%
We assume that the true system response is given by the Gaussian process
\begin{equation} \label{eq:zGP1dEx}
	z(x) \sim \set{GP} \left (\overline z(x), \,  g( x, \, x'') * c_z(x'', \, x''') *  g( x''', \,  x') \right )
\end{equation}
with the mean
\begin{equation}
	\overline z(x) = \frac{1}{5} \sin (\pi x) + \frac{1}{50} \sin (7 \pi x)
\end{equation}
and the squared exponential covariance kernel
\begin{equation}
	c_z (x, \, x') = 0.0225 \exp \left ( - 2 (x -x' )^2  \right ) \, .
\end{equation}
Of course, this true system response~$z(x)$  is in practice not known. Irrespective of~$z(x)$ we choose a finite element model with a deterministic diffusion coefficient \mbox{$\mu (x) =1$} and  a random source with a mean \mbox{$\overline{f}(x)= \pi^2/5$} and the squared-exponential covariance parameters~\mbox{$\sigma_f = 0.3$} and~\mbox{$\ell_f= 0.25$}.  The uniform finite element discretisation consists of~$32$ elements. The system response~$z(x)$ and the finite element solution~$u_h(x)$, i.e. their mean and $95\%$ confidence regions, are depicted in Figure~\ref{fig:trueVSsoln}.  The finite element solution is able to roughly capture the overall true system response but not its details.  

\begin{figure}[]  
	\centering
	\includegraphics[width=0.55\textwidth]{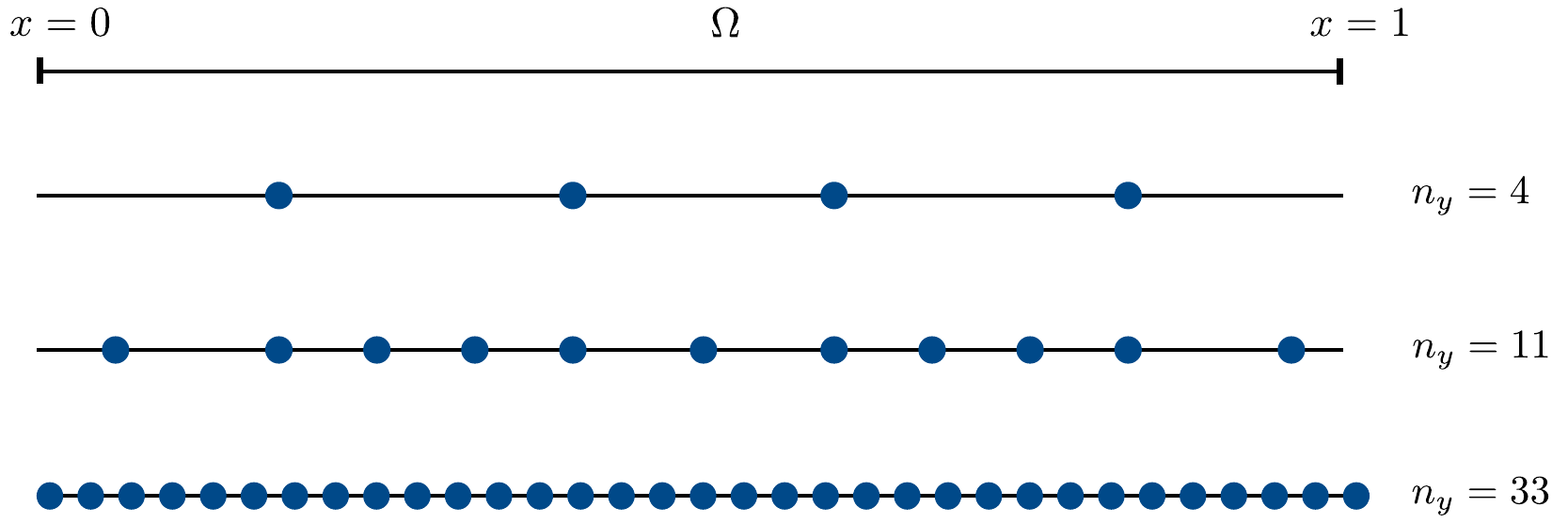}
	\caption{One-dimensional problem. Location of the~$n_y \in  \{ 4, \, 11,  \, 33 \}$ observations points for the data vector~$\vec y$. \label{fig:sensorPos}}
\end{figure}

In practice, the true response is not known and we can only observe~\mbox{$\vec y = \vec z + \vec e $} at the~$n_y$ observation points. We consider  the \mbox{$n_y \in \{ 4, \, 11, \, 33\} $ } observation points  shown in Figure~\ref{fig:sensorPos}. We sample at each of the $n_y$ observation points~\mbox{$n_o \in \{1, \, 10, \, 100, \, 1000 \} $}  repeated readings.  Hence, we sample a synthetic observation matrix~\mbox{$\vec Y \in \mathbb R^{n_y \times n_o}$} from the Gaussian process
\begin{equation} \label{eq:yGP}
	y(x) \sim  \set{GP} \left (\overline z(x), \, g( x, \, x'') * c_z(x'', \, x''') *  g( x''', \,  x')  + 2.5 \cdot 10^{-5} \delta_{x x'} \right ) \,  ,
\end{equation}
where~$ \delta_{x x'}$ is the Kronecker delta and~$\sigma_e^2 = 2.5 \cdot 10^{-5}$ is the observation noise. See~\ref{app:GPregress} for sampling from a Gaussian process. 

\begin{figure}[h!] 
	\centering
	\subfloat[$p(\rho | \vec Y)$]{
		\includegraphics[width=0.33\textwidth]{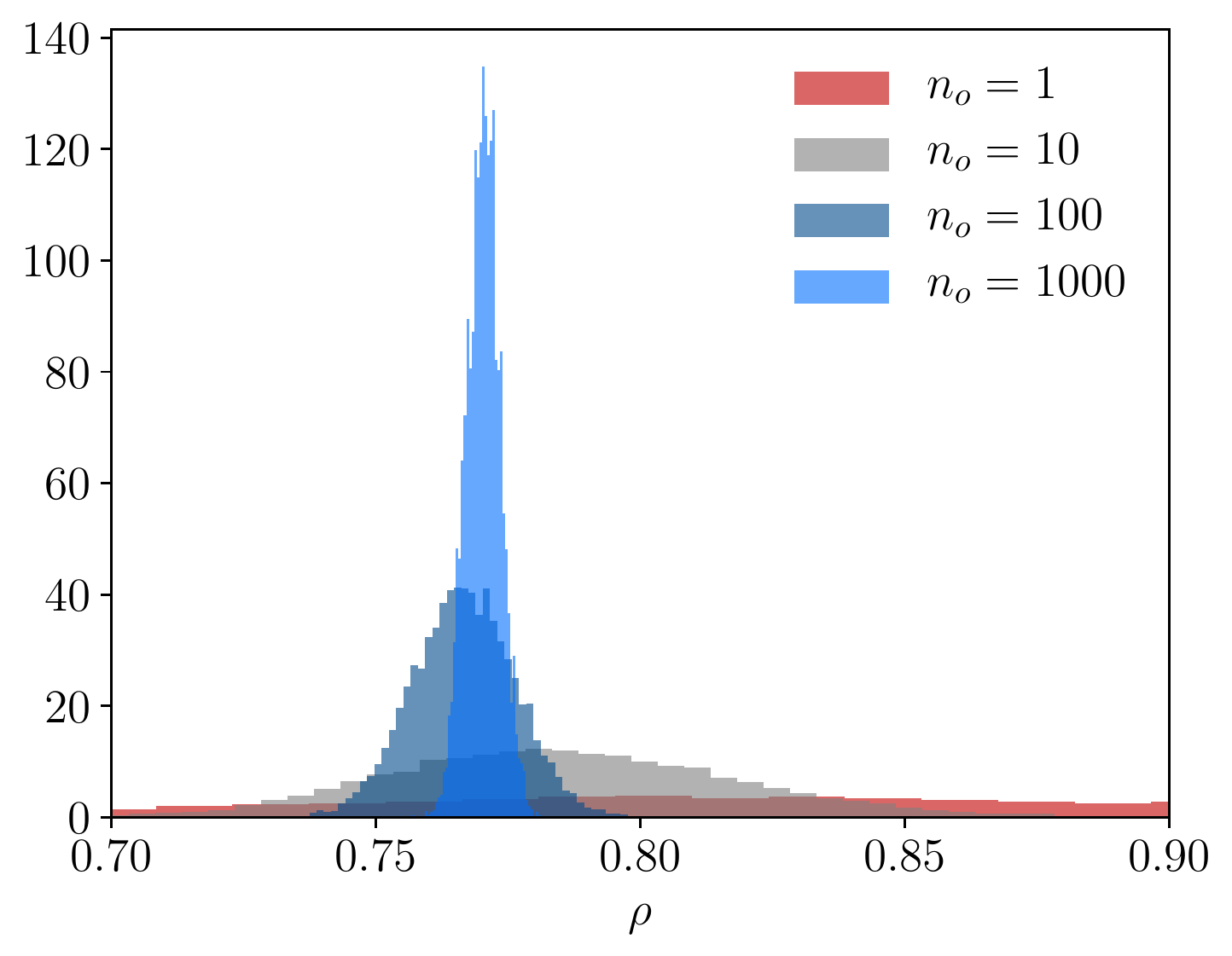}
	} 
	\subfloat[$p(\sigma_d | \vec Y)$]{
		\includegraphics[width=0.33\textwidth]{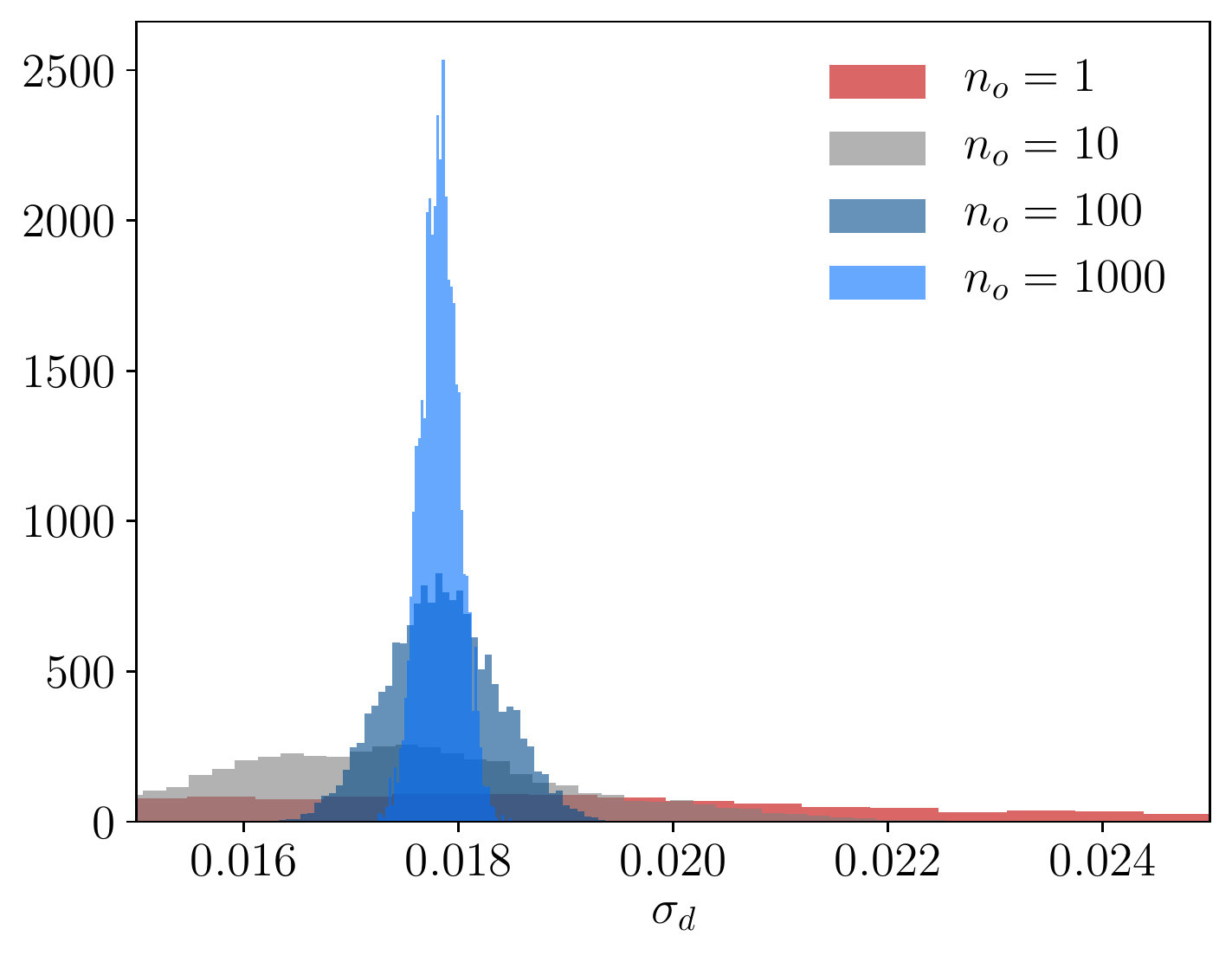}
	} 
	\subfloat[$p(\ell_d | \vec Y)$]{
		\includegraphics[width=0.33\textwidth]{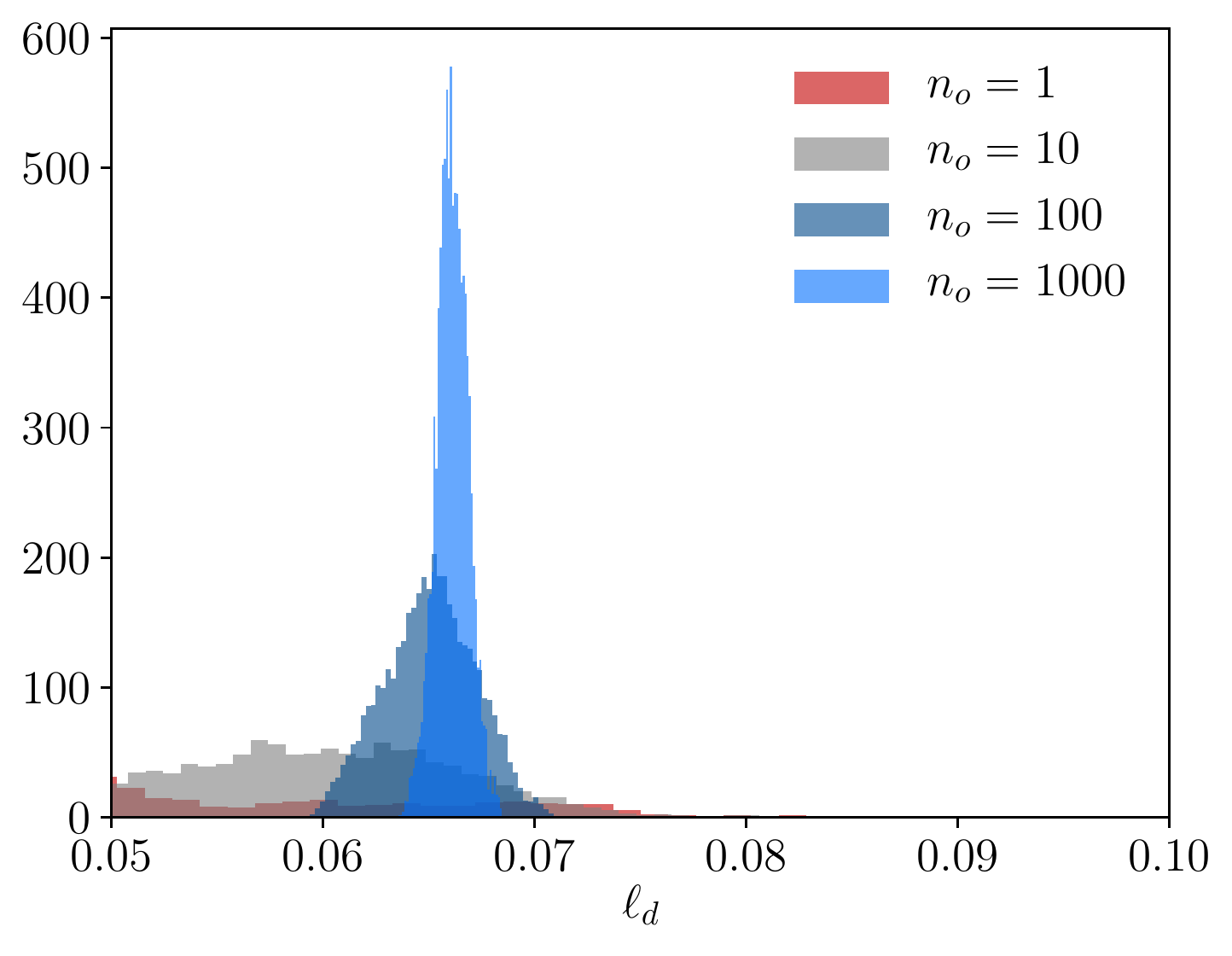}
	} 
	\caption{One-dimensional problem with random source. Posteriors of the parameters~$\rho$, $\sigma_d$ and $\ell_d$ for~$n_y=11$ and~$n_o \in \{1, \, 10,\, 100, \, 1000 \}$ sampled with MCMC. \label{fig:1dInferredHist}}
\end{figure}

Before computing the posterior finite element density~$p(\vec u | \vec Y)$, we first determine the (unknown) hyperparameters of the statistical generating model~\eqref{eq:decomposition}. In this example, amongst the parameters summarised in the graphical model in Figure~\ref{fig:graphStatFEM}, only the scaling parameter~$\rho$ and the mismatch covariance parameters~$\sigma_d$ and~$\ell_d$ are assumed to be unknown. We collect these hyperparameters in the vector~$\vec w= (\rho, \, \sigma_d, \, \ell_d)^\trans$. The posterior density of the hyperparameters is given by \mbox{$p(\vec w | \vec Y) \propto p (\vec Y | \vec w) p(\vec w)$}, see~\eqref{eq:bayesParams}.  To make the inference problem more challenging we assume a non-informative prior~$p(\vec w) \propto 1$ so that the posterior is proportional to the likelihood, i.e.~ \mbox{$p(\vec w | \vec Y) \propto p (\vec Y | \vec w)$}. We sample~$p(\vec w | \vec Y)$ using a standard MCMC algorithm while enforcing the positivity of the hyperparameters by sampling on the log scale, see~\ref{sec:appMCMC}. For each combination of $n_y$ and~$n_o$, we run $20000$ iterations with an average acceptance ratio of~$0.287$. In Figure~\ref{fig:1dInferredHist} the obtained normalised histograms for~$p(\rho | \vec Y)$, $p(\sigma_d | \vec Y )$ and~$ p( \ell_d | \vec Y) $  for $n_y=11$ are depicted. We can observe that the standard deviations become significantly smaller with increasing~$n_o$. In Tables~\ref{tab:1dRho}, \ref{tab:1dSigmaD} and \ref{tab:1dEllD} the empirical mean and standard deviations of these three plots and other~$n_y$ and~$n_o$ combinations are given. When either~$n_y$,~$n_o$ or both are increased, the standard deviations becomes smaller. Depending on the considered application it may be easier to increase either~$n_y$ or~$n_o$. 
\begin{table}[h!]
	\centering
	\rowcolors{2}{}{csmlLightBlue!30}
	\begin{tabular} { l c c c c }
	\toprule  
	& $n_{o} = 1$ & $n_{o} = 10$ & $n_{o} = 100$ & $n_{o} = 1000$ \\
	\midrule
	$n_y = 4$ & 
	$0.80118 \pm 0.16823$ 
	& $0.78730 \pm 0.03518$ 
	& $0.79015 \pm  0.01208$ 
	& $0.77836 \pm 0.00380$ 
	\\
	$n_y = 11$ & 
	$0.85740 \pm 0.11607$ 
	& $0.78549 \pm 0.03349$ 
	& $0.76662 \pm 0.00961$ 
	& $0.77024 \pm 0.00312$ 
	\\
	$n_y = 33$ & 
	$0.76473 \pm 0.11476$ 
	& $0.79153 \pm 0.03072$ 
	& $0.77200 \pm 0.00980$ 
	& $0.77135 \pm 0.00311$ 
	\\
	\bottomrule
	\end{tabular}
	\caption[]{One-dimensional problem with random source. Empirical mean and standard deviation of~$p(\rho | \vec Y)$. \label{tab:1dRho}}
	\vspace{1.2em}
	\centering
	\rowcolors{2}{}{csmlLightBlue!30}
	\begin{tabular} { l  c c c c }
		\toprule
	   & $n_{o} = 1$ & $n_{o} = 10$ & $n_{o} = 100$ & $n_{o} = 1000$ \\
		\midrule
	$n_y = 4$ & 
	$0.04610 \pm 0.02150$ 
	& $0.02292 \pm 0.00328$ 
	& $0.02310\pm 0.00128$ 
	& $0.02340 \pm 0.00046$ 
	\\
	$n_y = 11$ & 
	$0.01823 \pm 0.00471$ 
	& $0.01734 \pm 0.00164 $ 
	& $0.01784 \pm  0.00051$ 
	& $0.01782 \pm 0.00017$ 
	\\
	$n_y = 33$ & 
	$0.02976 \pm 0.01463$ 
	& $0.01969 \pm  0.00186$ 
	& $0.01860 \pm 0.00053$ 
	& $0.01874 \pm 0.00018$ 
	\\
	\bottomrule 
	\end{tabular}
	\caption[]{One-dimensional problem with random source. Empirical mean and standard deviation of $p(\sigma_d | \vec Y)$. \label{tab:1dSigmaD}}
		\vspace{1.2em}
	\centering
	\rowcolors{2}{}{csmlLightBlue!30}
	\begin{tabular} { l  c c c c }
	\toprule 
	&  $n_{o} = 1$ & $n_{o} = 10$ & $n_{o} = 100$ & $n_{o} = 1000$ \\
	\midrule	
	$n_y = 4$ & 
	$0.23291 \pm 0.06648$ 
	& $0.08071 \pm 0.03899$ 
	& $0.17373 \pm 0.01346$ 
	& $0.17703 \pm 0.00527$ 
	\\
	$n_y = 11$ & 
	$0.04447 \pm 0.01359$ 
	& $0.05925 \pm 0.00696$ 
	& $0.06505 \pm 0.00215$ 
	& $0.06610 \pm 0.00075$ 
	\\
	$n_y = 33$ & 
	$0.08967    \pm 0.01828$ 
	& $0.07577 \pm 0.00496$ 
	& $0.07309 \pm 0.00164$ 
	& $0.07436 \pm 0.00052$ 
	\\
	\bottomrule
	\end{tabular}
	\caption[]{One-dimensional problem with random source. Empirical mean and standard deviation of $p(\ell_d | \vec Y)$. \label{tab:1dEllD}}
\end{table}

\begin{figure}[]
\centering
	\includegraphics[width=0.24\textwidth]{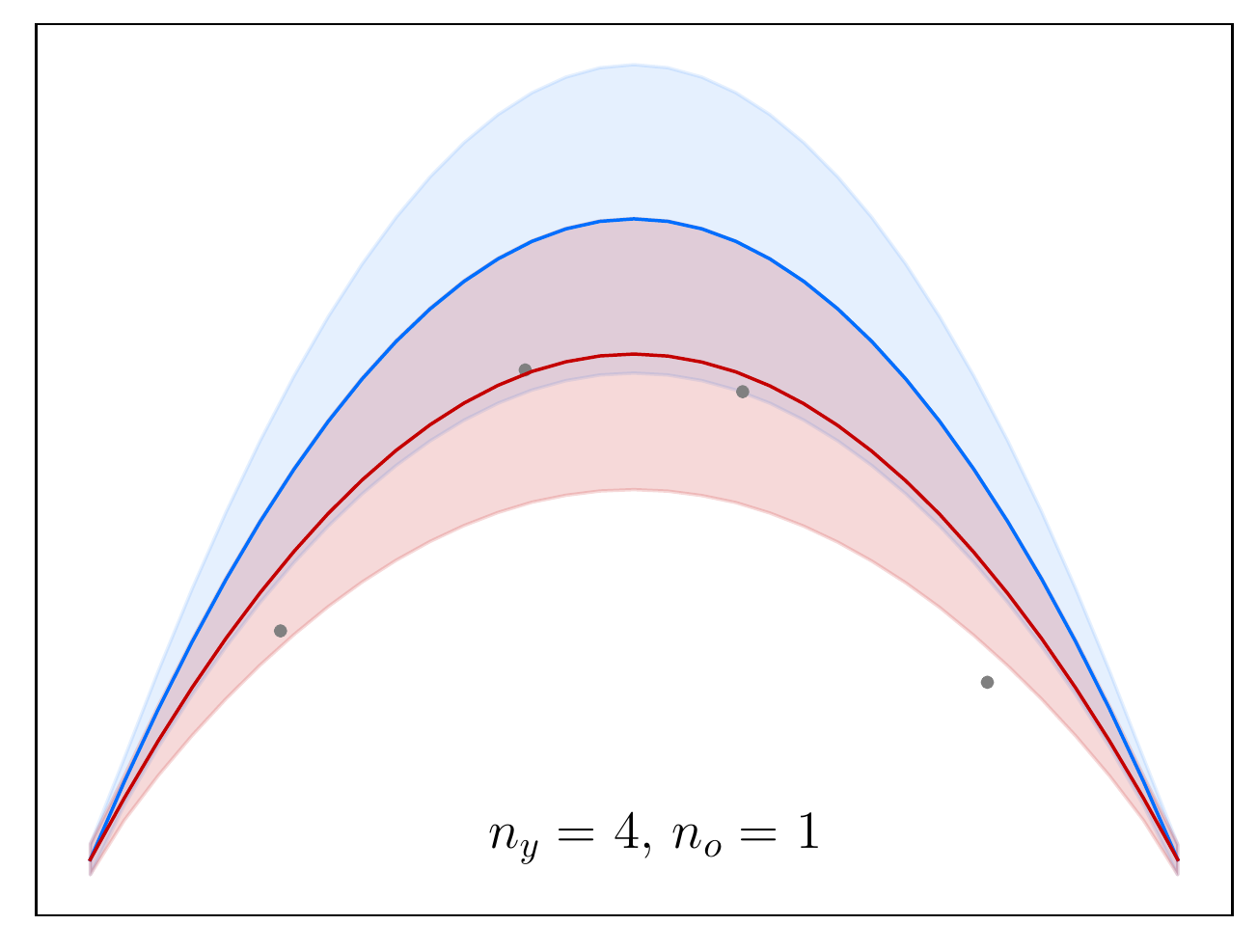}
	\includegraphics[width=0.24\textwidth]{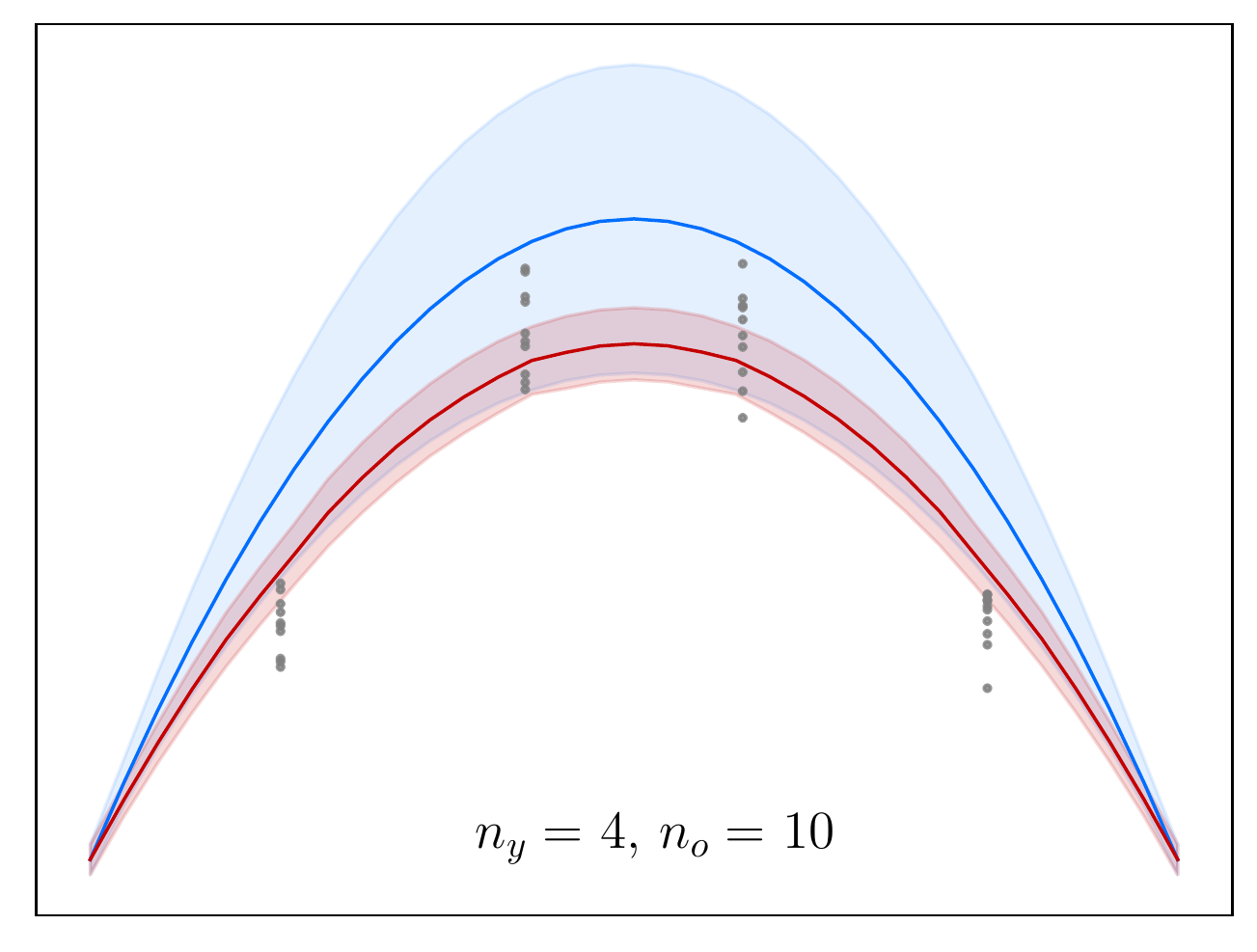}
	\includegraphics[width=0.24\textwidth]{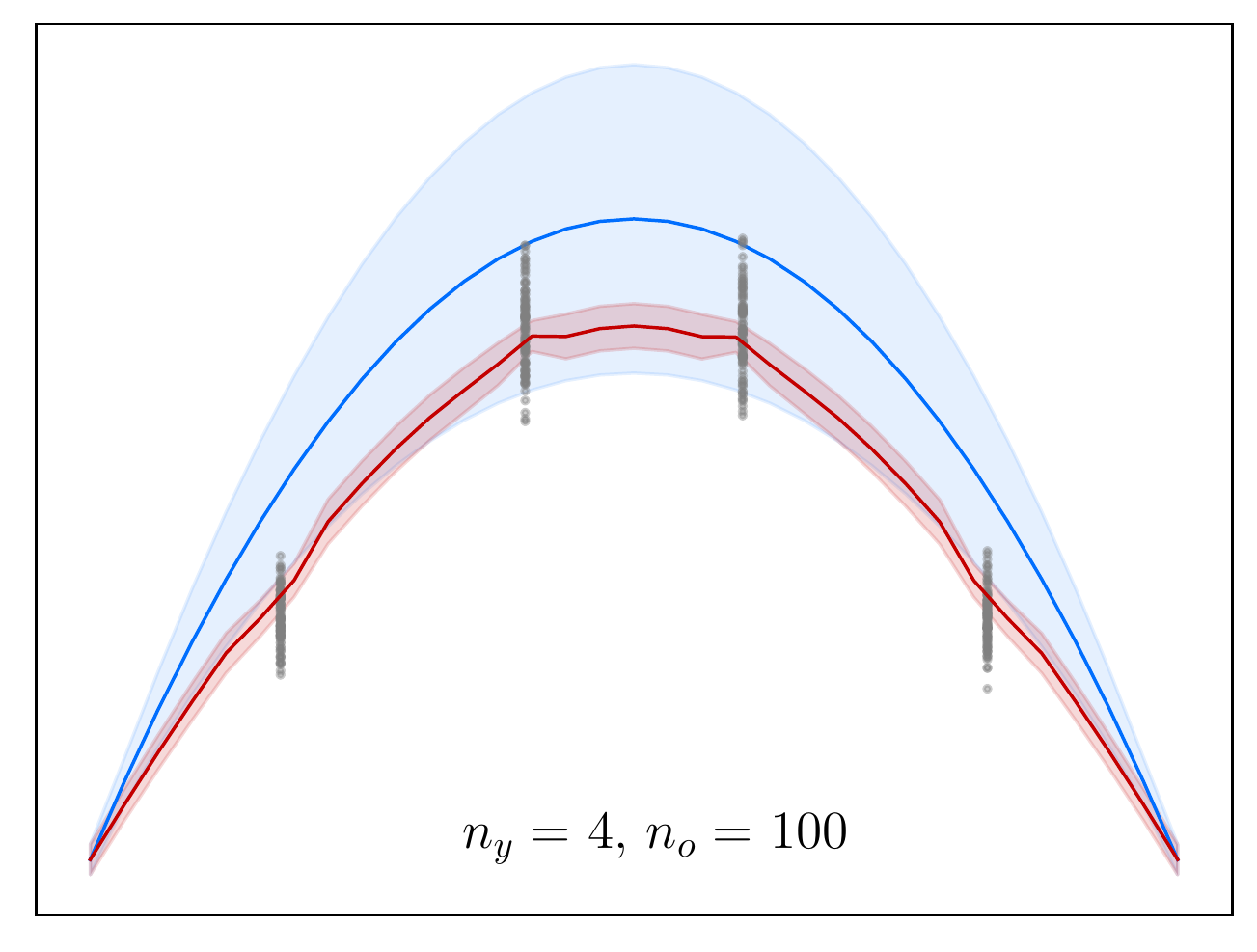}
	\includegraphics[width=0.24\textwidth]{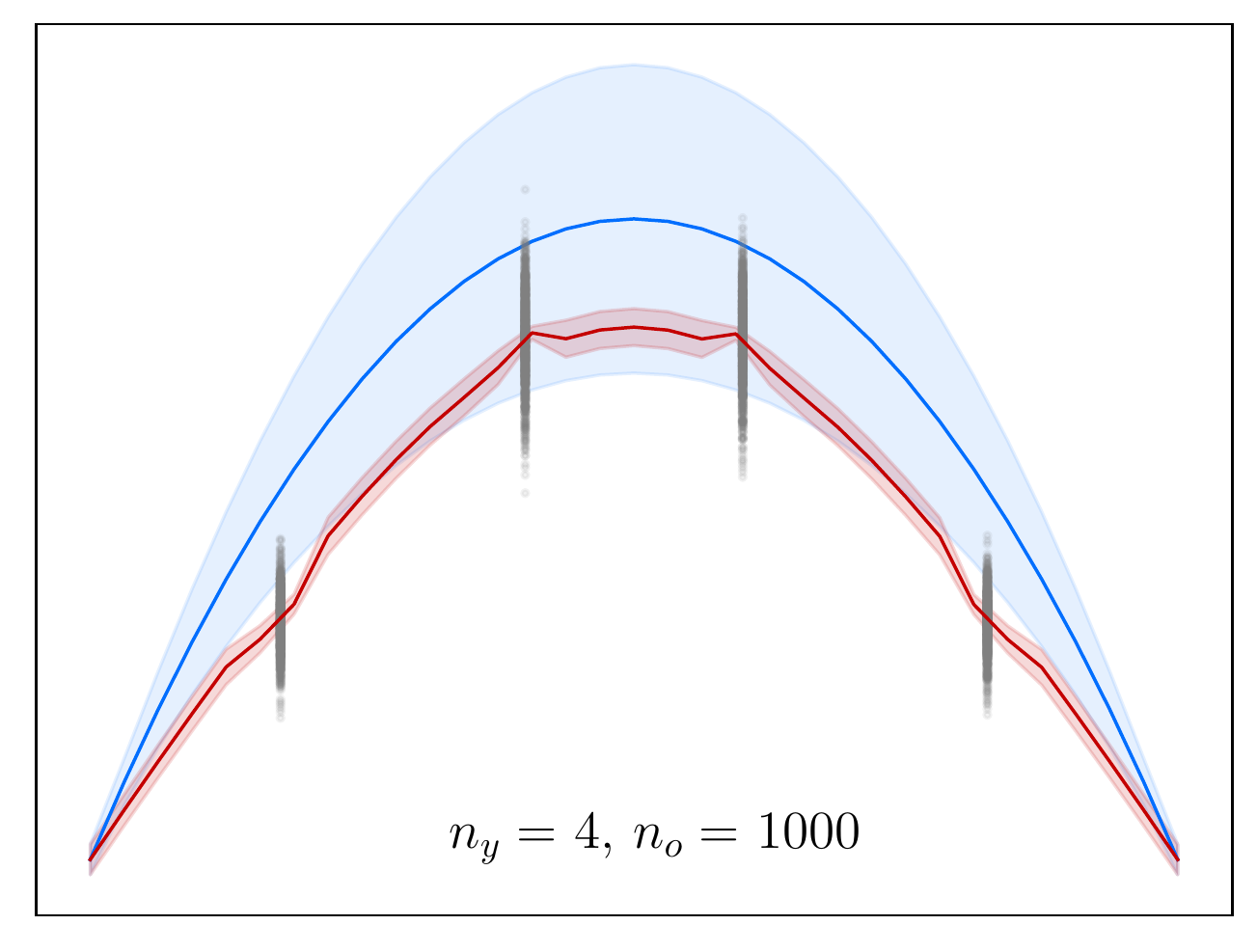} 
	\\
	\includegraphics[width=0.24\textwidth]{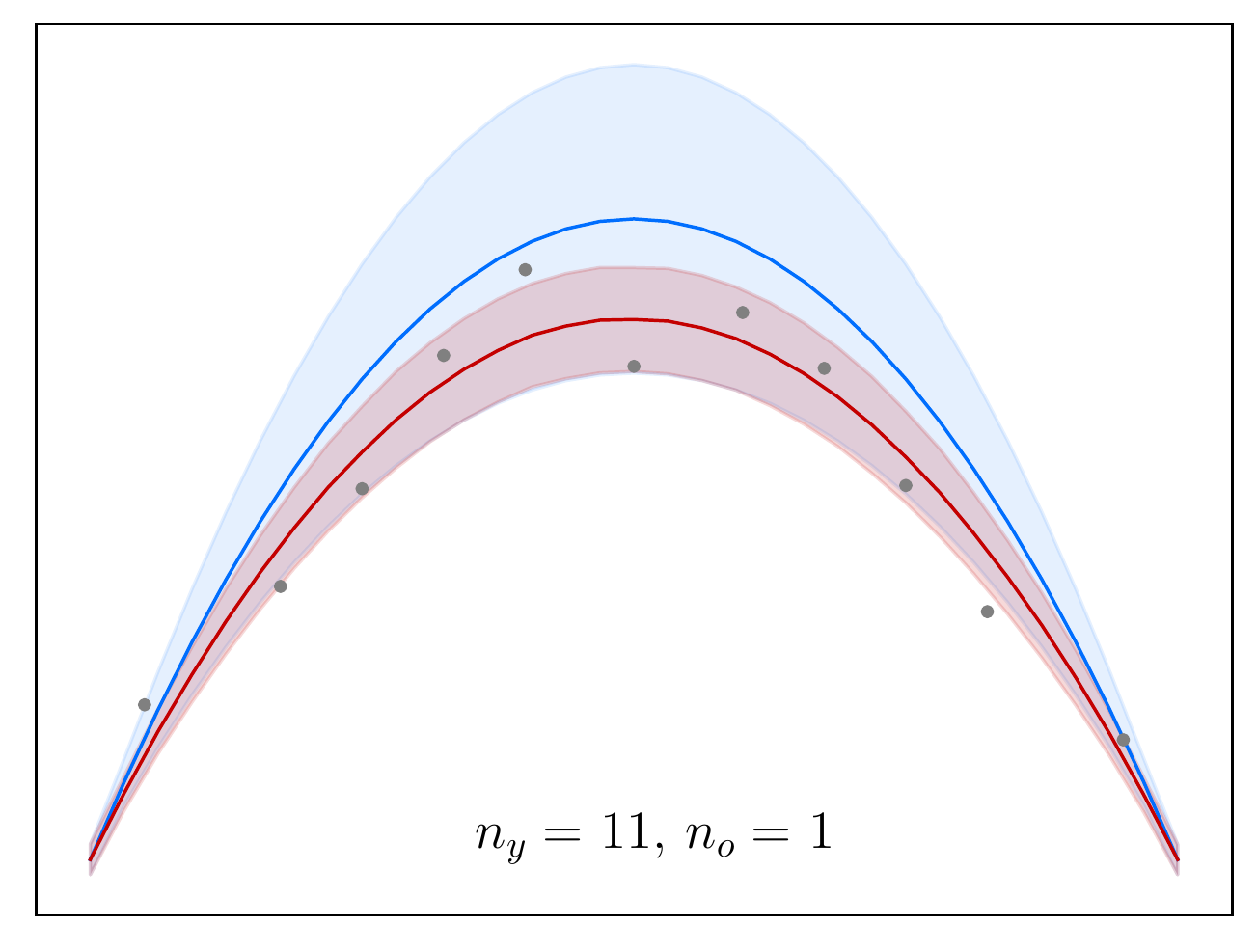}
	\includegraphics[width=0.24\textwidth]{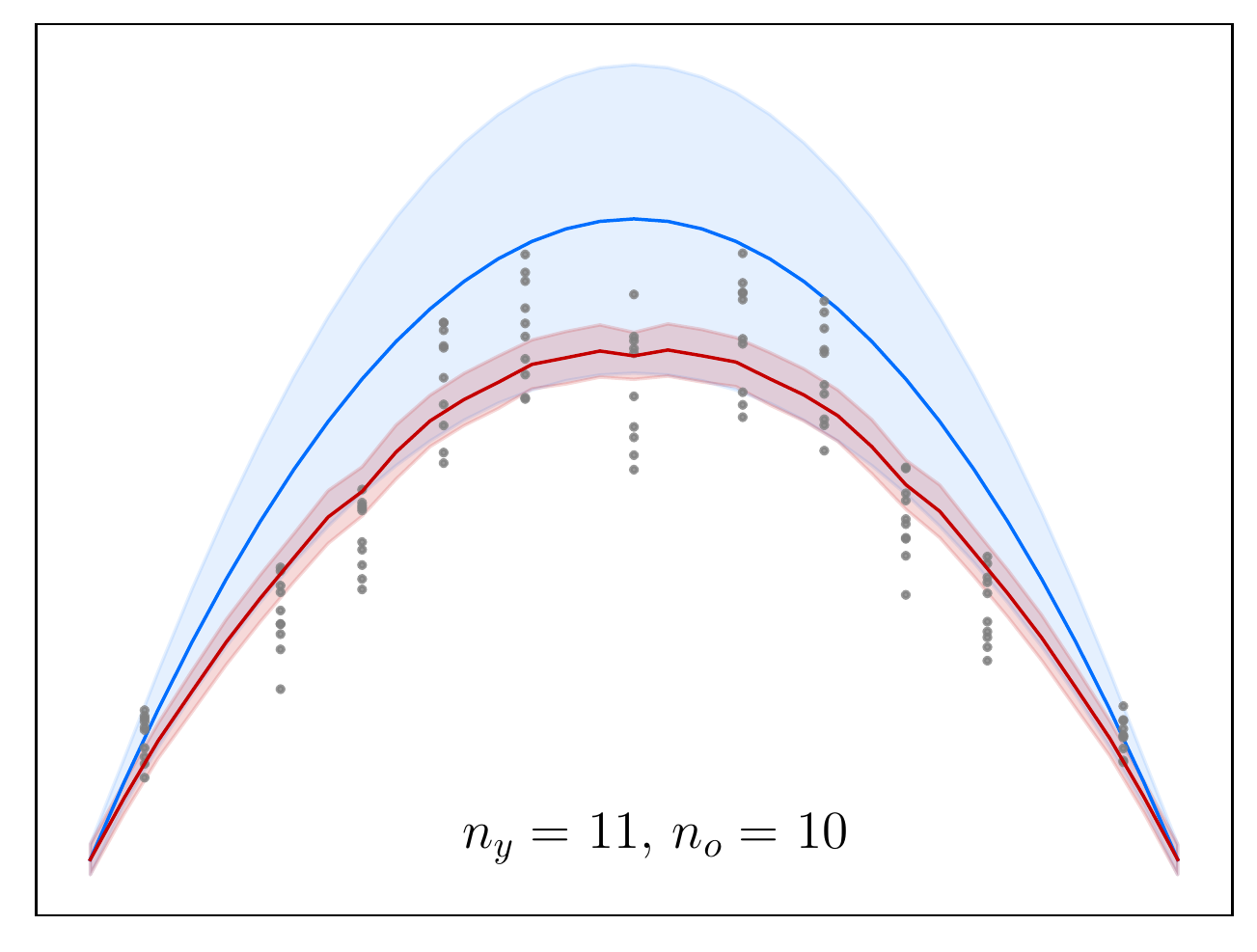}
	\includegraphics[width=0.24\textwidth]{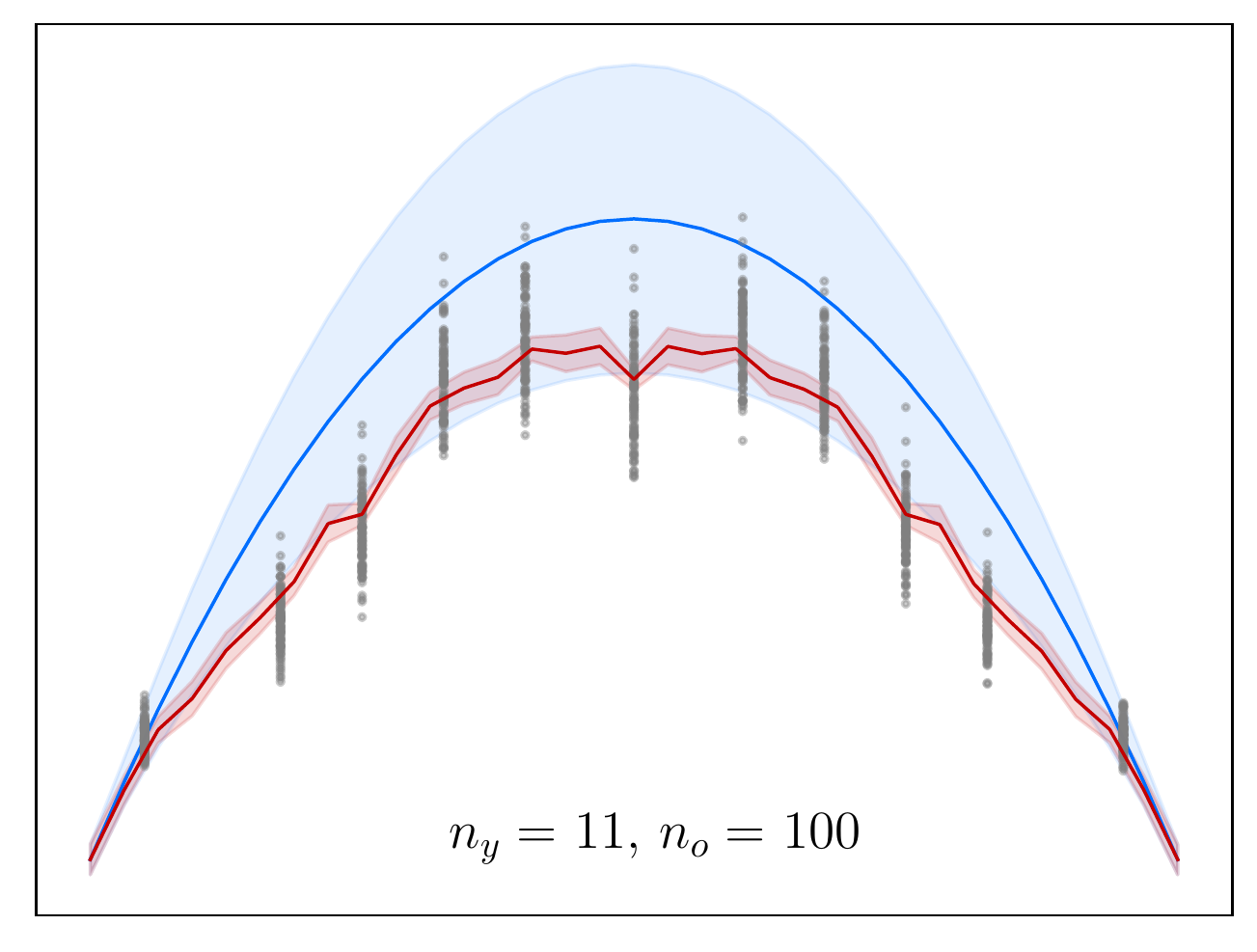}
	\includegraphics[width=0.24\textwidth]{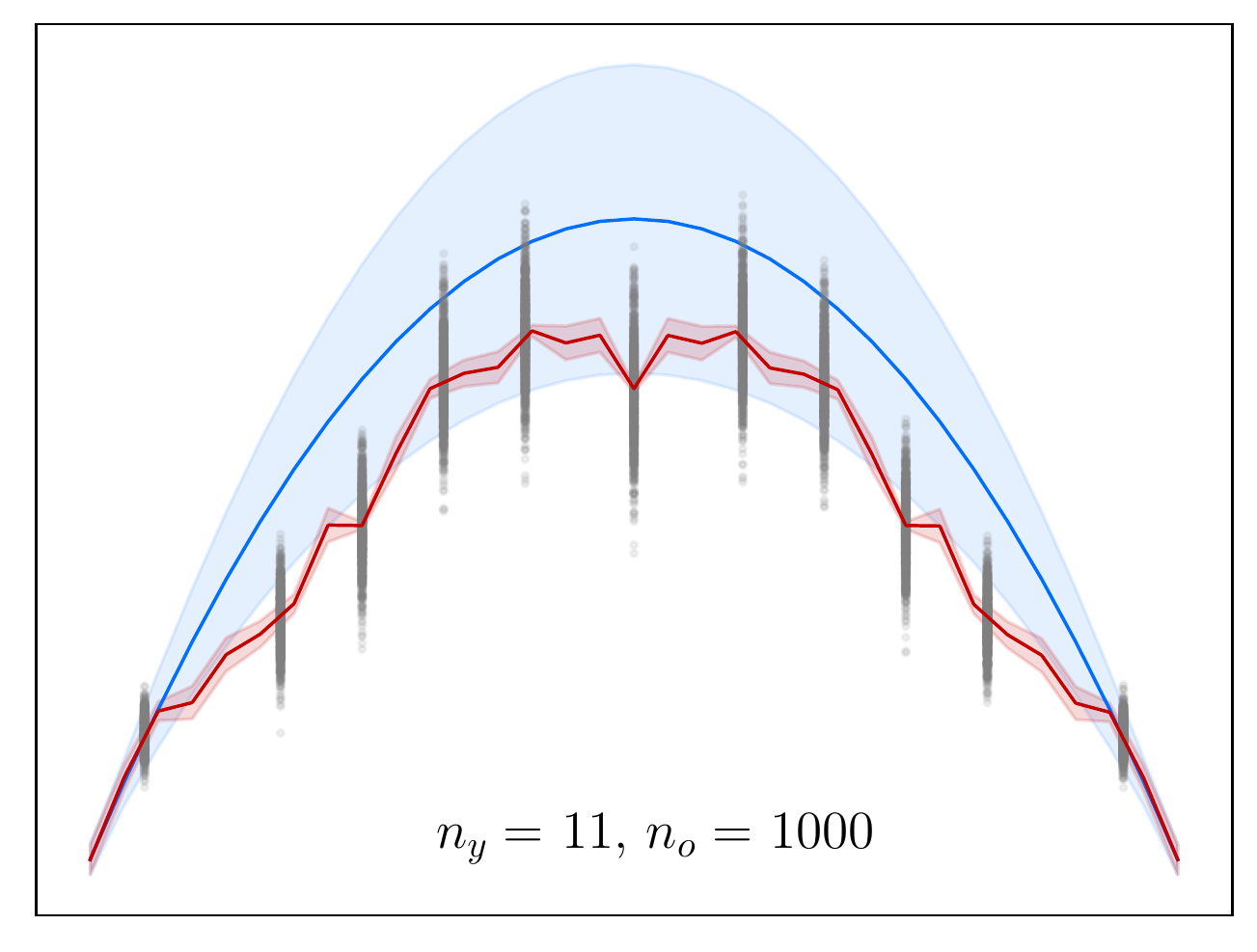} 
	\\
	\includegraphics[width=0.24\textwidth]{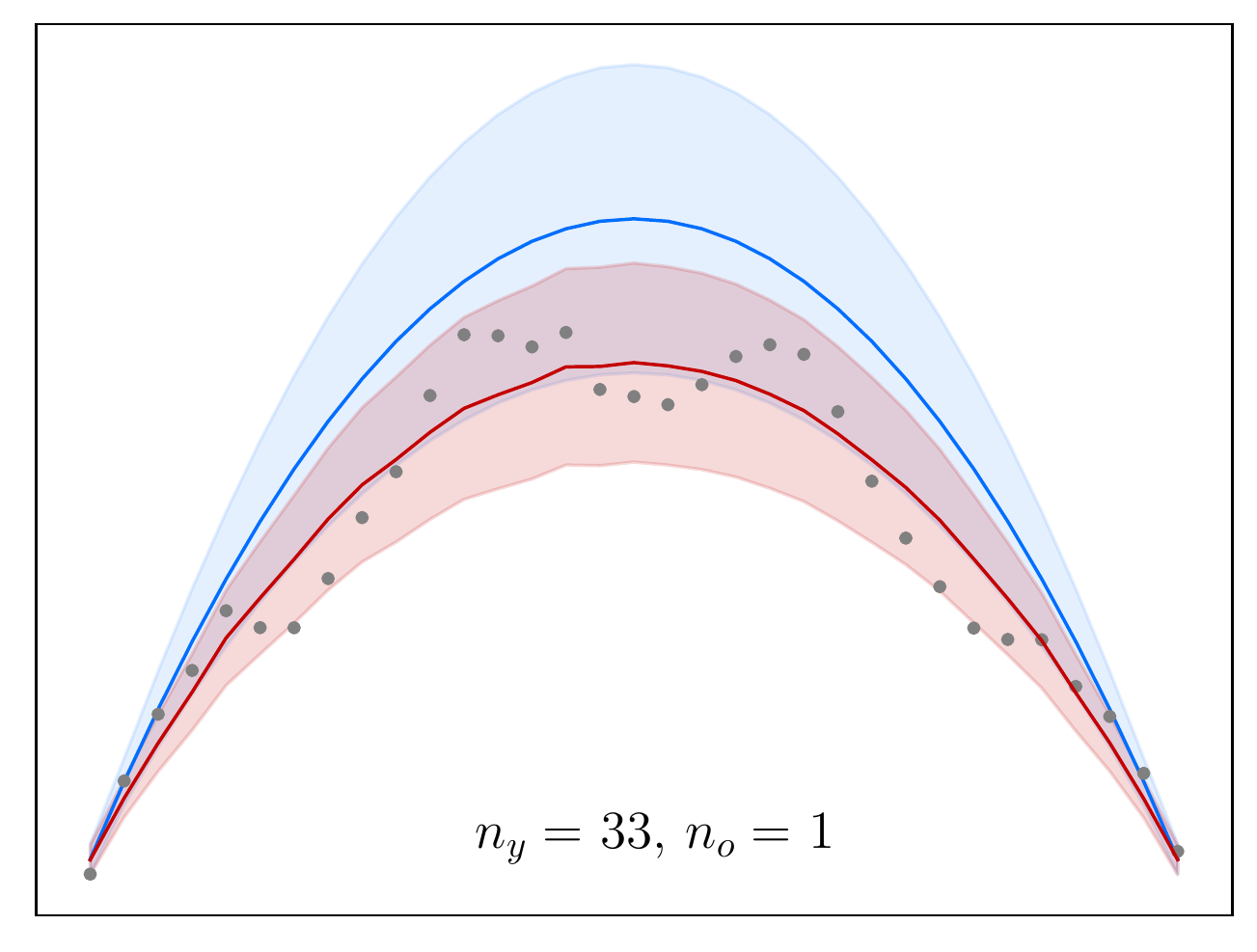}
	\includegraphics[width=0.24\textwidth]{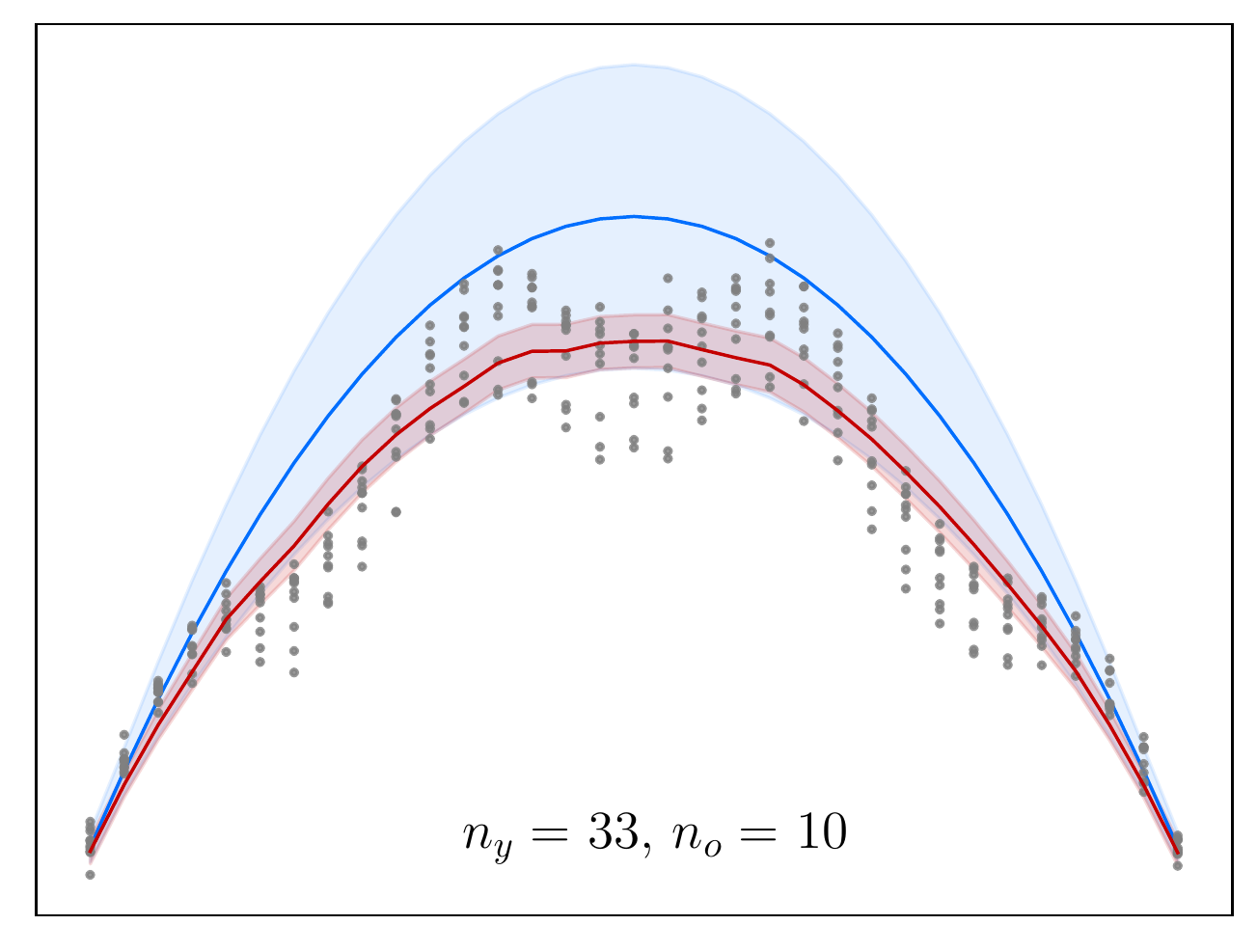}
	\includegraphics[width=0.24\textwidth]{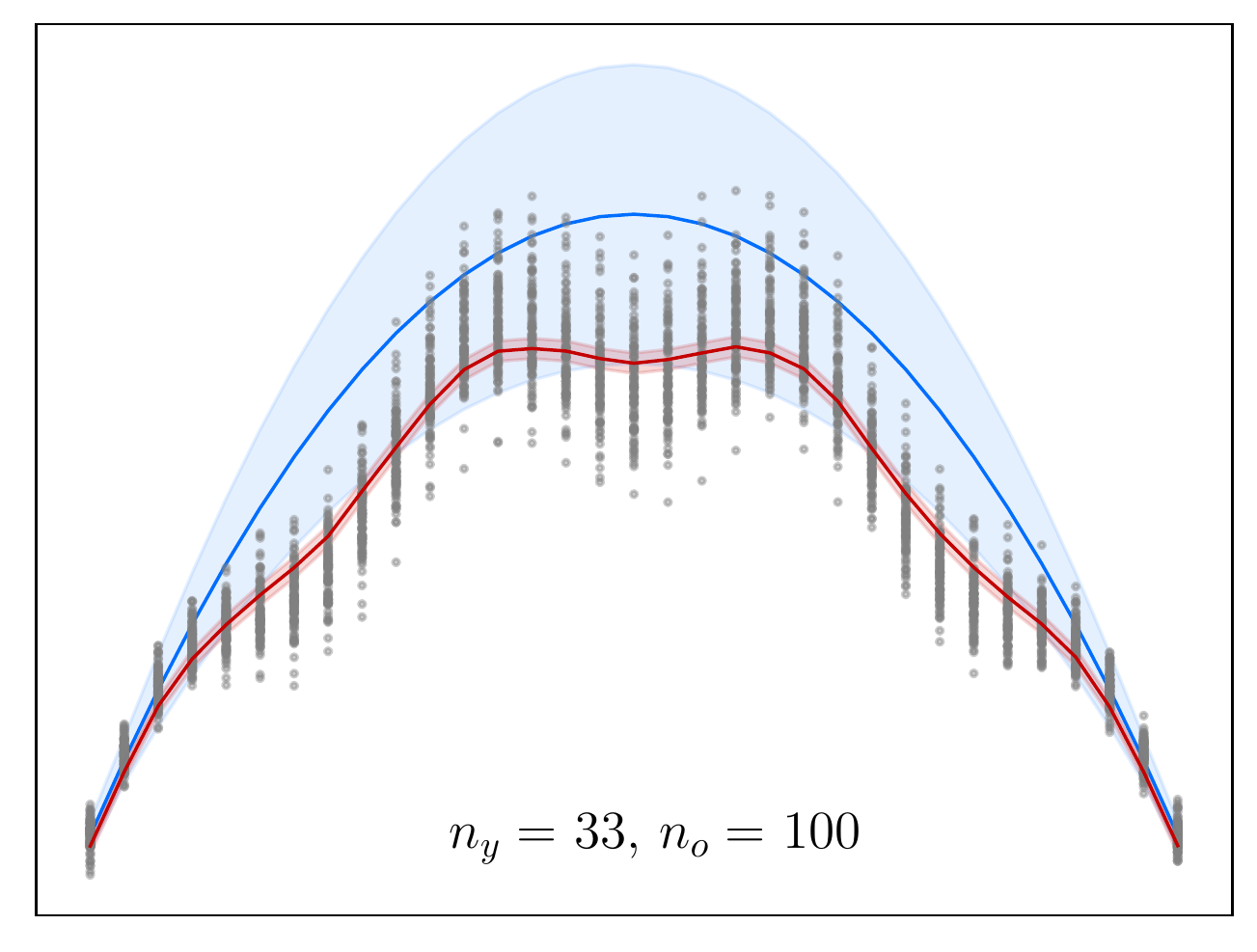}
	\includegraphics[width=0.24\textwidth]{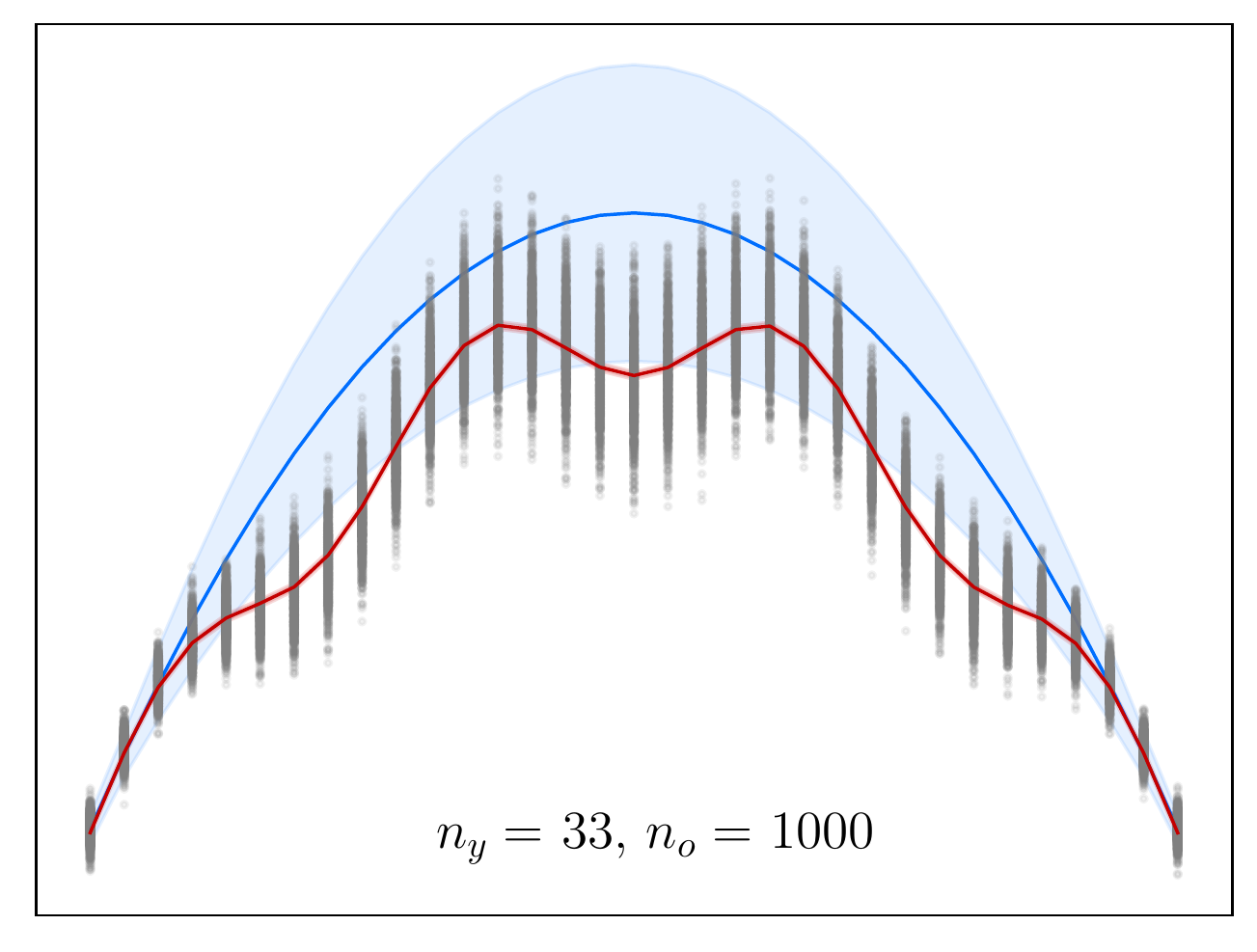} 
	\caption{One-dimensional problem with random source. Finite element density~$p(\vec u | \vec Y)$ conditioned on observation data (grey dots). The blue lines represent the mean~$\overline{\vec u}$ and the red lines the conditioned mean~$\overline {\vec u}_{|Y}$. The shaded areas denote the corresponding $95\%$ confidence regions. In each row the number of sensors~$n_y$ and each column the number of readings~$n_o$ is constant. \label{fig:1dPuhY}} 
\end{figure}

\begin{figure}
\vspace{2em}
\centering
	\includegraphics[width=0.24\textwidth]{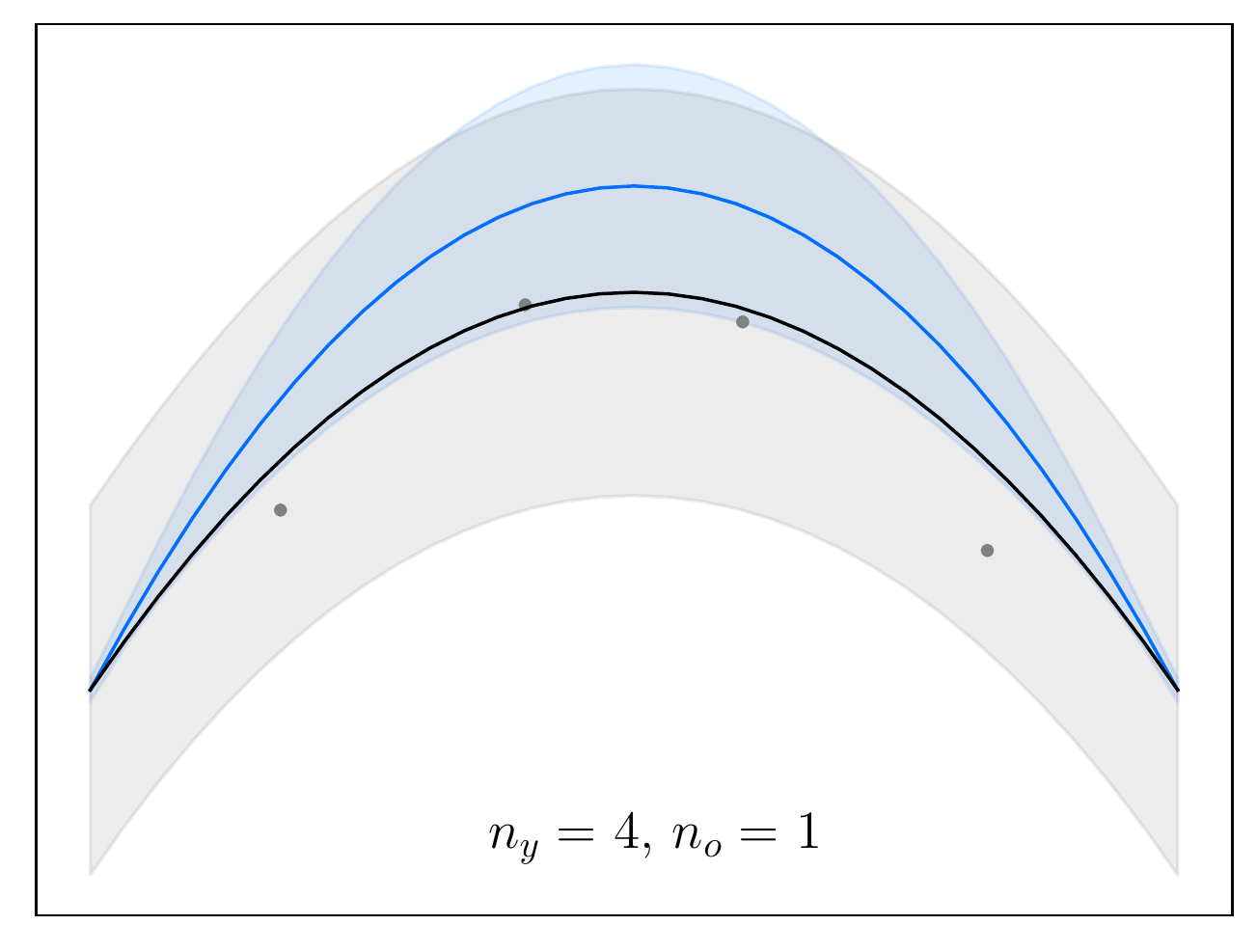}
	\includegraphics[width=0.24\textwidth]{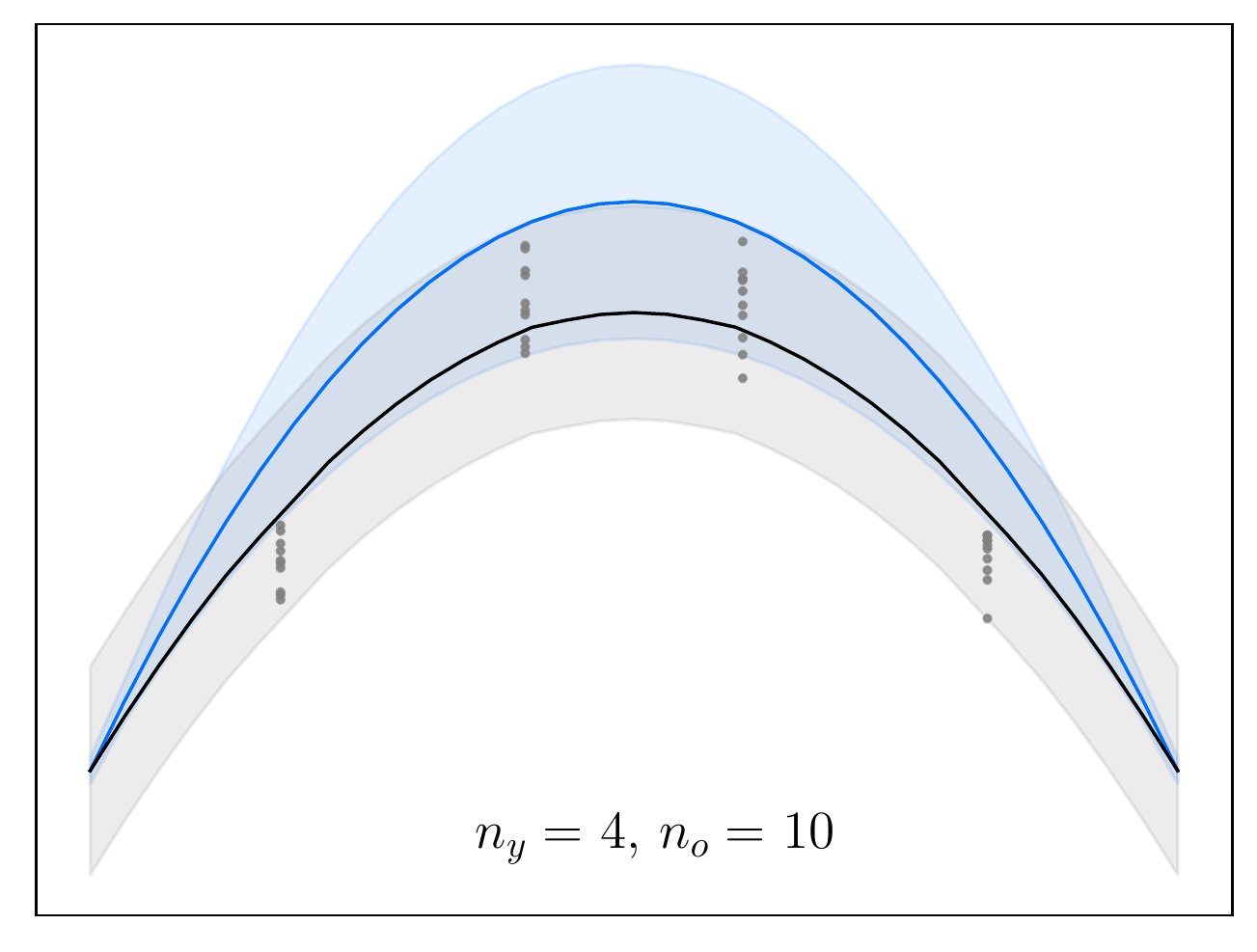}
	\includegraphics[width=0.24\textwidth]{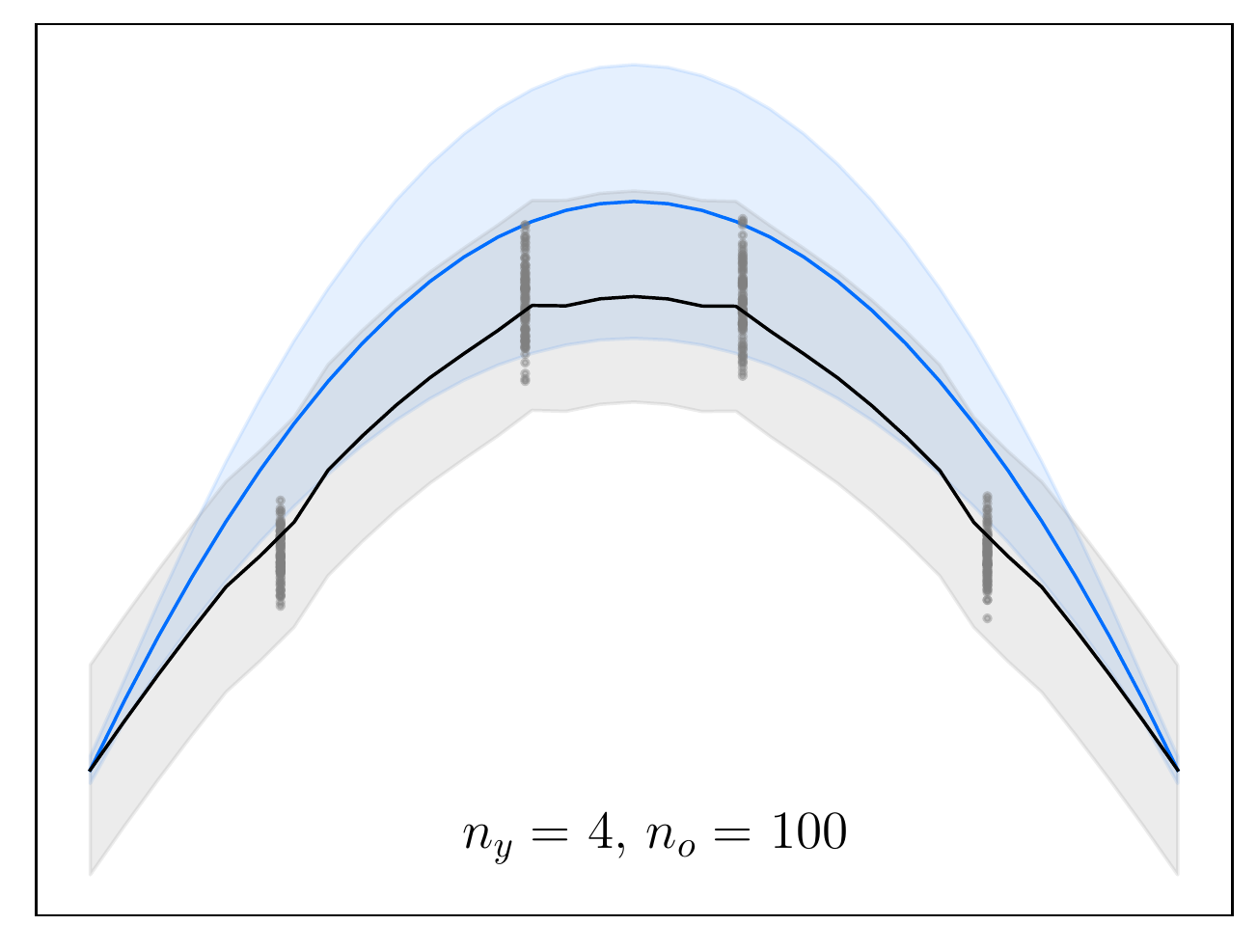}
	\includegraphics[width=0.24\textwidth]{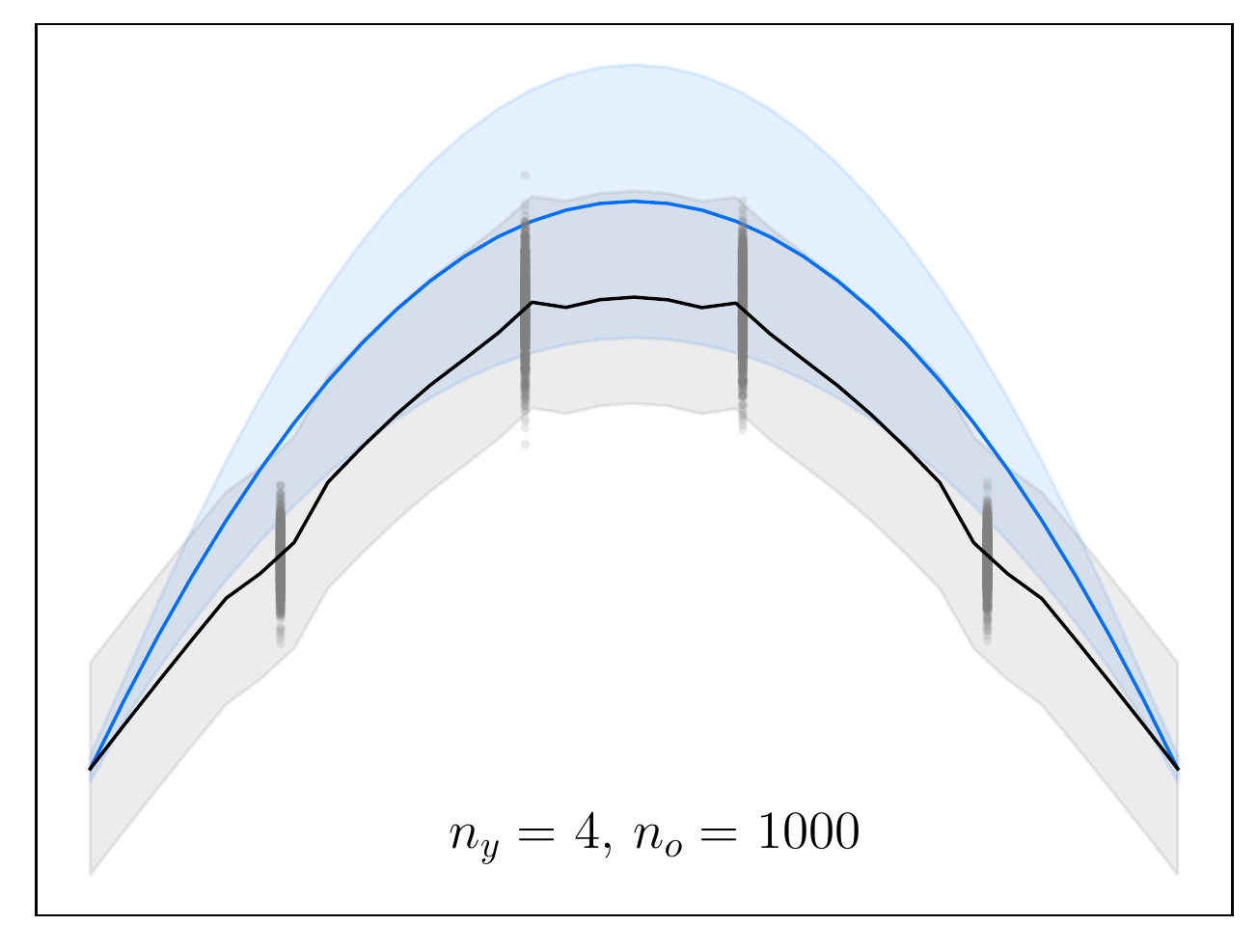} 
	\\
	\includegraphics[width=0.24\textwidth]{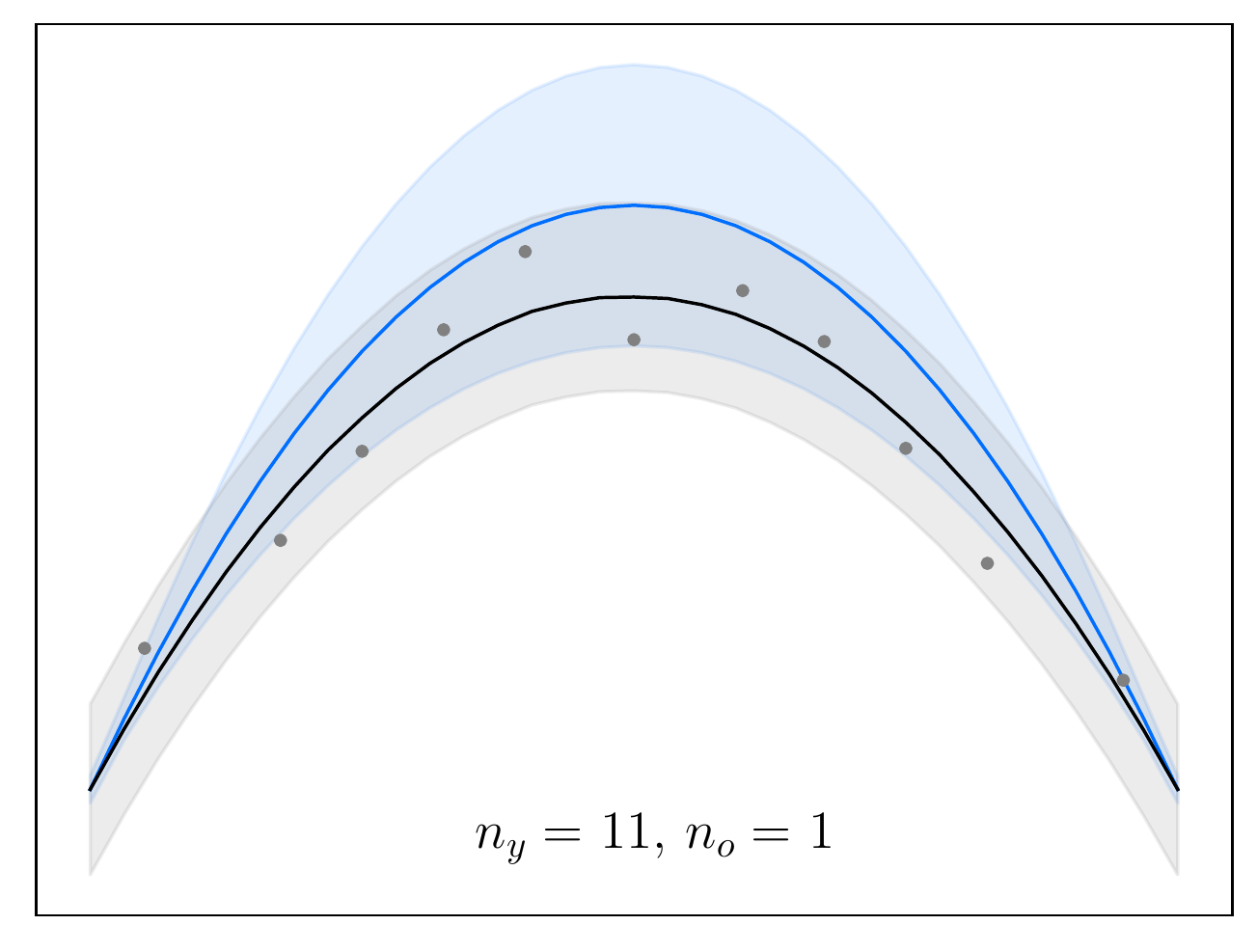}
	\includegraphics[width=0.24\textwidth]{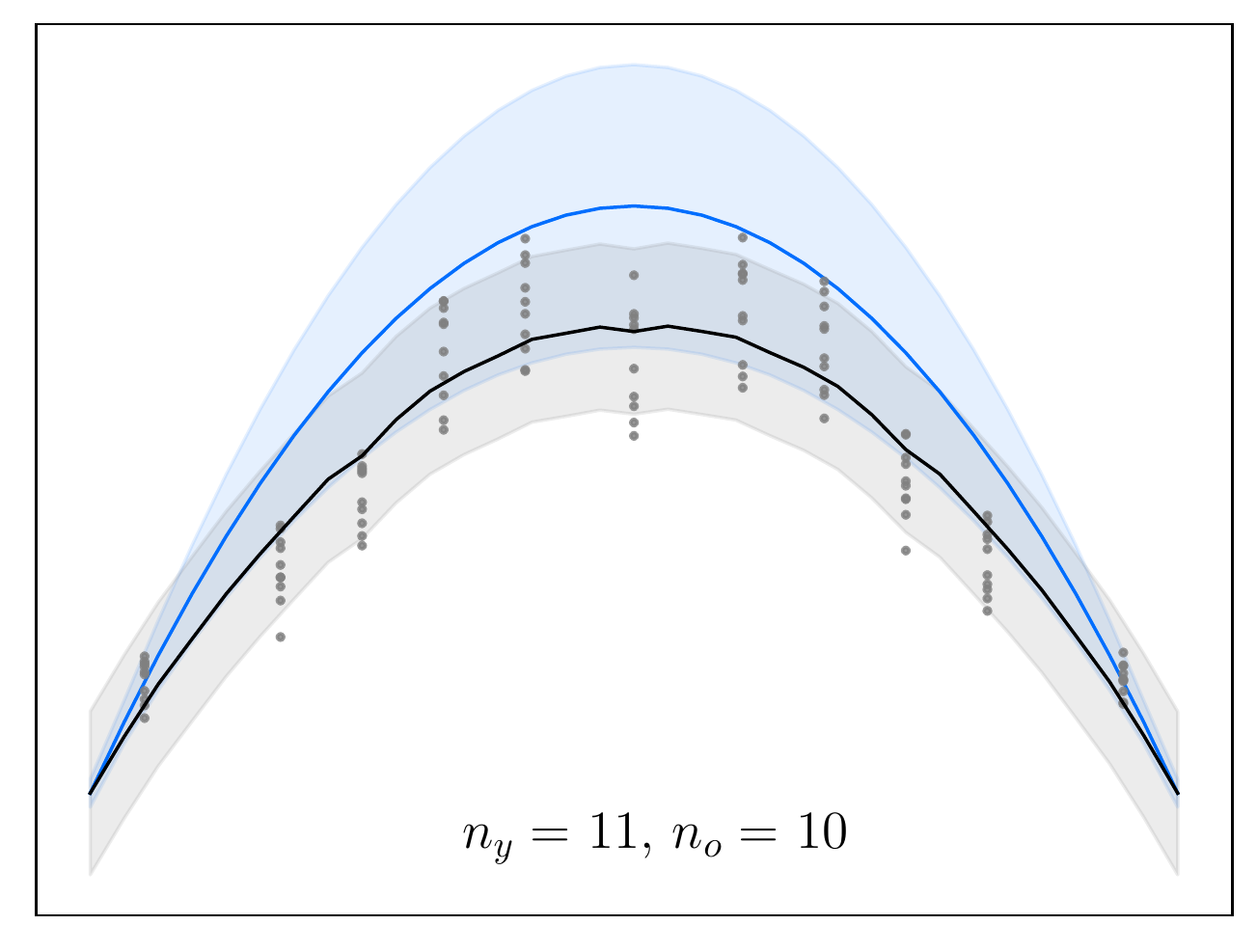}
	\includegraphics[width=0.24\textwidth]{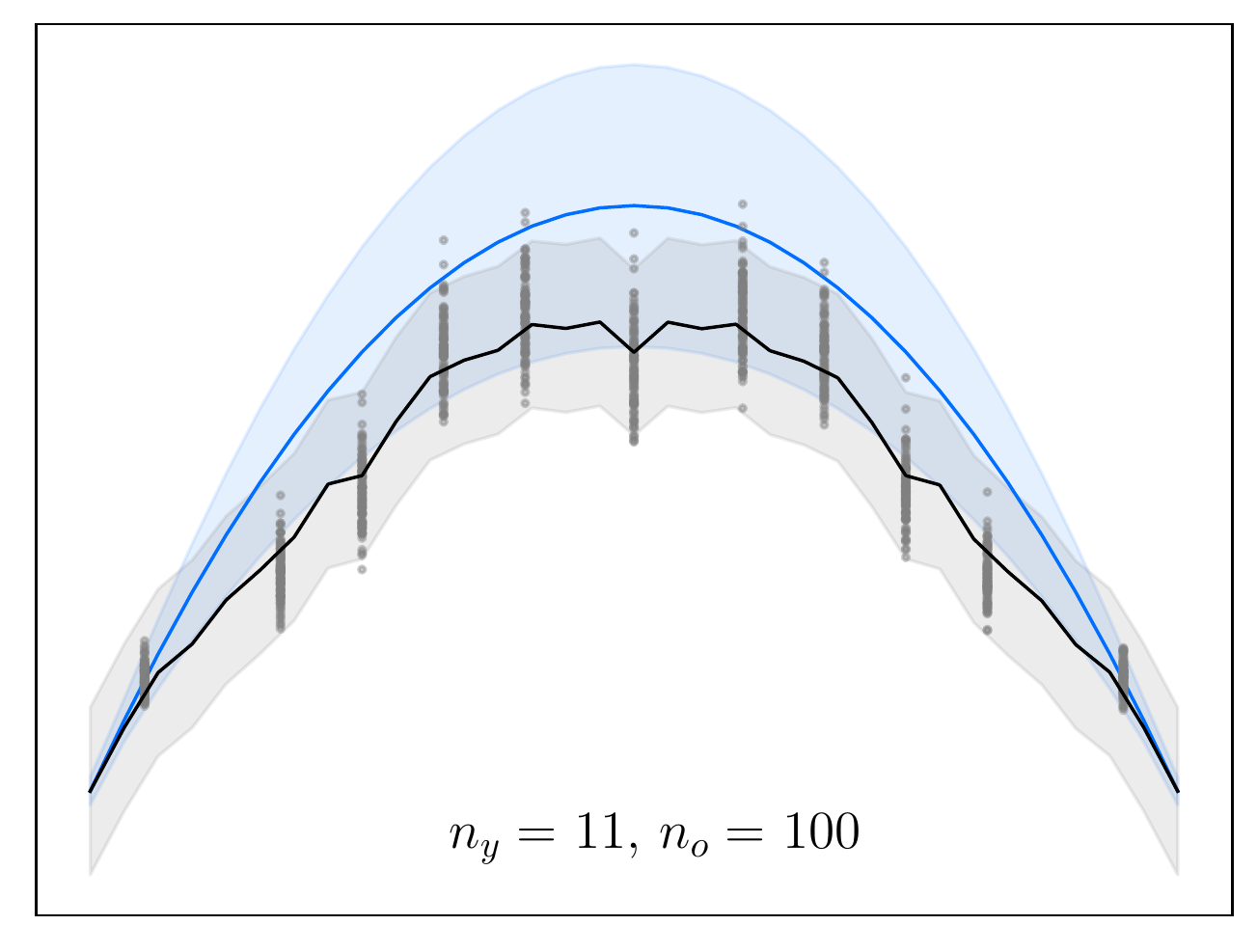}
	\includegraphics[width=0.24\textwidth]{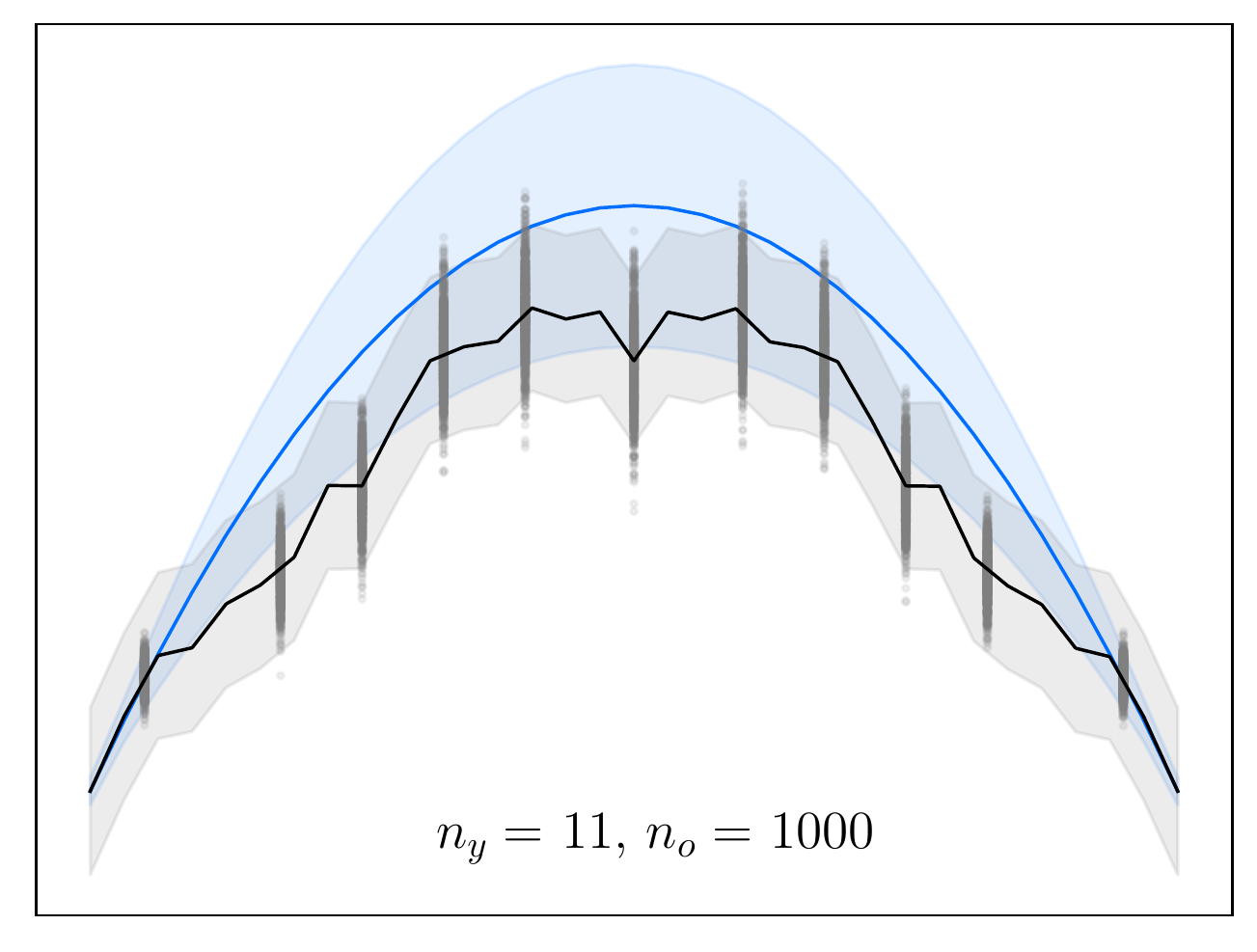} 
	\\
	\includegraphics[width=0.24\textwidth]{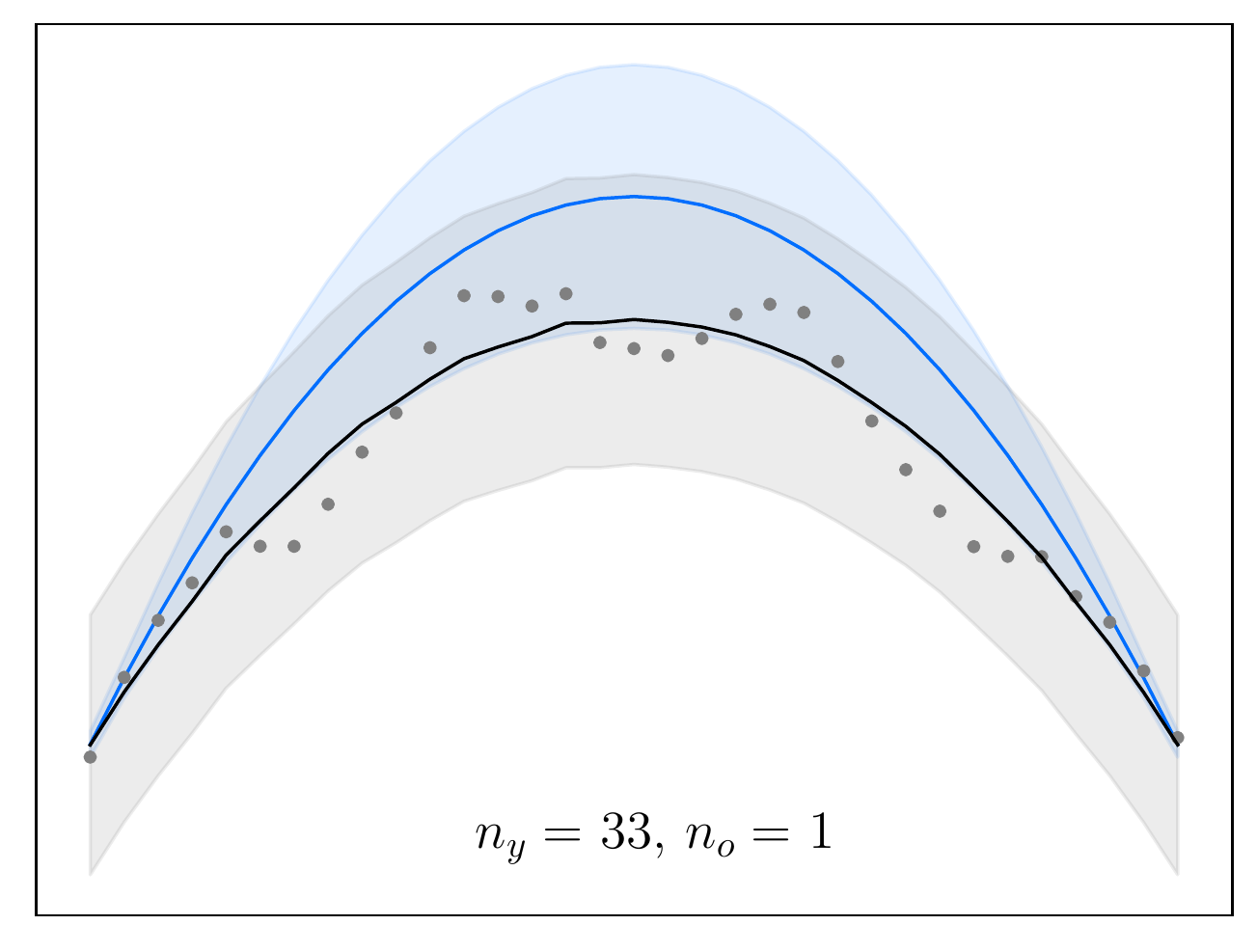}
	\includegraphics[width=0.24\textwidth]{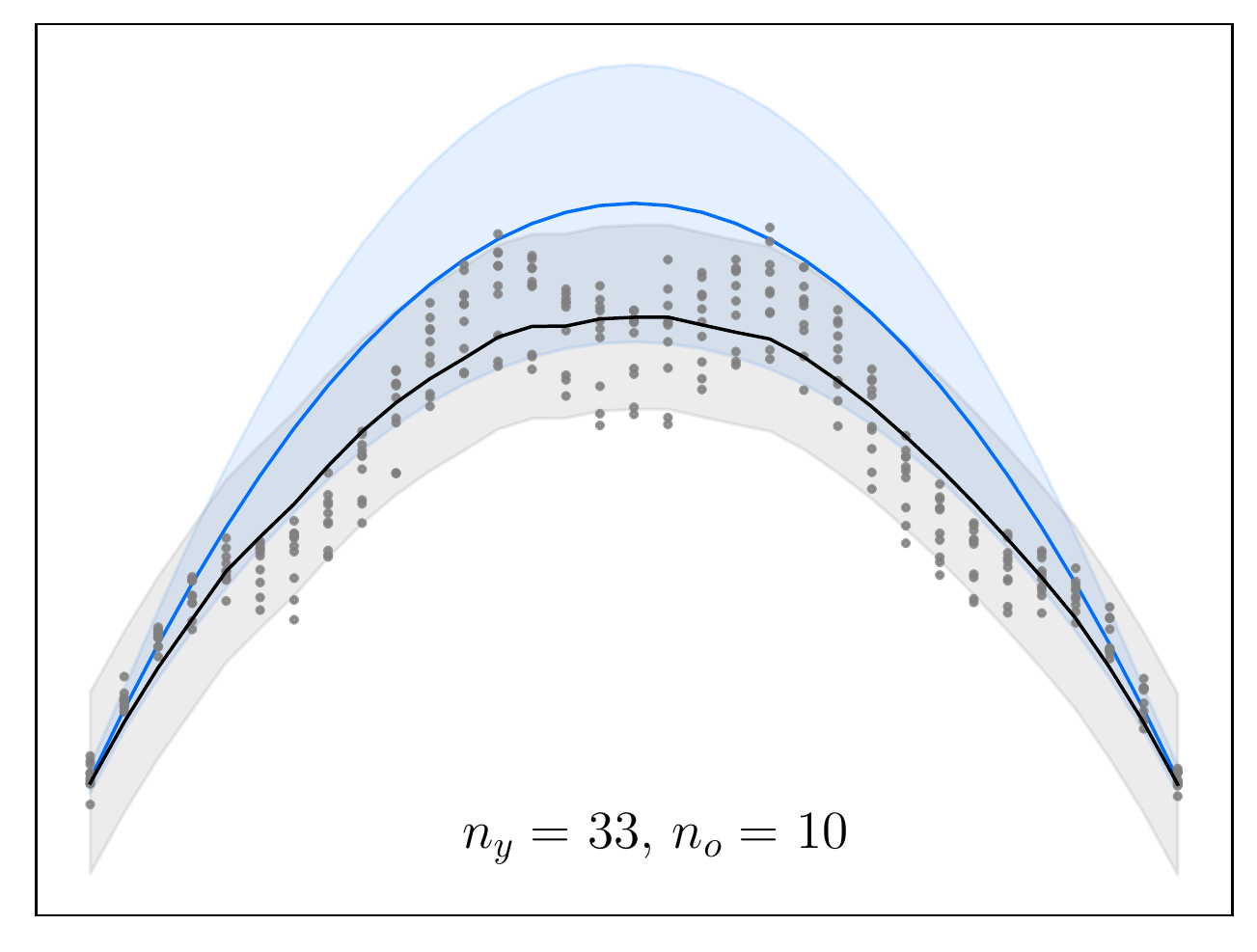}
	\includegraphics[width=0.24\textwidth]{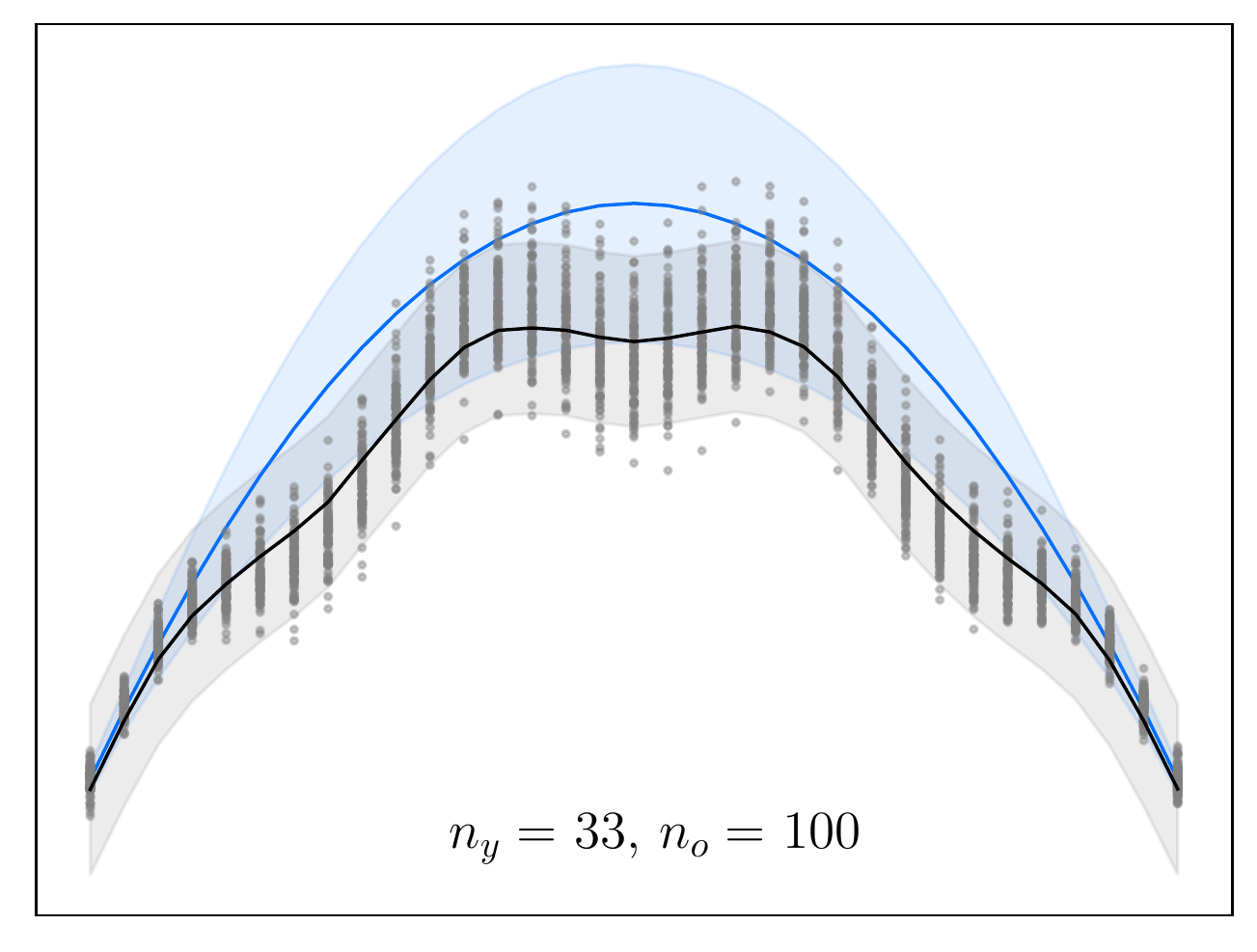}
	\includegraphics[width=0.24\textwidth]{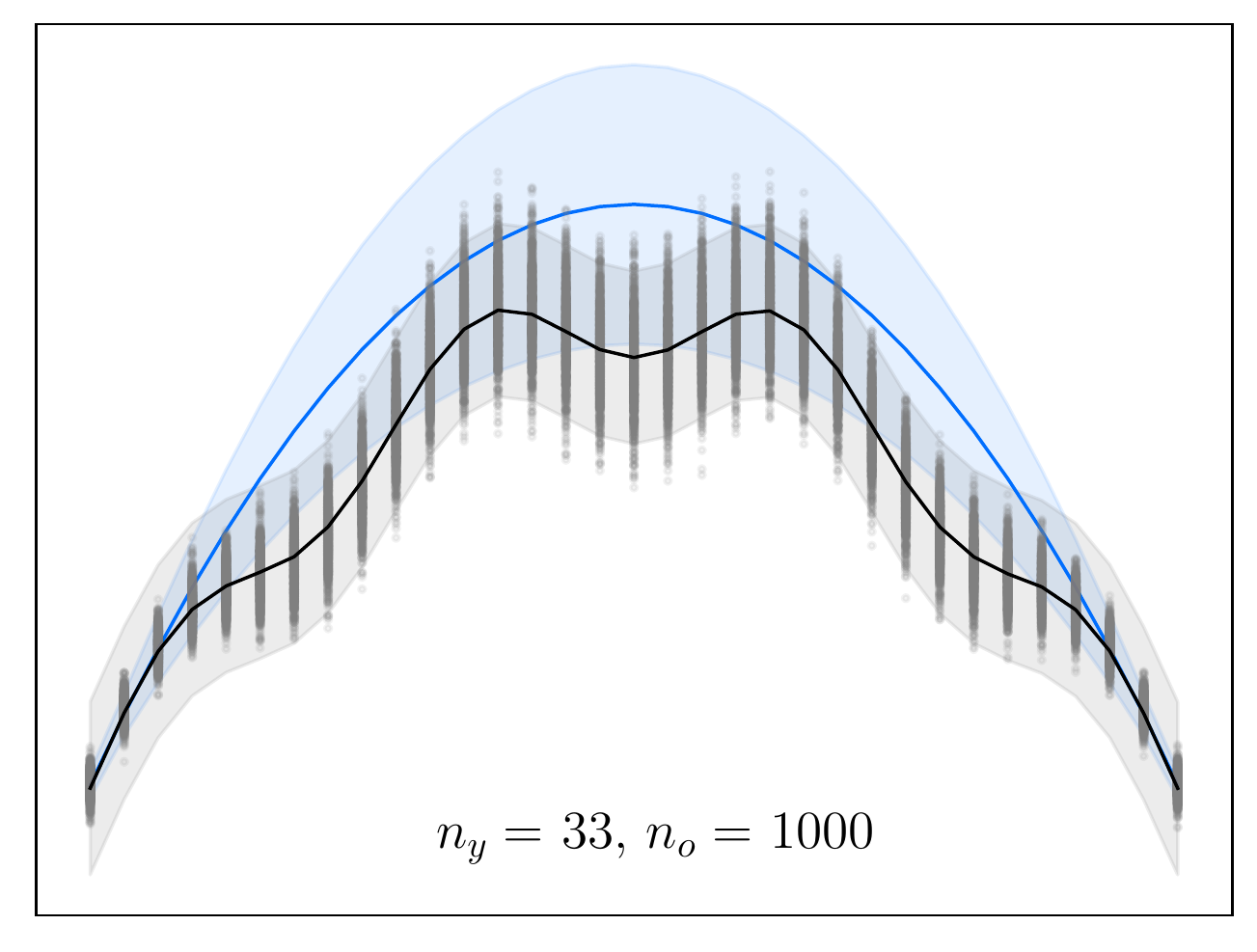} 
	\caption{One-dimensional problem with random source. Inferred true system density~$p(\vec z | \vec Y)$ conditioned on observation data (grey dots). The blue lines represent the mean~$\overline{\vec u}$ and the black lines the conditioned mean~$\overline {\vec z}_{|Y}$. The shaded areas denote the corresponding $95\%$ confidence regions. In each row the number of sensors~$n_y$ and each column the number of readings~$n_o$ is constant. \label{fig:1dPzY}} 
\end{figure}

With the density of the hyperparameters~$p(\vec w | \vec Y)$ at hand, it is possible to evaluate the posterior density of the finite element solution~$p(\vec u | \vec Y)$ given by~\eqref{eq:postUcYMult}. As discussed in Section~\ref{sec:hyperPlearning}, we use the empirical mean~\mbox{$\overline{\vec w} = \mathbb E [\vec w]$} of the hyperparameters~\mbox{$\vec w \sim p(\vec w | \vec Y)$} as a point estimate in evaluating the posterior density~$p(\vec u | \vec Y)$. The obtained posterior densities for different combinations of~$n_y$ and~$n_o$ are depicted in Figure~\ref{fig:1dPuhY}. Each observation sampled from the Gaussian process~\eqref{eq:yGP} is depicted with a dot.  It is remarkable that already a single set of readings can achieve a significant improvement of the finite element mean.  In all cases the~$95\%$ confidence regions become smaller with increasing number of readings~$n_o$. When both~$n_y$ and~$n_o$ are increased the posterior mean~$\overline {\vec u}_{|\vec Y}$ converges to the true process mean~$\overline z(x)$ and the covariance~$\vec C_{u|Y}$ converges to zero. In the corresponding Figure~\ref{fig:1dPzY} the obtained posterior densities~$ p(\vec z | \vec Y)$ for the inferred true system response according to~\eqref{eq:trueSystemPosterior} are shown. As expected with increasing~$n_y$ and~$n_o$  the inferred density~$ p(\vec z | \vec Y)$ converges to the, in this example known, true density given by the Gaussian process~\eqref{eq:zGP1dEx}. Only very few~$n_y$ and~$n_o$ yield a very good approximation to the true process mean~$\overline z(x)$. While the overall shape of the inferred  and the true process confidence regions are in good agreement there are some differences close to the boundaries.  These are related to the assumed covariance structure for the model mismatch vector~$\vec d$. The chosen squared exponential kernel is unable to provide a better approximation to the covariance of the true process.

%

%
\subsubsection{Posterior finite element density and system response for random diffusivity \label{sec:1dUcondYdiff}}
%

\begin{figure}[]
	\centering
	\subfloat[][Diffusion coefficient $\kappa(x)$ \label{fig:poisson1D-LHS-exA}] {
		\includegraphics[width=0.425\textwidth]{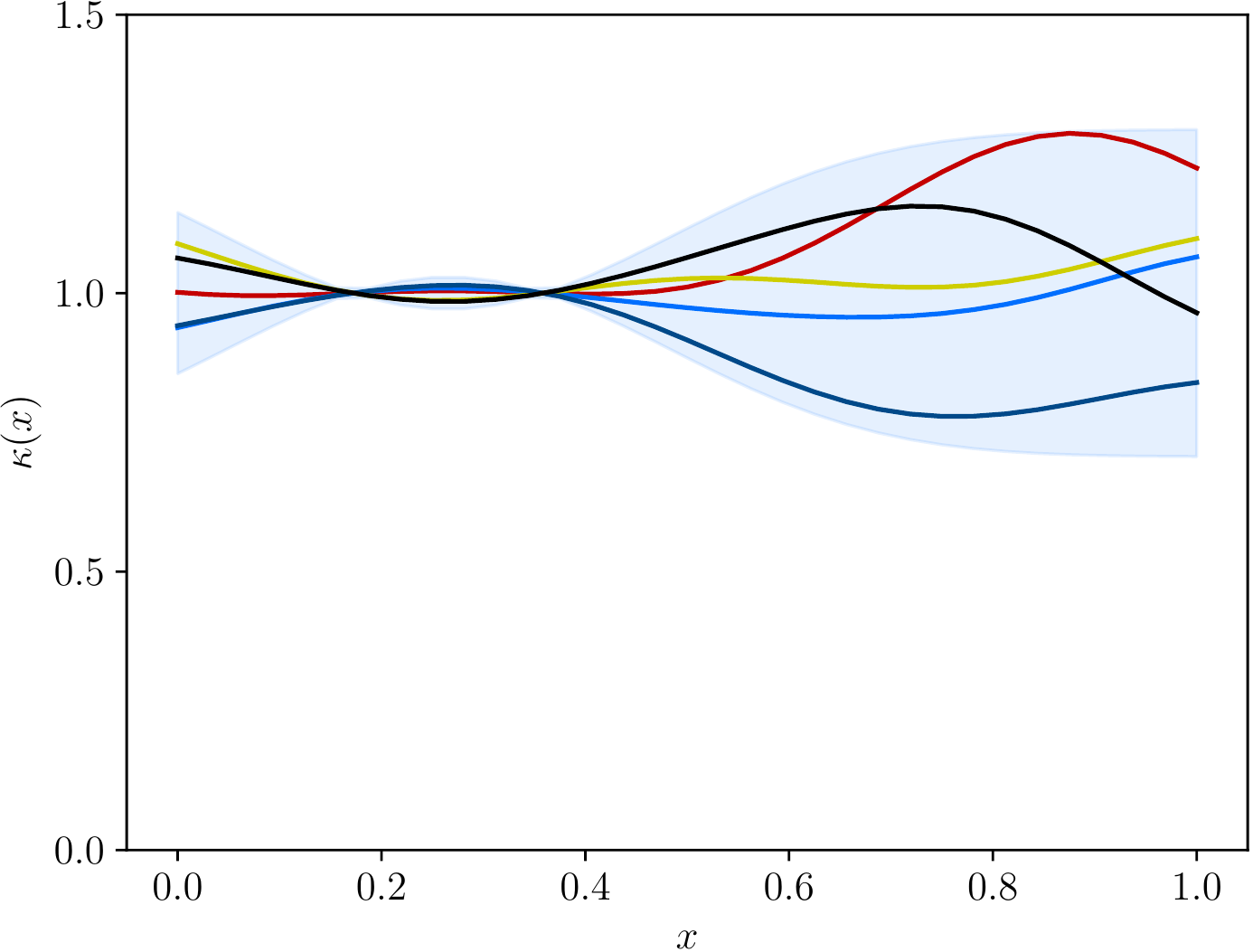}
	}
	\hspace{0.05\textwidth}
	\subfloat[][Solution $u_h(x)$] {
		\includegraphics[width=0.425\textwidth]{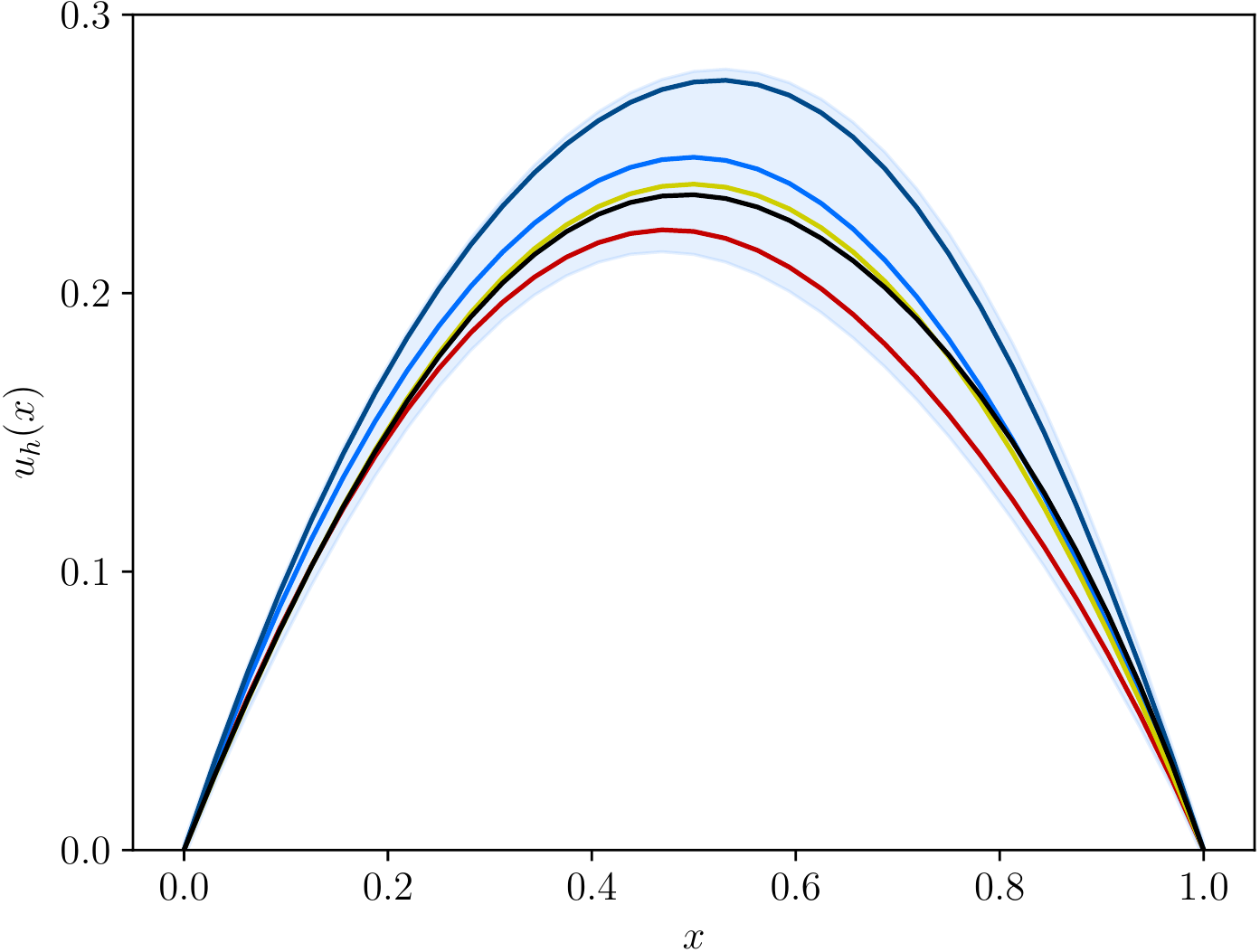}
	}
	\caption{One-dimensional problem with random diffusivity.  The diffusion coefficient is a Gaussian process with a mean~\mbox{$\overline \kappa (x) = 1 $} and covariance kernel parameters~\mbox{$\sigma_\kappa = 0.15 $} and \mbox{$\ell_\kappa= 0.25$.}  The five lines in (a) represent samples drawn from the Gaussian process~\eqref{eq:gprKappaExample} conditioned on \mbox{$\kappa(11/64) = \kappa(23/64) =  1.0$}. The corresponding lines in (b) show the solution. The shaded areas denote the corresponding  $95\%$ confidence regions. \label{fig:poisson1D-LHS-ex}} 
\end{figure}

We consider the case when the diffusion coefficient is random and the source is deterministic. We aim to compute the posterior finite element and true system densities~$p(\vec u | \vec Y)$ and ~$p(\vec z  | \vec Y)$  for observation matrices~$\vec Y$ sampled from the Gaussian process~\eqref{eq:yGP}.  The source is chosen to be~$f(x)= \pi^2 / 5$ and the diffusion coefficient is given by the Gaussian process 
\begin{equation} \label{eq:gprKappaExample}
	\kappa(x)  \sim \set{GP} \left ( 1.0, \, 0.0225 \exp \left ( - 8 (x -x' )^2  \right ) \right ) \, .
\end{equation}
The diffusion coefficient within each element is assumed to be constant. Hence, this Gaussian process is discretised by sampling the diffusion coefficient vector~$\vec \kappa \in \mathbb R^{n_e}$ at the element centres~$\vec X^{(c)}$ to yield the multivariate Gaussian density~\mbox{$p(\vec \kappa) = \set N(\overline{\vec \kappa}, \, \vec C_\kappa)$.} When the diffusion coefficient in some of the finite elements is known it can be taken into account by conditioning the density~$p(\vec \kappa)$ on those known values, see~\ref{app:GPregress}. In this example, we assume that the diffusion coefficient~\mbox{$\kappa(x=11/64) = 1.0$} and~\mbox{$\kappa(x= 23/64) =  1.0$} are  known. In Figure~\ref{fig:poisson1D-LHS-ex} five samples of the so conditioned diffusion coefficient and its~$95\%$ confidence region are shown. The corresponding finite element solutions are obtained by solving the forward problem with the given diffusion coefficient distribution. Evidently, the mapping between the diffusion coefficient~$\vec \kappa$ and the finite element solution~$\vec u$ is nonlinear.  As discussed in Section~\ref{sec:forwardFE}, we approximate this mapping with a first order perturbation method yielding the approximate density~\mbox{$p(\vec u) = \set N(\overline{\vec u}, \, \vec C_u)$,} see~\eqref{eq:forwardDensityFinal}.

As in Section \ref{sec:1dUcondY}, we first determine the unknown hyperparameters of the statistical generating model before computing the posterior densities~\mbox{$p(\vec u | \vec Y)$} and~\mbox{$p(\vec z | \vec Y)$.} The unknowns in this example are again the scaling parameter~$\rho$ and the mismatch covariance parameters~$\sigma_d$ and~$\ell_d$, which are collected in the vector~\mbox{$\vec w = (\rho, \, \sigma_d, \, \ell_d)^\trans$.} We sample the posterior~\mbox{$p(\vec w | \vec Y) \propto p(\vec Y, \vec w) p(\vec w)$} using standard MCMC and a non-informative prior~$p(\vec w) \propto 1$.  We consider \mbox{$n_o \in \{ 1, \, 10, \, 100, \, 1000 \}$} repeated readings sampled from~\eqref{eq:yGP} at the \mbox{$n_y \in \{ 4, 33 \}$} locations shown in Figure \ref{fig:sensorPos}. 

\begin{figure}[t] 
	\centering
	\includegraphics[width=0.24\textwidth]{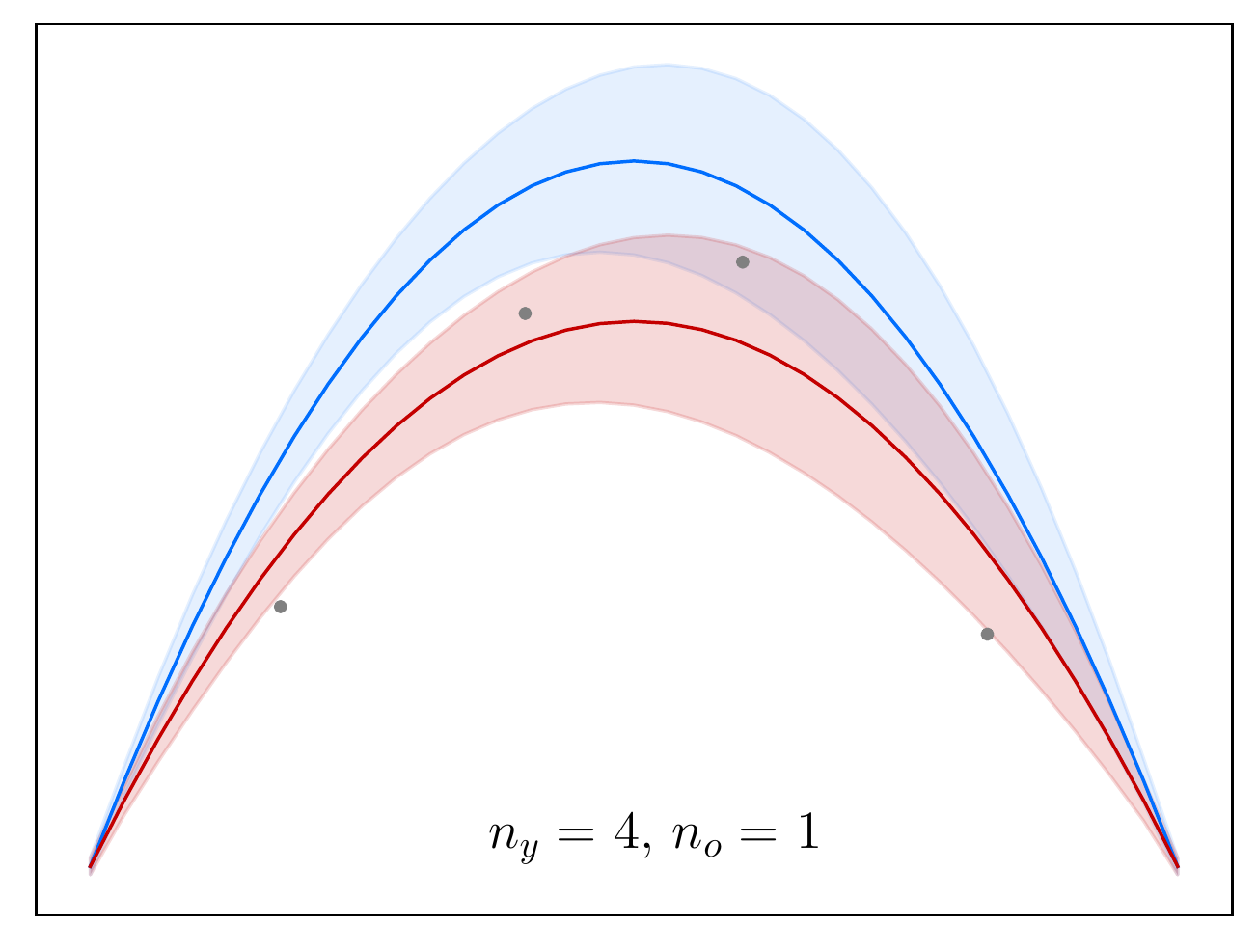}
	\includegraphics[width=0.24\textwidth]{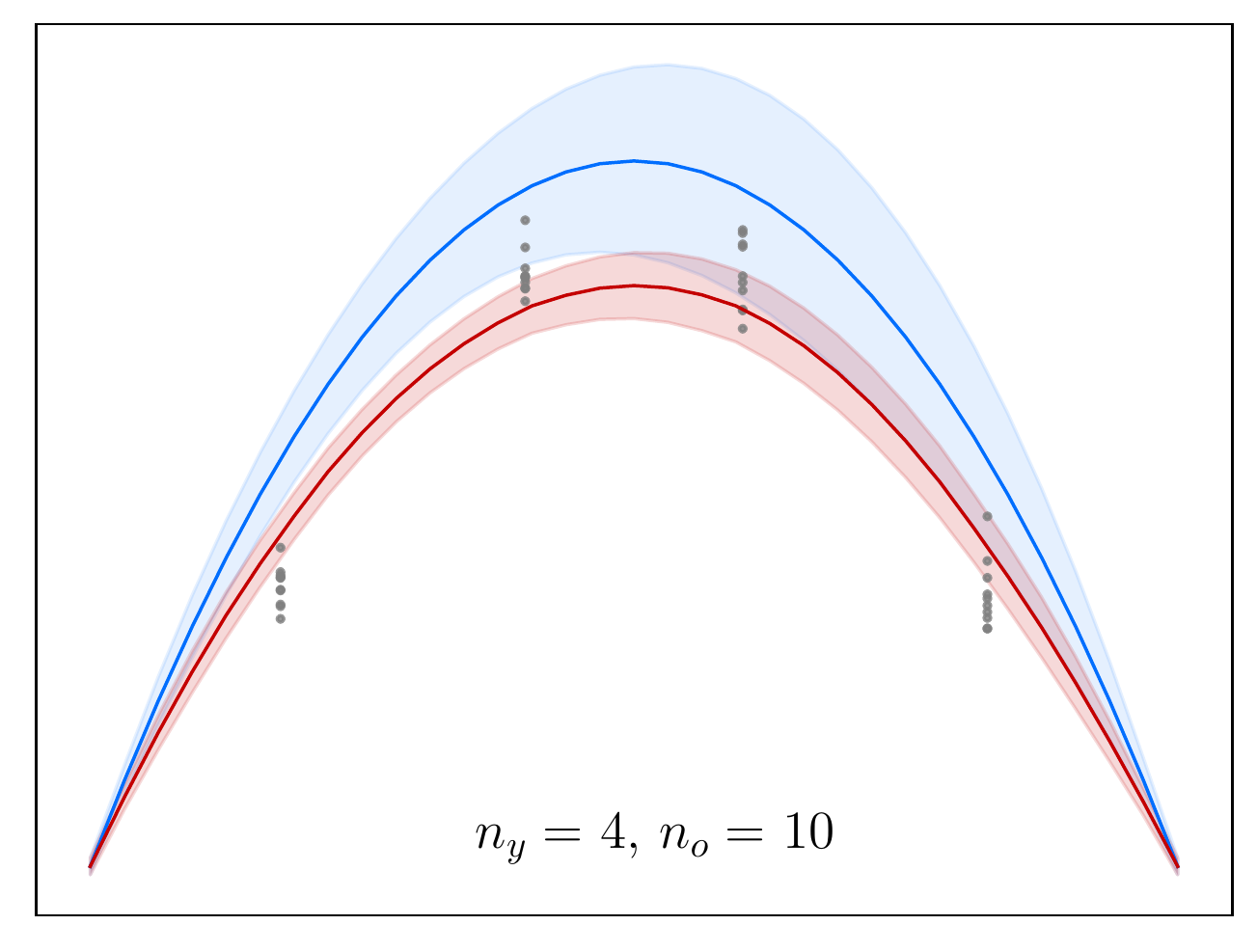}
	\includegraphics[width=0.24\textwidth]{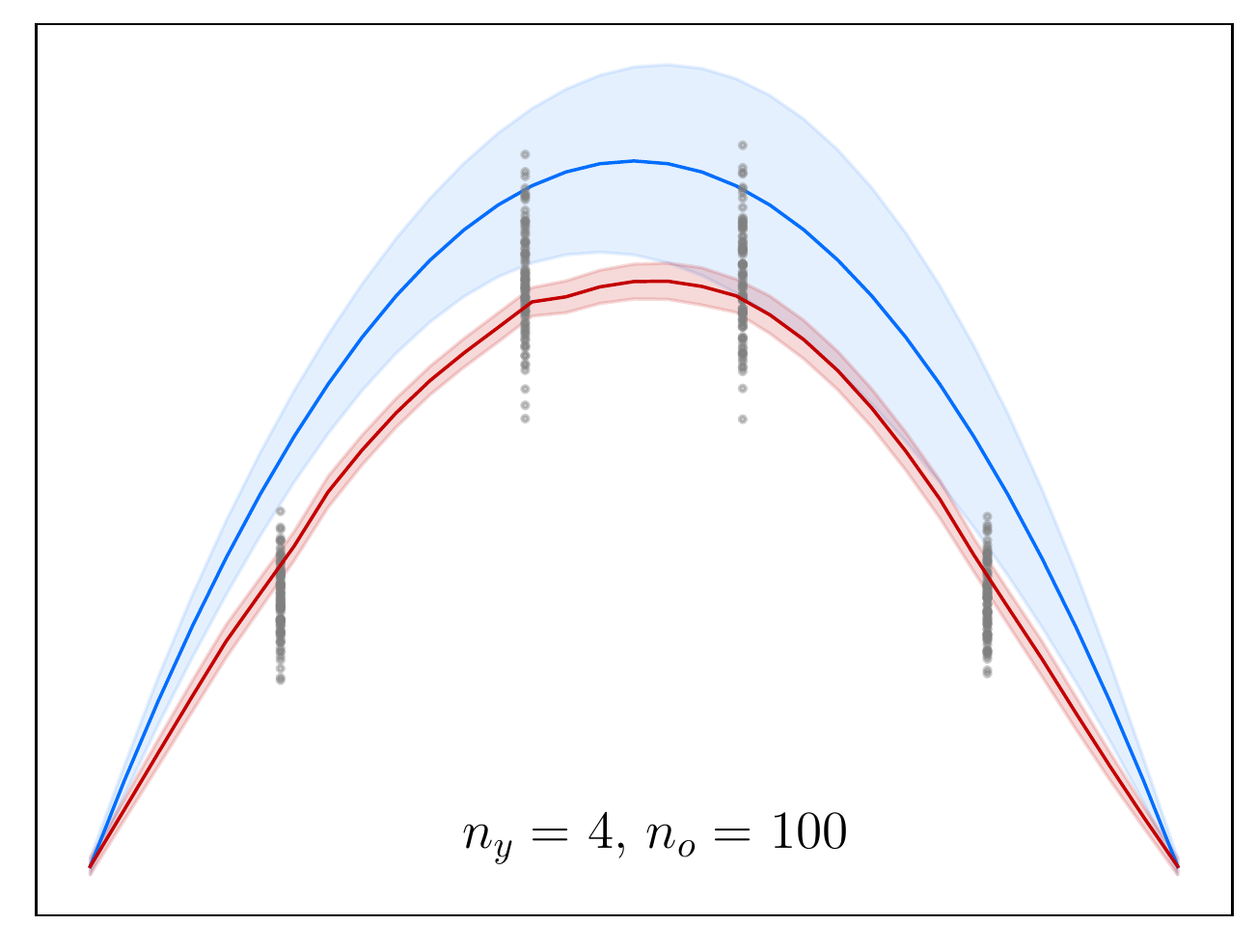}
	\includegraphics[width=0.24\textwidth]{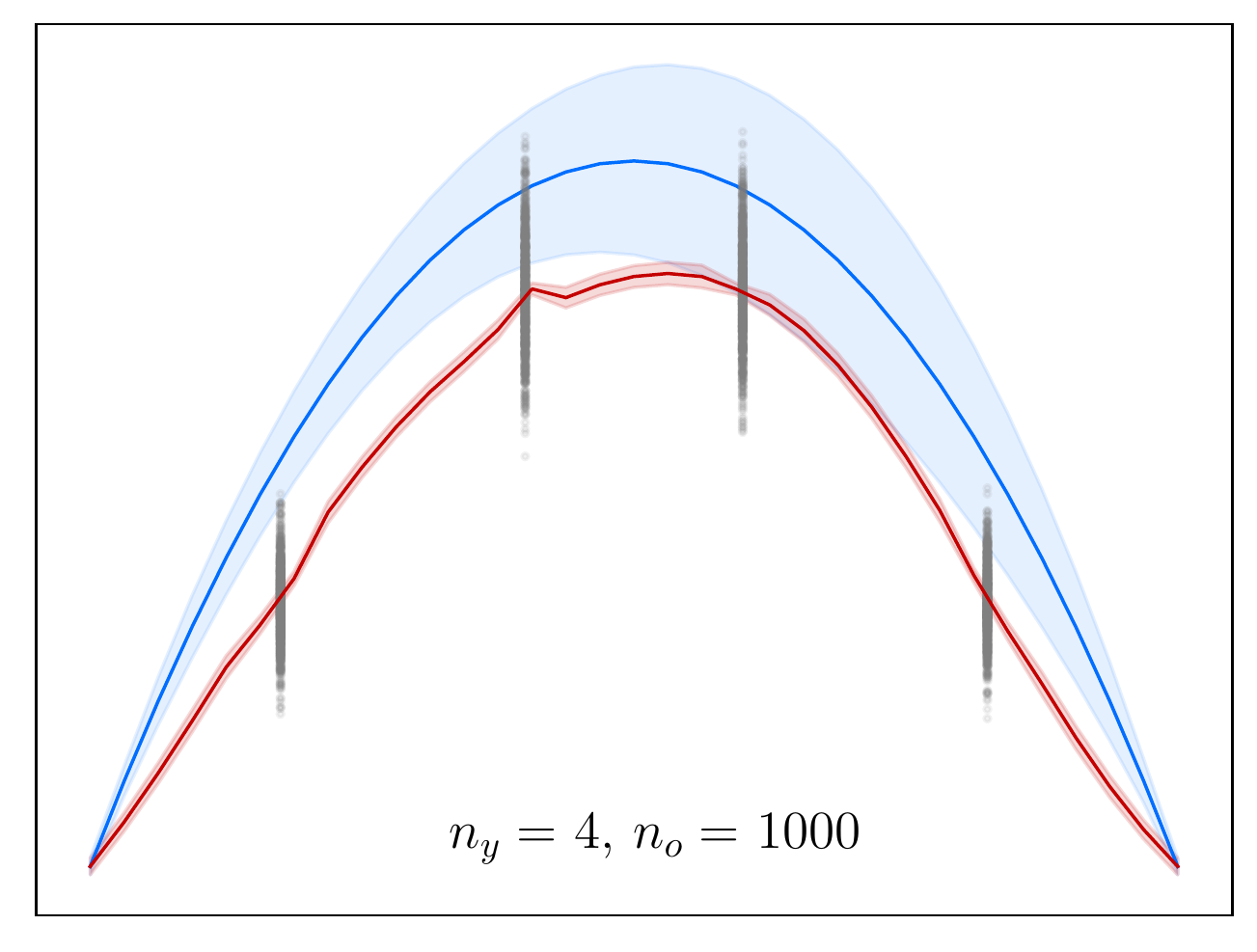} 
	\\
	\includegraphics[width=0.24\textwidth]{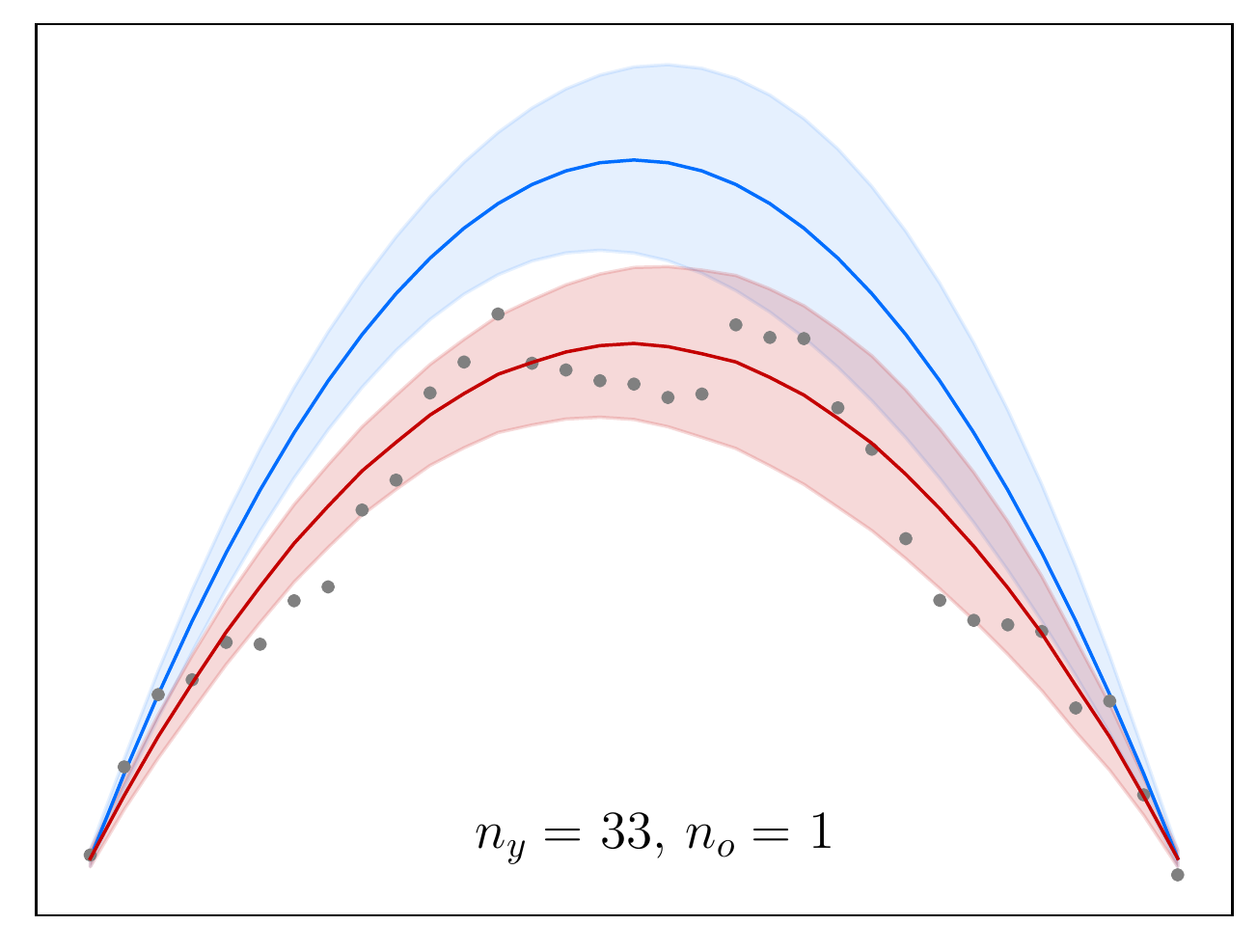}
	\includegraphics[width=0.24\textwidth]{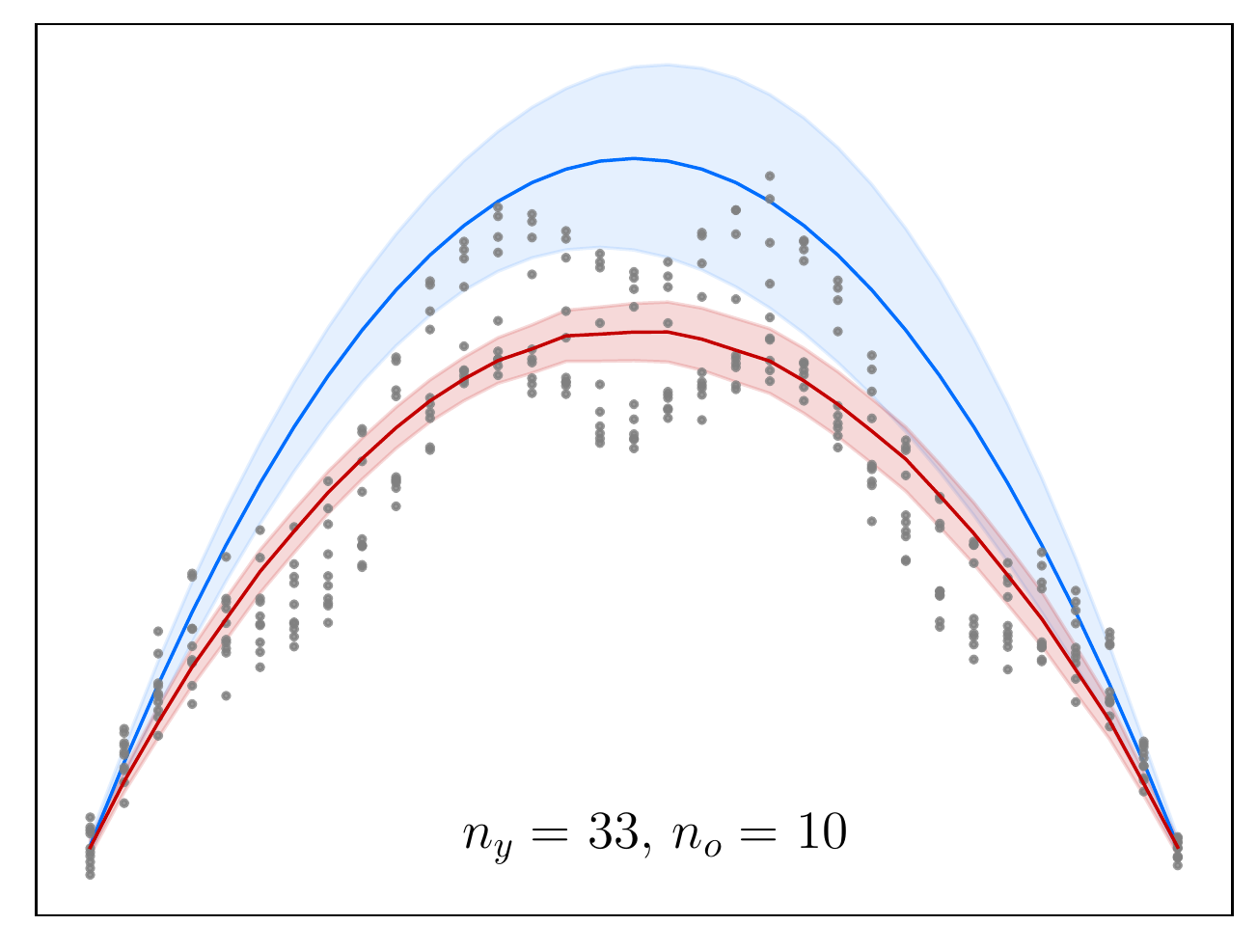}
	\includegraphics[width=0.24\textwidth]{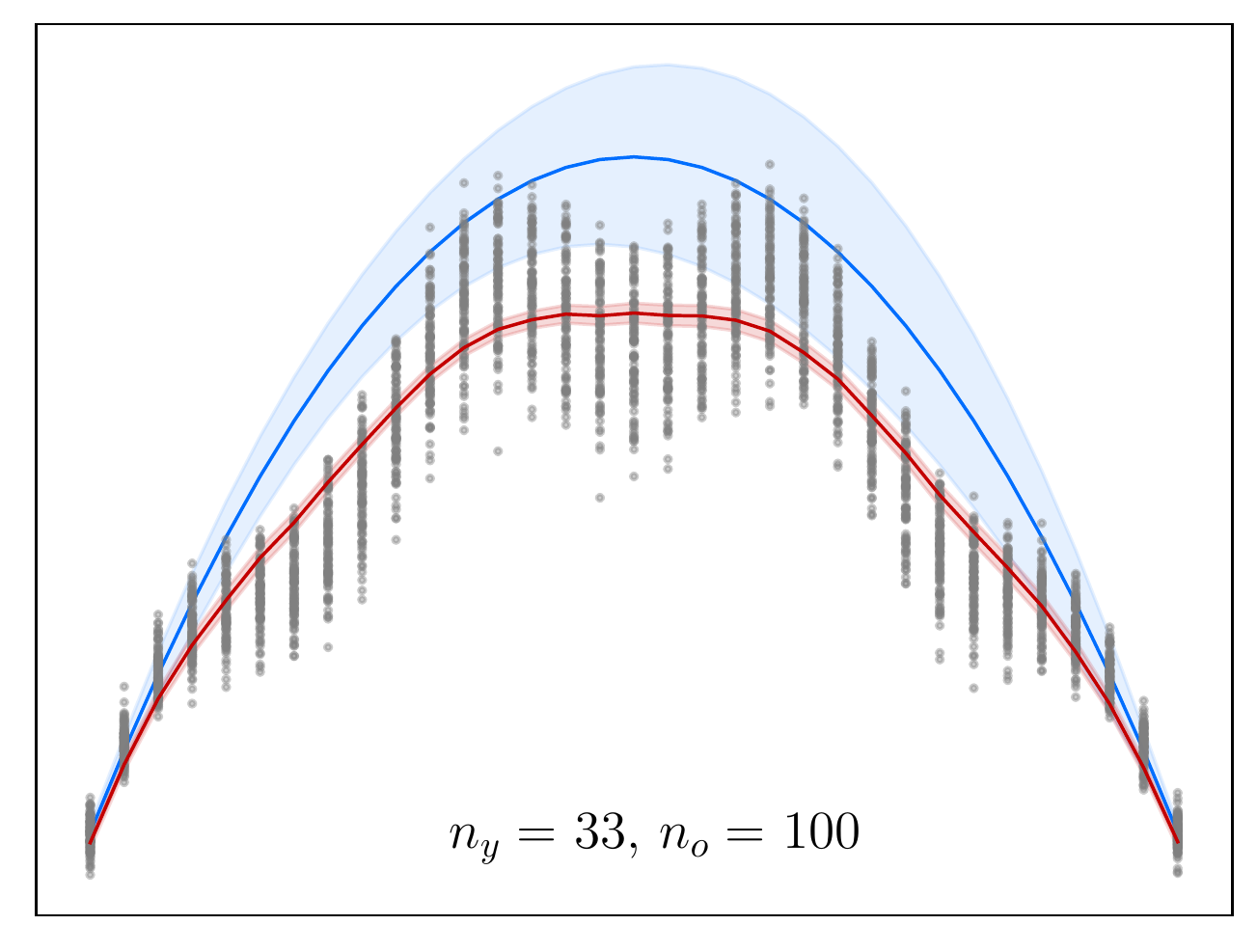}
	\includegraphics[width=0.24\textwidth]{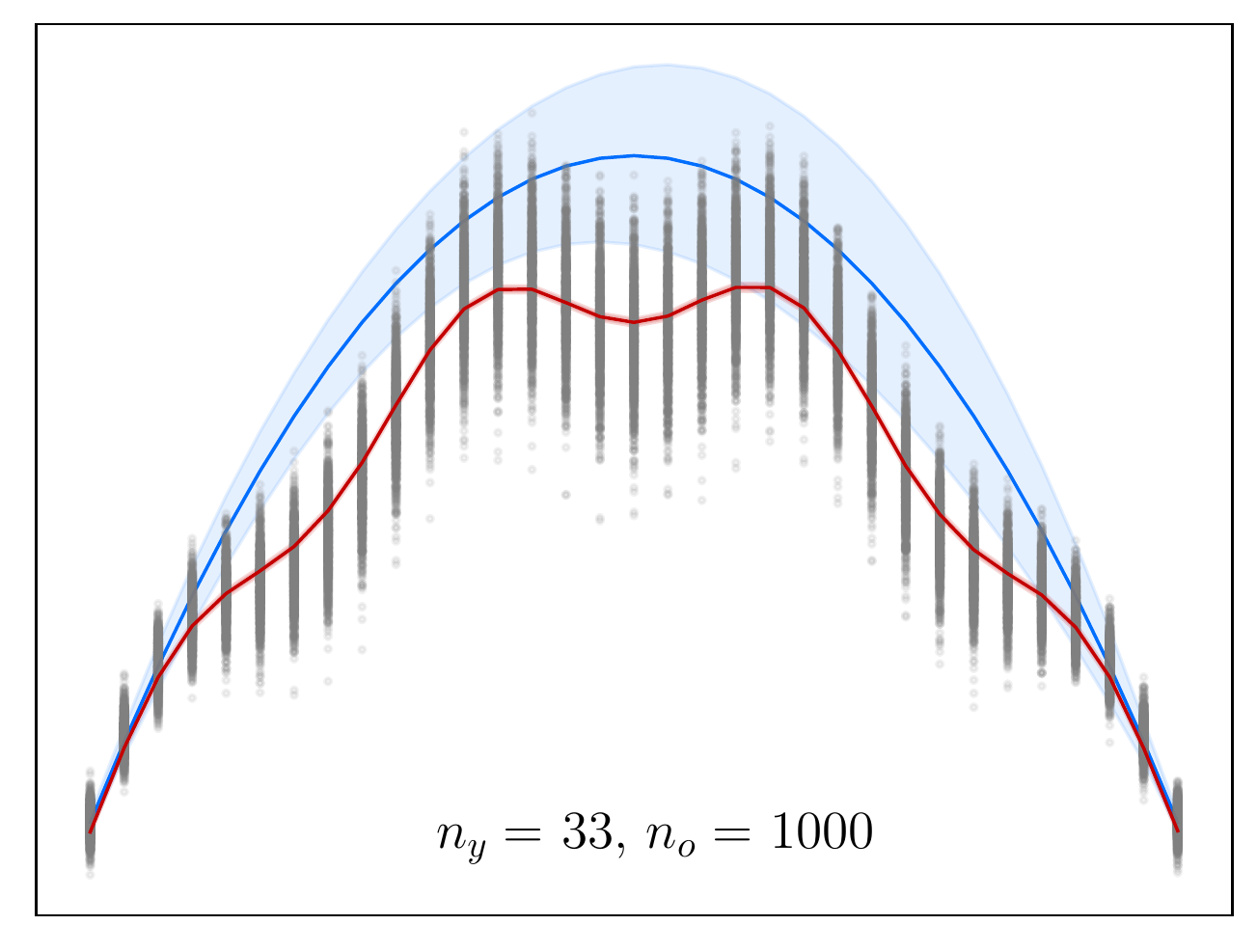} 
	\caption{One-dimensional problem with random diffusivity. Finite element density~$p(\vec u | \vec Y)$ conditioned on observation data (grey dots). The blue lines represent the mean~$\overline{\vec u}$ and the red lines the conditioned mean~$\overline {\vec u}_{|Y}$. The shaded areas denote the corresponding $95\%$ confidence regions. In each row the number of sensors~$n_y$ and each column the number of readings~$n_o$ is constant. \label{fig:1dPuhY-LHS}} 
%
\vspace{1.5em}
%
	\centering
	\includegraphics[width=0.24\textwidth]{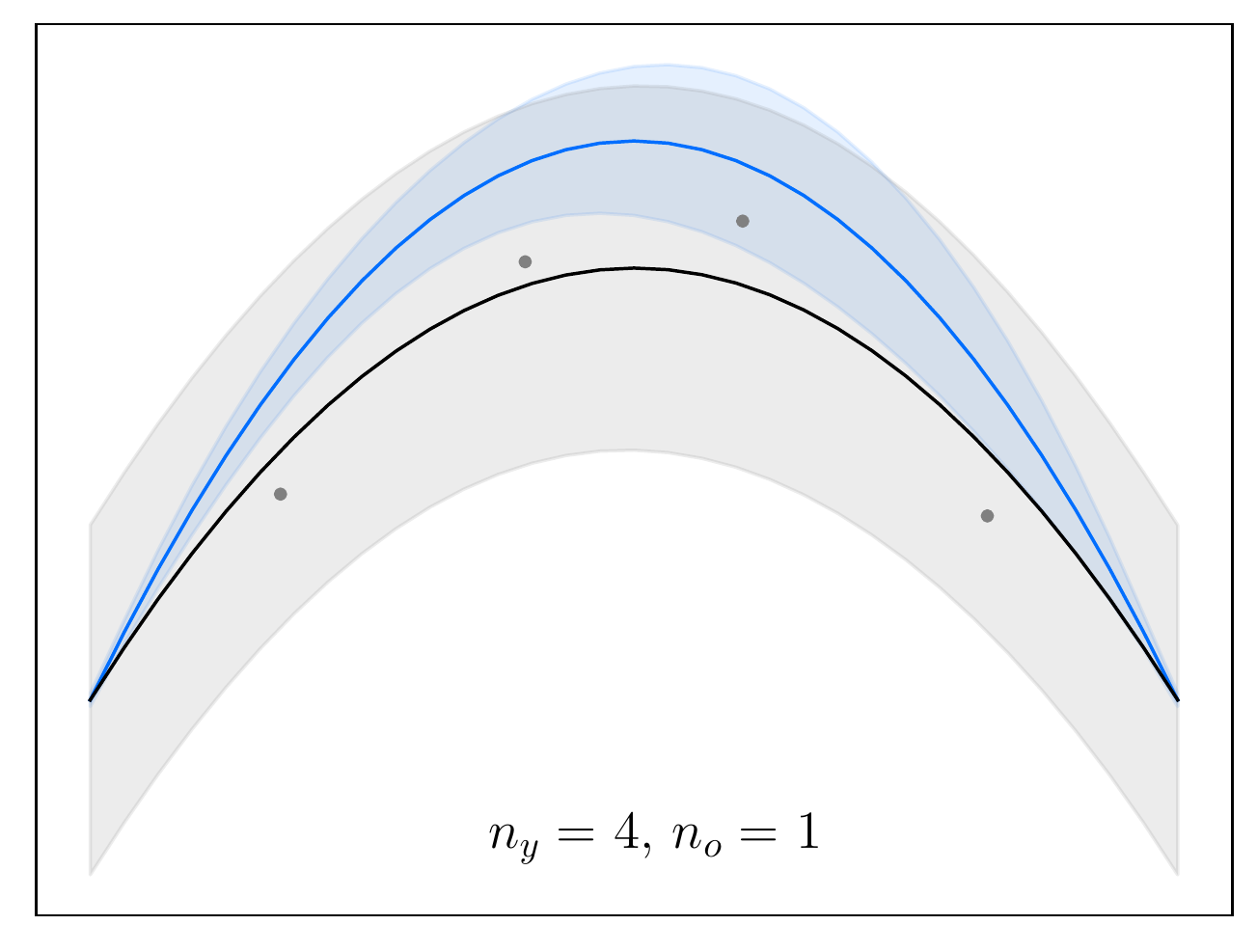}
	\includegraphics[width=0.24\textwidth]{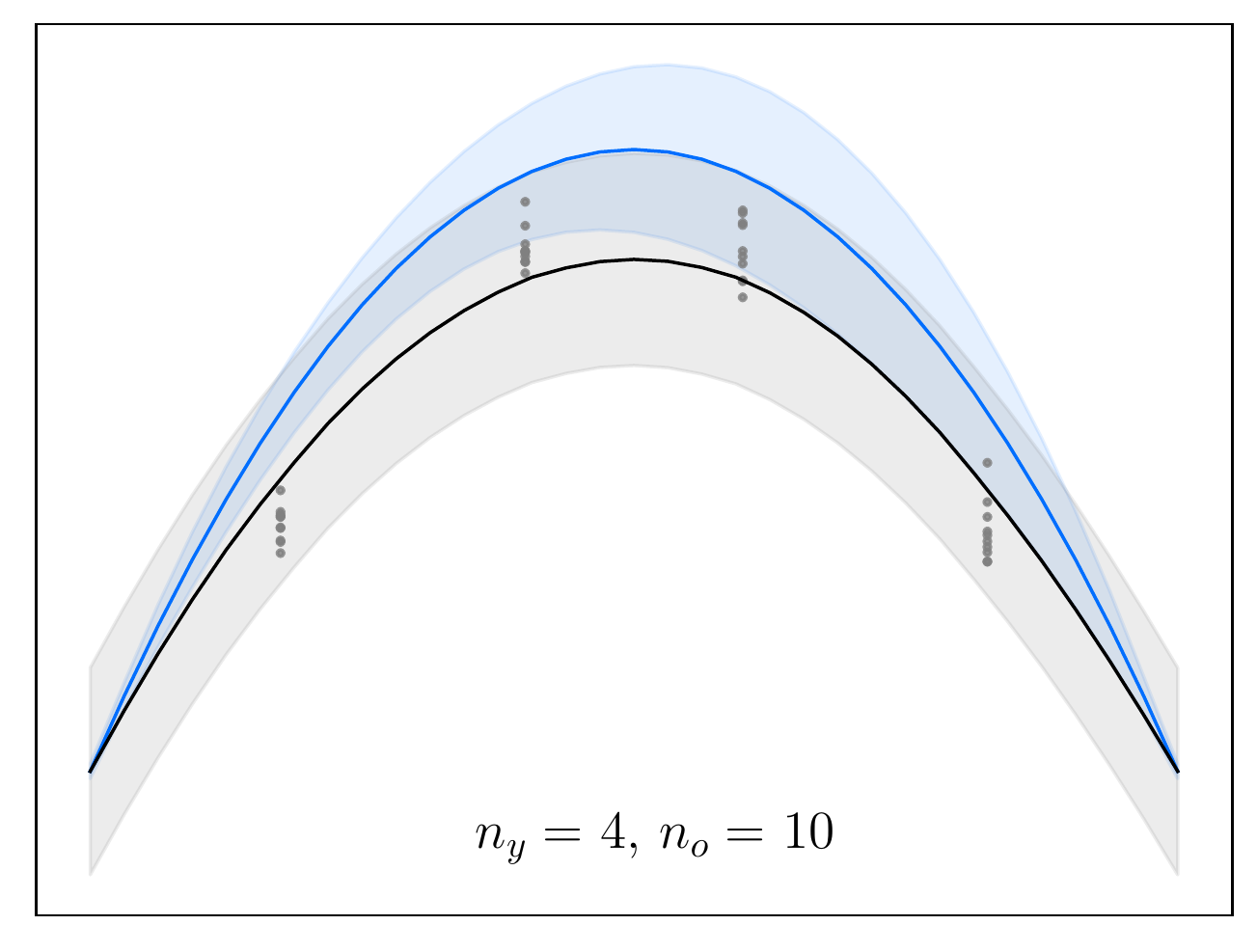}
	\includegraphics[width=0.24\textwidth]{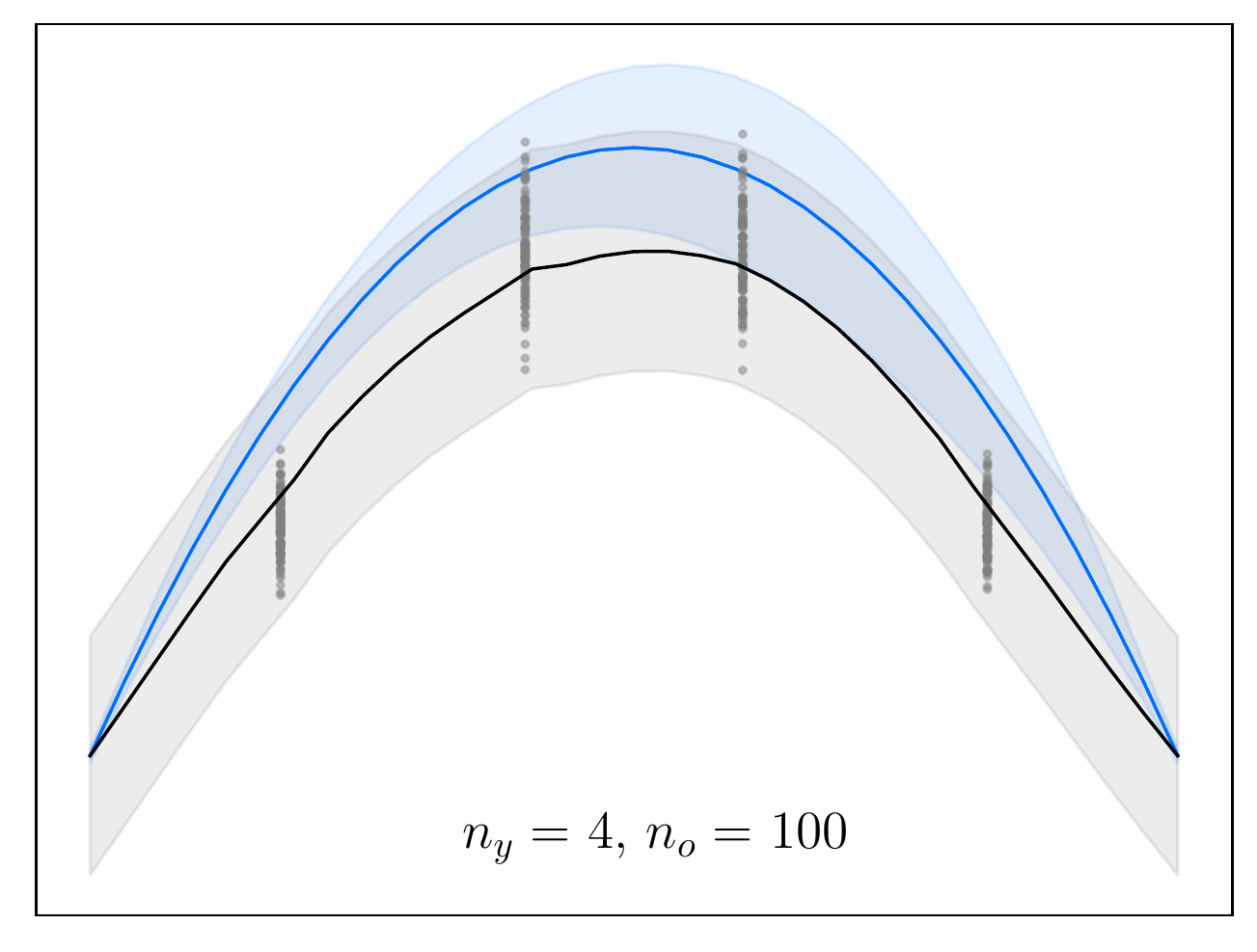}
	\includegraphics[width=0.24\textwidth]{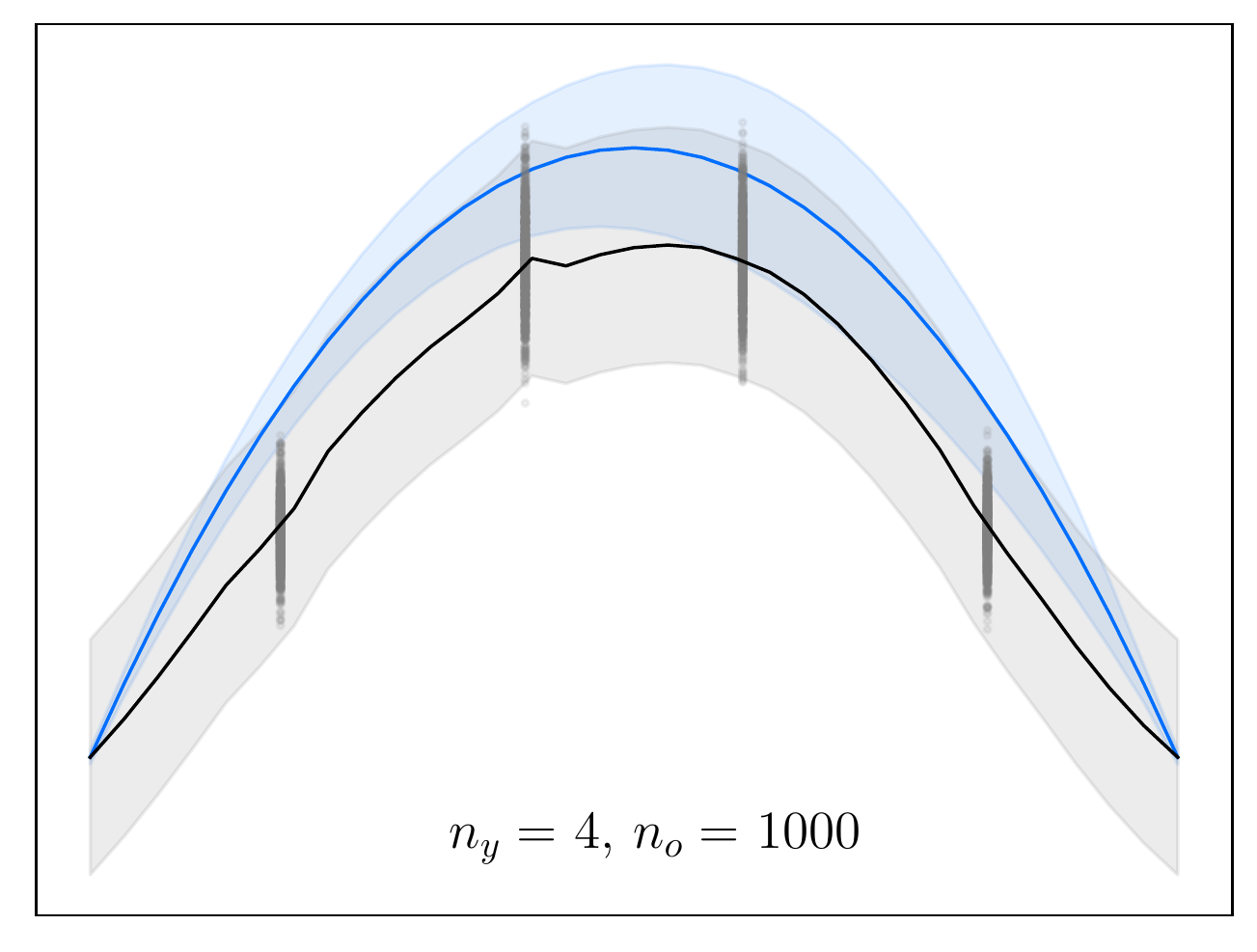} 
	\\
	\includegraphics[width=0.24\textwidth]{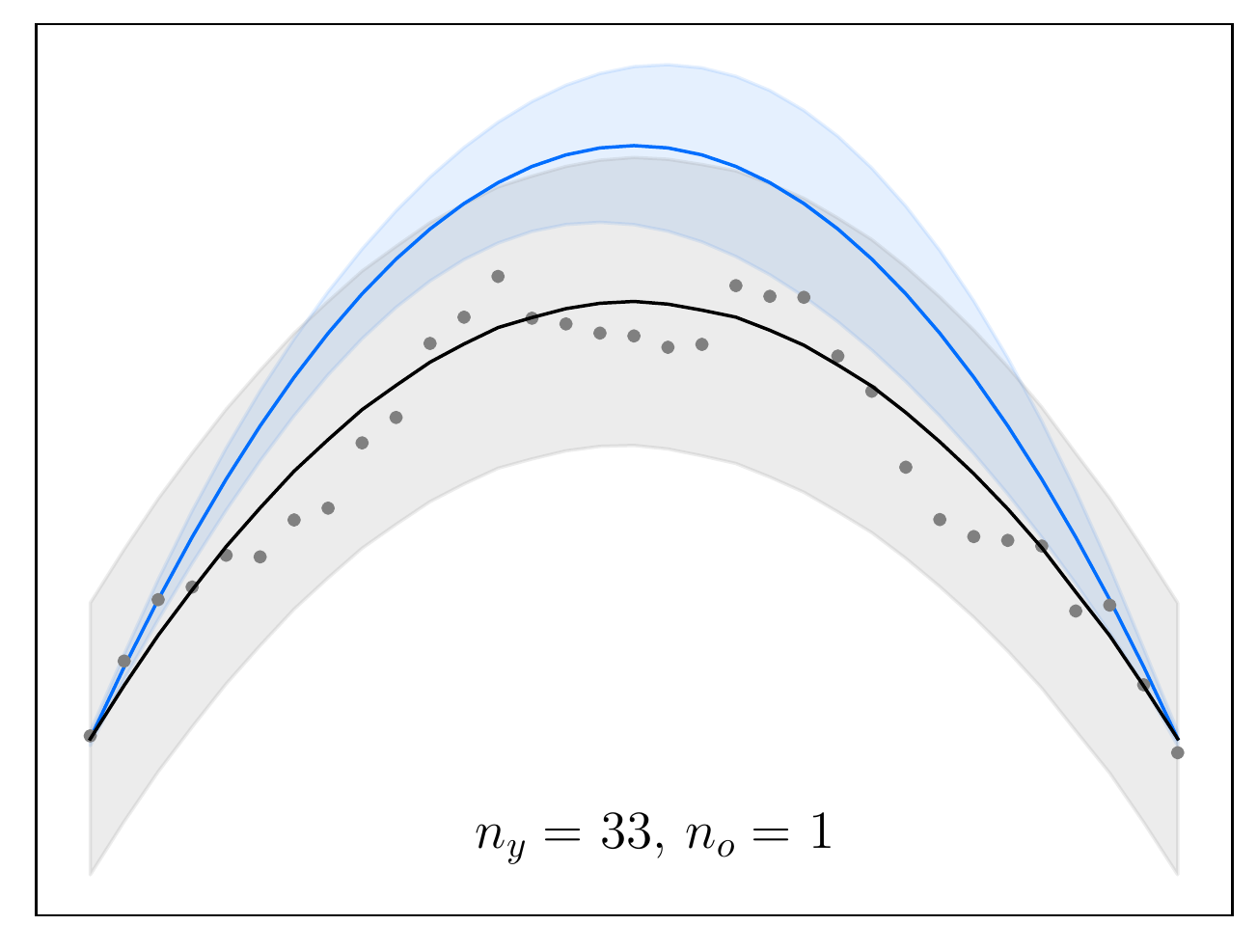}
	\includegraphics[width=0.24\textwidth]{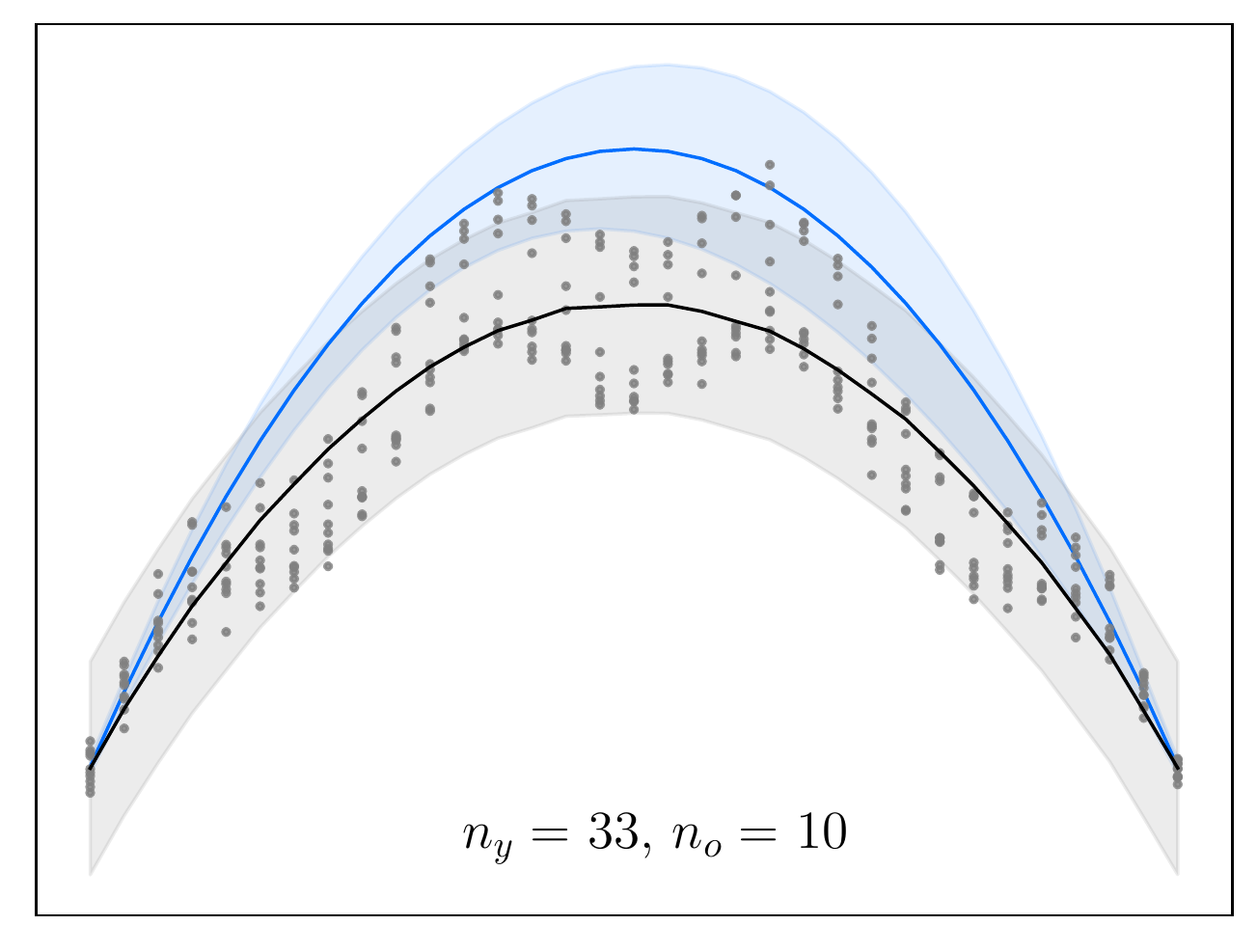}
	\includegraphics[width=0.24\textwidth]{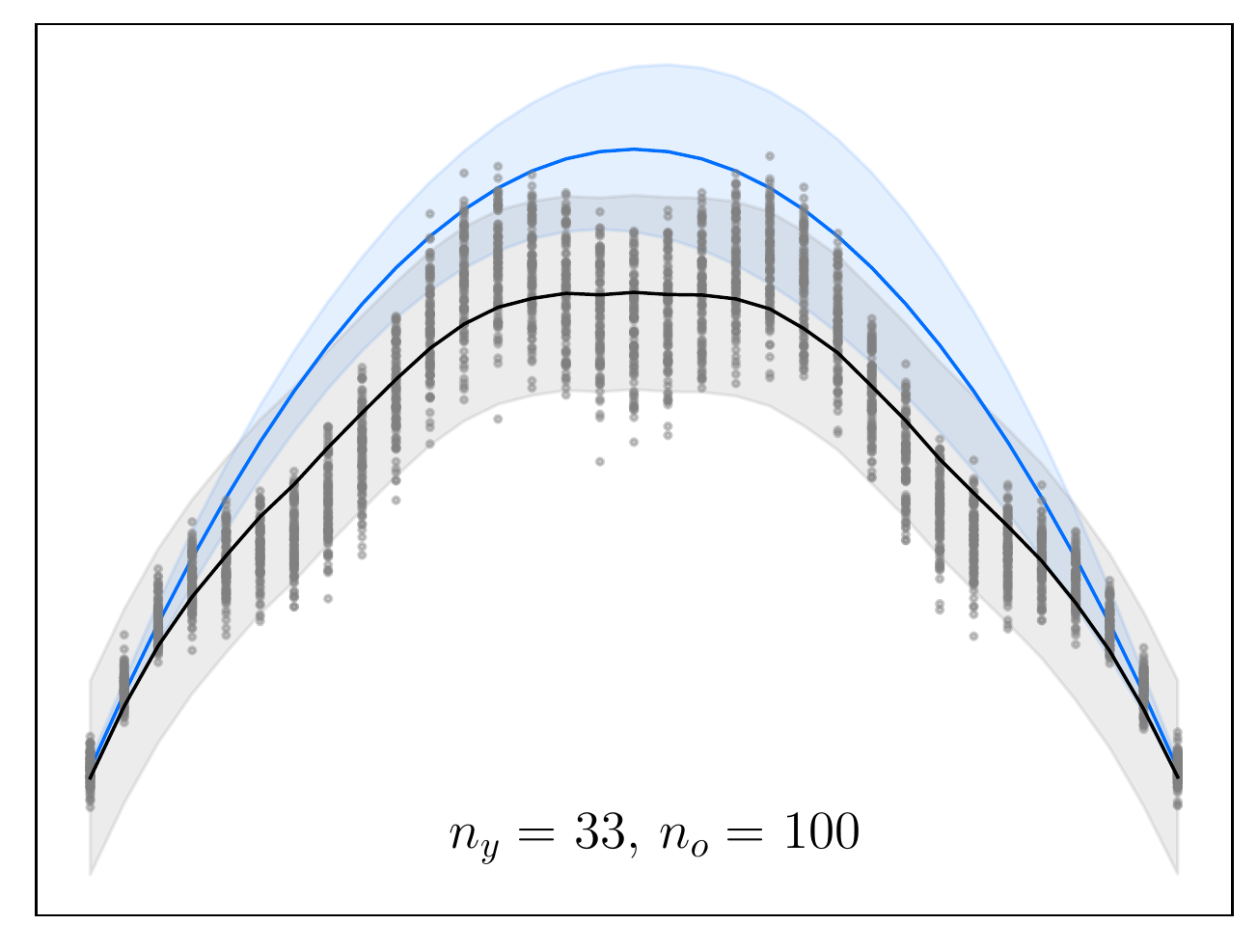}
	\includegraphics[width=0.24\textwidth]{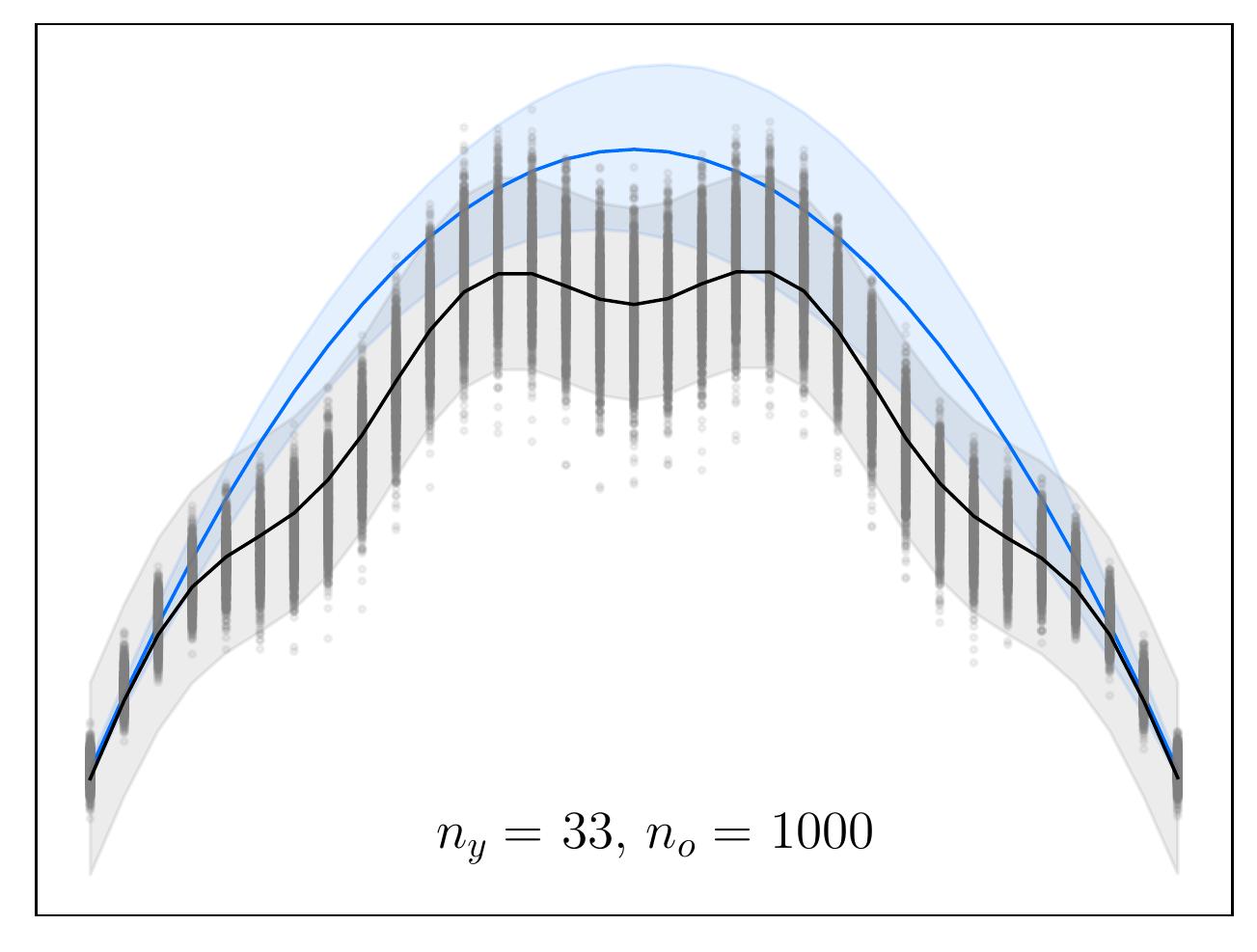} 
	\caption{One-dimensional problem with random diffusivity. Inferred true system density~$p(\vec z | \vec Y)$ conditioned on observation data (grey dots). The blue lines represent the mean~$\overline{\vec u}$ and the black lines the conditioned mean~$\overline {\vec z}_{|Y}$. The shaded areas denote the corresponding $95\%$ confidence regions. In each row the number of sensors~$n_y$ and each column the number of readings~$n_o$ is constant. \label{fig:1dPzY-LHS}} 
\end{figure}

We evaluate next the posterior finite element density~$p(\vec u | \vec Y)$.  In Figure \ref{fig:1dPuhY-LHS} the posterior and the prior finite element densities are compared for different~$n_y$ and~$n_o$ combinations. The prior mean and the~$95\%$ confidence region are slightly asymmetric due to the asymmetry of the diffusion coefficient, see Figure~\ref{fig:poisson1D-LHS-exA}. When~$n_y$ and~$n_o$ are low the posterior mean~$\overline{\vec u}|_{\vec Y}$ and the~$95\%$ confidence regions  are asymmetric as well. However, with increasing~$n_y$ and~$n_o$ the posterior mean converges to the symmetric true process mean~$\overline{z}(x)$ and the posterior covariance converges to zero. The inferred true system density~$p(\vec z | \vec Y)$ is shown in Figure \ref{fig:1dPzY-LHS}. It can be observed that for $n_y = 33$ observation locations and increasing number of readings~$n_o$  that both the mean~$\overline{\vec z}|_{\vec Y}$ and the covariance~$\vec C_{z|Y}$  of the inferred posterior show good agreement with the known true system density. These results assert the success of statFEM in inferring the true system density given a reasonable amount of observation data.

%
\subsubsection{Posterior diffusion coefficient density}
%
We contrast now statFEM to conventional Bayesian treatment of inverse problems. In statFEM our primary aim is to infer the posterior finite element and true system densities~$p(\vec u | \vec Y)$ and~$p(\vec z | \vec Y)$. While  in inverse problems the aim is to infer the posterior densities of certain model parameters, like the posterior diffusion coefficient density~$p(\vec \kappa^{(a)} |  \vec Y)$. The vector~$\vec \kappa^{(a)} $ represents the coefficients, or the parameters, used for discretising the diffusion coefficient. Commonly used approaches for discretising the diffusion coefficient can be expressed in the form
\begin{equation} \label{eq:kappaInterpol}
	\kappa(x) = \sum_i \psi_i  (x) \kappa_i^{(a)} = \vec \psi  (x)^\trans \vec \kappa^{(a)}	\, ,
\end{equation}  
where~$\vec \psi(x)$ is a set of basis functions and  $\vec \kappa^{(a)}$ are their coefficients. Usually, the~$\vec \psi(x)$ are chosen as either the Lagrange, B-spline, Karhunen-Loeve or Gaussian process basis functions, see~\cite{bui2013computational,vigliotti2018bayesian,marzouk2009dimensionality}. 

In this example, we discretise the diffusion coefficient using Gaussian process basis functions. Specifically,~$\vec \kappa^{(a)}$ are the diffusion coefficient values at the five {\em anchor} points with the coordinates \mbox{$\vec x^{(a)} =  (0, \, 0.25, \, 0.5, \, 0.75, \, 1 )^\trans$}. We obtain the basis functions $ \vec \psi^{(a)}$ by conditioning a Gaussian process on the anchor point coefficients~$\vec \kappa^{(a)}$, see~\eqref{eq:eqMeanGPregression} in~\ref{app:GPregress}.  The chosen Gaussian process has a zero mean and a squared exponential kernel with~\mbox{$\sigma_{\kappa^{(a)}} =1$} and~\mbox{$\ell_{\kappa^{(a)}} = 0.32$.} The obtained basis functions are infinitely smooth owing to the chosen squared exponential kernel and their shape is controlled by~$\ell_{\kappa^{(a)}}$. 

The vector of the element centre diffusion coefficients~$\vec \kappa(\vec X^{(c)})$ required for finite element analysis is obtained by evaluating~\eqref{eq:kappaInterpol}. When the source~$\vec f$ is deterministic, as is usually the case in inverse problems, the finite element density~\eqref{eq:feDensity2} conditioned on the anchor point coefficients~$\vec \kappa^{(a)}$ is given by  
\begin{equation} \label{eq:diracForward}
	p (\vec u | \vec \kappa^{(a)}  ) = \vec \delta \left  (\vec u - \vec A \left (\vec \kappa^{(a)}  \right)^{-1}  \vec f \right )  \, , 
\end{equation}
where~$\vec \delta (\cdot)$ denotes a Dirac measure. That is, the forward problem in conventional inverse problems is deterministic. On the contrary, in statFEM the forward problem has always a well defined distribution as given by~\eqref{eq:forwardDensityFinal}. The marginal likelihood corresponding to~\eqref{eq:diracForward} is according to~\eqref{eq:magrLikely} given by
\begin{equation} 
	p(\vec y |  \vec \kappa^{(a)}, \, \dotsc  ) = \int p(\vec y | \vec u, \, \dotsc  )  p (\vec u | \vec \kappa^{(a)}  ) \D \vec u =  \set N \left  ( \rho \vec P \vec A \left ( \vec {\kappa}^{(a)} \right )^{-1} \vec f , \, \vec C_d + \vec C_e  \right ) \, ,
\end{equation}
where we made in contrast to~\eqref{eq:marginalLevald} the conditioning of the marginal likelihood on the anchor point coefficients and other hyperparameters, represented by dots,  explicit.

In this example we choose a deterministic source~$f(x)=1$ and consider~$n_y = 33$ observation points located at each of the nodes of the uniform finite element mesh with~\mbox{$n_e=32$} elements and~\mbox{$n_o \in \{ 1, \, 5, \, 25, \, 50 \}$} repeated readings. We sample the synthetic observation matrix~$\vec Y \in \mathbb R^{n_y \times n_o}$ from the Gaussian
\begin{equation} \label{eq:yInversionSynthetic}
	p(\vec y | \vec \kappa^{(a)}, \, \dotsc  ) = \set N \left ( \vec P  \vec A \left  (  \vec \kappa_z^{(a)} \right )^{-1} \vec f , \,  0.01^2 \vec I  \right ) \, ,
\end{equation}
where~\mbox{${\vec \kappa}^{(a)}_z = (\ln 0.7, \, \ln 1, \, \ln 0.7, \, \ln 0.4, \, \ln 0.7 )^{\trans}$} are the true coefficients at the respective anchor points $\vec x^{(a)}$. In the underlying statistical generating model~\eqref{eq:decomposition}  the hyperparameters have been chosen with \mbox{$\rho=1$}, \mbox{$\sigma_d = 0$} and~\mbox{$\sigma_e  = 0.01$.} In addition,~$\rho$ and~$\sigma_d$ are chosen to be deterministic, i.e. their priors and posteriors are fixed to the given values, and~$\sigma_e$ is a random variable. Under these conditions the statistical generating model reduces to~\mbox{$\vec y = \vec u + \vec e$} as is usually assumed in inverse problems.
 
\begin{figure}[h!] 
	\centering
	\subfloat[$p( \kappa_0^{(a)} | \vec Y)$]{\includegraphics[width=0.32\textwidth]{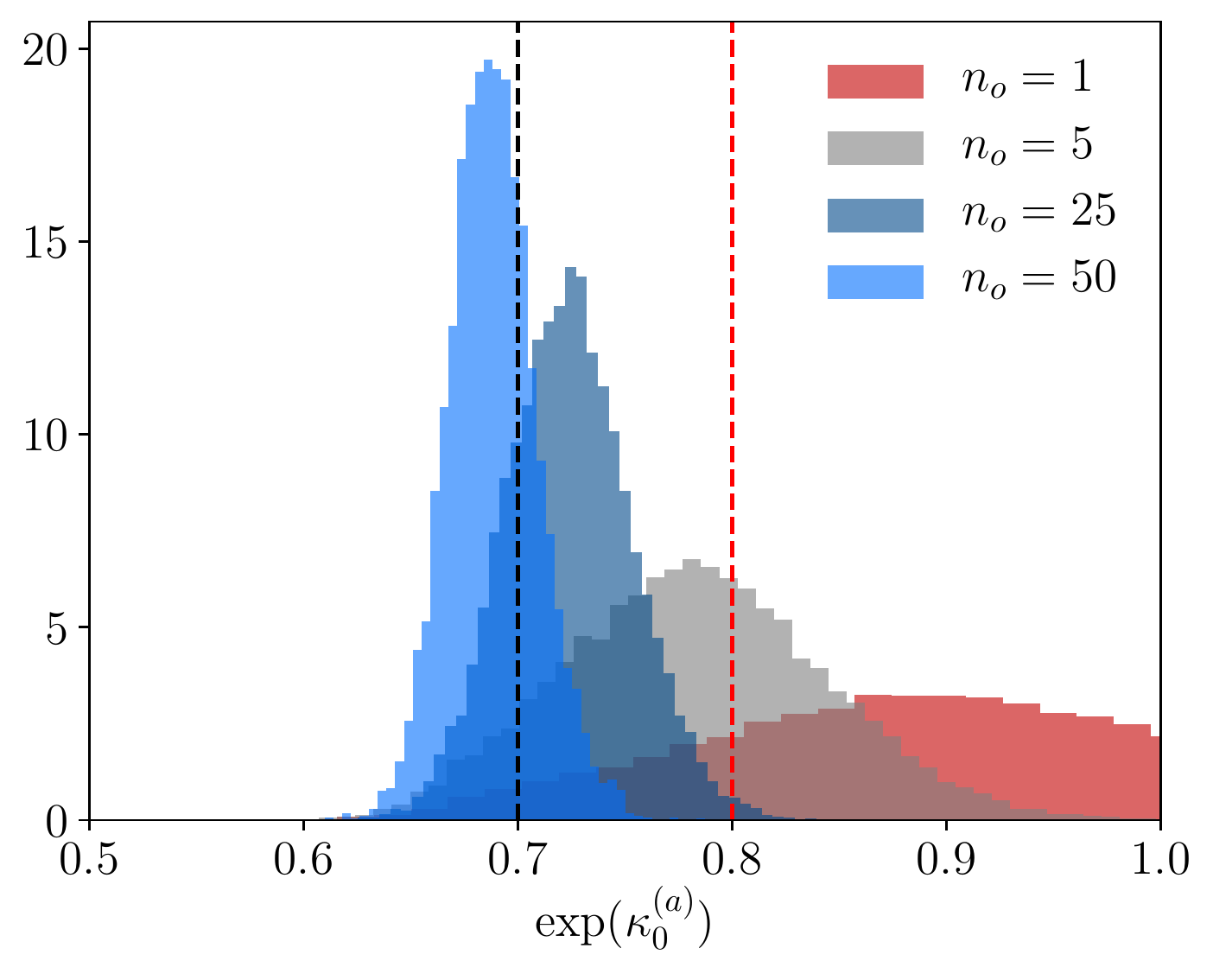}} \hfill
	\subfloat[$p( \kappa_1^{(a)} | \vec Y)$]{\includegraphics[width=0.32\textwidth]{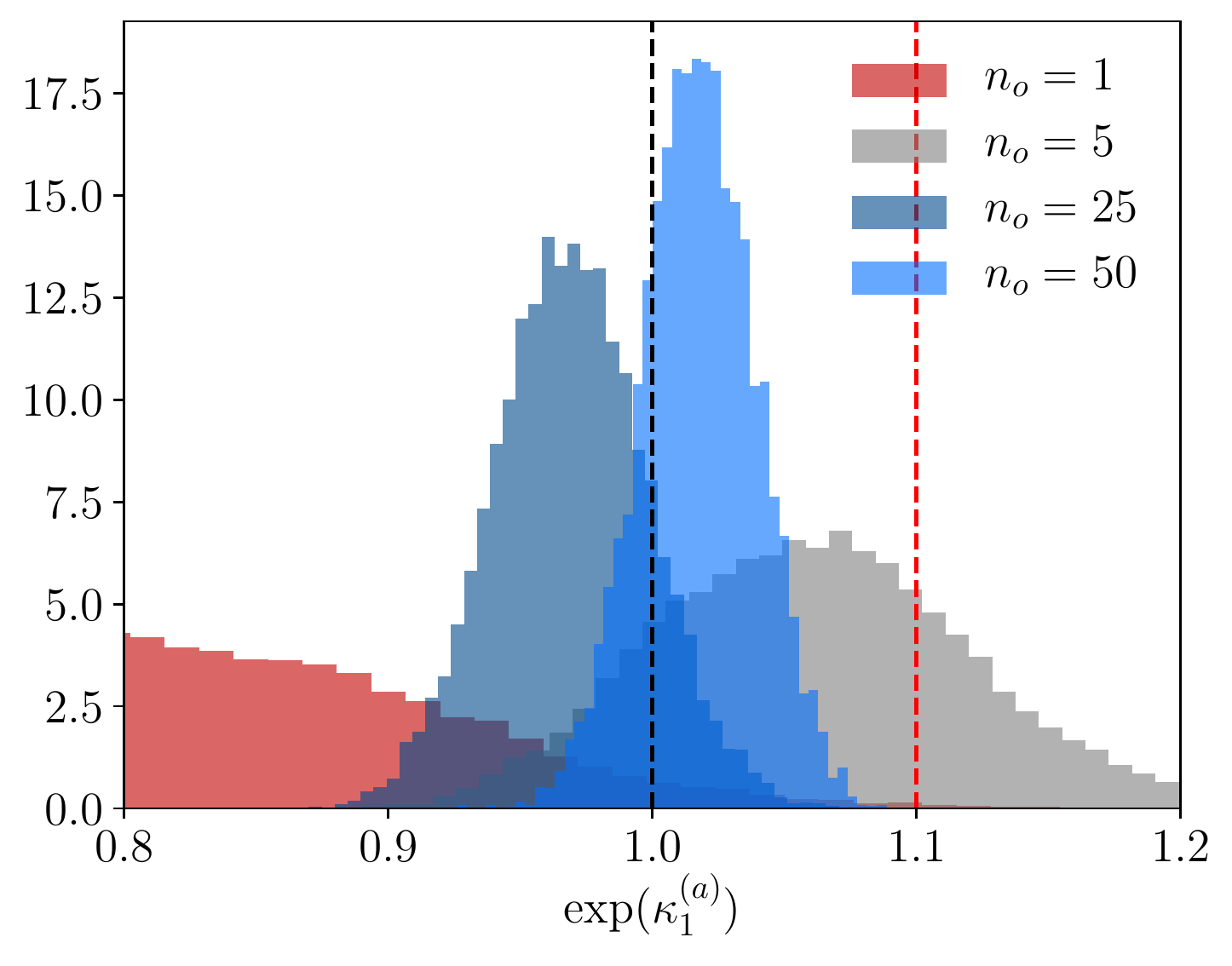}} \hfill
	\subfloat[$p( \kappa_2^{(a)} | \vec Y)$]{\includegraphics[width=0.32\textwidth]{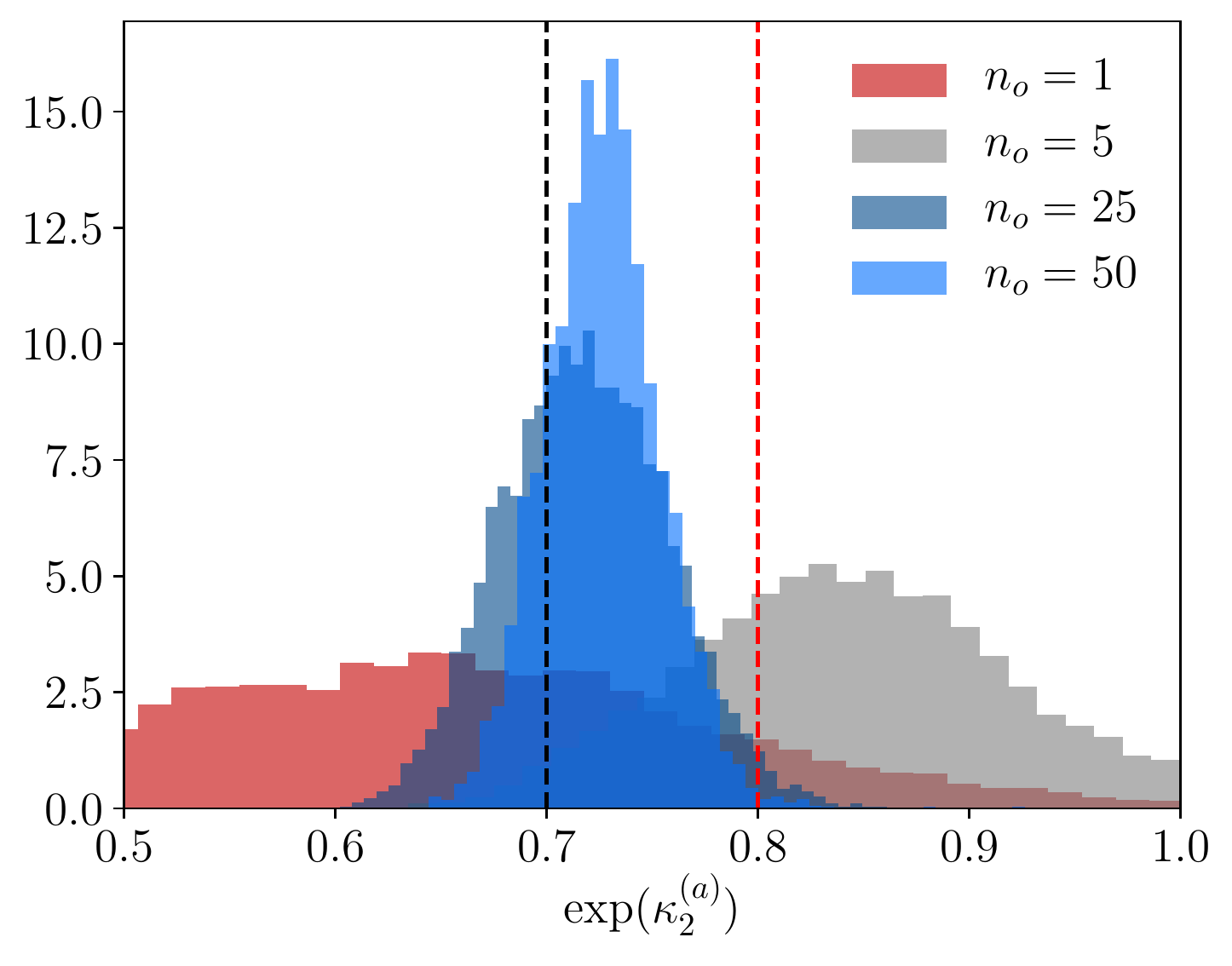}} \\
	\subfloat[$p( \kappa_3^{(a)}| \vec Y)$]{\includegraphics[width=0.32\textwidth]{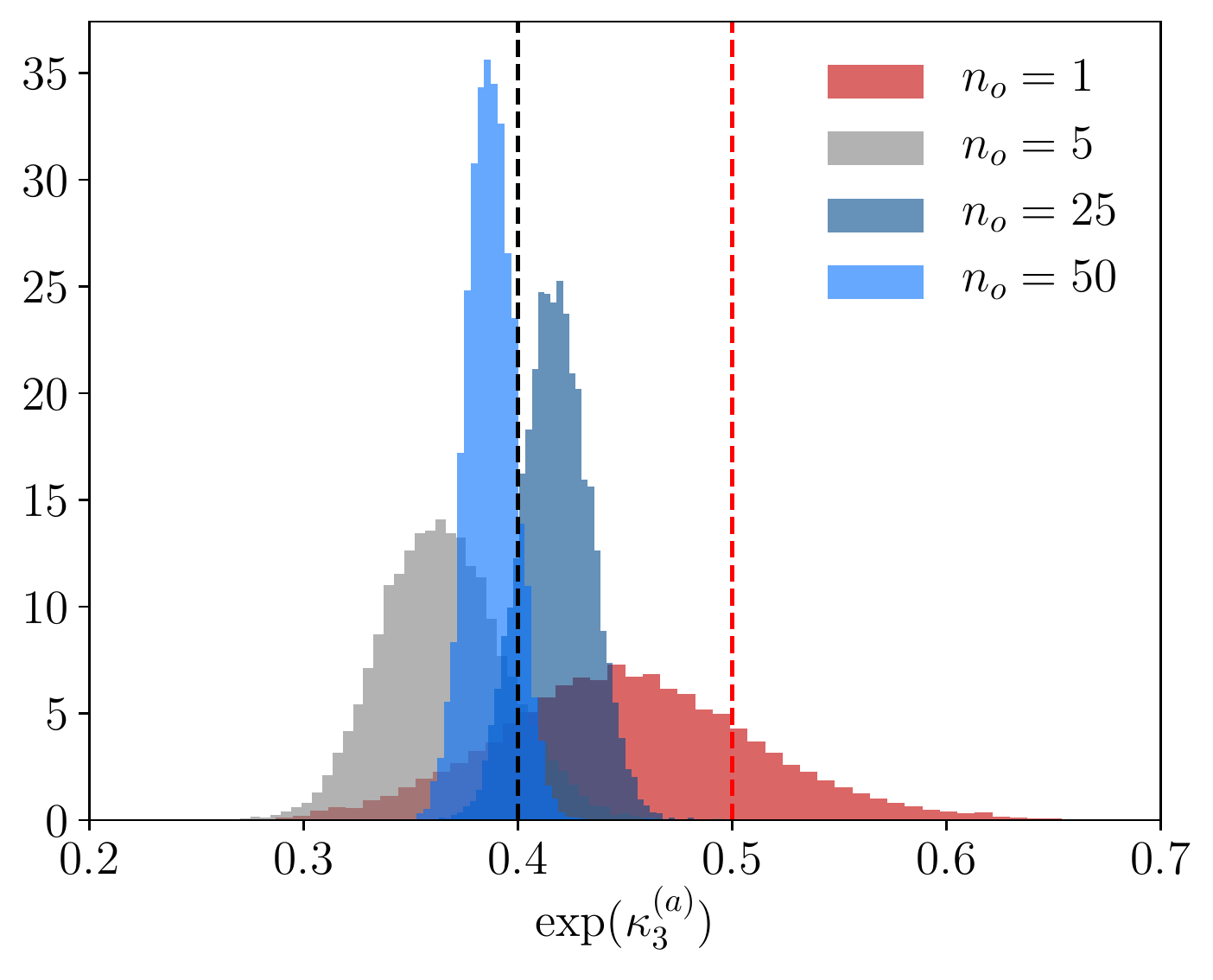}} \hfill
	\subfloat[$p( \kappa_4^{(a)} | \vec Y)$]{\includegraphics[width=0.32\textwidth]{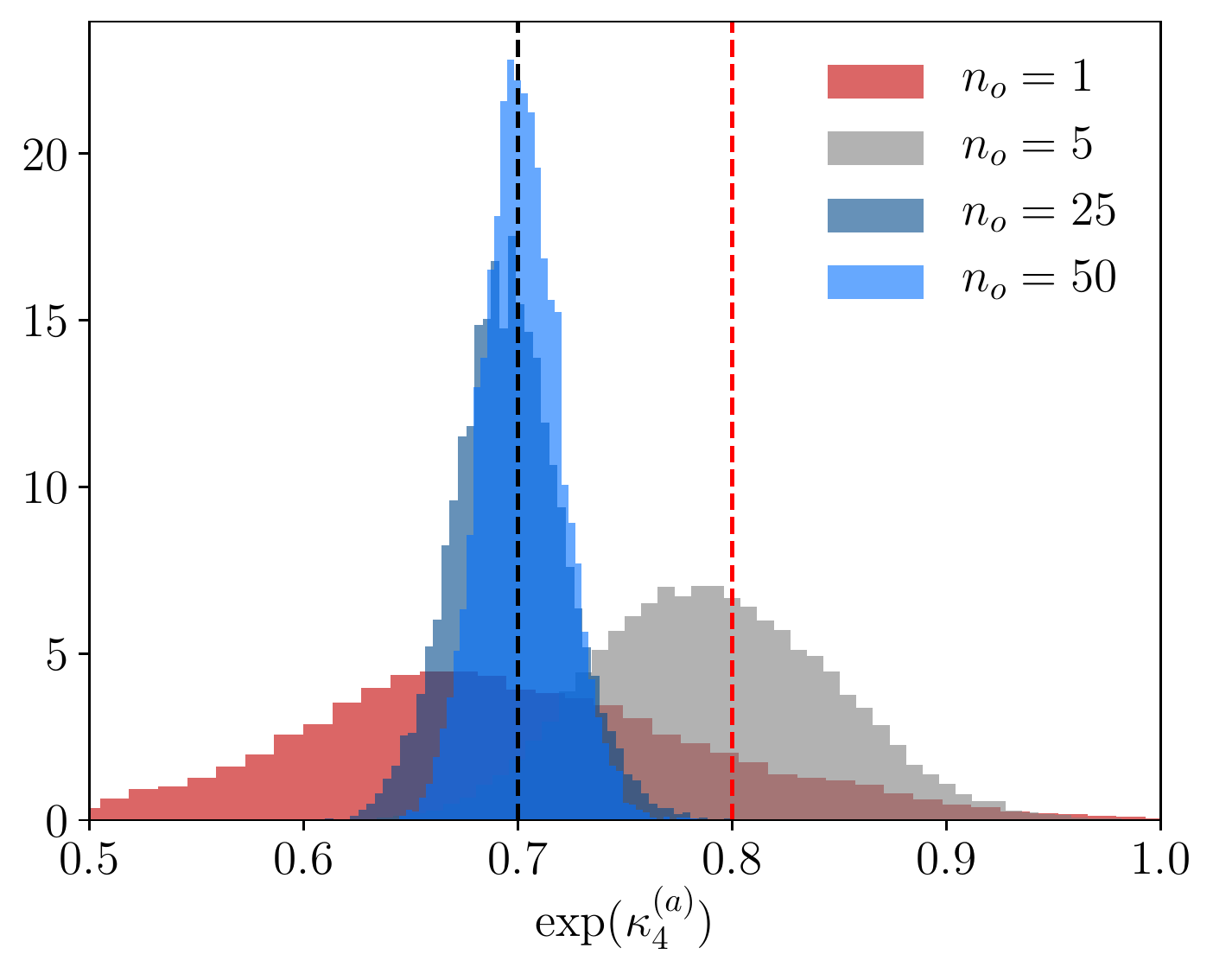}} \hfill
	\subfloat[$p(\sigma_e | \vec Y)$]{\includegraphics[width=0.32\textwidth]{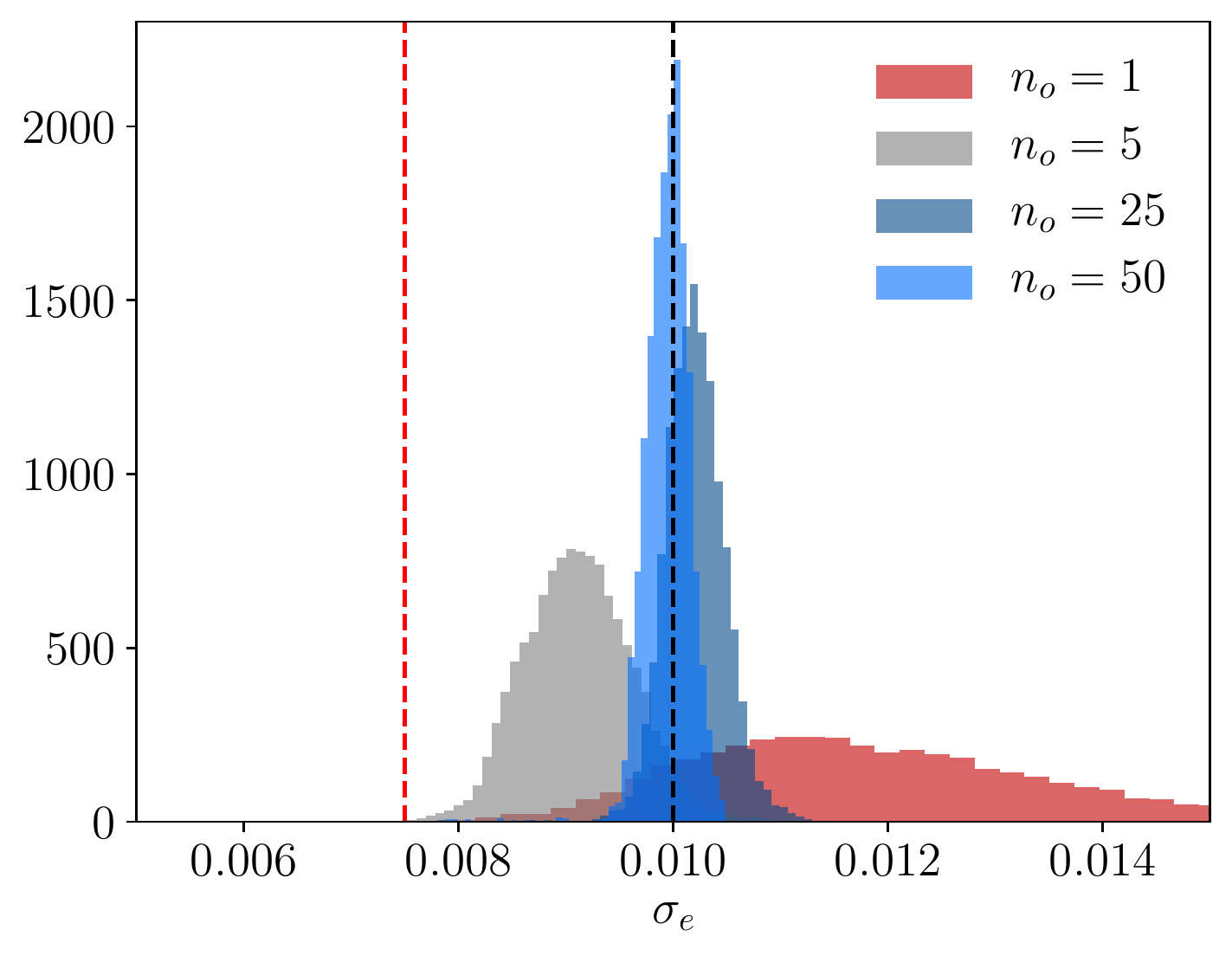}} 
	\caption{One-dimensional conventional Bayesian inverse problem. Posteriors of the parameters~$\vec w = ({ \vec \kappa^{(a)}}^\trans, \,  \sigma_e )^\trans$ for~$n_o \in \{ 1, \, 5, \,  25, \, 50 \}$  obtained with MCMC. The red vertical lines indicate the mean of the prior and the black lines the true generating process parameters.  }
	\label{fig:inferredParamLHS}
%
\vspace{1.5em}
%
	\centering
	\includegraphics[width=0.475\textwidth]{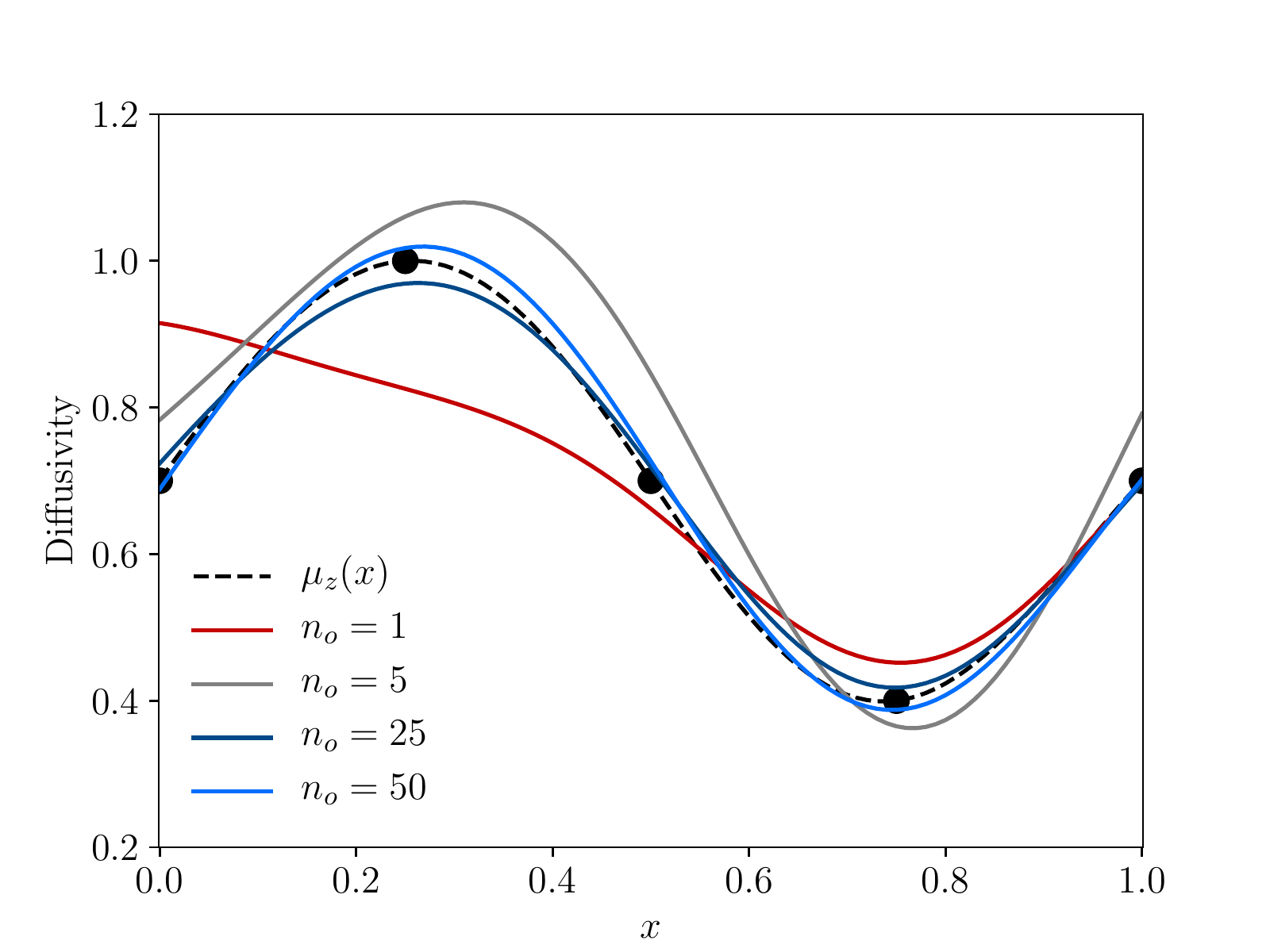}
	\caption{One-dimensional conventional Bayesian inverse problem. Comparison of the inferred diffusion coefficient $\mu(x)$ with the true diffusion coefficient $\mu_z(x)$ (dashed black line). The five dots denote the anchor points for~$\mu_z(x)$. \label{fig:trueDiff}}
\end{figure}

In hyperparameter learning we consider~$\vec w = ( {\vec \kappa^{(a)}}^\trans, \,  \sigma_e )^\trans$  as the unknown parameters to be inferred from the observation matrix~$\vec Y$ sampled from~\eqref{eq:yInversionSynthetic}. According to~\eqref{eq:bayesParams} the hyperparameter posterior is given by~\mbox{$p(\vec w| \vec Y) \propto p(\vec Y | \vec w) p (\vec w)$.} We choose as priors the Gaussian densities 
\begin{align}
	p (\vec \kappa^{(a)} )  = \set N \left ( (\ln 0.8, \, \ln 1.1, \, \ln 0.8,\, \ln 0.5, \, \ln 0.8)^\trans, \,0.0025 \,  \vec I  \right )  \, , \quad  p ( \sigma_e)  = \set N ( 0.0075  , \, 4 \cdot 10^{-6}\  ) \, .   
\end{align}
Notice that the mean of the priors are different from the true generating process parameters used in~\eqref{eq:yInversionSynthetic} for sampling the synthetic observations. We sample the posterior~$p(\vec w| \vec Y)$ with MCMC using $50000$ iterations with an acceptance ratio of around~$0.3$. Each~$n_o$ requires one MCMC run so that in total four runs are performed. The inferred densities~$p(\vec w  | \vec Y)$ are shown in Figure~\ref{fig:inferredParamLHS}. The red vertical lines indicate the mean of the priors and the black lines the mean of the true generating process.  Even with only one set of readings, $n_o=1$, the  mean of the posteriors are visually very different from the mean of the priors. When~$n_o$ becomes larger the mean of the posteriors converge indeed towards the true generating process parameter values and the standard deviation of the posteriors become smaller. In Figure~\ref{fig:trueDiff} the mean of the true diffusion coefficient~$\mu_z(x)$ and the inferred diffusion coefficient~$\mu(x)$ over the domain are shown. The anchor points and their coefficients are denoted by dots. As visually apparent with increasing number of readings the inferred diffusion coefficient converges towards the true diffusion coefficient. 
%

\subsubsection{Finite element mesh selection} 
%
We consider Bayesian model comparison for selecting a finite element mesh which can best explain the observed data. That is, for a given observation matrix~$\vec Y^{(i)}$ we are looking for the finite element mesh~$\set M_j$ with the highest posterior probability~$p(\set M_j | \vec Y^{(i)})$. In this example, we examine four different uniform element meshes~\mbox{$\{ \set M_j\}_{j=1}^4$} with the respective finite element sizes~$h_1=1/4$, $h_2= 1/8$, $h_3 = 1/16$ and $h_4 =1/32$. Furthermore, we choose a deterministic diffusion coefficient with~$\mu(x) = 1$ and a random source with the mean
\begin{align}
	\overline{f}(x) &= \frac{\pi^2}{5} \sin(\pi x) +  \frac{49 \pi^2}{50} \sin (7 \pi x)\, 
\end{align}
and the covariance parameters $\sigma_f = 0.2$ and $\ell_f= 0.25$. 

Four synthetic observation matrices~$\left \{ \vec Y^{(i)} \right \}_{i=1}^4$ are sampled from the marginal likelihood~\eqref{eq:marginalLevald} with the scaling parameter $\rho = 0.8$, the mismatch covariance parameters $\sigma_d = 0.005$ and $\ell_d = 0.3$, and the sensor noise $\sigma_e = 0.005$. Each mesh~$\set M_i$ has the corresponding observation matrix~$\vec Y^{(i)} \in \mathbb R^{n_y \times n_o}$  with  $n_y \in \{11, \, 33\} $ observation locations and $n_o = 100$ readings. See Figure~\ref{fig:sensorPos} for the location of the observation points. Thus, each column of~$\vec Y^{(i)}$  is sampled from the marginal likelihood   
\begin{equation} \label{eq:generatingModelSelect}
	p \left ( \vec y | \set M_{i}\right )  = \set N \left  ( 0.8 \vec P \overline { \vec u} ( \set M_{i}  ),  \, 0.64 \vec P\vec C_u  ( \set M_{i}  ) \vec P^{\top} + \vec C_d + \vec C_e \right ) \, ,
\end{equation}
where the finite element mean~$\overline { \vec u} (\set M_{i}) $  and covariance~$\vec C_u (\set M_{i})$ are obtained with the mesh~$\set M_i$.  

As in preceding sections, prior to computing the posterior probabilities~$p(\set M_j | \vec Y^{(i)})$ we first infer the hyperparameters \mbox{$\vec w = ( \rho, \, \sigma_d, \, \ell_d )^{\trans}$} of the statistical generating model. In doing so we choose a non-informative prior~$p(\vec w) \propto 1$. The determined hyperparameters have very similar values like the ones in the generating density~\eqref{eq:generatingModelSelect}, confirming the consistency of the proposed approach. Once the hyperparameters are known we compute the posterior probabilities~$p(\set M_j | \vec Y^{(i)}) \propto p( \vec Y^{(i)} | \set M_j) p(\set M_j) $ with the marginal likelihood~$p( \vec Y^{(i)} | \set M_j)$ given in~\eqref{eq:magrLikelyMult}. Assuming a non-informative prior~$p(\set M_j)  \propto 1$ we have~$p(\set M_j | \vec Y^{(i)}) \propto p( \vec Y^{(i)} | \set M_j) $.  
\begin{figure}[]
	\centering
	\subfloat[$\log p( \set M_j | \vec Y^{(1)})$]{\includegraphics[width=0.4\textwidth]{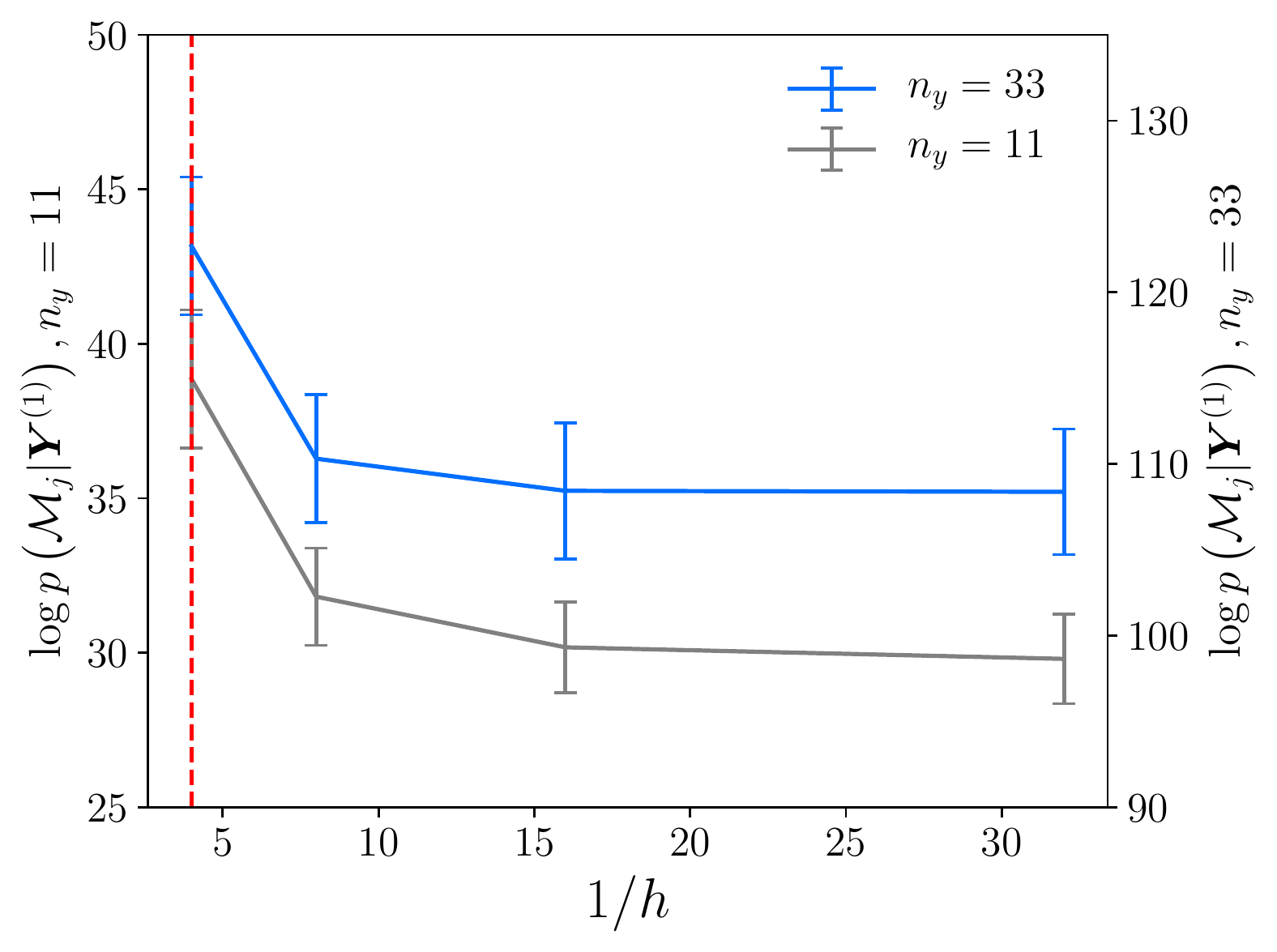}} \hspace{0.1\textwidth}
	\subfloat[$\log p( \set M_j | \vec Y^{(2)})$]{\includegraphics[width=0.4\textwidth]{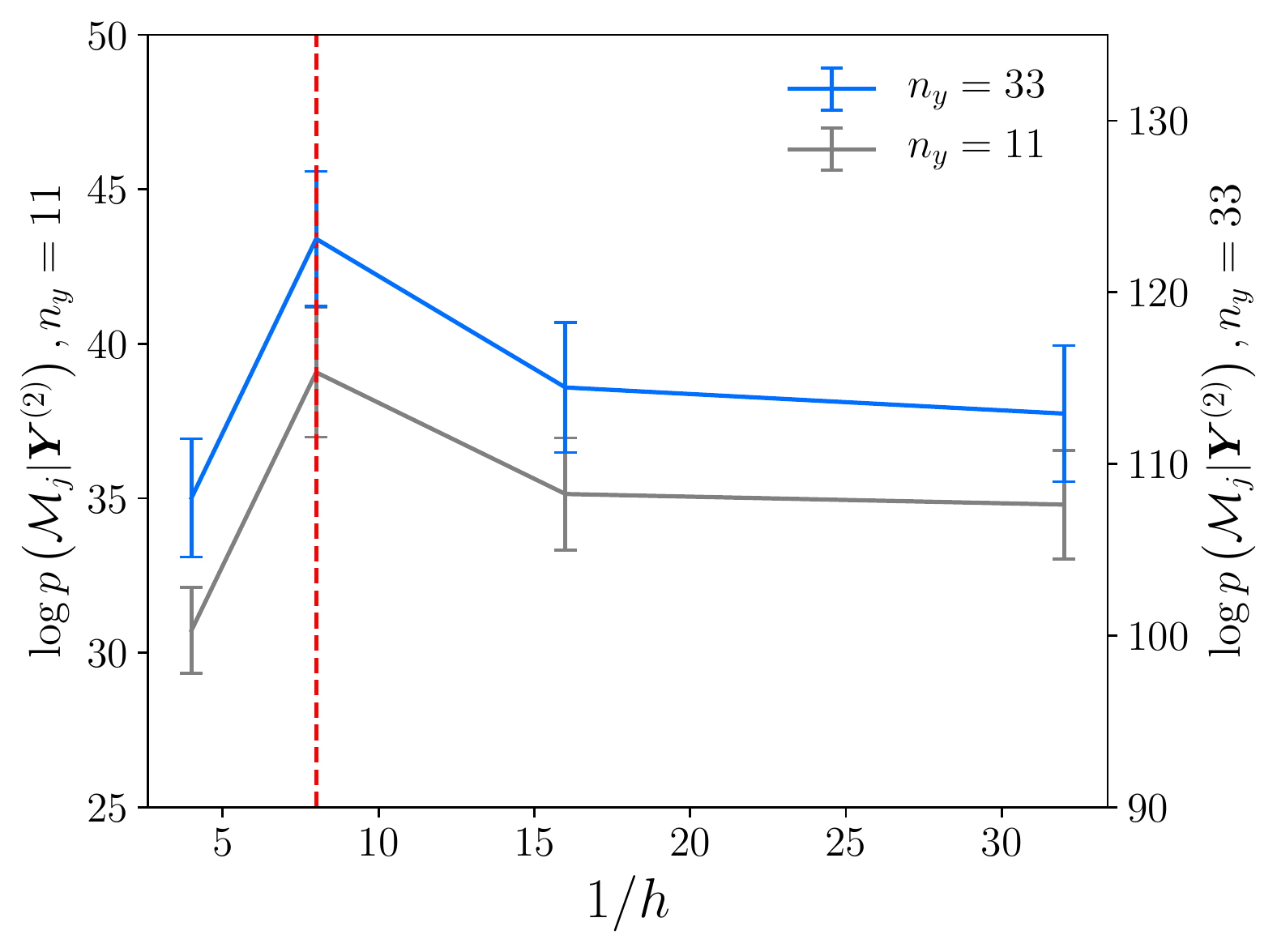}} \\
	\subfloat[$\log p( \set M_j | \vec Y^{(3)})$]{\includegraphics[width=0.4\textwidth]{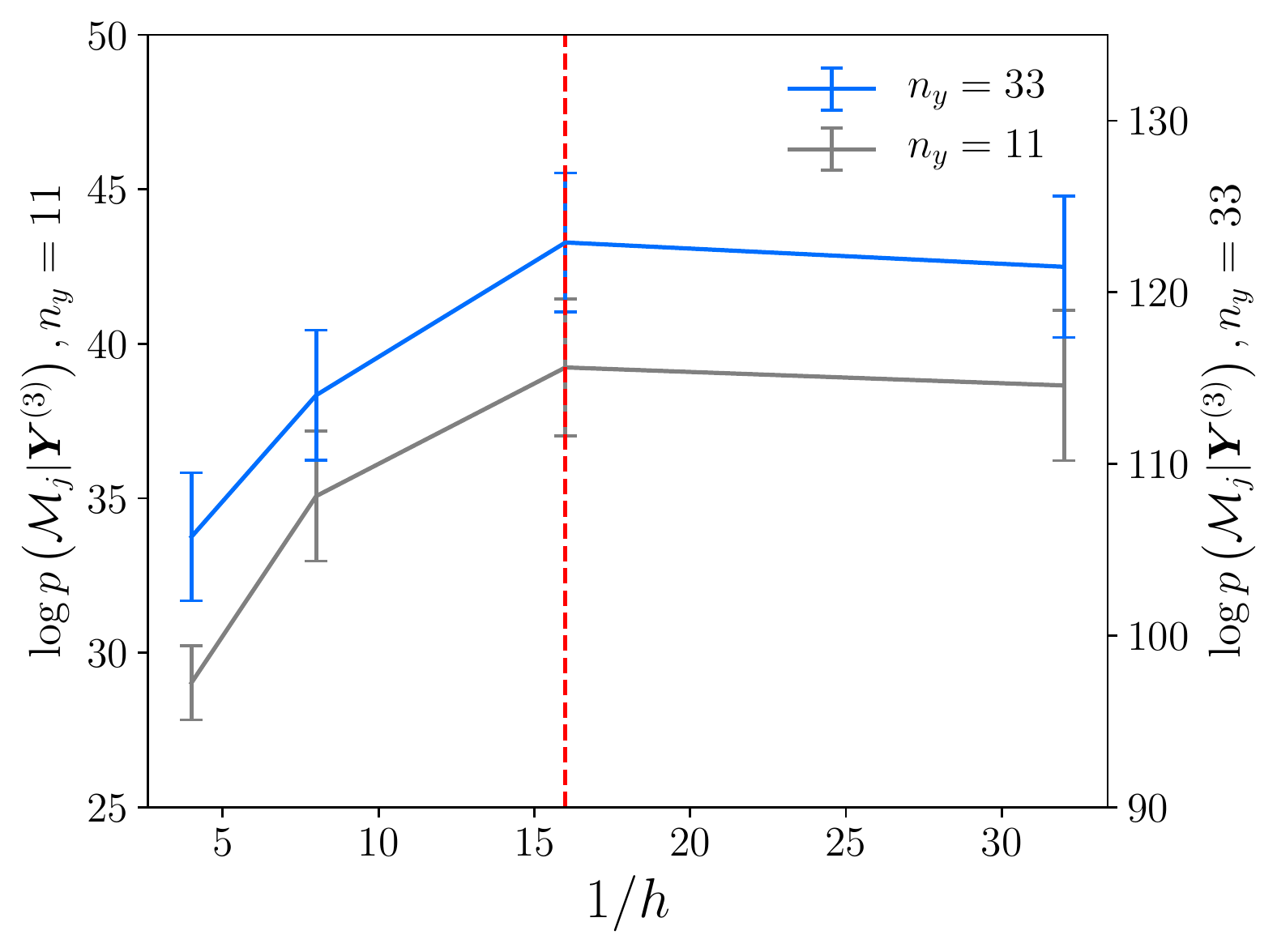}}  \hspace{0.1\textwidth}
	\subfloat[$\log p( \set M_j | \vec Y^{(4)})$]{\includegraphics[width=0.4\textwidth]{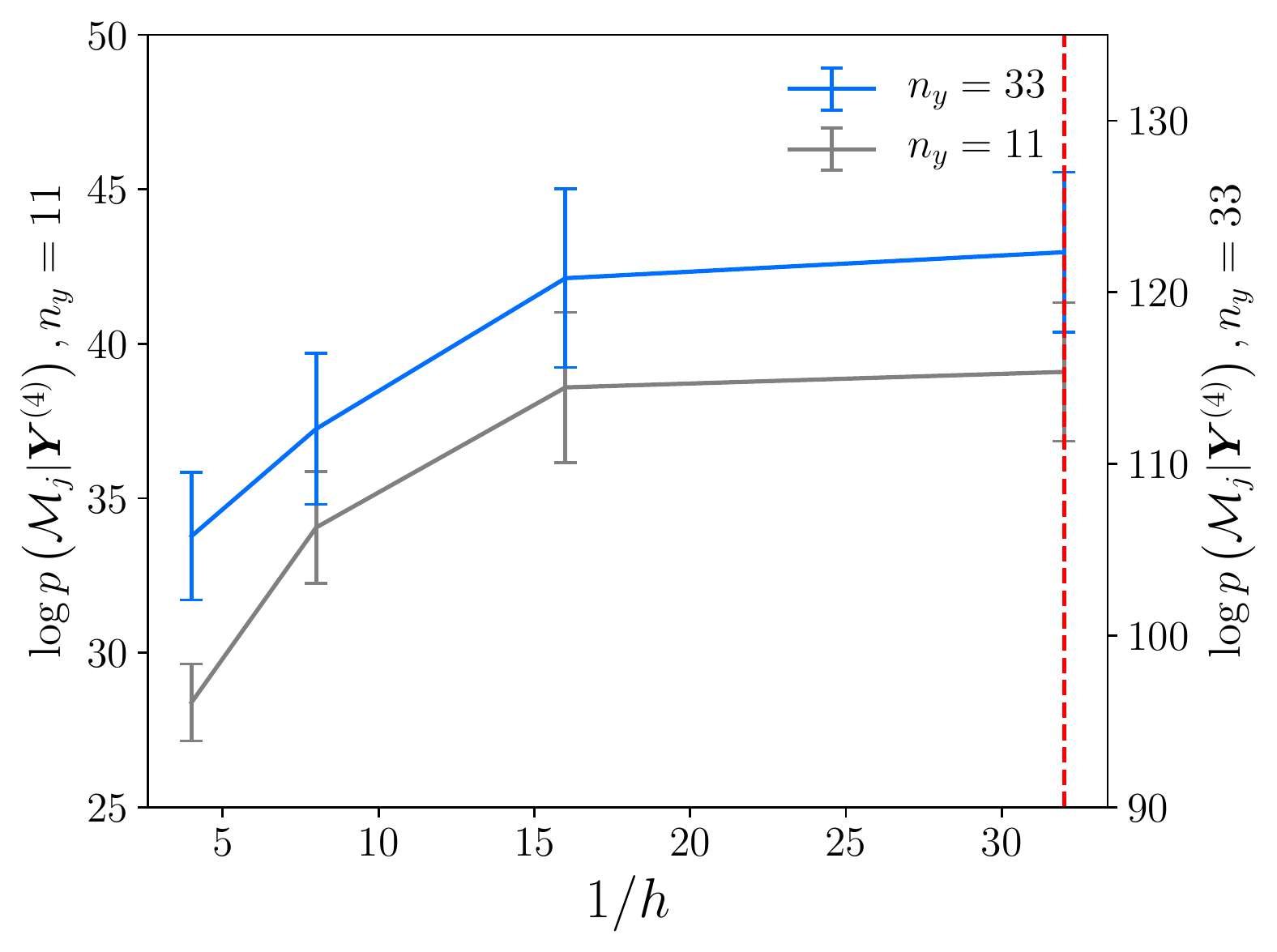}} 
	\caption{Comparison of one-dimensional models. Mesh log-posterior probability~\mbox{$p( \set M_j | \vec Y^{(i)}) $} over the inverse element size~$1/h$ for the four different meshes \mbox{$\{ \set M_j \}_{j=1}^4$} and data sets \mbox{$\left \{ \vec Y^{(i)} \right \}_{i=1}^4$.} In each plot the legend on the left refers to~$n_y=11$ and on the right to~$n_y=33$ observation locations. }
	\label{fig:modelSelect}
\end{figure}

Figure \ref{fig:modelSelect} shows the log-posterior probability~$\log p \left (\set M_j | \vec Y^{(i)} \right )$ of the four different meshes and the four different observation matrices. Considering that there are two different observation arrangements~$n_y  \in \{ 11, \, 33\}$,  there are in total $2 \times 16$ $\set M_j$ and~$ \vec Y^{(i)}$ combinations.  The bars in Figure~\ref{fig:modelSelect} indicate the standard deviations of~$\log p \left (\set M_j | \vec Y^{(i)} \right )$ obtained by sampling each~$50$ times.  Clearly, the maximum of~$\log p(\set M_j | \vec Y^{(i)})$ is always where~$j$ is equal to~$i$. That is, for a given data set the most probable mesh is the one with which the data has been generated. We can use this information to choose the most suitable mesh for a given data set. When two meshes have a similar log-posterior probability we can choose for computational efficiency the coarser one. To explain Figure \ref{fig:modelSelect}, note that an observation matrix generated with a coarse mesh will lack the higher frequencies of the solution field. For such an observation matrix there  is no need to use a complex computational model with a finer mesh. Consequently, Bayesian inference allows us to identify the simplest possible model as stipulated by the well-known Occam's razor principle~\cite[Ch. 28]{mackay2003information}.

%
\subsection{Plate with a hole \label{sec:twoD}}
\begin{figure}[tb]
          \centering
          \subfloat[][Geometry and boundary conditions \label{fig:2dDomainA}] {
            	 \includegraphics[angle=0, height=0.27\textwidth]{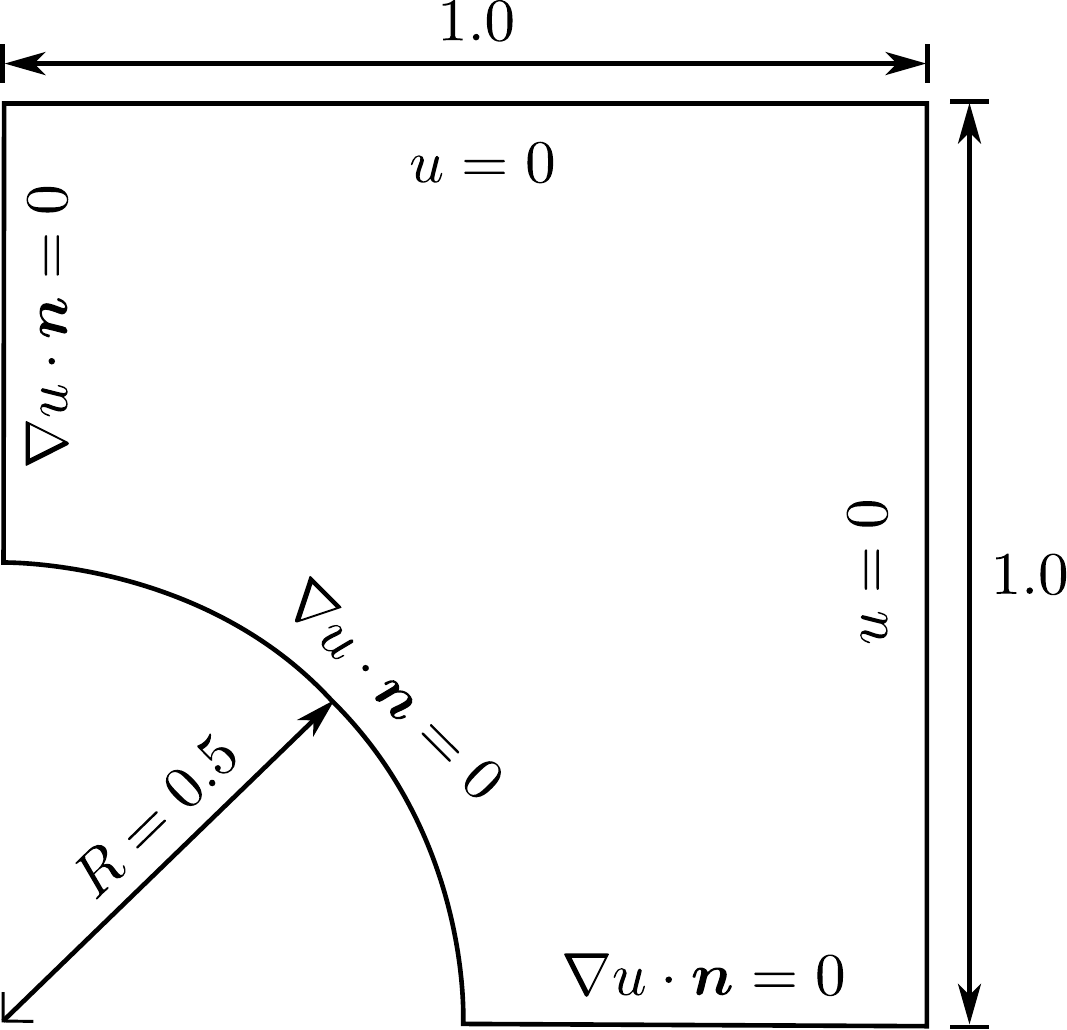} 
	 }
	 \hfill
	      \subfloat[][Isocontours of source~$f(\vec x)$ drawn from a Gaussian process with~$\overline{f}(\vec x) = 1.0$, $\sigma_f = 0.3$ and $\ell_f =0.15$ and its corresponding solution~$u_h(x)$ \label{fig:2dDomainB}] {   
	      \includegraphics[height=0.19\textheight]{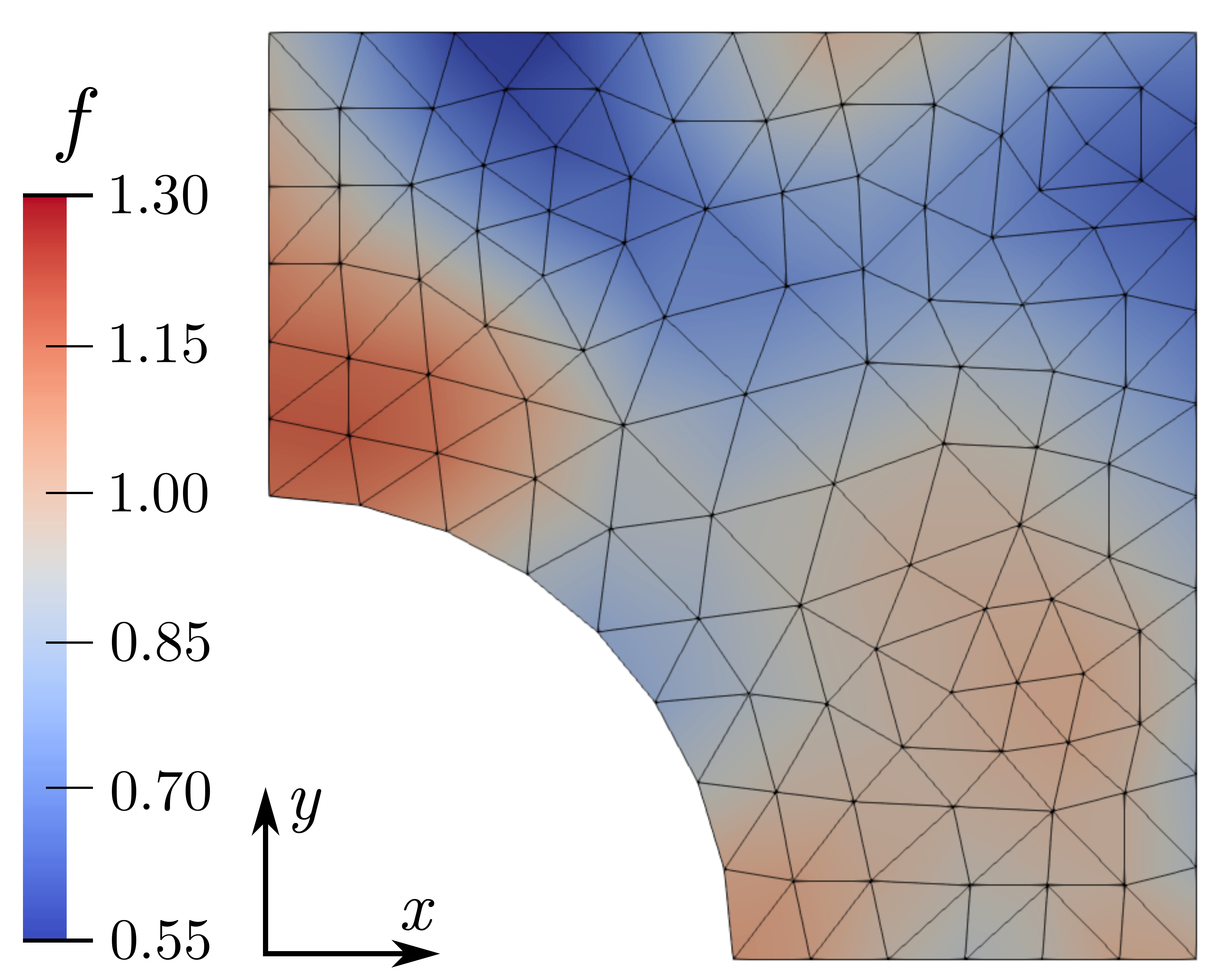} \hspace{0.005\textwidth}
	      \includegraphics[height=0.19\textheight]{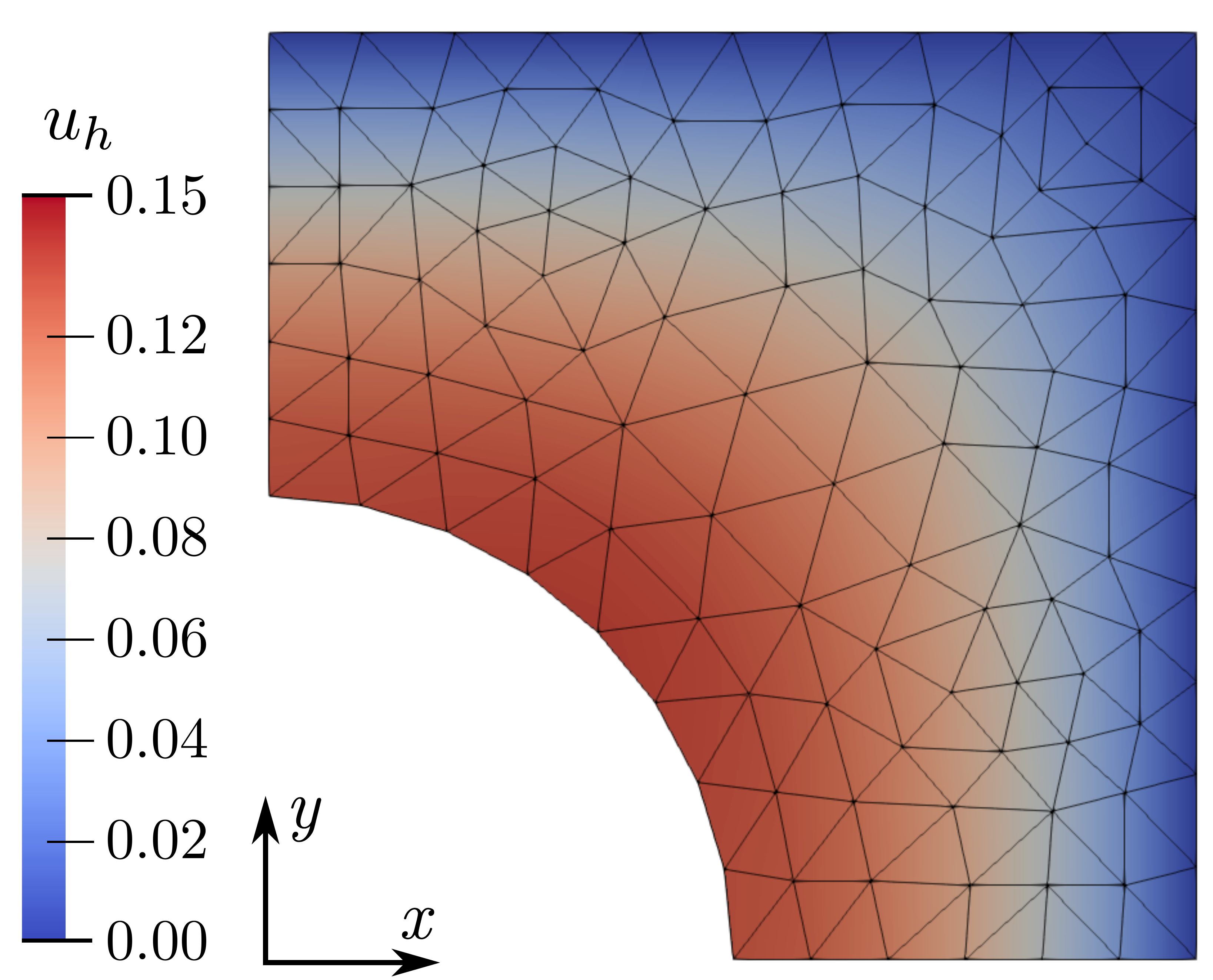}  \hspace{0.005\textwidth}	      
     	     }      
    \caption{Plate with a hole.}
\label{fig:2dDomain}
\end{figure}

%
As a two-dimensional example, we study a Poisson problem on a unit-square with a circular hole shown in Figure~\ref{fig:2dDomainA}. The boundary conditions on the five edges are chosen as indicated in the figure. In this example only the source~$f(\vec x)$ is chosen as random. We discretise the weak form with the finite element mesh shown in Figure~\ref{fig:2dDomainB} consisting of~$208$ standard linear triangular elements and $125$ nodes. 

%
\subsubsection{Posterior finite element density and system response for random source \label{sec:2dUcondY}}
%
The deterministic diffusion coefficient is assumed to be~$\mu(\vec x) = 1.0$ and the random source~$f(\vec x)$ is a Gaussian process with a mean \mbox{$\overline{f}(\vec x) = 1.0$} and a squared exponential covariance kernel with the parameters~$\sigma_f = 0.3$ and~$\ell_f =0.15 $. As introduced in Section~\ref{sec:forwardFE}, the density of the source vector~$\vec f$ is given by the multivariate Gaussian~\mbox{$p (\vec f) = \set N(\overline{\vec f}, \, \vec C_f)$} and the density of the finite element solution by~\mbox{$p (\vec u) = \set N(\vec A^{-1} \overline{\vec f}, \,  \vec A^{-1} \vec C_f \vec A^{-\trans} )$}. In Figure~\ref{fig:2dDomainB} a representative source distribution and its respective finite element solution are depicted. As visually apparent and discussed in Section~\ref{sec:forwardFE} the solution field~$u_h(\vec x)$ is significantly smoother than the source field $f(\vec x)$ owing to the smoothing property of the inverse differential, i.e. Laplace, operator. 

\begin{figure}
          \centering
                      	 \includegraphics[angle=0, height=0.35\textwidth]{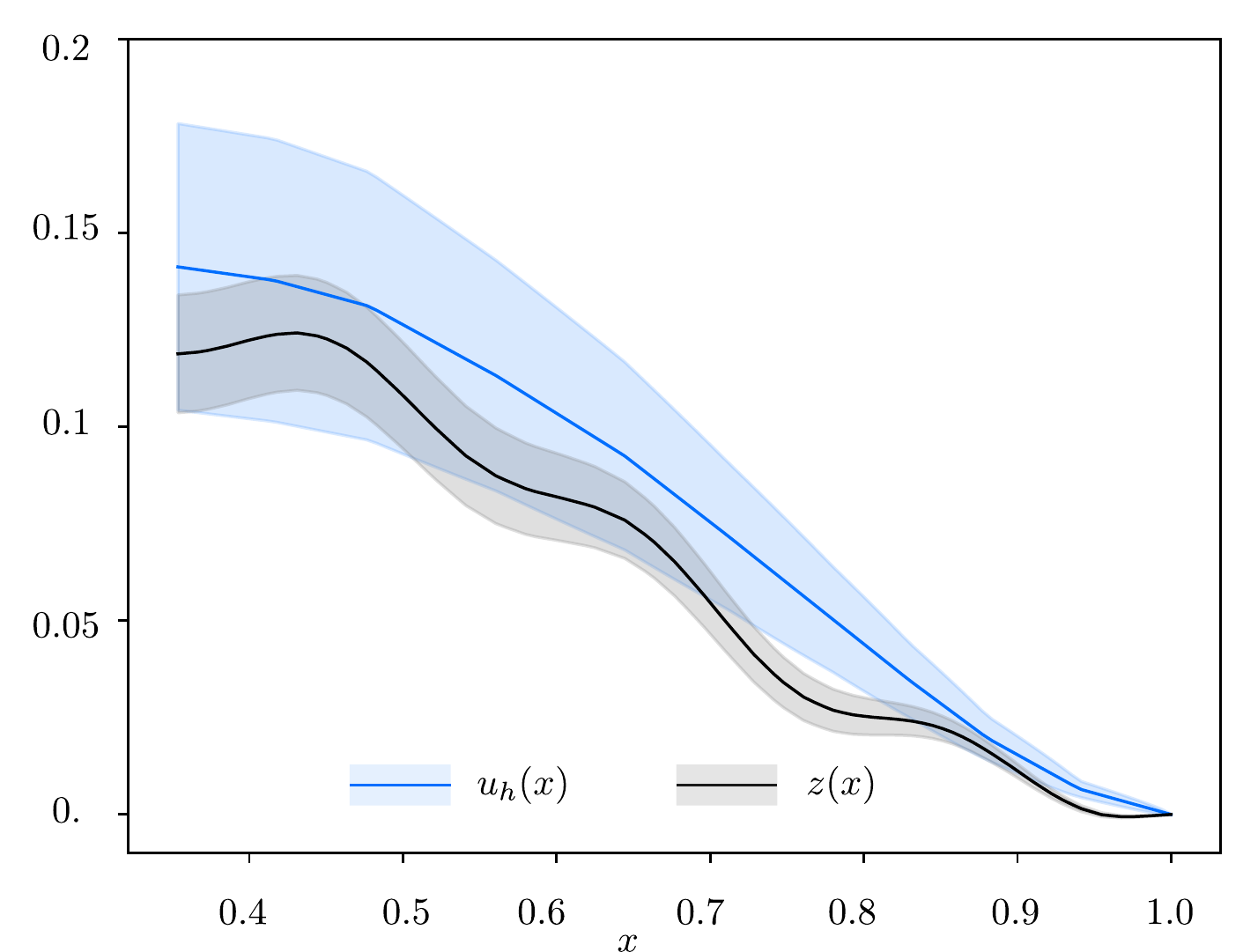} 
        \caption{Plate with a hole.  Finite element solution $u_h(x)$ and the true system response $z(x)$ along the diagonal of the domain with $x^{(2)} = x^{(1)}$.  The solid lines represent the respective means and the shaded areas the~$95\%$ confidence regions.}
\label{fig:zUcomparison}
\end{figure}

We consider as the true system response~$z(x)$ the solution of a second much finer finite element model.  The fine mesh is obtained by repeated quadrisection of the coarse mesh shown in Figure~\ref{fig:2dDomainB} and has  $53248$ elements. The random source of the fine finite element model is a Gaussian process 
\begin{equation}
g(\vec x) \sim \set {GP} \left ( \overline g (\vec x), \, c_{g} (\vec x, \, \vec x')   \right )
\end{equation}
with the mean 
\begin{equation}
	\overline g (\vec x) = \frac{1}{2}  + \frac{1}{2} \sin \left ( \pi \|\vec x\|  \right ) + 3 \sin \left (7 \pi \| \vec x \| \right ) 
\end{equation}
and the squared exponential covariance kernel with the parameters~\mbox{$\sigma_g = 0.1$}and~\mbox{$\ell_g = 0.2$.} The density of the corresponding finite element solution on the fine mesh is given by
\begin{equation}
	 \vec z =   \set N \left ( \overline {\vec z}, \, \vec C_z  \right )  = \set N \left (\vec A_g^{-1} \overline {\vec g}, \, \vec A_g^{-1}  \vec C_{u_g} \vec A_g^{-\trans} \right )  \, , 
\end{equation}
The true system response~$z(\vec x) = \sum_i \phi_i (\vec x) z_i$ and the finite element solution~$u_h(\vec x) = \sum_i \phi_i (\vec x) u_i$ are compared in Figure~\ref{fig:zUcomparison}. Both fields are plotted along the diagonal of the domain, i.e. the line with $x^{(2)} = x^{(1)}$. As described above the source terms and meshes chosen for~$z(\vec x)$ and $u_h(\vec x)$ are very different. Their difference represents the model mismatch. 

\begin{figure}
	\centering
		\includegraphics[height=0.19\textheight]{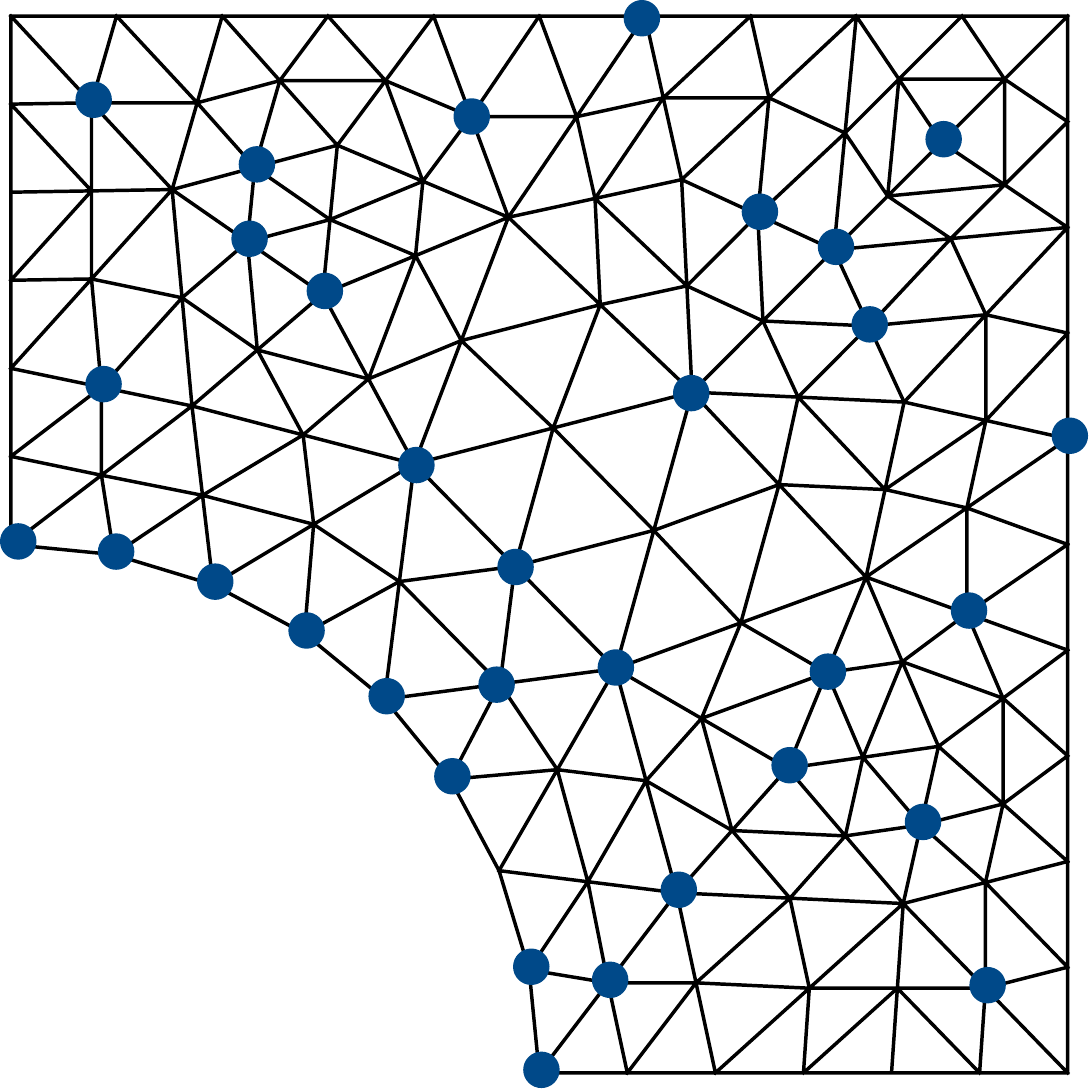}\hspace{0.075\textwidth}
		\includegraphics[height=0.19\textheight]{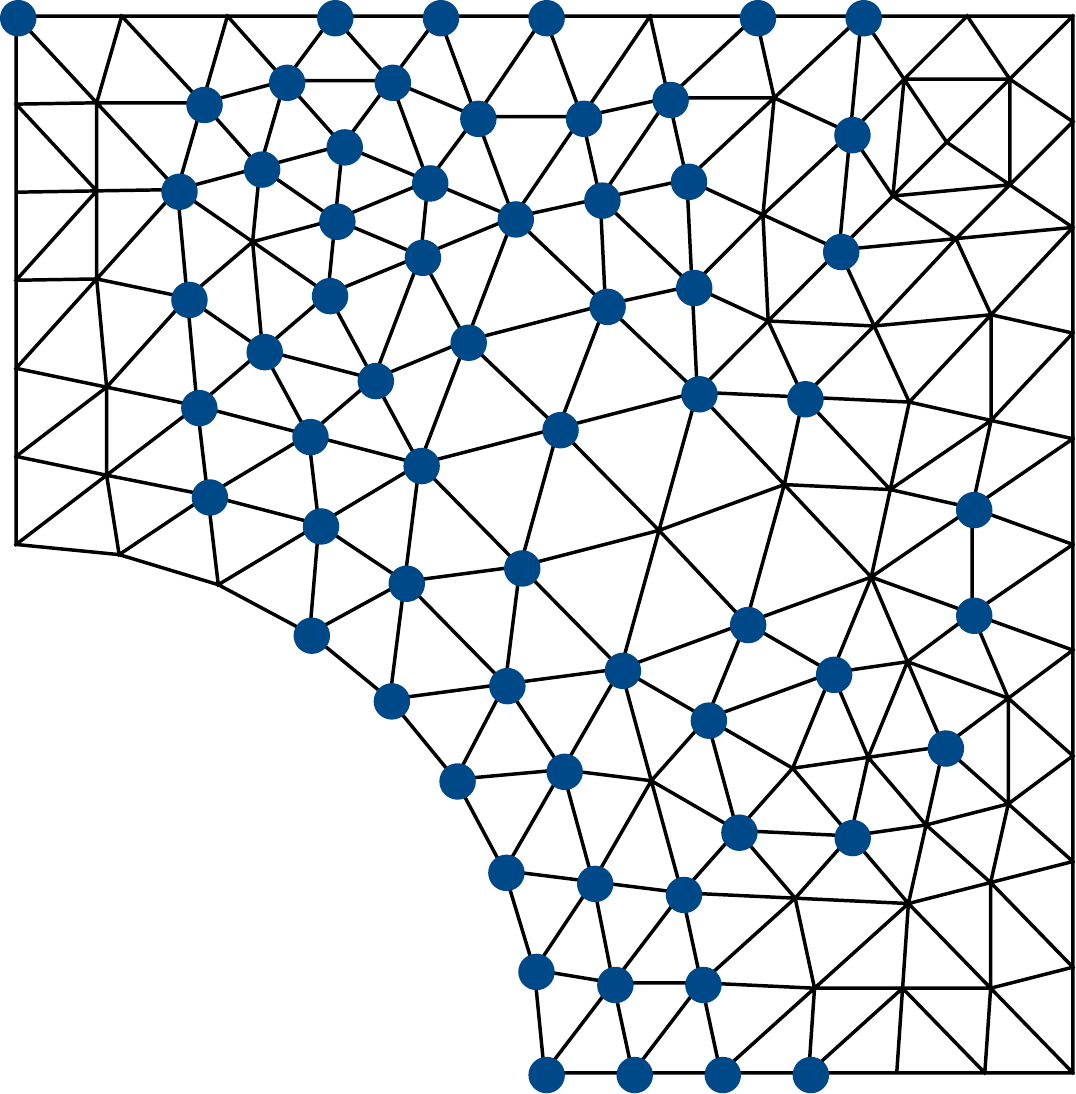}\hspace{0.075\textwidth}
		\includegraphics[height=0.19\textheight]{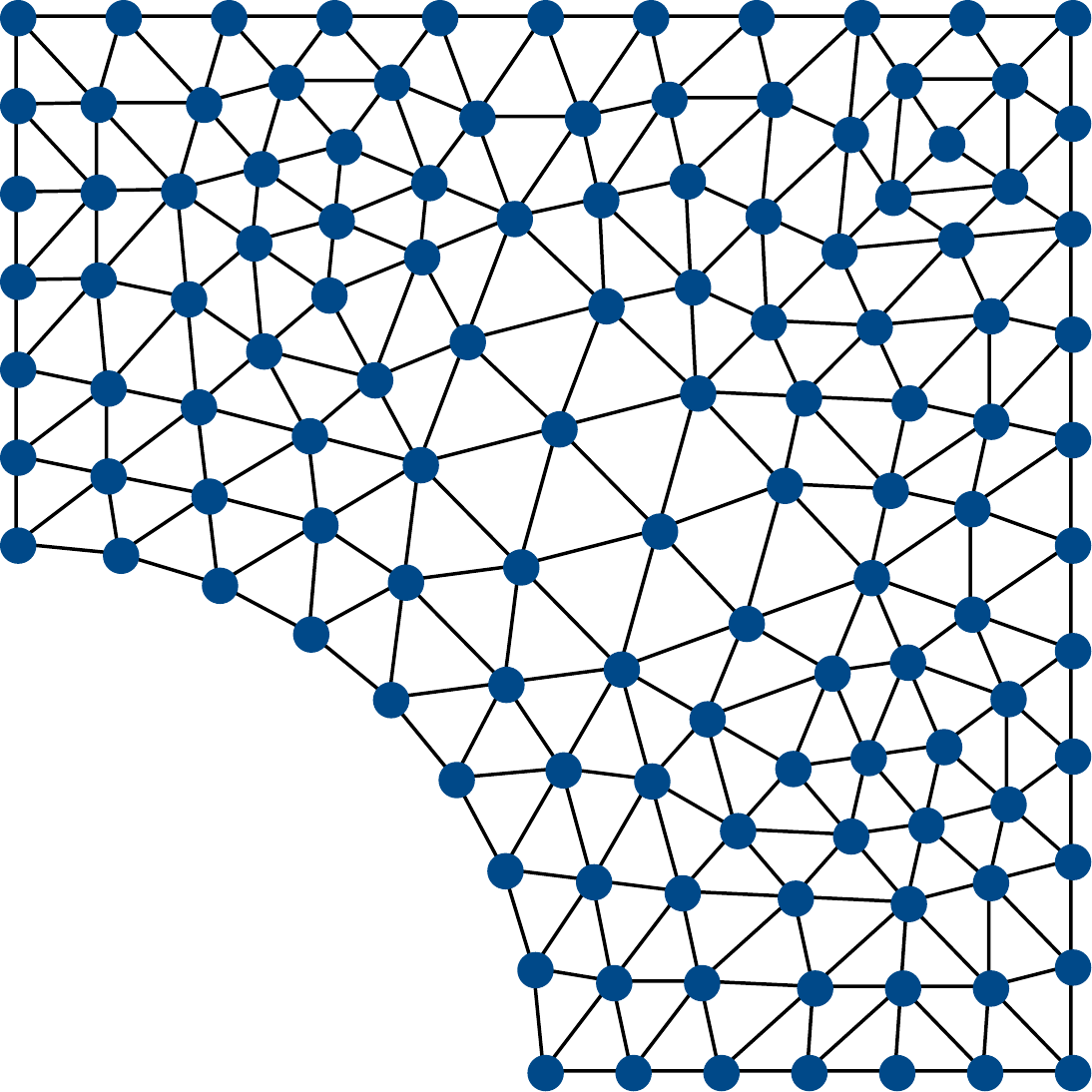}
	\caption{Plate with a hole. Location of the~$n_y \in  \{ 32, \, 64, \, 125 \}$ observations points for the data vector~$\vec y$ chosen according to a Sobol sequence.  The observation points are located at the finite element nodes marked with dots. 
	\label{fig:2dSensorPos}}
\end{figure}

We sample the synthetic observation matrix~$\vec Y \in \mathbb R^{n_y \times n_o}$ from the multivariate Gaussian 
\begin{equation}  \label{eq:observMultivariateGauss}
	 \vec y    =  \set N \left (\vec A_g^{-1} \overline {\vec g}, \, \vec A_g^{-1}  \vec C_g \vec A_g^{-\trans} + 2.5 \cdot 10^{-5} \vec I  \right )  
\end{equation}
with the observation noise~$\sigma_e^2 = 2.5 \cdot 10^{-5}$. The selected~$n_y \in \{32, \, 64, \, 125 \}$ observation points are all located at the finite element nodes, see Figure~\ref{fig:2dSensorPos}. They are distributed according to a Sobol sequence so that the set observation points are nested~\cite{sobol1976uniformly}. We sample at each of the~$n_y$ sample points~$n_o = \{1,\, 10, \, 100, \, 1000\}$ repeated readings.

\begin{figure}[h!] 
	\centering
	\subfloat[$p(\rho | \vec Y)$]{
		\includegraphics[height=.24\textwidth]{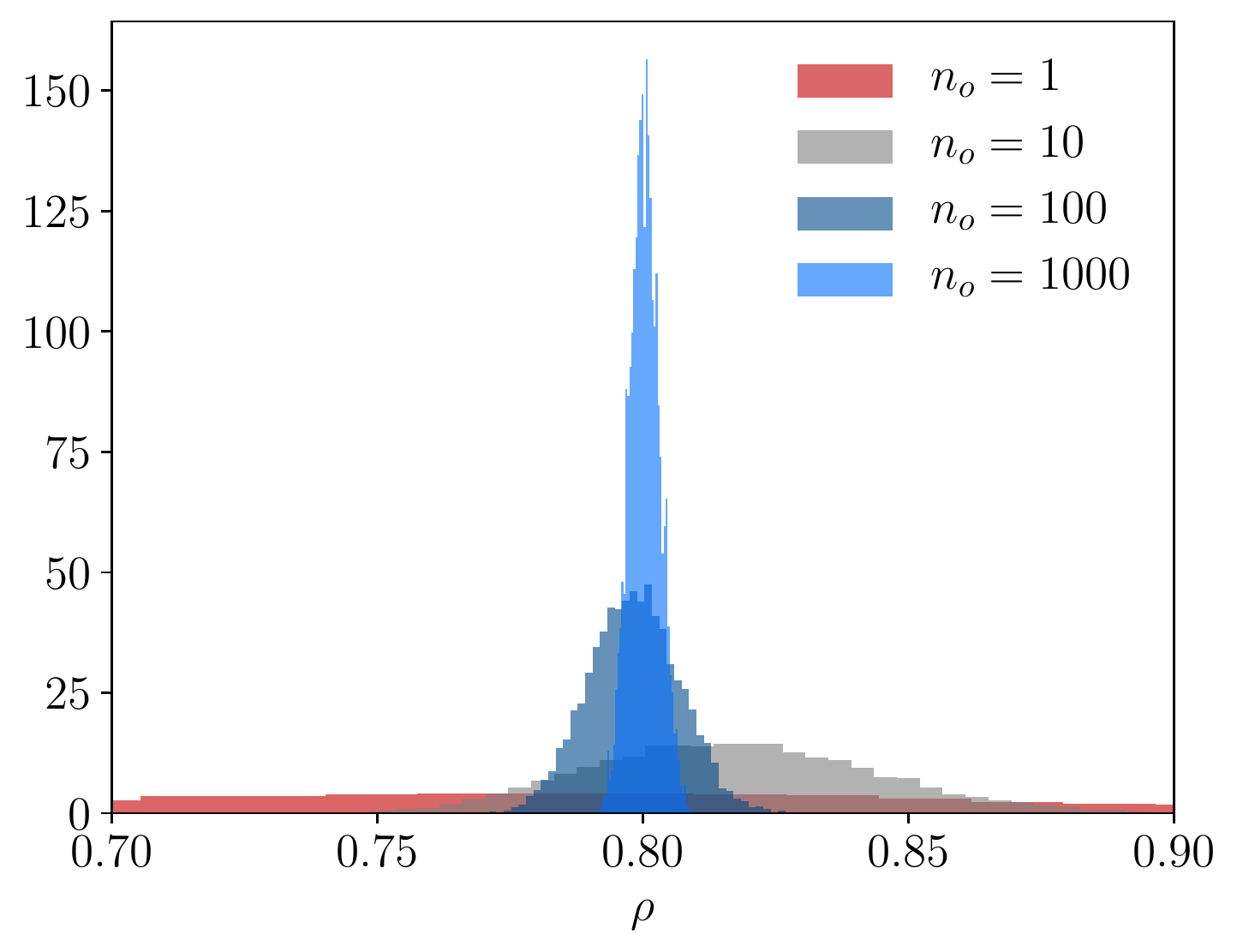}
	} 
	\subfloat[$p(\sigma_d | \vec Y)$]{
		\includegraphics[height=.24\textwidth]{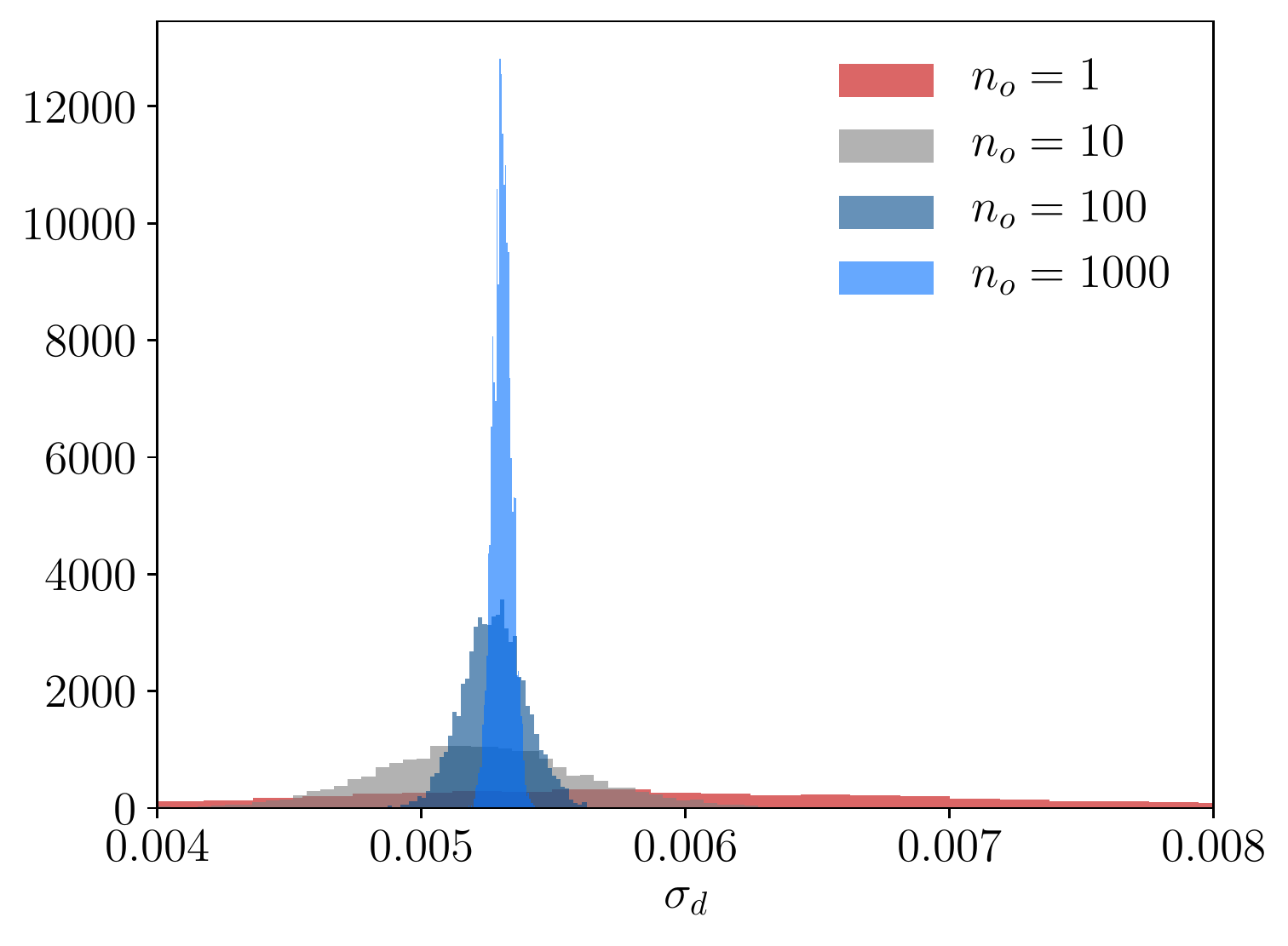}
	} 
	\subfloat[$p(\ell_d | \vec Y)$]{
		\includegraphics[height=.24\textwidth]{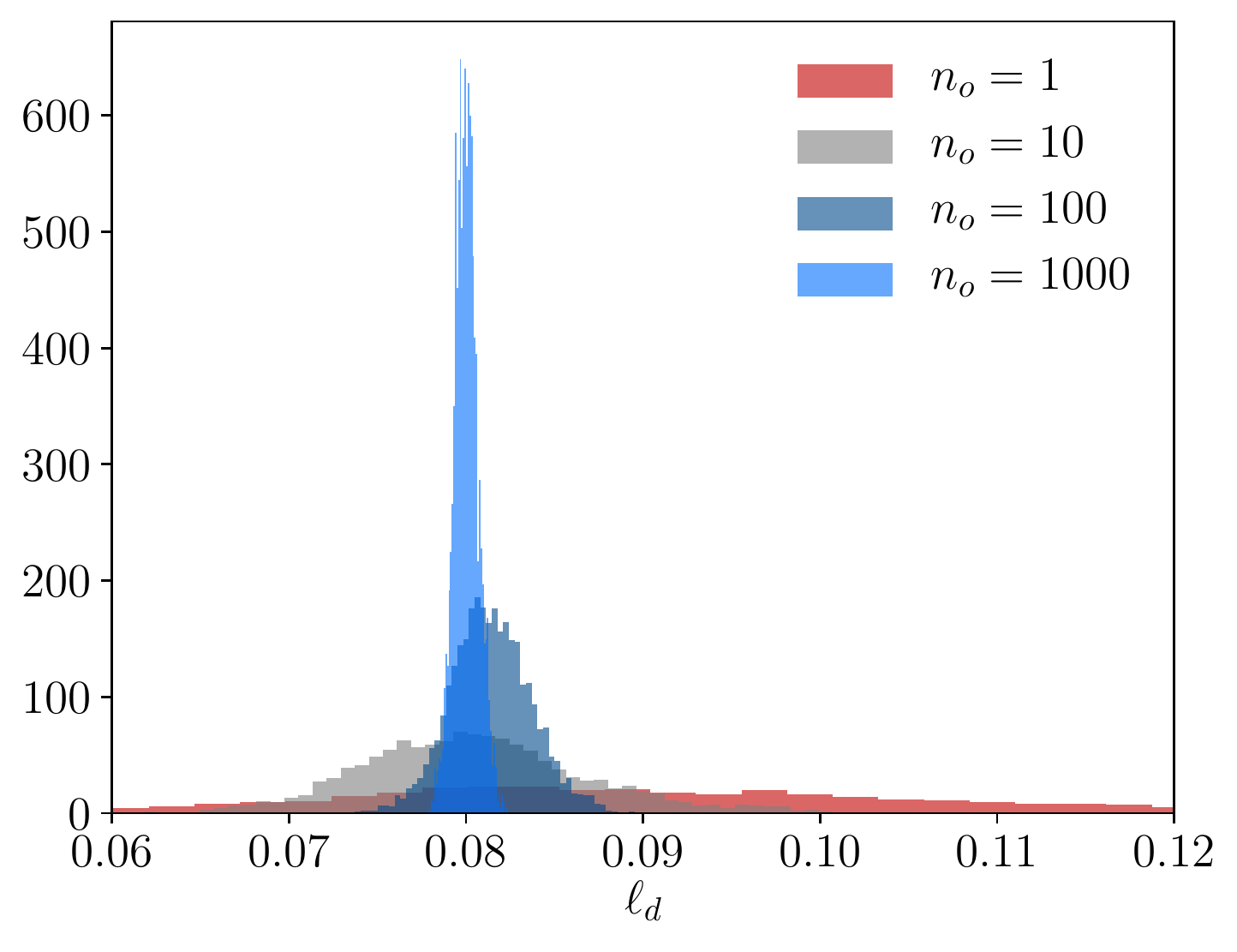}
	} 
	\caption{Plate with a hole. Posteriors of the parameters~$\rho$, $\sigma_d$ and $\ell_d$ for~$n_y=64$ and~$n_o \in \{1, \, 10,\, 100, \, 1000 \}$ obtained with MCMC. \label{fig:2dInferredHist}}
\end{figure}

As for the one-dimensional example in Section~\ref{sec:1dUcondY}, only the parameters~$\vec w= (\rho, \, \sigma_d, \, \ell_d)^\trans$ of the statistical generating model are assumed to be unknown. Choosing an uninformed prior~$p(\vec w) \propto 1$ we sample the posterior density of the hyperparameters~$p(\vec w | \vec Y)$ using a standard MCMC algorithm, see~\ref{sec:appMCMC}. For each combination of $n_y$ and~$n_o$ we run $50000$ iterations with an average acceptance ratio of~$0.254$. In Figure~\ref{fig:2dInferredHist} the obtained histograms for~$p(\rho | \vec Y)$, $p(\sigma_d | \vec Y )$ and~$ p( \ell_d | \vec Y) $  for $n_y=64$ are depicted. The observed overall trends are very similar to the one-dimensional example. The standard deviations become smaller with increasing~$n_o$. In Tables~\ref{tab:2dRho}, \ref{tab:2dSigmaD} and \ref{tab:2dEllD} the mean and standard deviations of these three plots and other~$n_y$ and~$n_o$ combinations are given. As to be expected when either~$n_y$,~$n_o$ or both are increased the standard deviation becomes smaller. 

\begin{table}[h!]
\centering
\rowcolors{2}{}{csmlLightBlue!30}
\begin{tabular}{ l   c   c   c   c }
\toprule
 & $n_o=1$ & $n_o=10$ & $n_o=100$ & $n_o=1000$\\
\midrule
$n_y = 32$ & 
$0.81665 \pm 0.09841$ 
& $0.81056 \pm 0.02826$ 
& $0.80928 \pm  0.00943$ 
& $0.81085 \pm 0.00289$ 
\\
$n_y = 64$ & 
$0.81051 \pm 0.10648$ 
& $0.81710 \pm 0.02713$ 
& $0.79831 \pm 0.00853$ 
& $0.80026 \pm 0.00278$ 
\\
$n_y = 125$ & 
$0.78432 \pm 0.08124$ 
& $0.82240 \pm 0.02761$ 
& $0.79888 \pm 0.00861$ 
& $0.79806 \pm 0.00271$ \\
\bottomrule
\end{tabular}
\caption{Plate with a hole. Empirical mean and standard deviation of $p(\rho|\vec Y)$. \label{tab:2dRho}} 
\vspace{1.2em}
\centering
\rowcolors{2}{}{csmlLightBlue!30}
\begin{tabular}{ l  c  c  c  c  }
\toprule
 & $n_o=1$ & $n_o=10$ & $n_o=100$ & $n_o=1000$\\
\midrule
$n_y = 32$ & 
$0.00678 \pm 0.00229$ 
& $0.00498 \pm 0.00043$ 
& $0.00546 \pm 0.00016$ 
& $0.00545 \pm 0.00005$ 
\\
$n_y = 64$ & 
$0.00595 \pm 0.00145$ 
& $0.00521 \pm 0.00038 $ 
& $0.00527 \pm  0.00011$ 
& $0.00530 \pm 0.00003$ 
\\
$n_y = 125$ & 
$0.00444 \pm 0.00058$ 
& $0.00503 \pm  0.00024$ 
& $0.00463 \pm 0.00008 $ 
& $0.00469 \pm 0.00002$  \\
\bottomrule
\end{tabular}
\caption{Plate with a hole. Empirical mean and standard deviation of $p(\sigma_d|\vec Y)$. \label{tab:2dSigmaD}}
\vspace{1.2em}
\centering
\rowcolors{2}{}{csmlLightBlue!30}
\begin{tabular}{ l   c   c   c   c }
\toprule
 & $n_o=1$ & $n_o=10$ & $n_o=100$ & $n_o=1000$\\
\midrule
$n_y = 32$ & 
$0.28759 \pm 0.17127$ 
& $0.06962 \pm 0.01399$ 
& $0.08344\pm 0.00447$ 
& $0.08644 \pm 0.001188$ 
\\
$n_y = 64$ & 
$0.08726 \pm 0.02194$ 
& $0.08049 \pm 0.00637$ 
& $0.08141 \pm 0.00228$ 
& $0.080011 \pm 0.00066$ 
\\
$n_y = 125$ & 
$0.06002    \pm 0.00752$ 
& $0.06870 \pm 0.00429$ 
& $0.06949 \pm 0.00133$ 
& $0.06949 \pm 0.00041$  \\
\bottomrule 
\end{tabular}
\caption{Plate with a hole. Empirical mean and standard deviation of $p(\ell_d|\vec Y)$. \label{tab:2dEllD}}
\end{table}

\begin{figure}
	\centering
		\includegraphics[width=.24\textwidth]{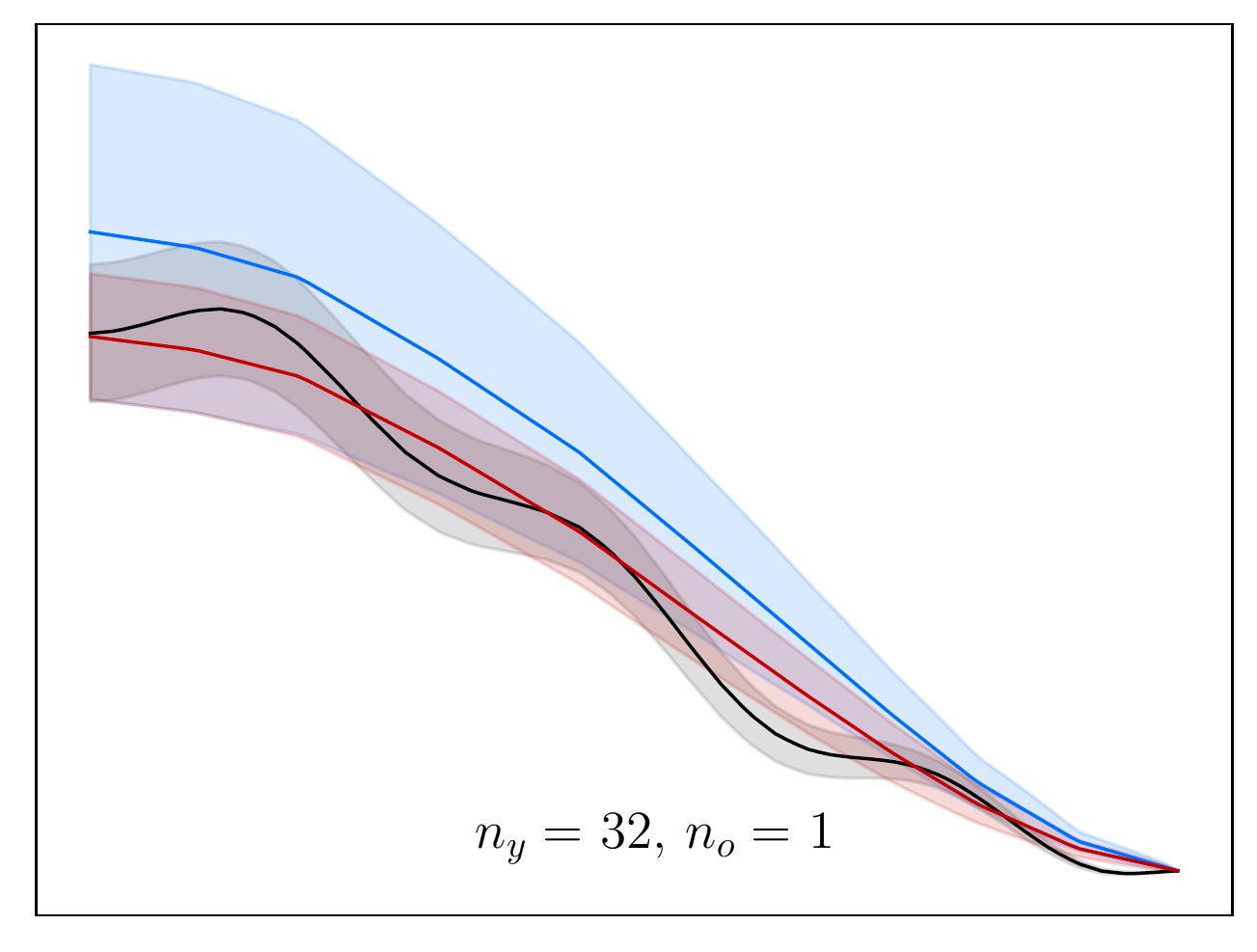}
		\includegraphics[width=.24\textwidth]{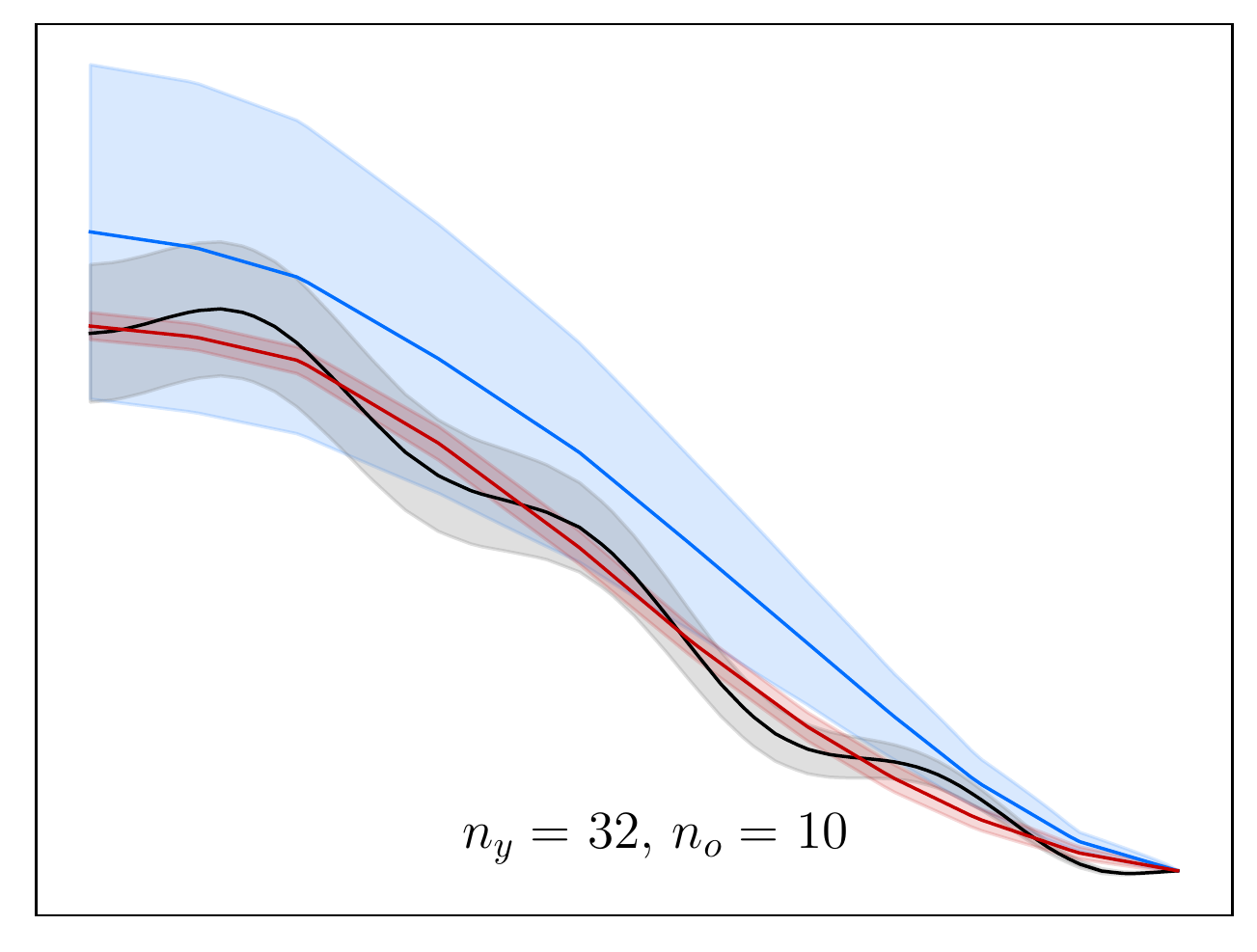}
		\includegraphics[width=.24\textwidth]{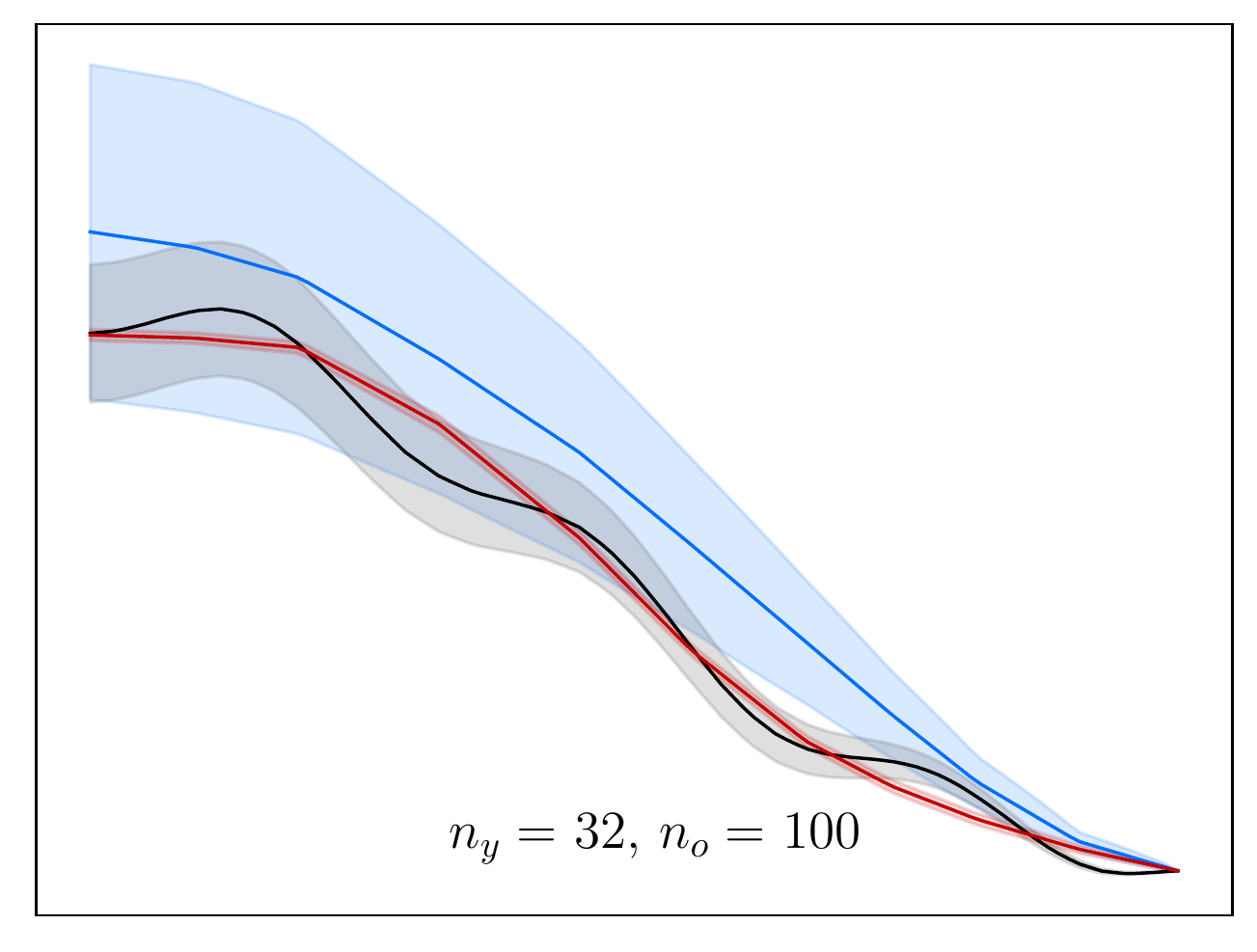}
		\includegraphics[width=.24\textwidth]{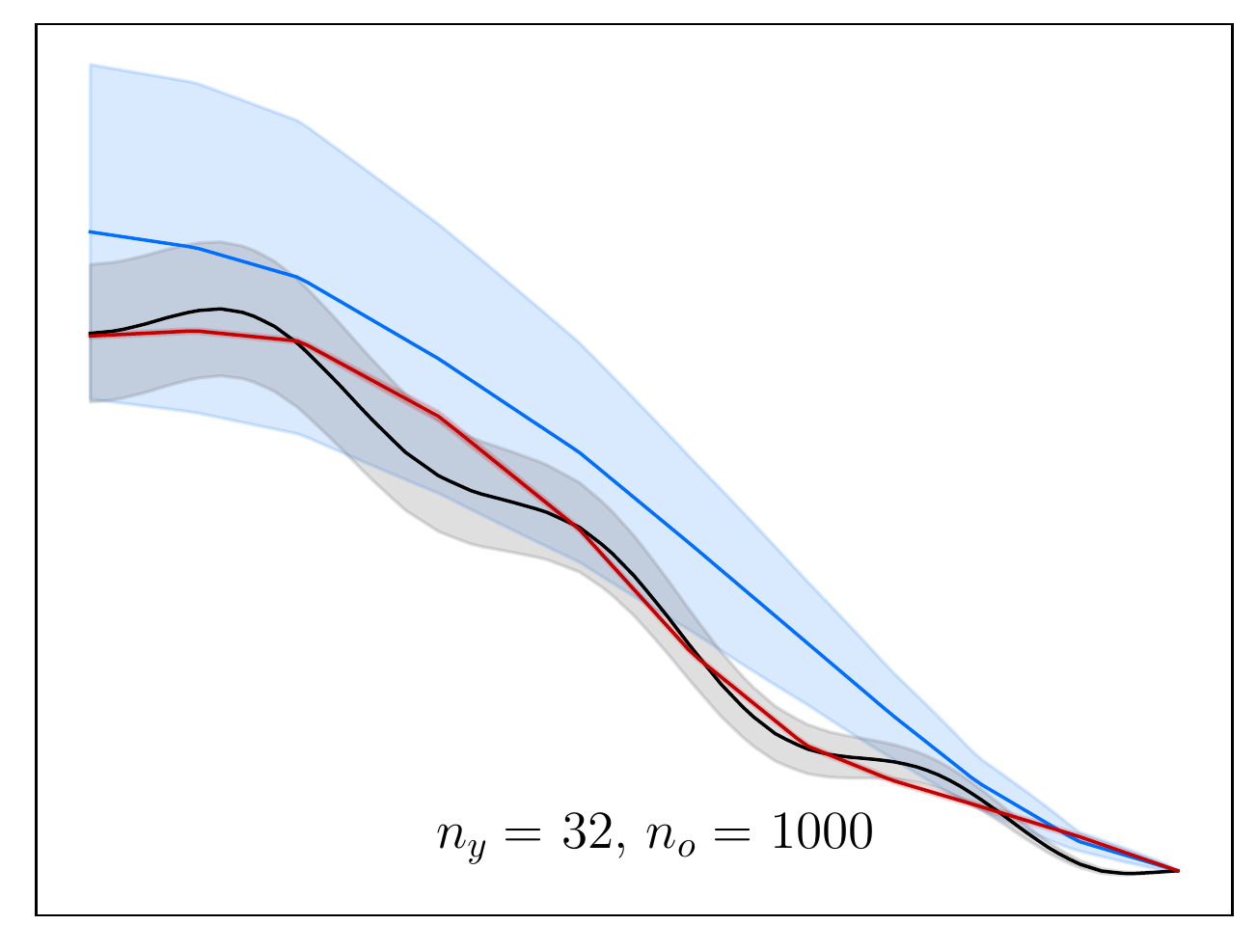}
	\\\
		\includegraphics[width=.24\textwidth]{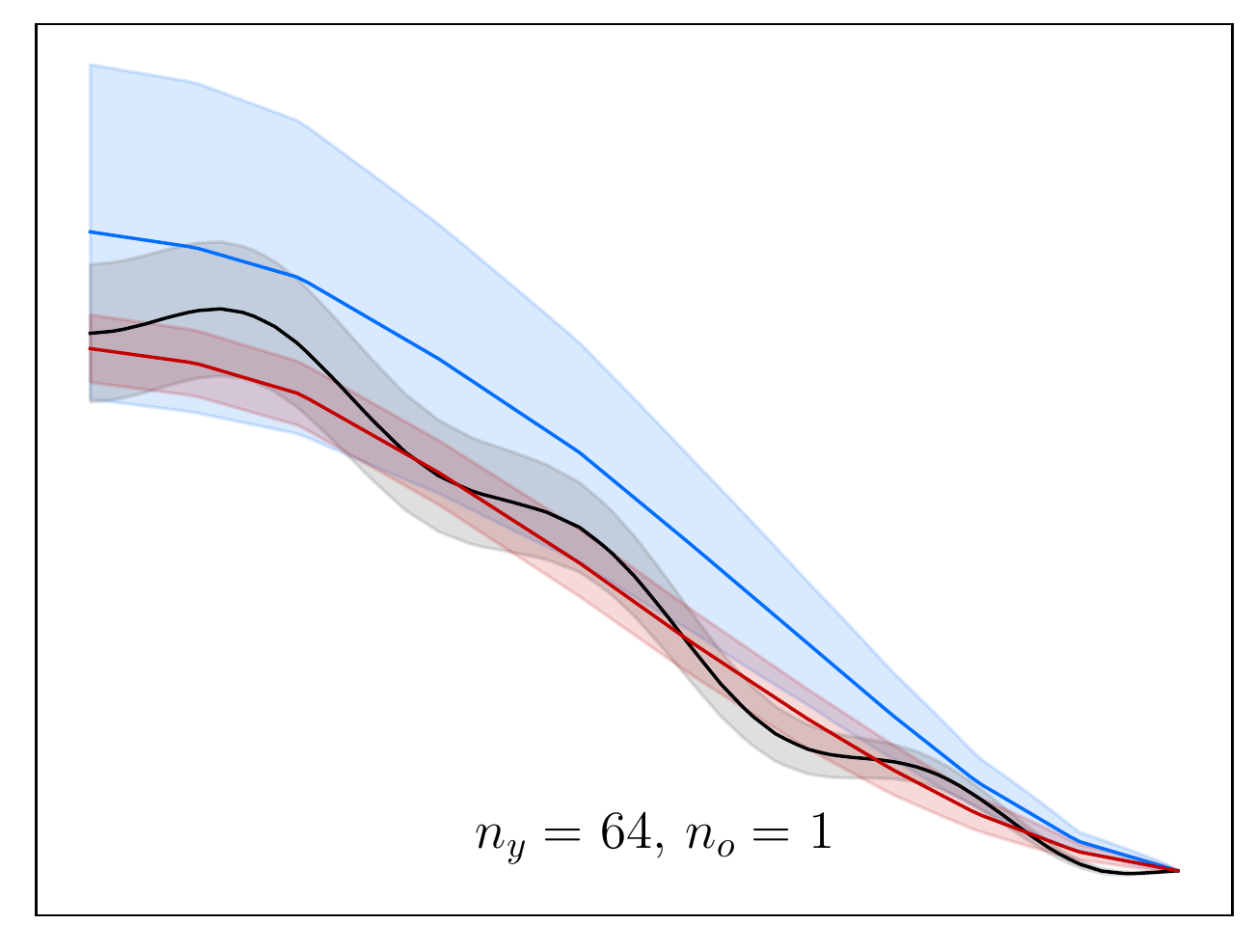}
		\includegraphics[width=.24\textwidth]{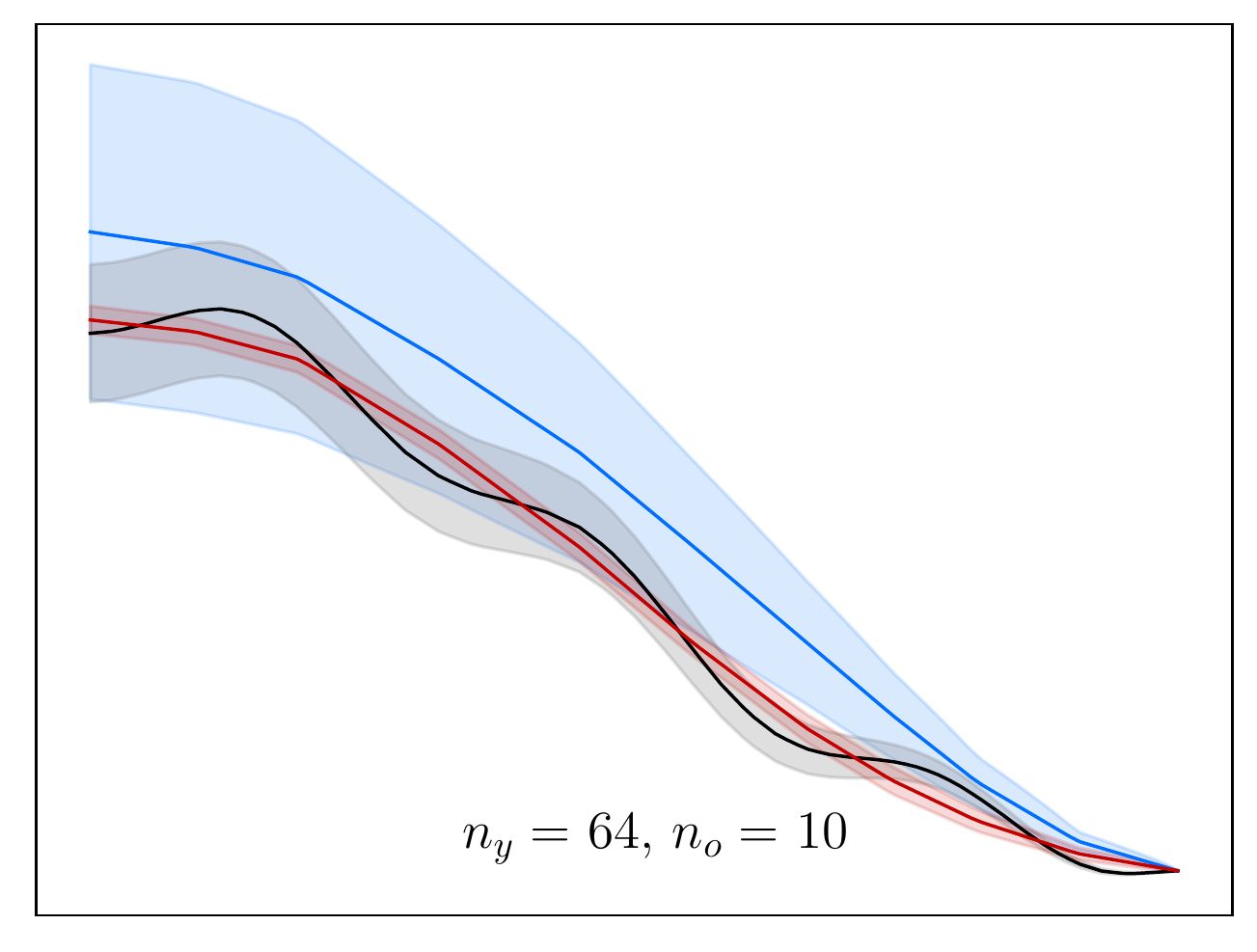}
		\includegraphics[width=.24\textwidth]{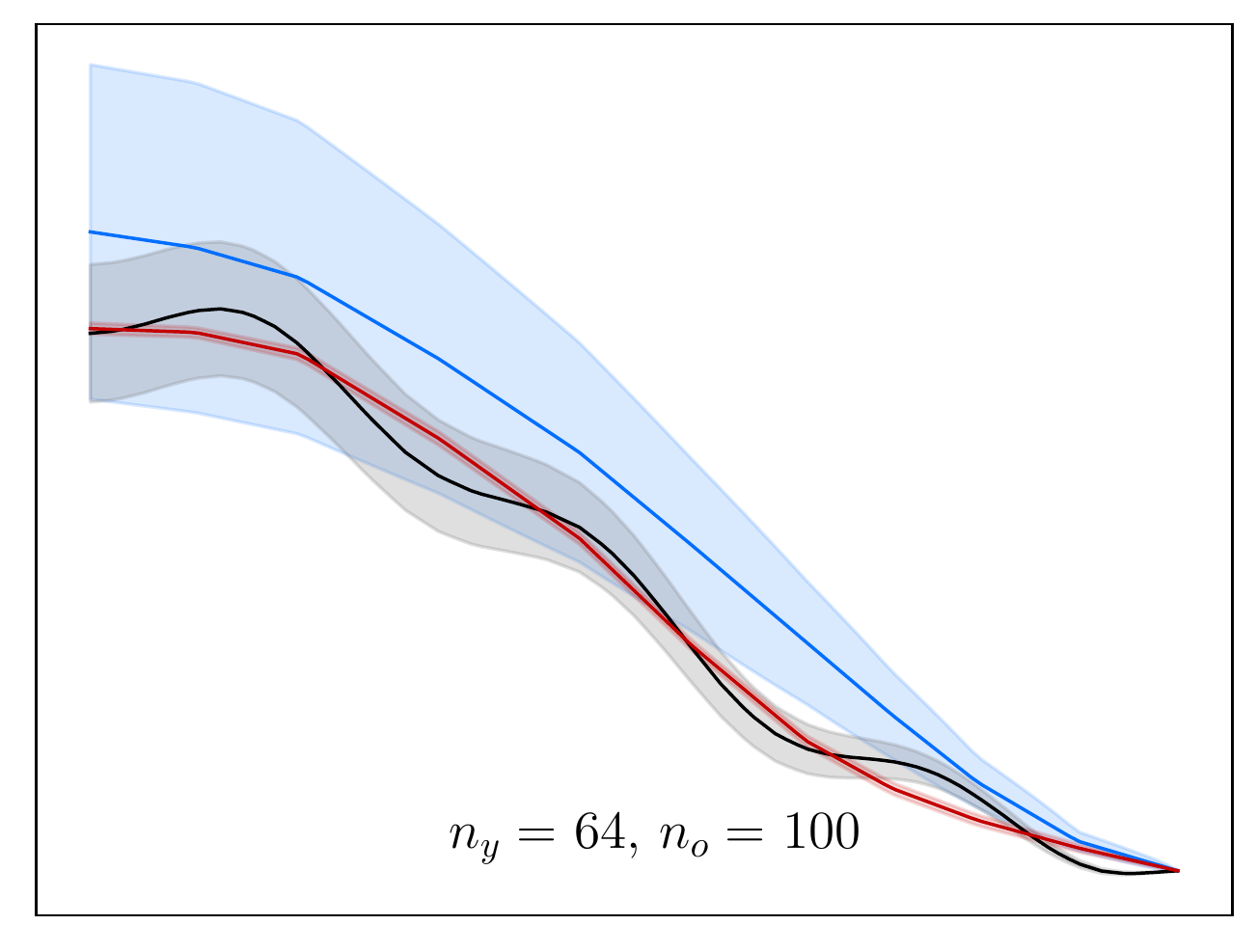}
		\includegraphics[width=.24\textwidth]{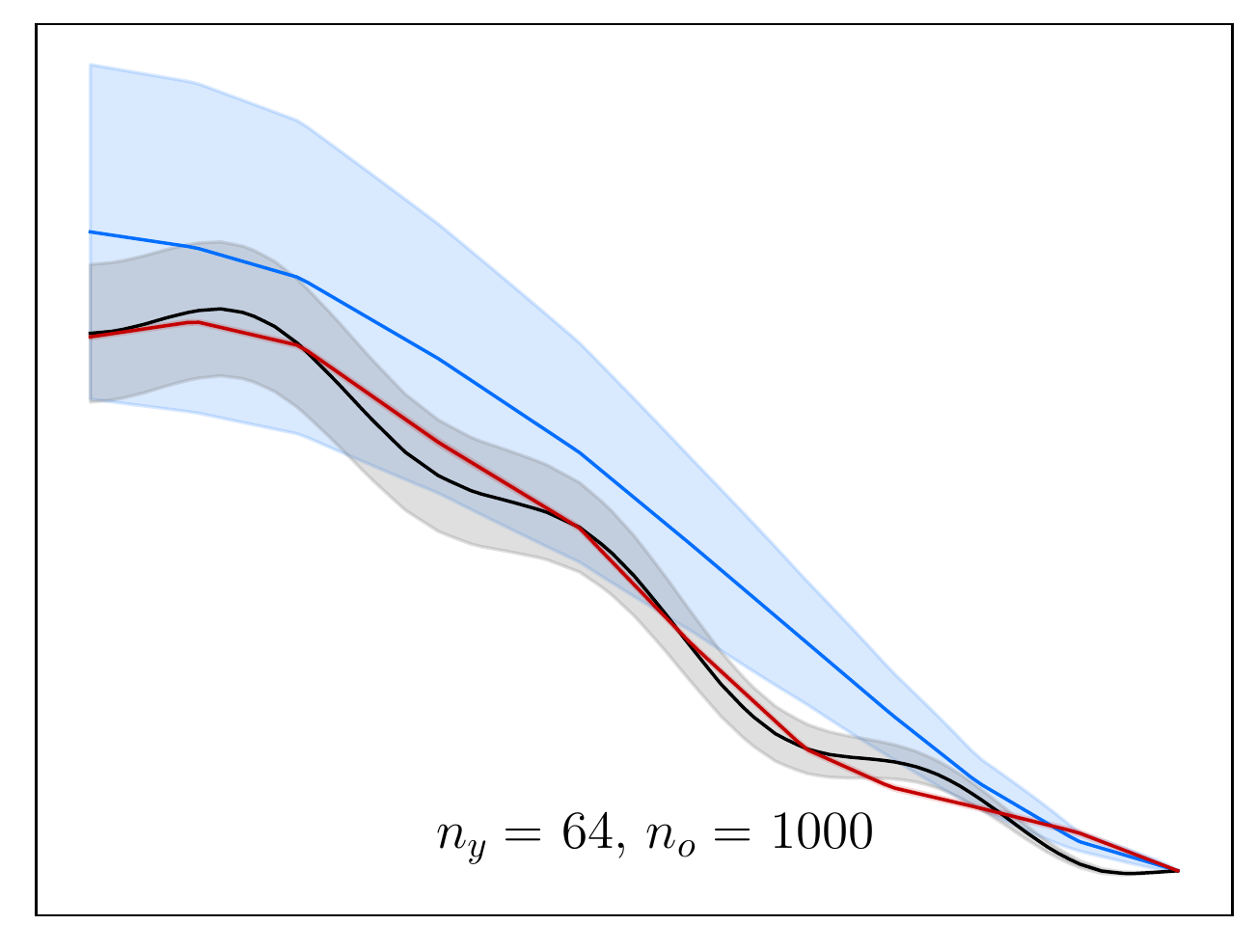}
	\\\
		\includegraphics[width=.24\textwidth]{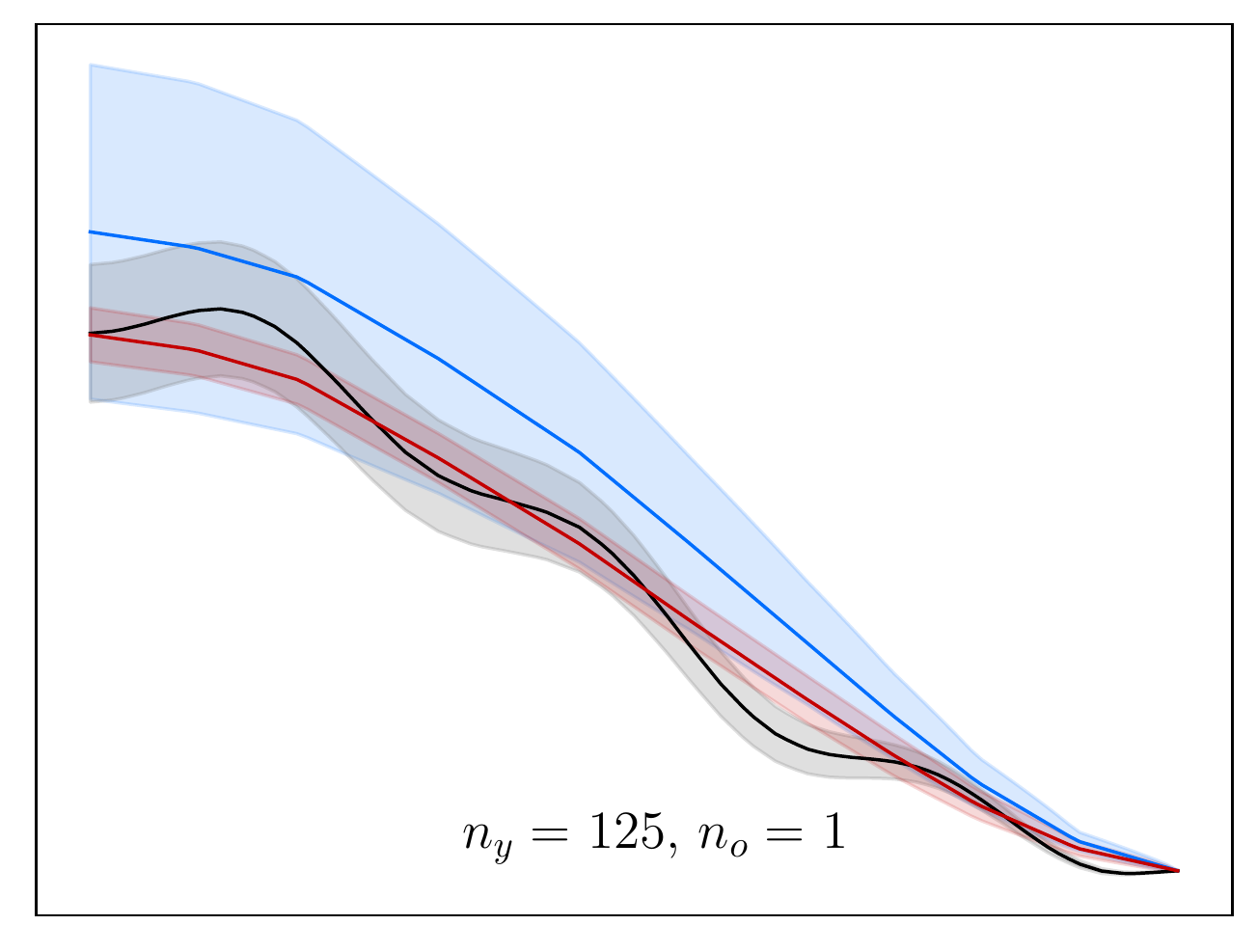}
		\includegraphics[width=.24\textwidth]{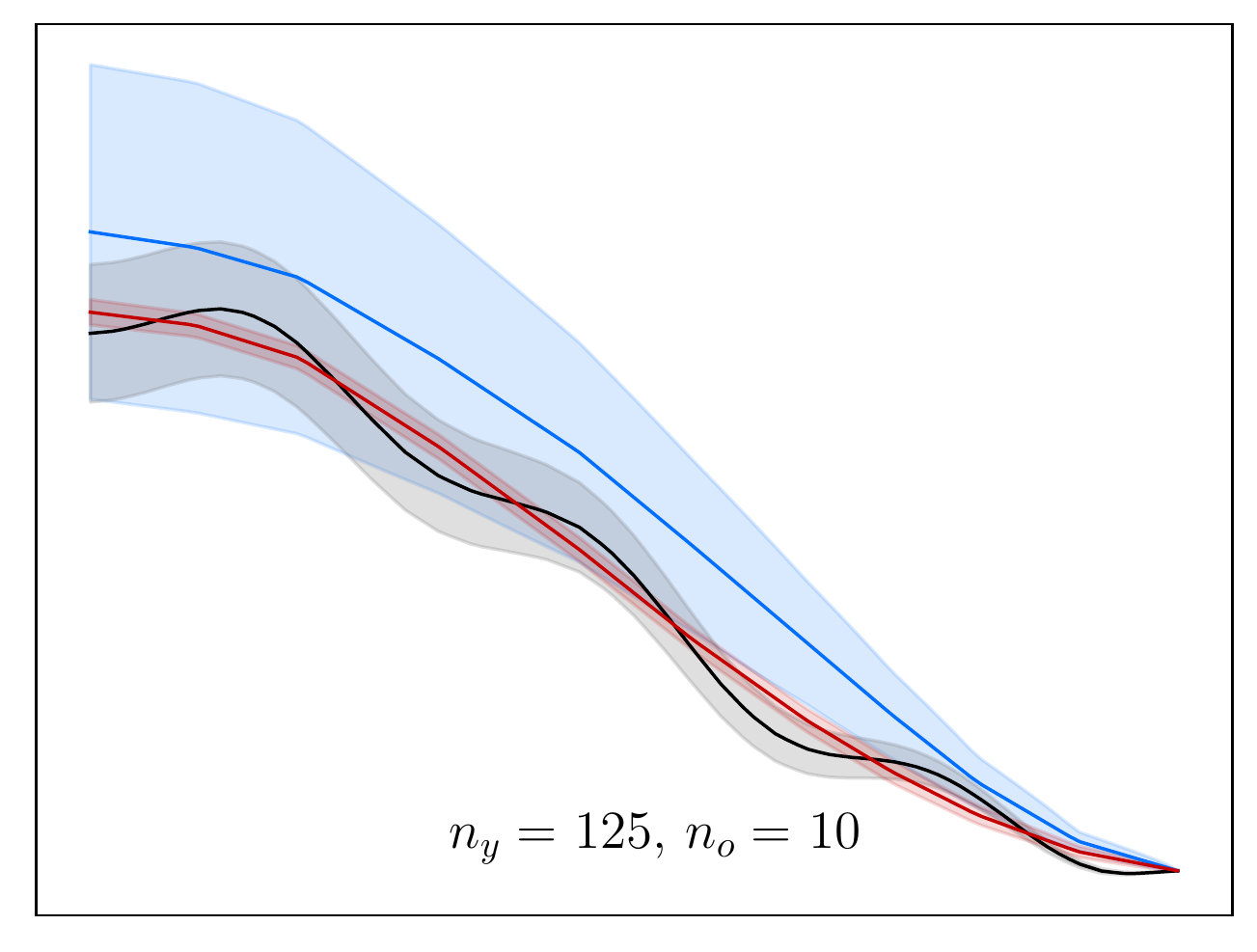}
		\includegraphics[width=.24\textwidth]{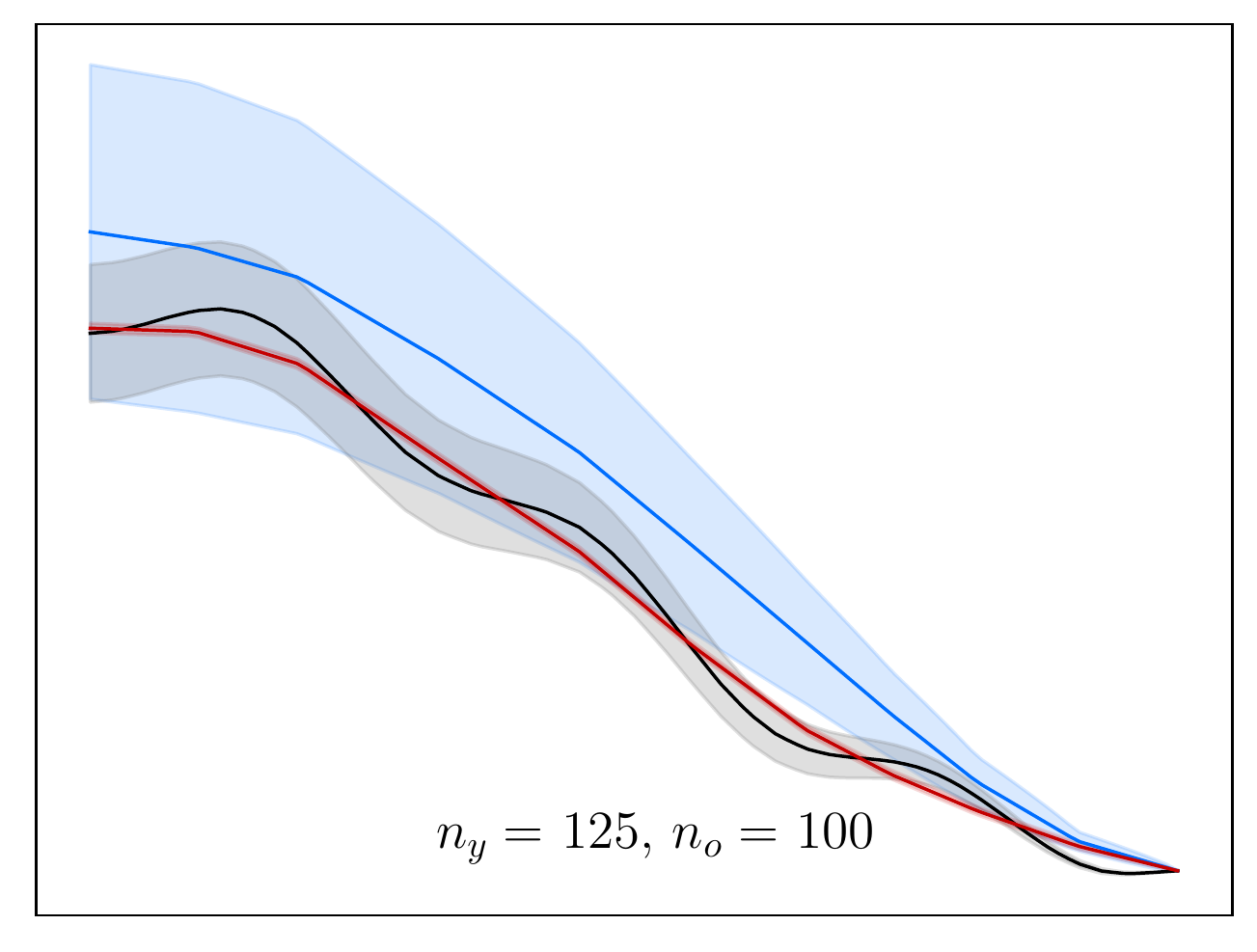}
		\includegraphics[width=.24\textwidth]{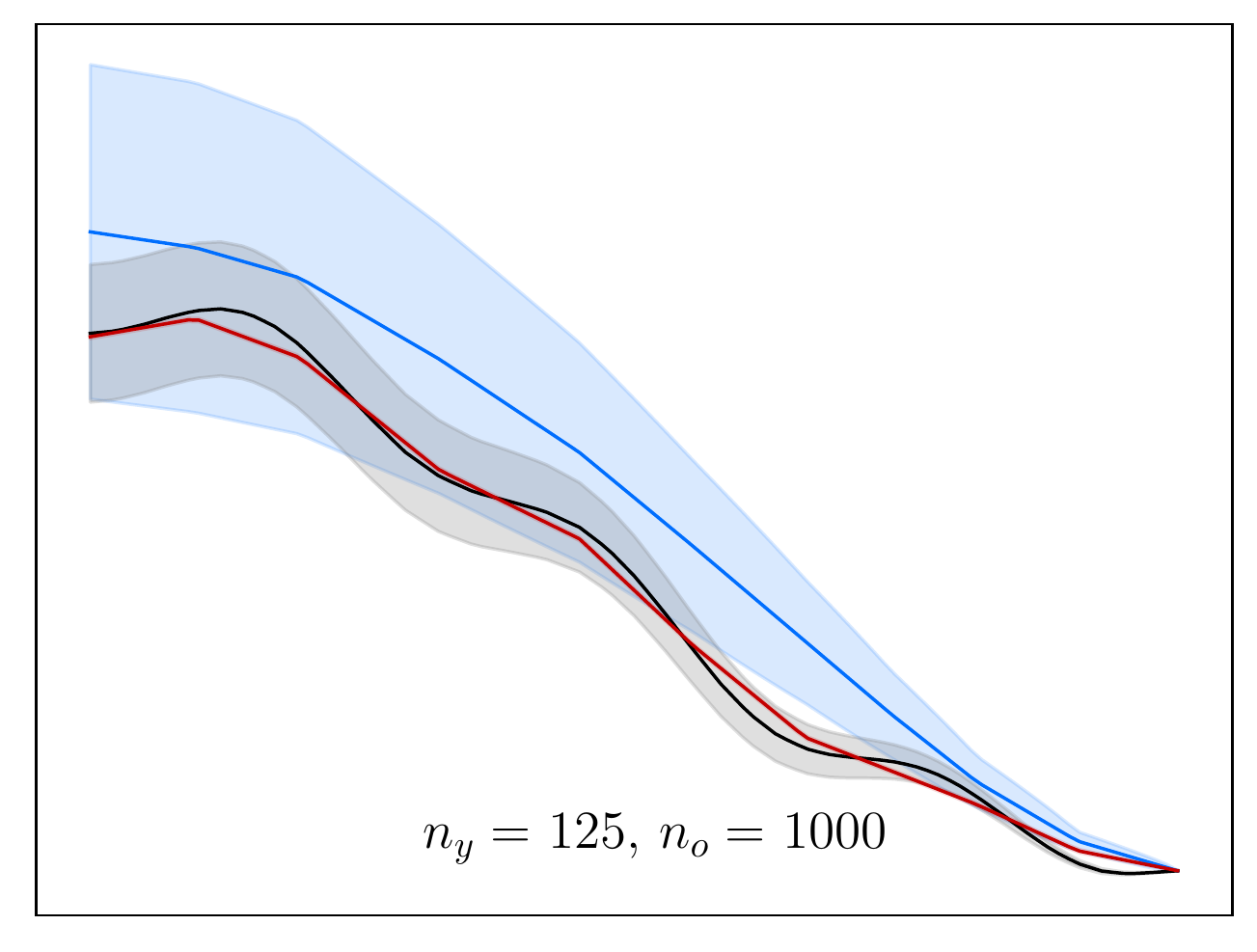}
	\caption{Plate with a hole. Finite element density~$p(\vec u | \vec Y)$ conditioned on observation data sampled from the multivariate Gaussian density~\eqref{eq:observMultivariateGauss} (black line). The blue lines represent the mean~$\overline{\vec u}$ and the red lines the conditioned mean~$\overline {\vec u}_{|Y}$. The shaded areas denote the corresponding $95\%$ confidence regions. In each row the number of sensors~$n_y$ and each column the number of readings~$n_o$ is constant. All plots along the diagonal of the domain~with $x^{(2)} = x^{(1)}$}
	\label{fig:2DpostUh}
\end{figure}

\begin{figure}
	\centering
		\includegraphics[width=.24\textwidth]{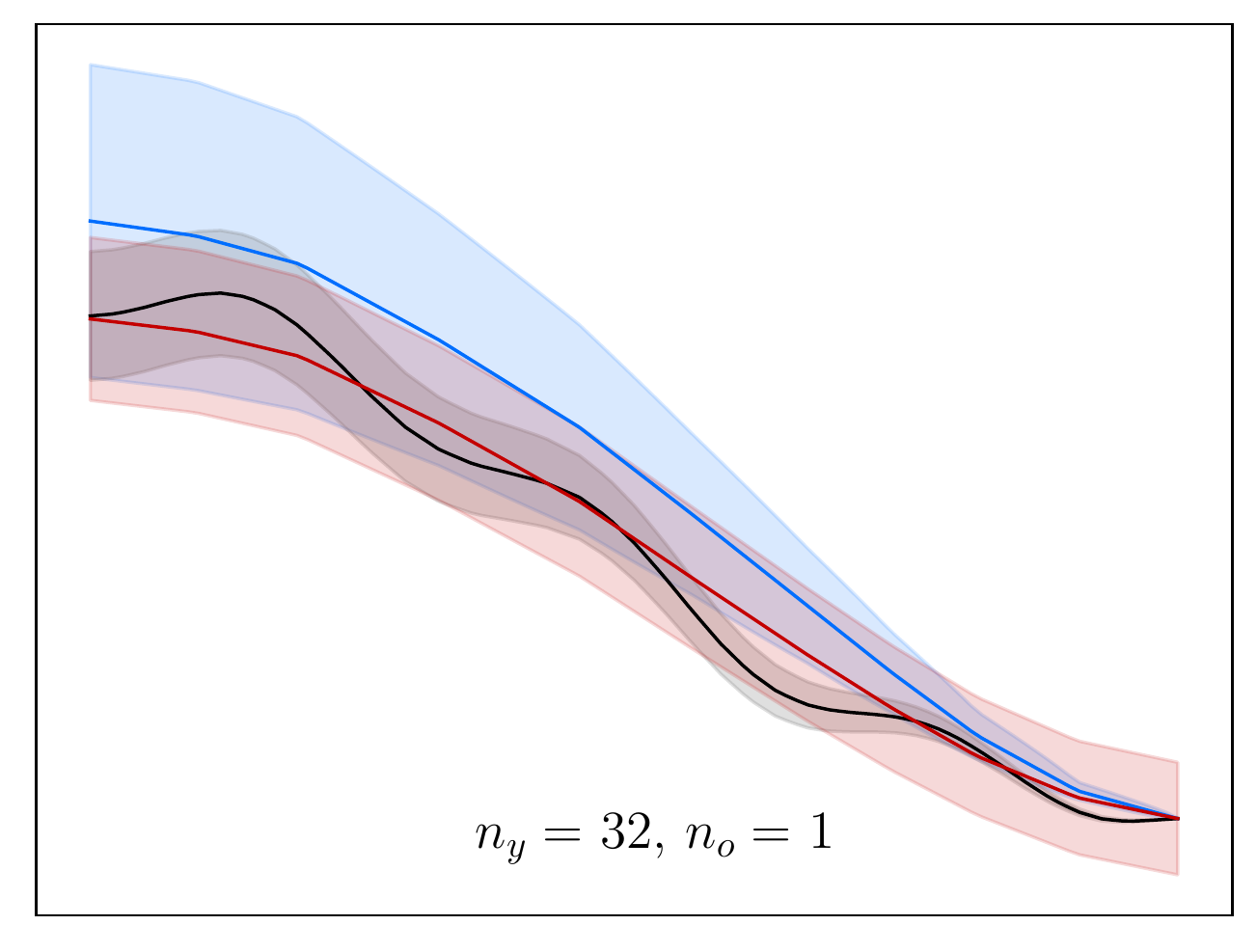}
		\includegraphics[width=.24\textwidth]{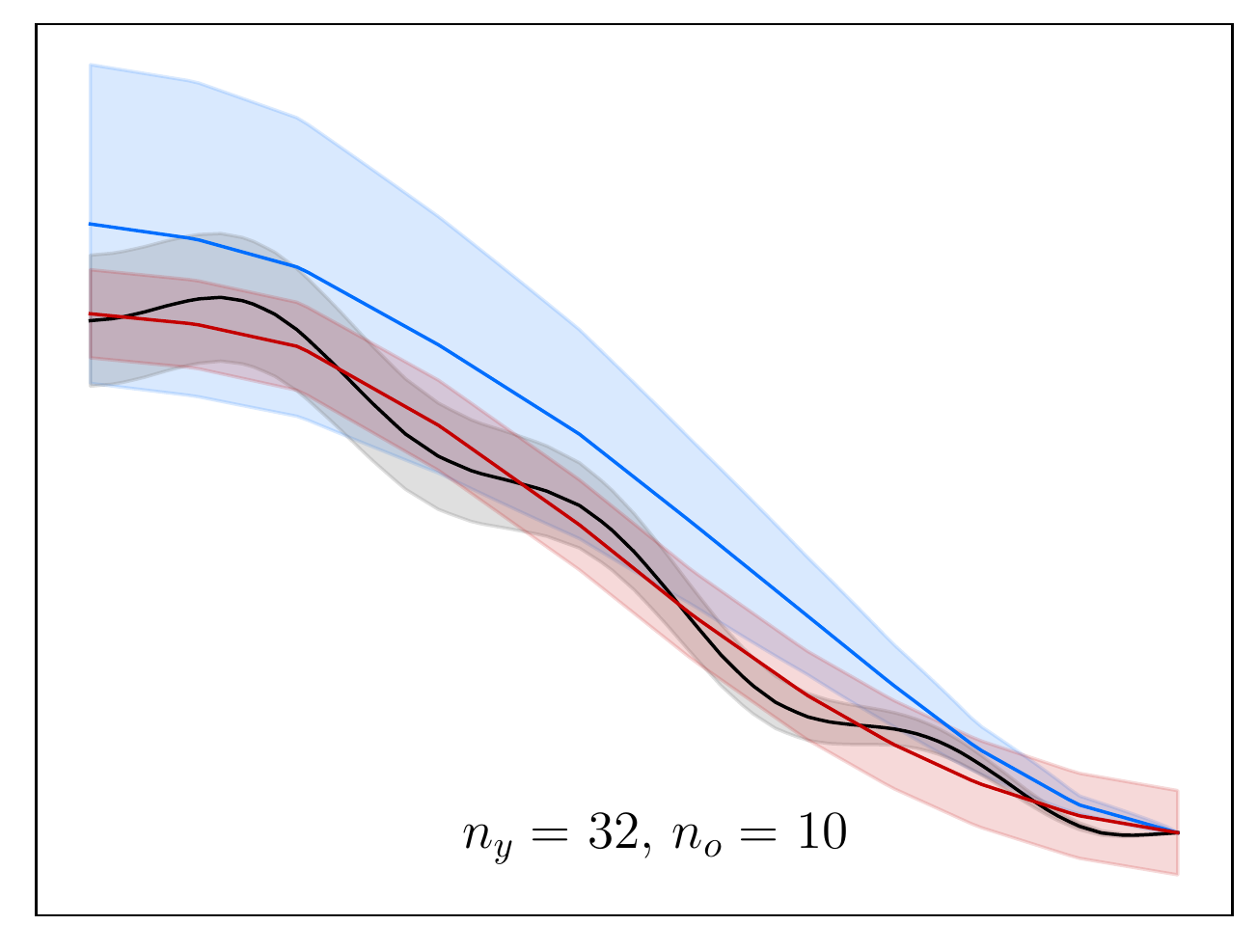}
		\includegraphics[width=.24\textwidth]{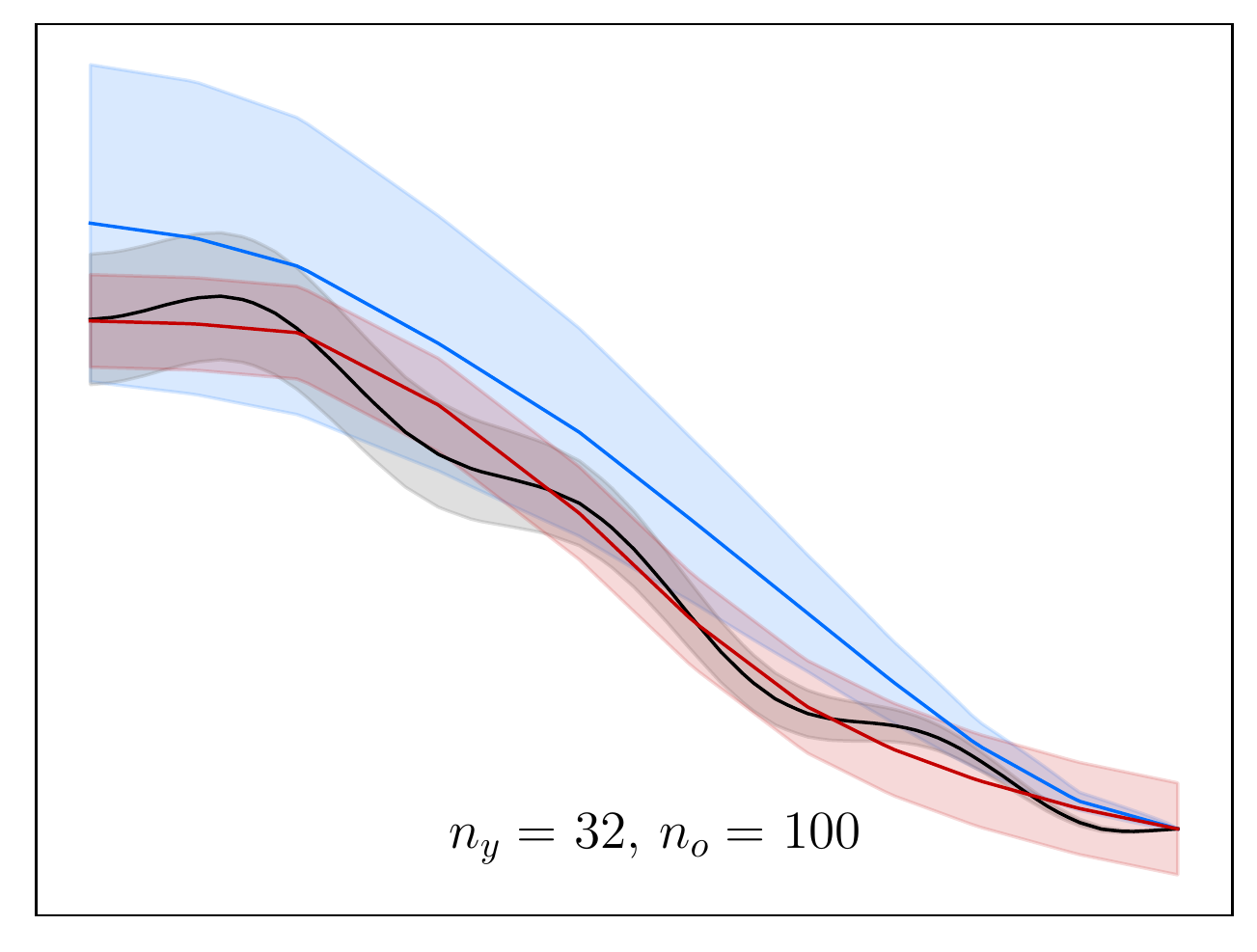}
		\includegraphics[width=.24\textwidth]{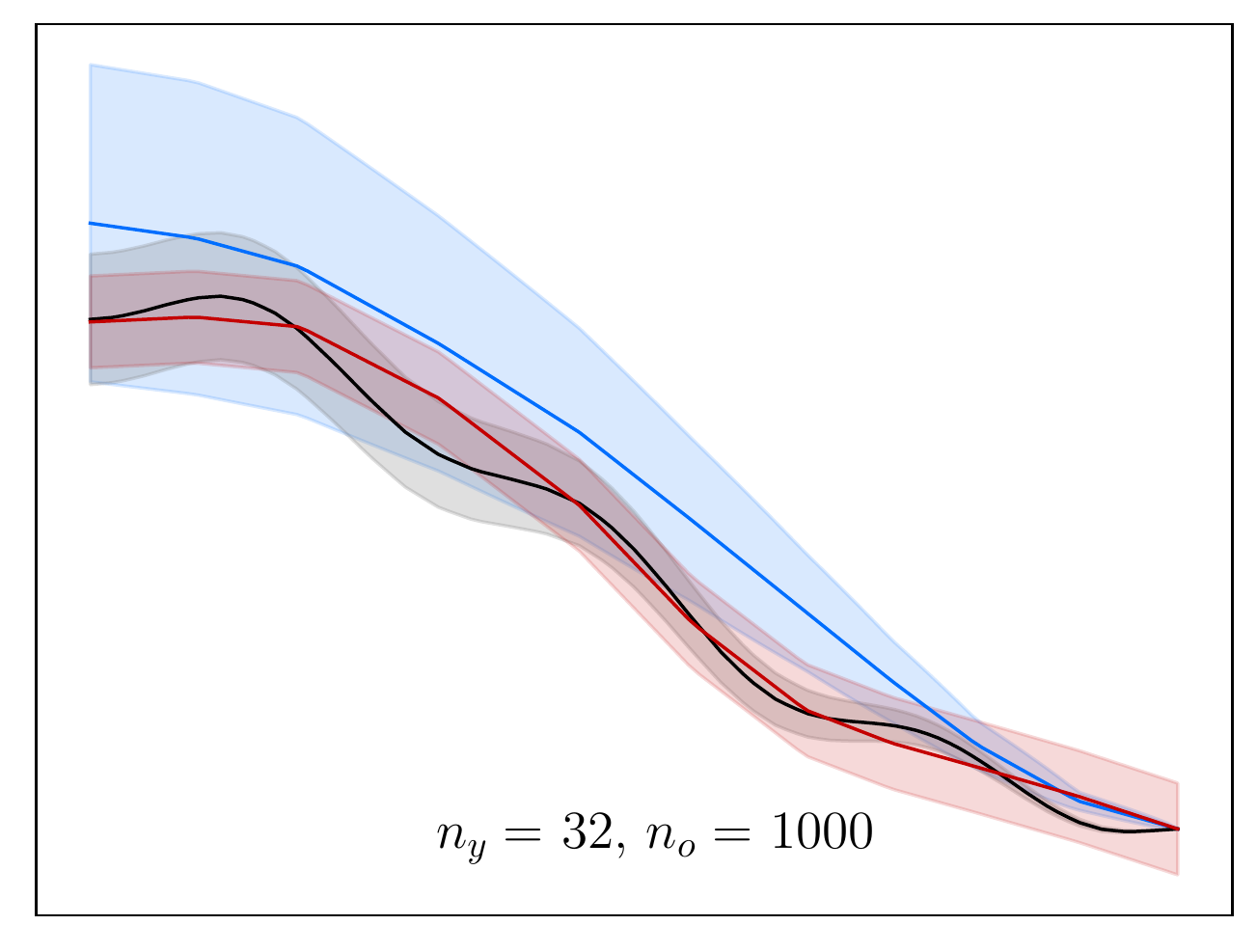}
	\\\
		\includegraphics[width=.24\textwidth]{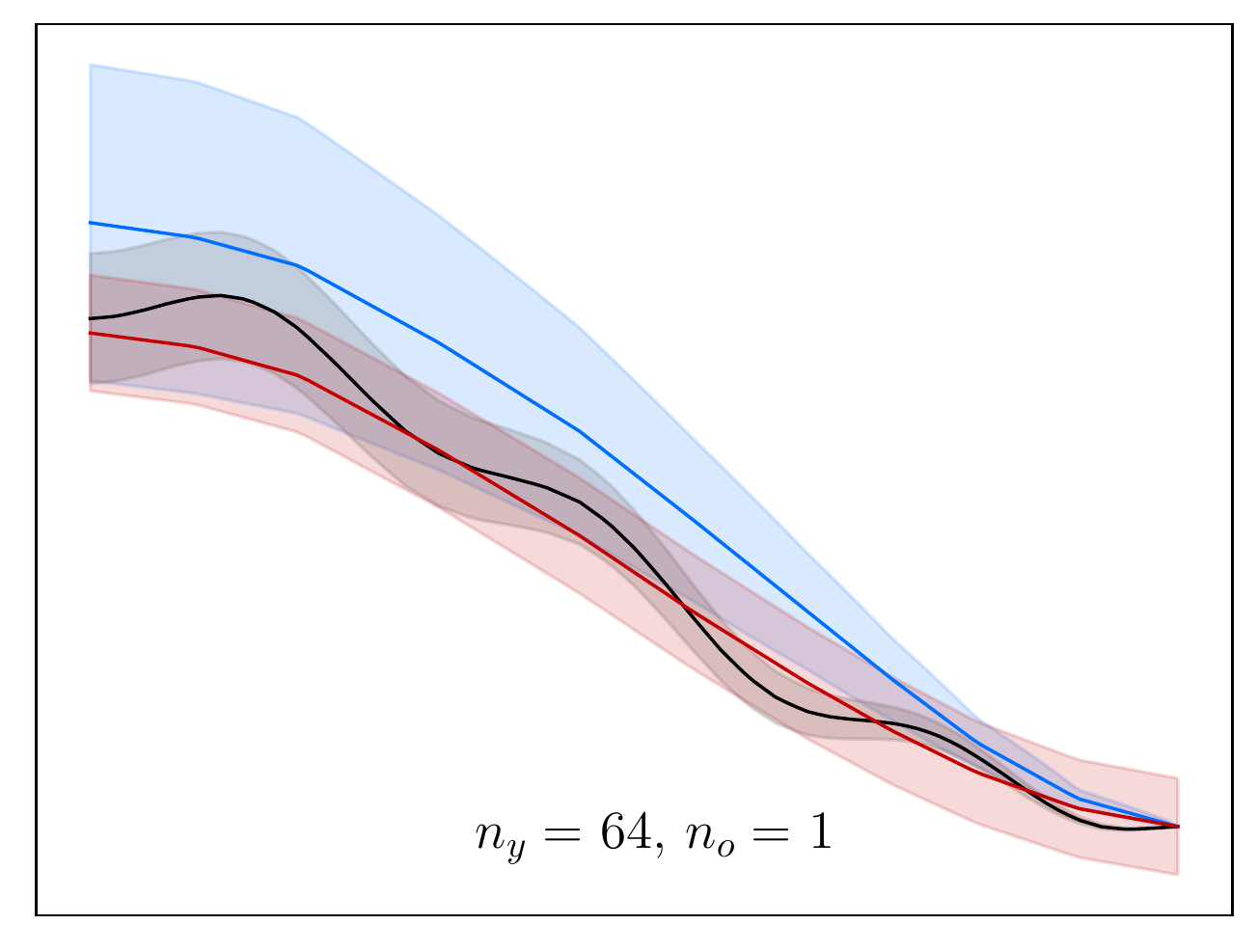}
		\includegraphics[width=.24\textwidth]{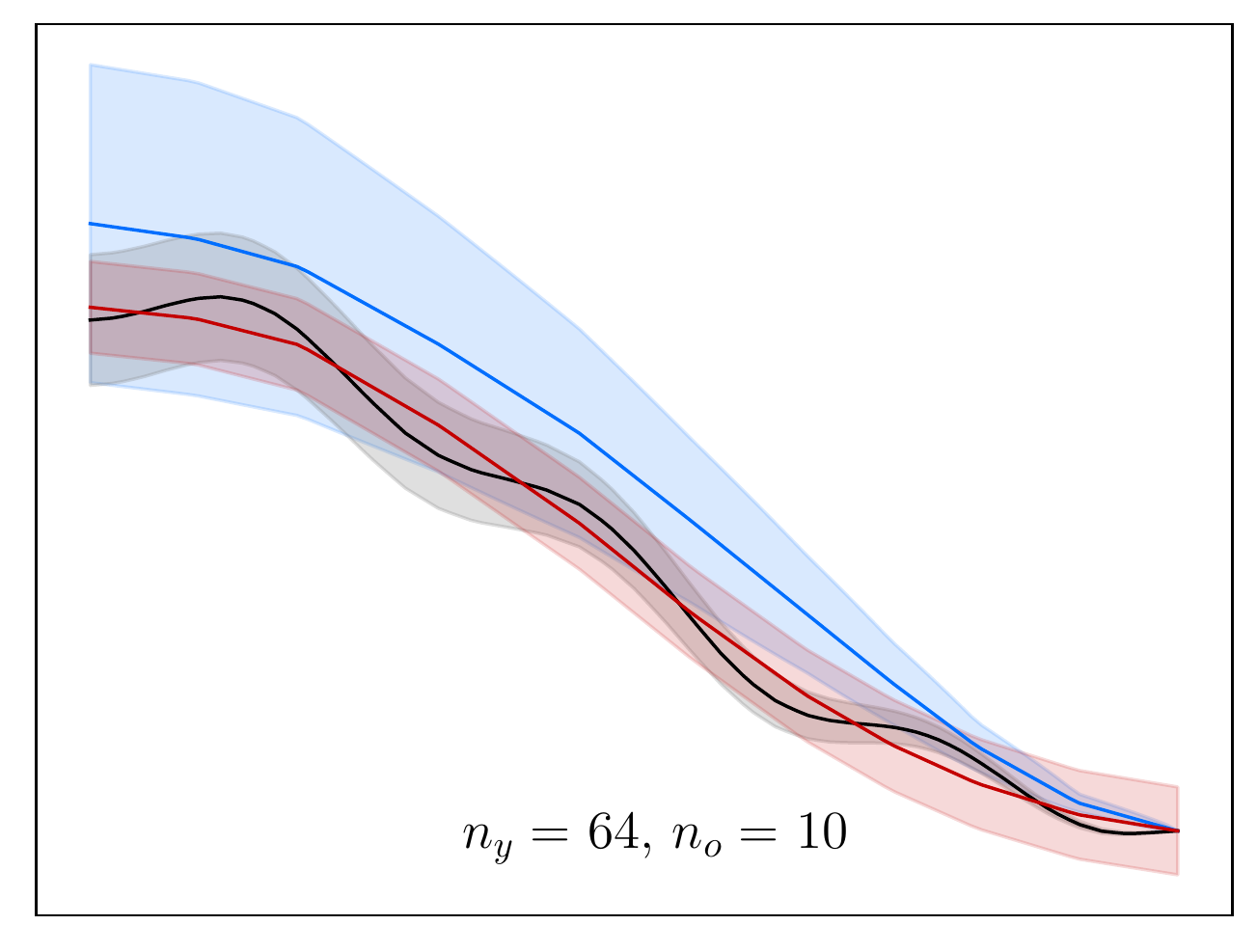}
		\includegraphics[width=.24\textwidth]{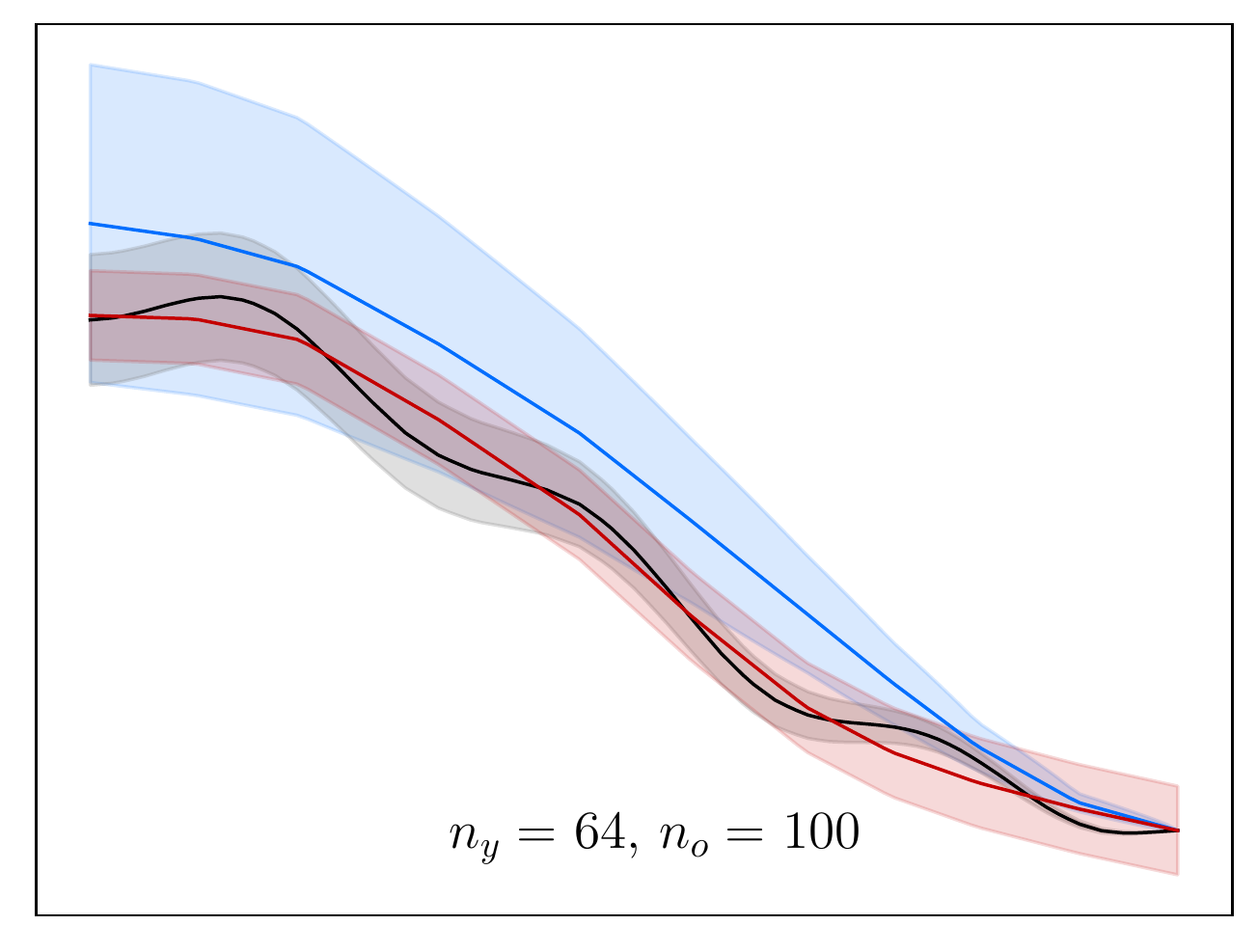}
		\includegraphics[width=.24\textwidth]{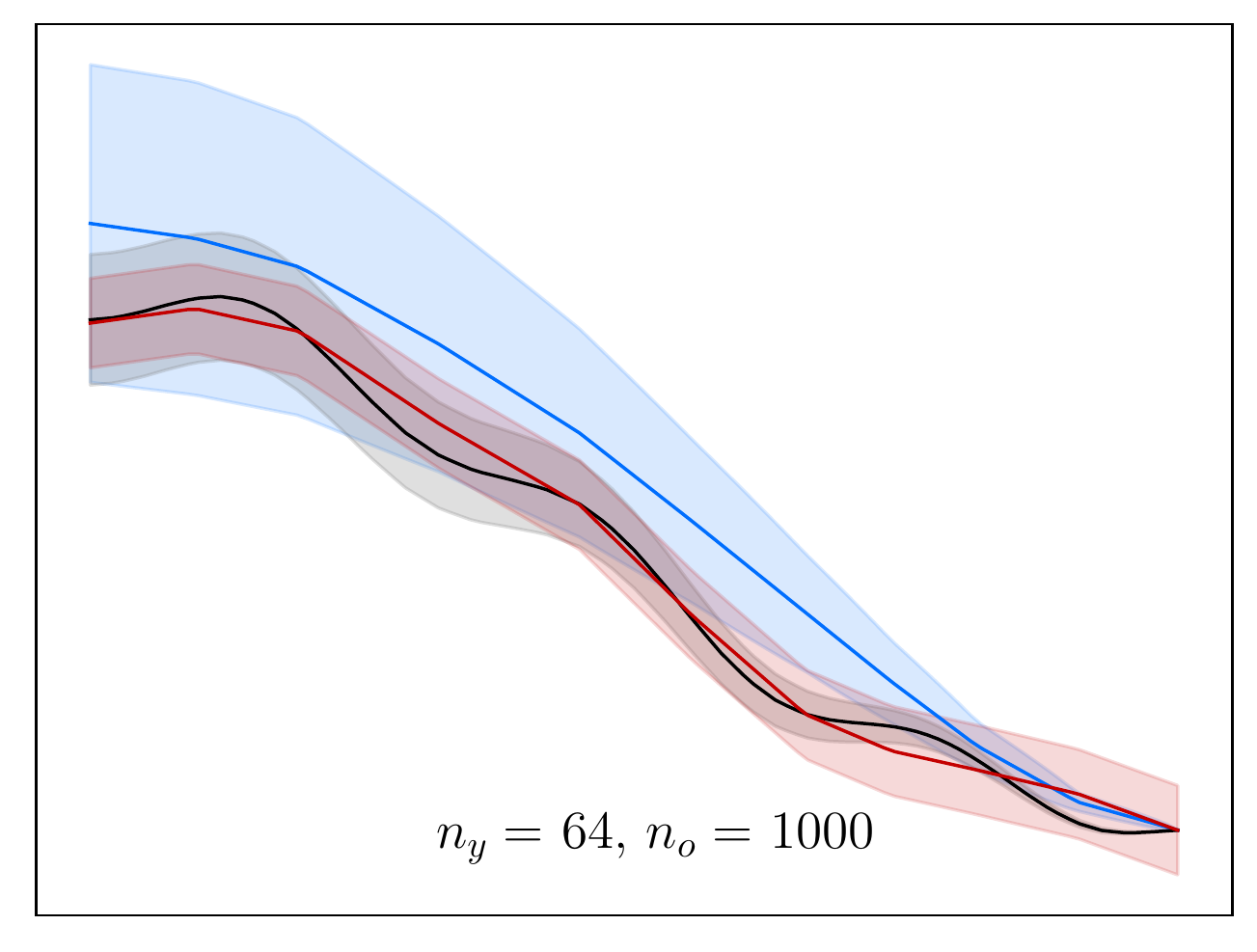}
	\\\
		\includegraphics[width=.24\textwidth]{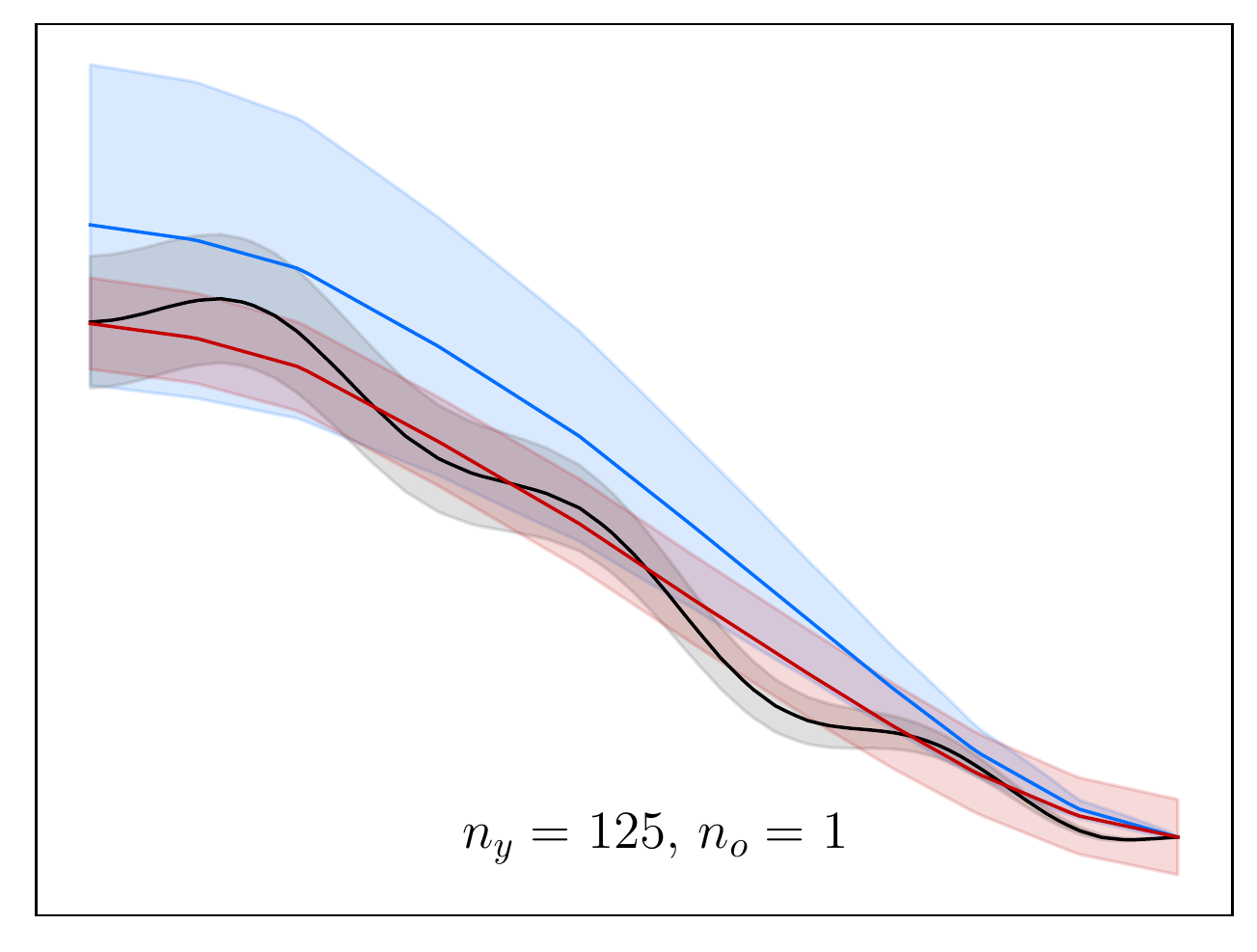}
		\includegraphics[width=.24\textwidth]{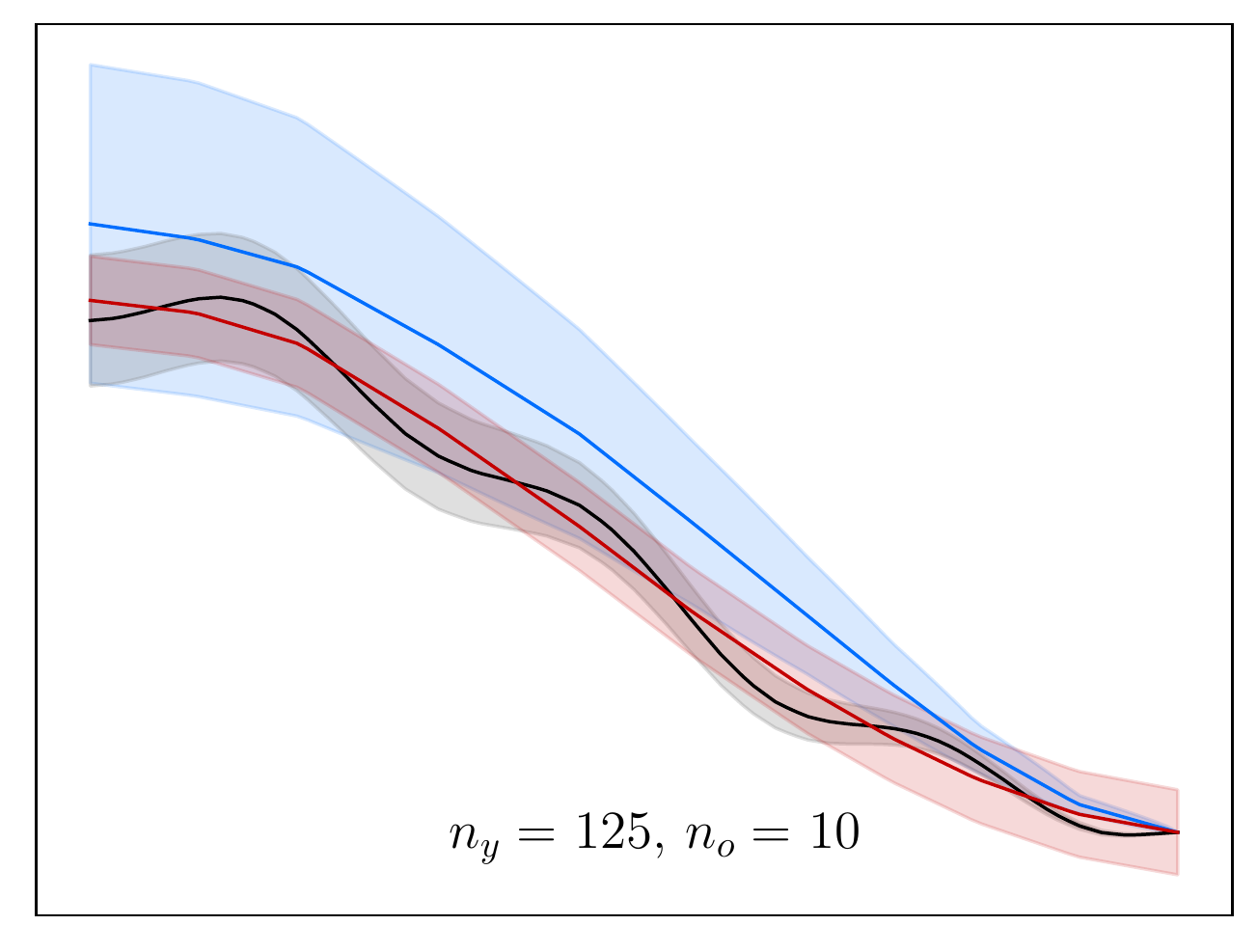}
		\includegraphics[width=.24\textwidth]{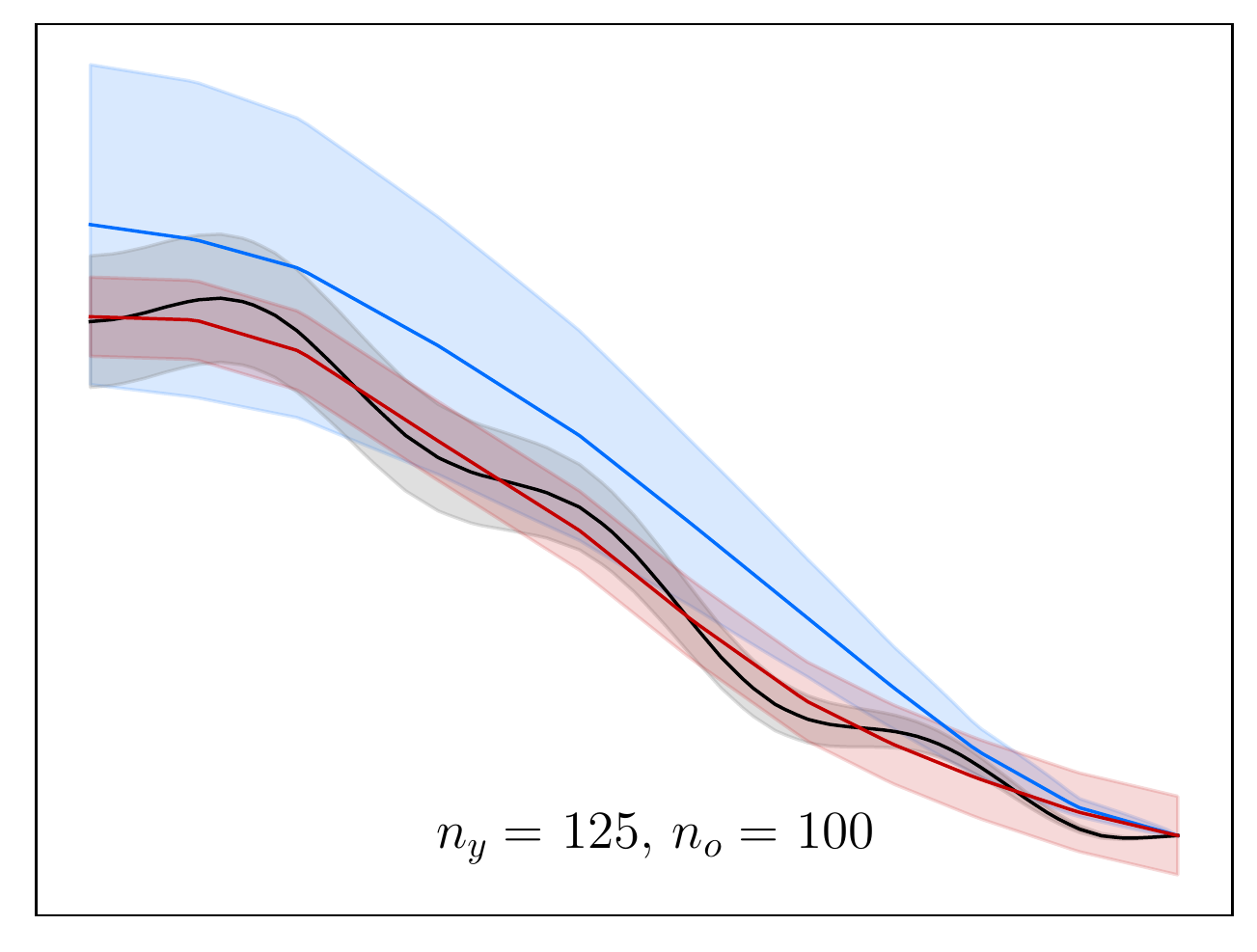}
		\includegraphics[width=.24\textwidth]{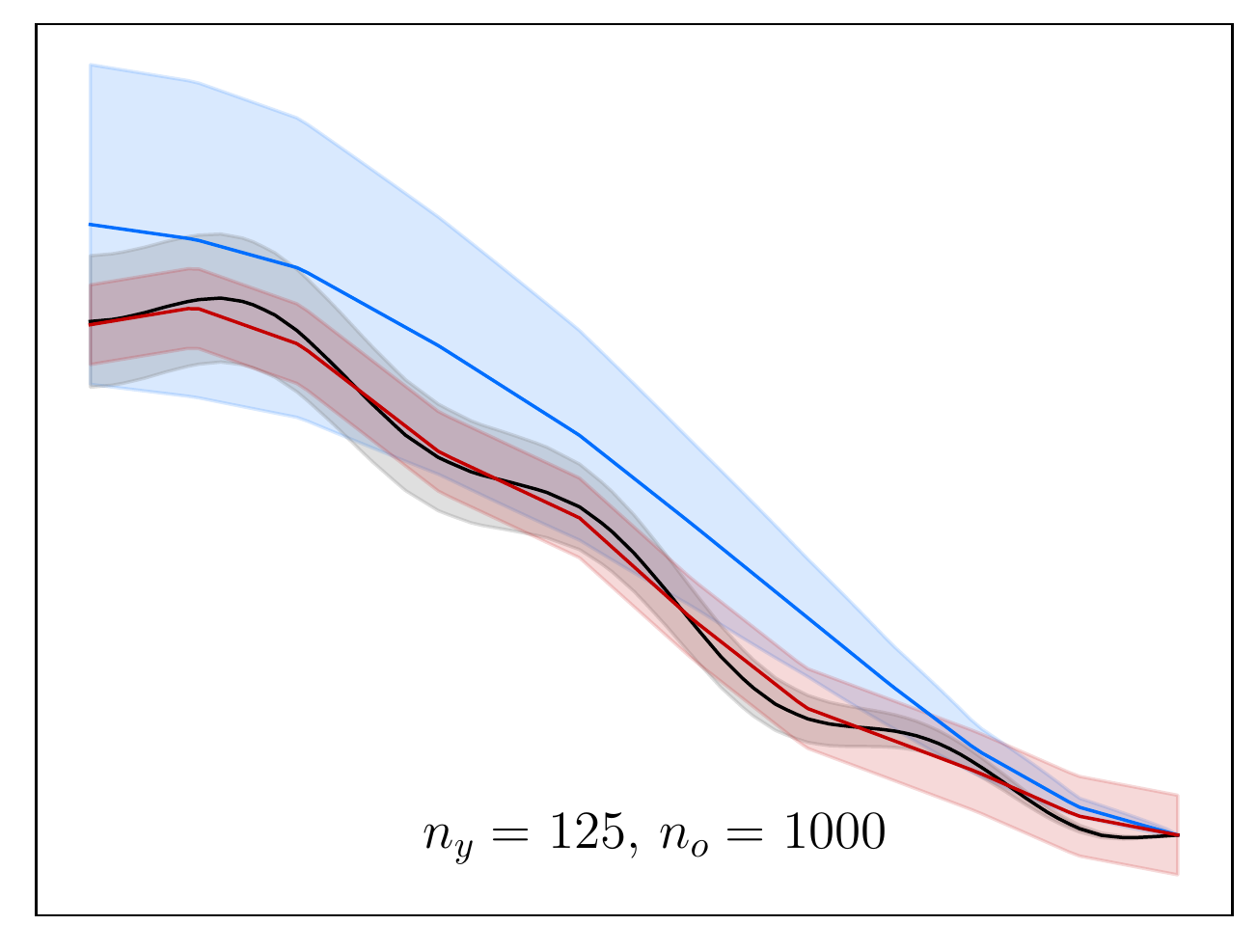}
	\caption{Plate with a hole. Inferred true system density~$p(\vec z | \vec Y)$  conditioned on observation data sampled from the multivariate Gaussian density~\eqref{eq:observMultivariateGauss} (black line). The blue lines represent the mean~$\overline{\vec u}$ and the red lines the conditioned mean~$\overline{\vec z}_{| \vec Y}$. The shaded areas denote the corresponding $95\%$ confidence regions. In each row the number of sensors~$n_y$ and each column the number of readings~$n_o$ is constant.  All plots along the diagonal of the domain with~$x^{(2)} = x^{(1)}$}
	\label{fig:2DtrueZ}
\end{figure}

The inferred posterior finite element densities~$p(\vec u | \vec Y)$ computed according to~\eqref{eq:postUcYMult} for different number of observation points~$n_y$ and readings~$n_o$ are shown in Figure~\ref{fig:2DpostUh}. We used again the empirical averages~$\overline{\vec w} = \expect[ {\vec w}]$ of the hyperparameters as point estimates. As for the one-dimensional example, the posterior mean~$\overline {\vec u}_{|Y}$ converges with increasing~$n_y$ and~$n_o$ to the true system response mean~$\overline{\vec z}$. At the same time, the posterior covariance~$\vec C_{u|Y}$ converges to zero. Obviously, the finite element solution space cannot represent the true system response so that some differences between~$\overline{\vec u}_{|Y}$ and~$\overline{\vec z}$ remain, see, e.g., the plot for~$n_y=125$ and~$n_o=1000$. In Figure~\ref{fig:2DtrueZ}  the inferred true system response densities~$p(\vec z | \vec Y)$ according to~\eqref{eq:trueSystemPosterior} are shown. The two means~$\overline{\vec z}_{|Y}$  and~$\overline{\vec u}_{|Y}$  are identical. In comparison to the true system covariance~$\vec C_z$, the obtained posterior covariance~$\vec C_{z|Y}$ is slightly smaller towards the centre and larger towards the boundary of the plate. To achieve a better match between~$\vec C_z$ and~$\vec C_{z|Y}$ it is necessary to model the mismatch covariance differently. The used squared exponential kernel~\eqref{eq:mismatchCovKernel} depends on a scalar scaling parameter~$\sigma_d$. A possible remedy would involve the modelling of~$\sigma_d$ as a spatially varying field.

%
\section{Conclusions \label{sec:conclusions}}
%
We introduced the statistically constructed finite element method, statFEM, that provides a means for coherent synthesis of observation data and finite element discretised mathematical models. Thus, statFEM can be interpreted as a physics-informed machine learning or Bayesian learning technique.  The mathematical models used in engineering practice are highly misspecified due to inherent uncertainties in loading, material properties, geometry, and the many inevitable modelling assumptions. StatFEM enables the fusion of observation data from in-situ monitoring of engineering systems with the possibly severely misspecified finite element model. This is conceptually different from traditional Bayesian inversion or calibration, which aim to learn the parameters, like the material and geometry, of the model. In statFEM the probabilistic finite element model provides the prior density which is used in determining the posterior densities of the random variables and hyperparameters of the postulated statistical generating model.  As numerically demonstrated, with increasing number of observations the obtained posterior densities converge towards the true system density. Informally, choosing a more informative prior, or in some sense better finite element model, enables us to approximate the true system density with less data. In engineering practice observational data is usually scarce so that informative priors are important. When data is abundant instead of statFEM one could argue that a purely data-driven approach, like Gaussian process regression, may be sufficient if the system under study is relatively simple, see e.g.~\cite{bessa2019bayesian}. Building upon the Bayesian statistics framework, statFEM provides a wide range of techniques to compare and interrogate models and data of different fidelity and resolution, which we only partially exploited in this paper.

In closing, a number of possible extensions of statFEM are noteworthy. We used in this paper, purely for illustrative purposes,  the squared exponential kernel as a covariance kernel. Especially, for the mismatch variable the use of other kernels or linear combination of kernels from Gaussian process regression literature appears promising, see e.g.~\cite[Ch. 4]{williams2006gaussian}. Moreover, we obtained the prior density of the finite element solution by approximating the forward problem in the random domain by a first-order perturbation method. Although this method has certain advantages, like ease of extensibility to nonlinear and nonstationary problems~\cite{liu1988transient}, the random perturbations must be relatively small. Advanced approaches for solving stochastic partial differential equations do not have this limitation and may provide more informative priors. Furthermore, we apply in statFEM Bayes rule on several levels in turn which requires that hyperparameter densities are replaced with point estimates. Alternatively, it is possible to obtain the density of random variables and hyperparameters by marginalisation from a high-dimensional joint density. The joint density can be sampled, for instance, with the Metropolis-within-Gibbs algorithm; see~\cite{marzouk2009dimensionality} for an application of this approach in the Bayesian inversion context.  A possible further extension of statFEM concerns the refinement of the postulated statistical model to incorporate computational models of different fidelity by incorporating ideas from recursive co-kriging~\cite{kennedy2000predicting, perdikaris2015multi, perdikaris2017nonlinear}. This would allow, for instance, the synthesis of observation data with a detailed expensive to evaluate 3D elasticity model and an empirical engineering formula. Finally, the extension of statFEM to time-evolving linear and nonlinear Korteweg-de Vries equation has been considered in~\cite{duffin:2020}.

\section*{Acknowledgement}
The authors gratefully acknowledge support by the Lloyd's Register Foundation Programme on Data Centric Engineering and by The Alan Turing Institute under the EPSRC grant EP/N510129/1. In addition, MG acknowledges support by the EPSRC (grants EP/R034710/1, EP/R018413/1, EP/R004889/1 and EP/P020720/1), the Lloyd's Register Foundation and the Royal Academy of Engineering Research Chair in Data Centric Engineering.

\appendix
%
\section{Appendix \label{sec:appendix}}
%

\subsection{Gaussian process regression \label{app:GPregress}}
%
The  diffusion coefficient~$\kappa(\vec x) \sim \set{GP} ( \overline {\kappa} (\vec x), \, c_{\kappa}(\vec x, \, \vec x'))$ is assumed to be a Gaussian process and is approximated as constant in each finite element giving the multivariate Gaussian density~\eqref{eq:kappaMvG}, repeated here for convenience,  
\begin{equation*}
	\vec \kappa \sim p( \vec \kappa) = \set N \left (\overline{\vec \kappa} \left(  \vec X^{(c)} \right ), \, \vec C_{\kappa} \left ( \vec X^{(c)} , \, \vec X^{(c)} \right ) \right ) \, .
\end{equation*}
A common operation is to condition this density on a set of prescribed coefficients~\mbox{$\vec \kappa^{(a)} = \{ \kappa_i^{(a)} \}_{i=1}^{n_a}$} at  the respective anchor points~$\vec X^{(a)} = \{ \vec x_i^{(a)} \}_{i=1}^{n_a}$.  The joint probability of the anchor point diffusion vector~$ \vec \kappa^{(a)} \in \mathbb R^{n_a}$ and the element diffusion vector~$ \vec \kappa \in \mathbb R^{n_e}$ is given by
\begin{equation} \label{eq:kappaGP}
	p( \vec \kappa^{(a)}, \, \vec \kappa ) = \set N \left (  
	\begin{bmatrix}
	\overline{\vec \kappa} \left ( \vec X^{(a)} \right )    \\[0.6em]
	\overline {\vec \kappa} \left (\vec X^{(c) } \right )
	\end{bmatrix}, 
	\begin{bmatrix}
	\vec C_\kappa \left (\vec X^{(a)}, \, \vec X^{(a)} \right ) &  \vec C_\kappa \left ( \vec X^{(a)}, \, \vec X^{(c)} \right )  \\[0.6em] 
	\vec C_\kappa \left (\vec X^{(c)} , \, \vec X^{(a)} \right ) & \vec C_\kappa  \left ( \vec X^{(c)}, \,  \vec X^{(c)} \right ) \, 
	\end{bmatrix}  
	\right ) \, . 
\end{equation} 
The mean and the covariance matrix are obtained by evaluating the prescribed mean~$\overline {\kappa} (\vec x)$ and covariance kernel~$ c_{\kappa}(\vec x, \, \vec x')$ at the respective points. According to  standard results, see e.g. \cite[Chapter~2]{williams2006gaussian}, the density of~$\vec \kappa$ conditioned on~$ \vec \kappa^{(a)}$ is given by 
\begin{equation} \label{eq:kappaConditioned}
	 {\vec \kappa}_{|  \kappa^{(a)}}   \sim p \left (\vec \kappa  | \vec \kappa^{(a)} \right ) =   \set N  \left  (  \overline {\vec \kappa}_{|  \kappa^{(a)}}, \, \vec C_{\kappa | \kappa^{(a)}}   \right )  \, ,
\end{equation}
where
\begin{subequations}
\begin{align} 
	\overline {\vec \kappa}_{|  \kappa^{(a)}}  &=  \overline {\vec \kappa} \left (\vec X^{(c) } \right ) +  \vec C_\kappa \left (\vec X^{(c)}, \,  \vec X^{(a)} \right )  \vec C_\kappa \left (  \vec X^{(a)}, \,  \vec X^{(a)}  \right )^{-1} \left ( {\vec \kappa}^{(a)}  - \overline{\vec \kappa} \left ( \vec X^{(a)} \right )  \right )  \, , \label{eq:kappaConditionedMean} \\
  	 \vec C_{\kappa | \kappa^{(a)}} &=  \vec C_\kappa \left ( \vec X^{(c)}, \,  \vec X^{(c)}  \right ) -  \vec C_\kappa \left ( \vec X^{(c)}, \,  \vec X^{(a)}\right )   \vec C_\kappa \left (  \vec X^{(a)}, \,  \vec X^{(a)} \right )^{-1}  \vec C_\kappa \left ( \vec X^{(a)}, \,  \vec X^{(c)} \right )    \, .  \label{eq:kappaConditionedVar}
\end{align}
\end{subequations}
If needed, we sample from this density by first computing the Cholesky decomposition~$ \vec C_{\kappa | \kappa^{(a)}} = \vec L \vec L^\trans$  and then sampling the Gaussian white noise~\mbox{$\vec e \sim \set N (\vec 0, \, \vec I)$} to obtain 
\begin{equation}
	 \vec \kappa_{|  \kappa^{(a)}} = \overline {\vec \kappa}_{|  \kappa^{(a)}} + \vec L \vec e \, .
\end{equation}

\begin{figure} 
	\centering
	\includegraphics[scale=0.45]{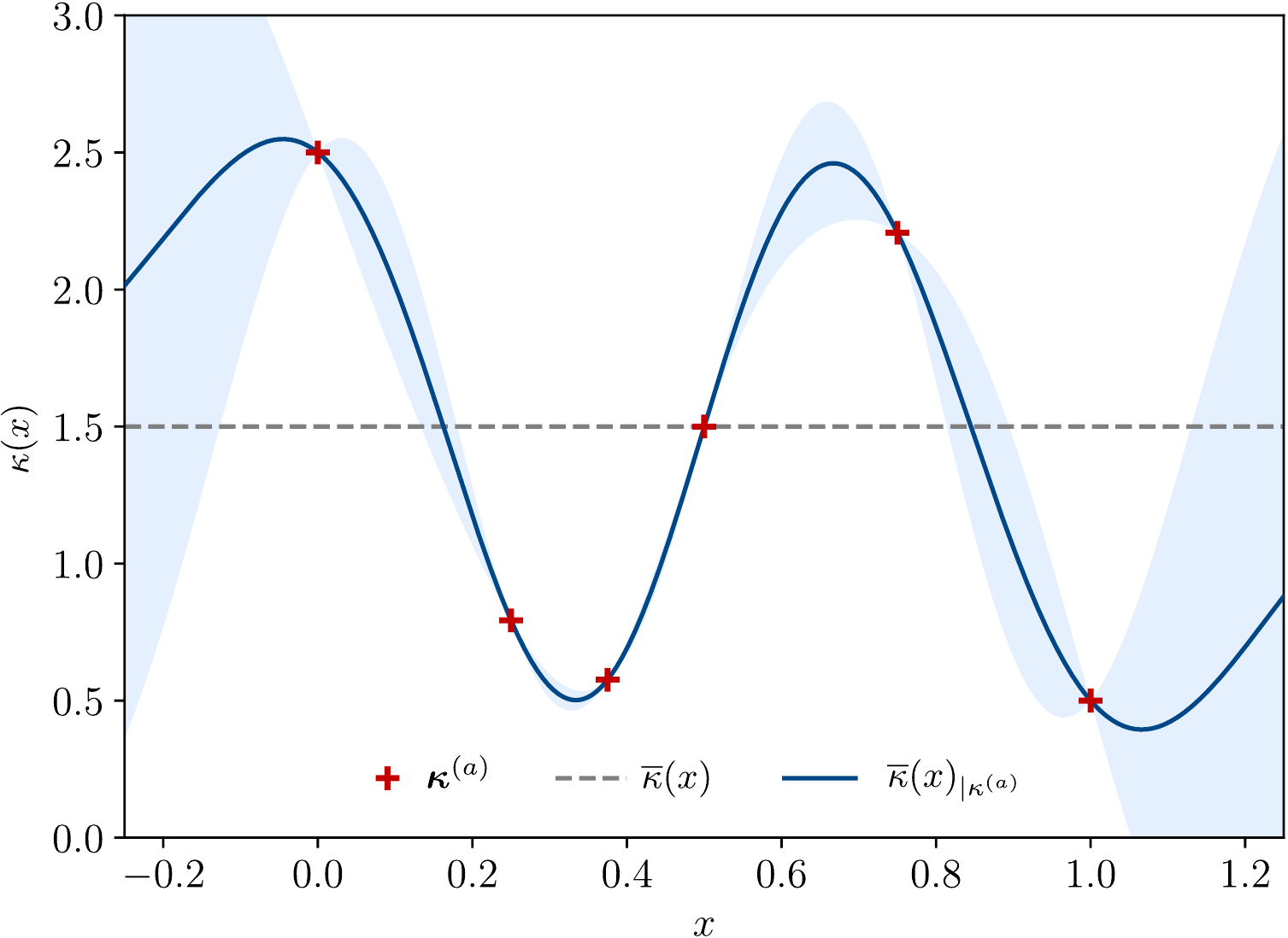}
	\caption{Illustrative Gaussian process regression example. The coefficients~$\vec \kappa^{(a)}$ at the six anchor points (in red) are prescribed. The solid line represents the conditioned mean~$\overline {\vec \kappa}_{|  \kappa^{(a)}}$ and the shaded area the~$95\%$ confidence region obtained from the covariance $\vec C_{\kappa | \kappa^{(a)}}$.  \label{fig:gpDiffusion}} 
\end{figure}

As an example for Gaussian process regression, in Figure~\ref{fig:gpDiffusion} the approximation of~$\kappa(\vec x)$ over a one-dimensional domain~\mbox{$\Omega =(0, \, 1)$} is illustrated. The six prescribed anchor point coefficients~$\vec \kappa^{(a)}$ lie on the curve~\mbox{$1.5 + \cos \left( 3 \pi x \right)$}.  The prescribed mean is~\mbox{$\overline \kappa(x) = 1.5$} and the parameters of the squared exponential kernel are~$\sigma_\kappa =1 $ and~$\ell_\kappa=0.2$. The depicted conditioned mean and the $95\%$ confidence region are obtained from~\eqref{eq:kappaConditionedMean} and~\eqref{eq:kappaConditionedVar}. 

Finally, the conditioned mean~\ref{eq:kappaConditionedMean} can be used to define the interpolating basis functions~$\vec \psi(\vec x)$. Specifically, choosing the mean of the Gaussian process as~$\overline \kappa(\vec x) = 0$ we can define
\begin{equation} \label{eq:eqMeanGPregression}
	\overline {\kappa} (\vec x)_{|  \kappa^{(a)}}  =   \vec C_\kappa \left (\vec x, \,  \vec X^{(a)} \right )  \vec C_\kappa \left (  \vec X^{(a)}, \,  \vec X^{(a)}  \right )^{-1}  {\vec \kappa}^{(a)}  = \vec \psi (\vec x)^\trans  {\vec \kappa}^{(a)}  \, .
\end{equation}
%

%
\subsection{Computation of the posterior finite element density \label{sec:appPuCy}}
%
In deriving the posterior finite element density~$p(\vec u | \vec y)$ and several other places we make use of the fact that the product of two Gaussian densities is again a Gaussian, see e.g.~\cite{bishop2006pattern,murphy2012machine}. To see this, it is sufficient to focus on the argument of the respective exponential function and to bring it into a quadratic form. The normalisation constant for the so obtained Gaussian can be determined by inspection. The steps in obtaining the quadratic form are as follows
\begin{align}  \label{eq:derivPuY}
\begin{split}
	 p(  \vec u  | \vec y ) &= \frac{ p(\vec y | \vec u)  p  (\vec u)}{ p (\vec y)} \propto p(\vec y | \vec u)  p  (\vec u)  \\
	&
	 \propto  \exp \left (  \left ( \rho \vec P \vec u - \vec y \right )^\trans ( \vec C_d + \vec C_e)^{-1}    \left ( \rho \vec P \vec u - \vec y \right ) \right )  \exp \left ( \left ( \overline{\vec u} - \vec u \right )^\trans \vec C_u^{-1}  \left ( \overline{\vec u} - \vec u \right )   \right ) 
	 \\
	& = \exp \left (    \vec u^\trans \vec B  \vec u -  2 \vec a^\trans  \vec u + \dotsc \right ) \, ,
\end{split}	
\end{align} 
where
\begin{align*}
	\vec B &=  \rho^2 \vec P^\trans (\vec C_d + \vec C_e)^{-1} \vec P  + \vec C_u^{-1}   \\ 
	\vec a & =   \rho  \vec P^\trans  \left ( \vec C_d + \vec C_e \right )^{-1} \vec y   +  \vec C_u^{-1} \overline{\vec u}   \, .
\end{align*}
By completion of the square, the last expression in~\eqref{eq:derivPuY} is evidently proportional to  
\begin{equation}
	 p(  \vec u | \vec y ) \propto \exp \left ( \left ( \vec B^{-1} \vec a - \vec u \right )^\trans \vec B \left ( \vec B^{-1} \vec a - \vec u \right )  \right )  \, .
\end{equation}
After determining the normalisation constant by inspection the posterior is given by 
\begin{equation}	
	 p(  \vec u | \vec y ) =  \frac{1}{ \sqrt{ (2 \pi)^{n_u} | \vec B|}  } \exp \left ( \left ( \vec B^{-1} \vec a - \vec u \right )^\trans \vec B \left ( \vec B^{-1} \vec a - \vec u \right )  \right )  \, .
\end{equation}
By comparison with~\eqref{eq:postUcY} we can conclude 
\begin{equation}
	\vec C_{u|y} = \vec B^{-1} \, , \quad  \overline {\vec {u}}_{|y} = \vec C_{u|y} \vec a \, .
\end{equation}

To avoid the inversion of large dense matrices, i.e.~$\vec C_u$, in evaluating~$ p(\vec u | \vec y) $, we use the Sherman-Morrison-Woodbury identity to obtain 
\begin{subequations}
\begin{align}
	\vec C_{u|y} &=  \vec C_u - \vec C_u \vec P^\trans \left ( \frac{1}{\rho^2} \left ( \vec C_d + \vec C_e \right )  + \vec P \vec C_u \vec P^\trans   \right )^{-1} \vec P    \vec C_u  \label{eq:shermanCov}\\  
	\overline {\vec {u}}_{|y} &=   \vec C_{u|y}   \left(  \rho  \vec P^\trans  \left ( \vec C_d + \vec C_e \right )^{-1} \vec y   +   \vec C_u^{-1}  \overline{\vec u}  \right)  \label{eq:shermanMean} \, .
\end{align}
\end{subequations}
Note that $  \vec C_u \in \mathbb R^{n_u \times n_u}$,   $\vec P \in  \mathbb R^{n_y \times n_u}$ and $\vec C_{d} , \,  \vec C_e \in \mathbb R^{n_y \times n_y}$  so that the bracket expression to be inverted is a dense matrix of dimension~$n_y \times n_y$, where~$n_y \ll n_u$ in most applications. When a direct solver is used, it is possible to speed up the computation of~$\vec P \vec C_u \vec P^\trans$, which involves according to~\eqref{eq:cUperturb}, the evaluation of terms like  
\begin{equation} \label{eq:tmpOps}
	\vec P \vec A^{-1} \vec C_f \vec A^{-\trans} \vec P^\trans = \vec R \vec C_f \vec R^\trans  \, . 
\end{equation}
Here, it is sufficient to factorise the sparse system matrix~$\vec A$ only once and to obtain $ \vec R$ by column-wise back and forward substitution of~$\vec P$. 

For completeness, we note that the posterior density~$p(\vec u | \vec y)$  can be alternatively obtained from the joint density of~$p( \vec u, \, \vec y)$, see~\cite{kaipio2006statistical,kennedy2001bayesian}. The statistical model introduced in Section~\ref{sec:generatingModel} is given by
\begin{equation}
	\vec y = \rho \vec P \vec u + \vec d + \vec e \, ,
\end{equation}
and the random vectors have the densities $\vec u \sim N(\overline{\vec u}, \, \vec C_u)$,~$\vec d \sim N (\vec 0, \, \vec C_d)$ and~$\vec e \sim N (\vec 0, \, \vec C_e)$. The joint density is defined by
\begin{equation}
	p (\vec u, \, \vec y) =
	 \set N \left ( 
	\begin{pmatrix}
		\overline {\vec u } \\ \overline { \vec y} 
	\end{pmatrix}
	, \,
	\begin{pmatrix}
		\expect \left ( ( \vec u -   \overline {\vec u } ) \otimes  ( \vec u -   \overline {\vec u } ) \right )   &  \expect \left (  ( \vec u -   \overline {\vec u } ) \otimes  (\vec y -  \overline { \vec y}   )  \right )  \\ 
		\expect \left ( ( \vec y -   \overline {\vec y } ) \otimes  ( \vec u -   \overline {\vec u } ) \right )   &  \expect \left (  ( \vec y -   \overline {\vec y } ) \otimes  (\vec y -  \overline { \vec y}   )  \right ) 
	\end{pmatrix}	
	\right ) \, ,
\end{equation}
where the expectations must be taken over the random vectors~$\vec u$,~$\vec d$ and~$\vec e$. After introducing the statistical model and the densities of the random vectors we obtain the joint density 
\begin{equation}
	p (\vec u, \, \vec y)
	= \set N \left ( 
	\begin{pmatrix}
		\overline {\vec u } \\ \rho \vec P \overline { \vec u} 
	\end{pmatrix}
	, \,
	\begin{pmatrix}
		\vec C_u  &  \rho  \vec C_u \vec P^\trans   \\ 
		  \rho  \vec P \vec C_u  & \rho^2 \vec P \vec C_u \vec P^\trans  + \vec C_d + \vec C_e    
	\end{pmatrix}	
	\right ) \, .
\end{equation}
The respective conditional density~$p(\vec u | \vec y)$ according to Section~\ref{app:GPregress} yields for the covariance~$\vec C_{u|y}$ the same expression as~\eqref{eq:shermanCov} and for the mean~$\overline {\vec {u}}_{|y}$ a slightly different expression than~\eqref{eq:shermanMean}.

%
\subsection{Markov chain Monte Carlo method (MCMC) \label{sec:appMCMC}}
%
We obtain the posterior of the statistical generating model parameters, i.e.~\mbox{$p(\vec w | \vec Y) \propto p(\vec Y  | \vec w) p(\vec w)$}, by sampling with the MCMC Metropolis algorithm. To avoid numerical stability issues with  products of small densities in the likelihood~$p(\vec Y  | \vec w)$ the logarithm of the posterior is considered. As the proposal density we choose a normal distribution~$q(\vec w | \vec v ) = \set N(\vec v, \, \sigma_q^2 \vec I)$ with the algorithmic parameter~$\sigma_q$. 

For a given sample~$\vec w^{(i)}$ one MCMC iteration consists of the following steps:
\begin{enumerate}
	\item[S1.] Sample  $\vec w \sim q (\vec w | \vec w^{(i)})$\,.
	\item[S2.] Compute acceptance probability  
	\begin{equation}\label{eq:mcmcA}
		\alpha = \min \left ( 0, \, \ln \left (p (\vec w | \vec Y ) \right ) - \ln ( p \left (\vec w^{(i)} | \vec Y) \right ) \right )  \, . 
	\end{equation}
	\item[S3.] Generate a uniform random number $u \sim  \set U [0, \,1]$\,.
	\item[S4.] Set new sample to 
	\begin{equation}\label{eq:mcmcB}
		\vec w^{(i+1)} = 
		\begin{cases}
			\vec w  \, \, \,  & \text{ if } \ln (u) < \alpha \\
			\vec w^{(i)} \, \, \,  & \text{ if } \ln(u) \ge  \alpha 
		\end{cases}\, .
	\end{equation}
\end{enumerate}

The non-negativity of the parameters collected in~$\vec w$ can be enforced by choosing a prior density with non-negative support or applying a parameter transformation. When the prior has a non-negative support any negative proposal~$\vec w$ will yield a zero posterior~$p (\vec w | \vec Y )$ and will be rejected in Step S4. As a result,  all the collected samples~$\{ \vec w^{(i)} \} $ will be non-negative. The many redundant samples make this approach, however, inefficient.
To avoid this, we consider the parameter  transformation~$\widetilde {\vec w} = \ln (\vec w)$, where the logarithm is applied component-wise, and sample in the transformed domain. Rather than rewriting the above algorithm, the transformation can be taken into account by simply replacing the acceptance probability in Step~S2 with
\begin{equation}
	\alpha = \min \left ( 0, \, \ln \left (p \left ( \exp( \vec w) | \vec Y \right ) \right ) +   \sum_j w_j   - \ln \left ( p \left ( \exp( \vec w^{(i)} ) | \vec Y \right )  \right )  -   \sum_j w_j^{(i)}   \right )  \, ,
\end{equation}
where the two additional terms represent the Jacobian of the transformation of the posterior density. The samples~$\{\vec w^{(i)} \}$ are now in the transformed domain and have to be transformed back. The  so obtained samples~$\{ \exp(\vec w^{(i)}) \}$ are all positive and have the desired distribution.  

Moreover, in our computations we choose the algorithmic parameter $\sigma_q$ for the proposal density~$q(\vec w | \vec v ) $ so that the acceptance ratio in Step S4 is around $0.25$.  There are a number of efficient algorithms available to automate the selection of~$\sigma_q$~\cite{andrieu2008tutorial}. We discard the the first $25\% - 30\%$ of the obtained samples to account for the burn-in phase. For further details on MCMC algorithms see~\cite{robert2013monte}.

\bibliographystyle{elsarticle-num-names}
\bibliography{statFEMrv}

\end{document}